%% file: main.tex
\documentclass[journal=jacsat, manuscript=article]{achemso}
\usepackage{graphicx} 
\usepackage{array}
\usepackage{booktabs}
\usepackage{hypdoc}
\usepackage{listings}
\usepackage{lmodern}
\usepackage{mathpazo}
\usepackage{microtype}
\usepackage{threeparttable}

\usepackage{amsmath}
\usepackage{subcaption} 

\title{Deep Learning-Optimized, Fabrication Error-Tolerant Photonic Crystal Nanobeam Cavities for Scalable On-Chip Diamond Quantum Systems}

\author{Sander van Haagen}
\affiliation{Department of Applied Sciences, Delft University of Technology, Delft 2628 CJ, NL}
\alsoaffiliation{Department of Quantum \& Computer Engineering, Delft University of Technology, Delft 2628 CD, NL}
\alsoaffiliation{QuTech, Delft University of Technology, Delft 2628 CJ, NL}
\email{svhaagen@gmail.com}

\author{Salahuddin Nur}
\affiliation{Department of Quantum \& Computer Engineering, Delft University of Technology, Delft 2628 CD, NL}

\author{Ryoichi Ishihara}
\affiliation{Department of Quantum \& Computer Engineering, Delft University of Technology, Delft 2628 CD, NL}
\alsoaffiliation{QuTech, Delft University of Technology, Delft 2628 CJ, NL}

\date{June 2024}

\begin{document}


\input{abstract}
\input{introduction}
\input{methodology}
\input{results}
\input{conslusions}

\nocite{Optical_microcavities_review,Optical_cavity_sensing,Cavity_sensing2,Optical_cavity_switch,Fabry_perot_applications,the_story_of_Q,the_story_of_V,Advanced_QM_book,QCEInfrastructure,GeLU}
\bibliography{references}

\input{acknowledgements}

\appendix
\renewcommand{\thefigure}{S\arabic{figure}} 
\renewcommand{\thetable}{S\arabic{table}} 
\renewcommand{\theequation}{S\arabic{equation}} 
\setcounter{figure}{0}  
\setcounter{table}{0}   
\setcounter{equation}{0}  
\input{A}
\input{B}
\input{C}
\input{D}
\input{E}
\input{F}
\input{G}
\input{H}

\end{document}

%% file: abstract.tex
\begin{abstract}
Cavity-enhanced diamond color center qubits can be initialized, manipulated, entangled, and read individually with high fidelity, which makes them ideal for large-scale, modular quantum computers, quantum networks, and distributed quantum sensing systems. However, diamond's unique material properties pose significant challenges in manufacturing nanophotonic devices, leading to fabrication-induced structural imperfections and inaccuracies in defect implantation, which hinder reproducibility, degrade optical properties and compromise the spatial coupling of color centers to small mode-volume cavities. A cavity design tolerant to fabrication imperfections—such as surface roughness, sidewall slant, and non-optimal emitter positioning—can improve coupling efficiency while simplifying fabrication. To address this challenge, a deep learning-based optimization methodology is developed to enhance the fabrication error tolerance of nanophotonic devices. Convolutional neural networks (CNNs) are applied to promising designs, such as L2 and fishbone nanobeam cavities, predicting Q-factors up to one million times faster than traditional finite-difference time-domain (FDTD) simulations, enabling efficient optimization of complex, high-dimensional parameter spaces. The CNNs achieve prediction errors below $3.99\%$ and correlation coefficients up to $0.988$. Optimized structures demonstrate a $52\%$ reduction in Q-factor degradation, achieving quality factors of $5 \times 10^4$ under real-world conditions and a two-fold expansion in field distribution, enabling efficient coupling of non-optimally positioned emitters. This methodology enables scalable, high-yield manufacturing of robust nanophotonic devices, including the cavity-enhanced diamond quantum systems developed in this study.
\end{abstract}

\newpage

%% file: introduction.tex
\section{Introduction}

Crystallographic defects in diamond known as color centers, such as nitrogen-vacancy (NV) centers and group IV defects (SiV, GeV, SnV and PbV centers), exhibit unique spin properties that make them promising building blocks for scalable and on-chip integrated quantum systems, including on-chip modular quantum computers \cite{3D_integration_ishihara}, distributed quantum sensors \cite{quantum_sensing} and secure quantum networks \cite{ruf2021quantum}. This promise is attributed to their host's solid-state characteristics, which enable the implementation of large-scale fabrication processes. Additionally, color centers in diamond have demonstrated the capability for initialization, manipulation, entanglement, and readout of individual spin-qubits with long coherence times and high fidelity \cite{ToenovdSar, PhD_cathryn, Rugar_PhD_thesis, debroux_2022}.
\par
The prospects of deploying color centers in diamond for solid-state quantum applications have driven the growth of diamond photonics. Photons serve as an effective medium for information transfer between qubits due to their fast propagation speed and weak interaction with the surrounding medium. Coherent spin-photon coupling emerges as the most promising approach for entangling such qubits, offering robust entanglement over extended distances \cite{Entanglement}. However, to fully leverage the potential of color centers in diamond, it is essential to enhance the zero-phonon line (ZPL) and thereby increase coherent light emission. \newline

Cavity quantum electrodynamics (cQED) offers a route to achieve this, as on-chip integrated optical resonators can act as interfaces that enhance emission into the ZPL \cite{QCED2, Review_groupIV}. This enhancement, known as Purcell enhancement \cite{Purcell_enhancement}, enables coherent coupling between the emitted light and the spin state of the spin-qubits, facilitating the possibility for on-chip quantum information processing tasks.
\par
One candidate of such optical resonators are photonic crystal (PhC) nanocavities. These cavities trap light by introducing defects in the periodicity of PhC lattices \cite{molding_the_flow_of_light}. Their demonstrated ability to achieve high quality factors (Q-factors) and small mode volumes at the ZPL results in large Purcell factors (see Supplementary Information \hyperlink{supinfA}{A}). These nanocavities can vary in terms of unit cell structure, PhC defects, and dimensionality, leading to a diverse range of design options, design parameters and optical characteristics (see Table \ref{tab:current_developments}). However, diamond's extreme hardness and chemical stability present significant challenges for precise nanofabrication of such nanocavities, often leading to fabrication imperfections that degrade the performance of diamond nanophotonic devices. Addressing these challenges is critical for fully realizing the potential of diamond-based solid-state quantum technologies.
\par
The most promising PhC nanocavities for scalable on-chip diamond quantum systems are 1D PhC nanocavities, also known as nanobeam cavities \cite{1D_PhC_cavitiy_PhD}. These cavities are characterized by periodic perturbations applied to a waveguide (see Figure \ref{fig:NC_schematic_geom} and \ref{fig:FB_schematic_geom}). They feature two Bragg mirrors positioned on either side of the resonant cavity. The cavity itself is typically formed by changing the distance between the two perturbations at the boundaries of the cavity region. These photonic structures are smaller than other 2D nanocavities, making them easier to fabricate using available techniques such as angled etching \cite{bhaskar2020, Bhaskar2019}, quasi-isotropic etching \cite{pregnolato2024, kuruma2021, rugar2021, mouradian2017}, or diamond film thinning \cite{ding2024, regan2021, lee2014, jung2019} (see Table \ref{tab:current_developments}). Additionally, their compact size is practical for scalable on-chip integration. However, fabricating these devices from diamond remains challenging and their performance is still highly susceptible to unavoidable fabrication imperfections. Table \ref{tab:current_developments} shows that the experimentally determined Q-factors are generally more than an order of magnitude lower than the simulated Q-factors, primarily due to fabrication imperfections.
\par
Hence, this work aims to address these fabrication challenges by developing an optimization methodology that enhances the tolerance of nanophotonic structures to common imperfections. More specifically, this study leverages deep learning (DL) techniques to optimize nanocavity designs for robustness against fabrication-induced imperfections while maintaining their optical performance metrics, such as high Q-factors and small mode volumes. This approach enables the creation of more resilient diamond nanocavities, which are critical for scalable, on-chip quantum technologies. \newline

Standard nanocavity design and optimization are facilitated by finite-difference time-domain (FDTD) simulations, which can compute the propagation of light in both the spatial and temporal domains. While these simulations offer high accuracy without relying on theoretical approximations, they demand significant computational resources due to their intensive grid-based numerical method \cite{Principles_of_photonic}. Previous optimization strategies have relied on techniques such as leaky mode visualization \cite{leaky_visualtization_1,leaky_visualtization_2}, Gaussian envelope approaches \cite{Gaussian_envelope_1, Gaussian_envelope_2, Gaussian_envelope_3}, and genetic algorithms \cite{Genetic_algorithm,genetic_algorithm_2}. These methods, whether trial-and-error or gradient-based, are often slow when dealing with large design parameter spaces like those encountered when optimizing nanobeam cavities \cite{Baba_CMA-ES, Baba_ML_300, Liu_genetic_algorithm, Renjie_CNN, Renjie_CNN_update, Takashi_CNN, Takashi_CNN_iter, Renjie_transformer, Renji_RL}.
\par
Recently, DL has emerged as an efficient alternative to computationally intensive methods based solely on FDTD simulations. Neural networks (NNs) acting as surrogate models can replace FDTD simulations, predicting optical properties up to a million times faster and thereby enabling the inverse design of complex photonic structures \cite{DL_photonic_design2, DL_photonics_general}. DL-based methods have achieved high accuracy in predicting key metrics of PhC nanocavities, such as Q-factor and mode volume, utilizing a range of NN architectures, including feed-forward neural networks (FNNs) \cite{Baba_CMA-ES, Baba_ML_300, Liu_genetic_algorithm}, convolutional neural networks (CNNs) \cite{Renjie_CNN, Renjie_CNN_update, Takashi_CNN, Takashi_CNN_iter}, transformers \cite{Renjie_transformer}, and reinforcement learning (RL) models \cite{Renji_RL}. However, these previous efforts focused on materials that are more accessible for fabrication, whereas diamond's extreme material properties introduce additional challenges. As a result, unlike prior work that focused solely on high Q-factors, this study prioritizes optimization efforts to enhance tolerance to fabrication errors, ensuring robust device performance and thereby reducing the gap between simulated and experimentally obtained Q-factors (see Table \ref{tab:current_developments}). \newline

This article focuses on optimizing the fabrication error tolerance of two base cavity designs, the L2 nanobeam (see Figure \ref{fig:NC_schematic_geom}) and the fishbone nanobeam (see Figure \ref{fig:FB_schematic_geom}), chosen for their inherent resilience against two specific challenges. Firstly, spin-qubits are susceptible to electric field fluctuations from charge variations, such as those occurring at nearby surfaces. Secondly, the defect implantation process can suffer from spatial inaccuracies, which degrade the spatial-spectral coupling between the emitter and the cavity mode \cite{Ion_implant_faberror, Ion_nonfaberror_coupling}. These base designs allow for large cavity regions, which helps counteract these issues by increasing the emitter's distance from nearby surfaces and distributing the mode more broadly at the defect implantation site.
\par
Three scenarios are compared for each base design, showcasing the proposed optimization technique. The two nanobeam base designs are each optimized under ideal conditions (no fabrication imperfections), against surface roughness \cite{Noda_faberror_roughness}, and against sidewall slant \cite{slanted_sidewalls}. These imperfections are common when fabricating such devices, even from thin film diamond, which is the most scalable fabrication technique among the three previously mentioned. A comparison of the simulated optical properties of ideally optimized nanobeam cavities with those optimized for fabrication error tolerance will reveal whether the latter perform better while suffering from these uncontrollable imperfections. \newline

In summary, the development of fabrication error-tolerant diamond nanobeam cavities represents a critical step towards the realization of scalable quantum systems. This research leverages DL techniques to optimize the design of two promising diamond nanobeam cavities, addressing challenges posed by fabrication imperfections. By doing so, it aims to advance the practical integration of color centers in diamond for quantum information processing and quantum network applications. While the methodology was demonstrated on diamond nanobeam cavities, its applicability extends to the design of other fabrication error-tolerant nanophotonic structures, making it a system-agnostic approach.

\newpage

\begin{table}[h!]
    \centering
    \caption{A table comparing the performance of different fabricated visible wavelength, suspended diamond nanocavities.}
    \begin{threeparttable}
        \scriptsize
        \resizebox{\textwidth}{!}{%
        \begin{tabular}{cccccccc}
        \hline
        \textbf{Cavity type}                                                 & \textbf{Unit cell} & \begin{tabular}[c]{@{}c@{}}\textbf{Wavelength} \\ \textbf{(nm)}\end{tabular} & \textbf{Q-factor}                                                                                                                                                 & \begin{tabular}[c]{@{}c@{}}\textbf{Mode volume}\\ $\boldsymbol{\left(\frac{\lambda}{n}\right)^3}$\end{tabular} & \textbf{Purcell factor}        & \textbf{Method}                                                            & \textbf{Reference}         \\ \hline
        \textbf{1D}                                                         &           &                                                            &                                                                                                                                                          &                                                                           &                       &                                                                   &                   \\ \hline
        \begin{tabular}[c]{@{}c@{}}Circular \\ holes\end{tabular}   &\raisebox{-0.38\totalheight}{\includegraphics[width=0.06\textwidth]{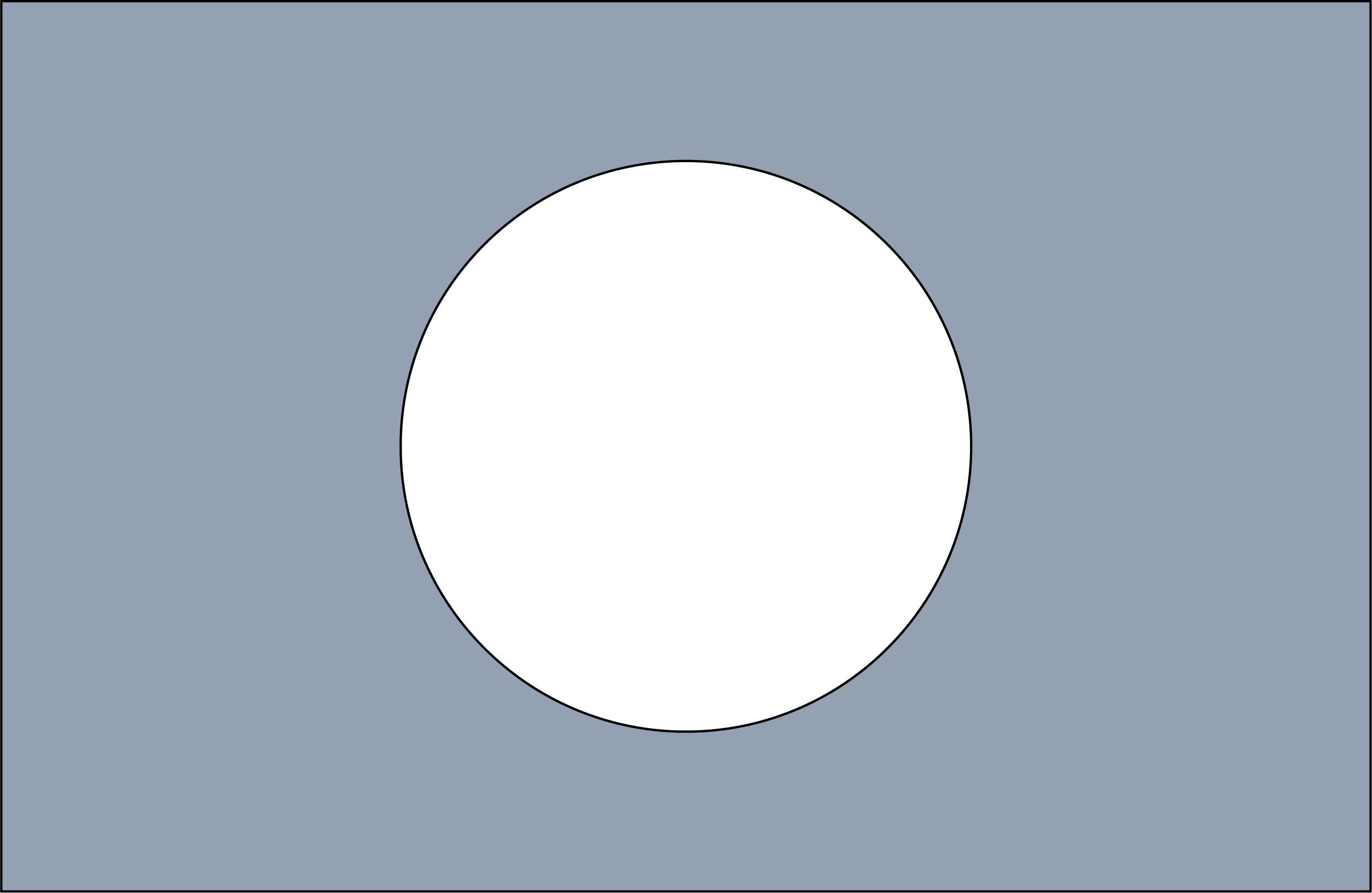}}            & $737$                                                        & \begin{tabular}[c]{@{}c@{}}$1.83\times10^5$ \\ ($10^{6*}$)\end{tabular}                                       & $0.5^*$                                                   & $13$                    & Thin film                                                         & Ding ($2024$) \cite{ding2024}       \\ \hline
        Sawfish                                                     & \raisebox{-0.38\totalheight}{\includegraphics[width=0.06\textwidth]{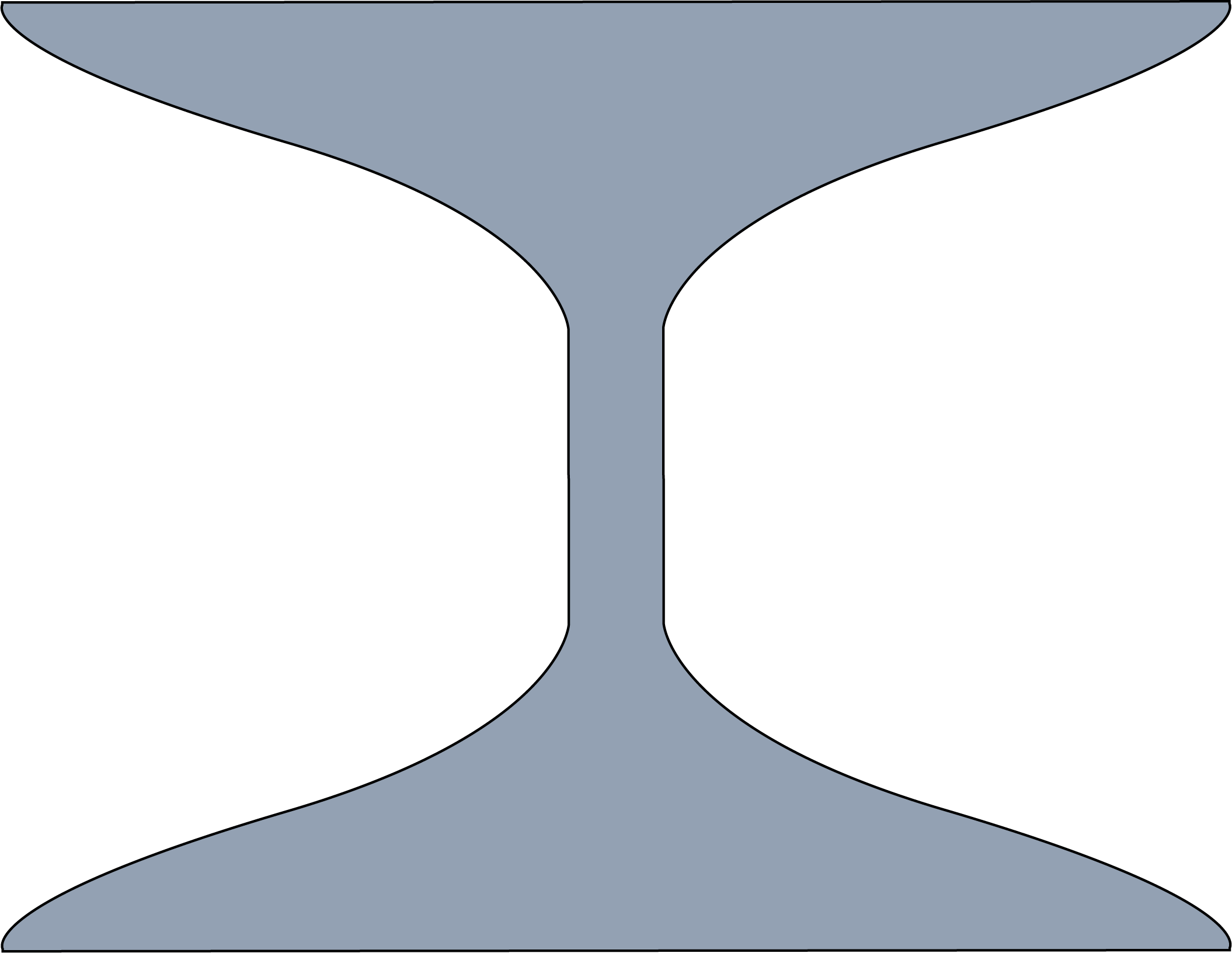}}          & $639$                                                        & \begin{tabular}[c]{@{}c@{}}$3825$\\ ($10^{6*}$)\end{tabular}                                                                & $0.8^*$                                                    & $46^*$ & \begin{tabular}[c]{@{}c@{}}Quasi-isotropic\\ etching\end{tabular} & Pregnolato ($2024$) \cite{pregnolato2024,sawfish2023} \\ \hline
        \begin{tabular}[c]{@{}c@{}}Circular\\ holes\end{tabular}    & \raisebox{-0.38\totalheight}{\includegraphics[width=0.06\textwidth]{Ciruclar_unitcell_schematic_top.png}}           & $618$                                                        & \begin{tabular}[c]{@{}c@{}}$1.1\times10^4$\\ ($6\times10^{5*}$)\end{tabular}   & $0.42^*$                                                   & $12$                    & \begin{tabular}[c]{@{}c@{}}Quasi-isotropic\\ etching\end{tabular} & Kuruma ($2021$) \cite{kuruma2021}    \\ \hline
        \begin{tabular}[c]{@{}c@{}}Rectangular\\ holes\end{tabular} &\raisebox{-0.38\totalheight}{\includegraphics[width=0.06\textwidth]{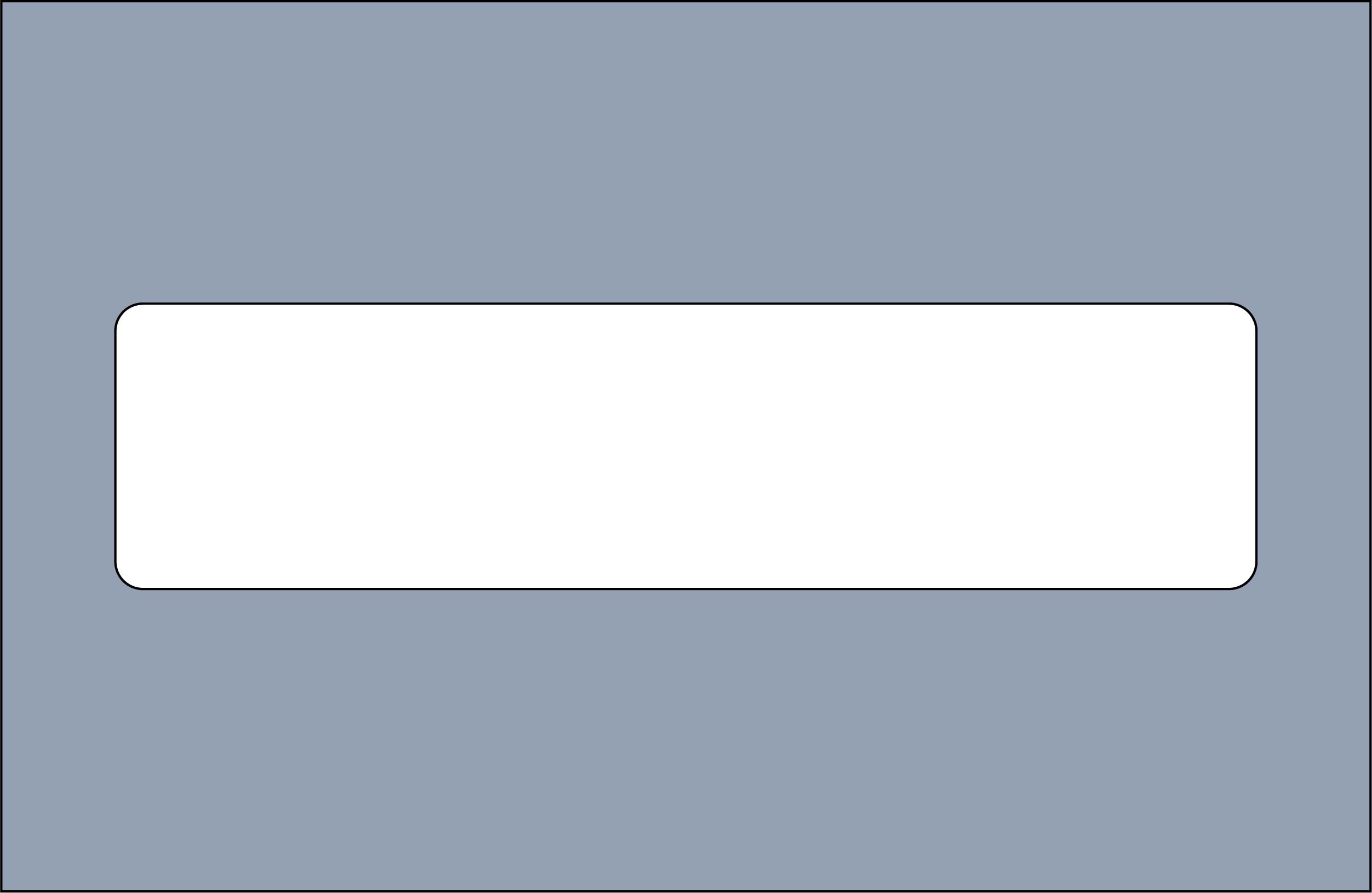}}           & $620$                                                        & \begin{tabular}[c]{@{}c@{}}$2800$\\ ($2\times10^{5*}$)\end{tabular}                                           & $2.8^*$                                                    & N/A                   & Thin film                                                         & Regan ($2021$) \cite{regan2021}     \\ \hline
        \begin{tabular}[c]{@{}c@{}}Rectangular\\ holes\end{tabular} &\raisebox{-0.38\totalheight}{\includegraphics[width=0.06\textwidth]{Rectangular_unitcell_schematic_top.png}}           & $775$                                                        & \begin{tabular}[c]{@{}c@{}}$8400$\\ ($2\times10^{5*}$)\end{tabular}                                           & $2.8^*$                                                    & N/A                   & Thin film                                                         & Regan ($2021$) \cite{regan2021}      \\ \hline
        \begin{tabular}[c]{@{}c@{}}Circular\\ holes\end{tabular}    &\raisebox{-0.38\totalheight}{\includegraphics[width=0.06\textwidth]{Ciruclar_unitcell_schematic_top.png}}            & $617$                                                        & \begin{tabular}[c]{@{}c@{}}2135\\ ($2\times10^{5*}$)\end{tabular}                                           & $0.56^*$                                                   & $25$                    & \begin{tabular}[c]{@{}c@{}}Quasi-isotropic\\ etching\end{tabular} & Rugar ($2021$) \cite{rugar2021}      \\ \hline
        \begin{tabular}[c]{@{}c@{}}Elliptical\\ holes\end{tabular}  & \raisebox{-0.38\totalheight}{\includegraphics[width=0.06\textwidth]{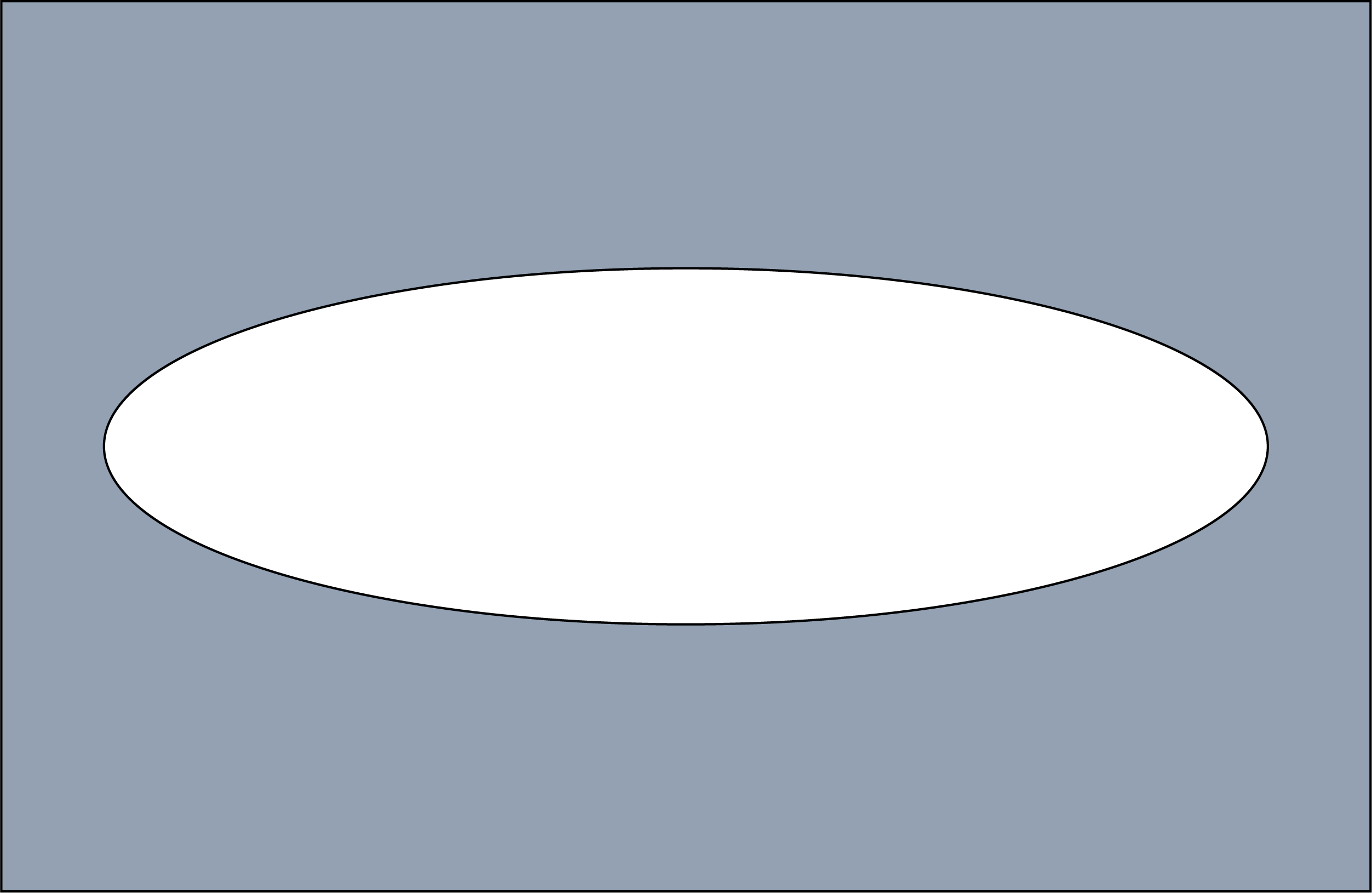}}          & $737$                                                        & \begin{tabular}[c]{@{}c@{}}$2\times10^4$\\ ($5\times10^{5*}$)\end{tabular}                                                                                                             & $0.5^*$                                                    & N/A                   & Angled etching                                                     & Bhaskar ($2020$) \cite{bhaskar2020,Bhaskar2019}   \\ \hline
        \begin{tabular}[c]{@{}c@{}}Circular\\ holes\end{tabular}    &\raisebox{-0.38\totalheight}{\includegraphics[width=0.06\textwidth]{Ciruclar_unitcell_schematic_top.png}}            & $637$                                                        & \begin{tabular}[c]{@{}c@{}}$1.4\times10^4$\\ ($10^{6*}$)\end{tabular}                                           & N/A                                                                       & N/A                   & \begin{tabular}[c]{@{}c@{}}Quasi-isotropic\\ etching\end{tabular} & Mouradian ($2017$) \cite{mouradian2017}  \\ \hline
        \begin{tabular}[c]{@{}c@{}}Circular\\ holes\end{tabular}    &\raisebox{-0.38\totalheight}{\includegraphics[width=0.06\textwidth]{Ciruclar_unitcell_schematic_top.png}}            & $660$                                                        & \begin{tabular}[c]{@{}c@{}}$2.4\times10^4$\\ ($2.7\times10^{5*}$)\end{tabular}                    & $0.47^*$                                                   & $20$                    & Thin film                                                         & Lee ($2014$) \cite{lee2014}        \\ \hline
        \textbf{2D}                                                          &           &                                                            &                                                                                                                                                          &                                                                           &                       &                                                                   &                   \\ \hline
        \begin{tabular}[c]{@{}c@{}}Circular\\ holes\end{tabular}    &\raisebox{-0.38\totalheight}{\includegraphics[width=0.06\textwidth]{Ciruclar_unitcell_schematic_top.png}}            & $746$                                                        & \begin{tabular}[c]{@{}c@{}}$1.6\times10^5$\\ ($7.6\times10^{5*}$)\end{tabular} & $2.18^*$                                                                      & N/A                   & Thin film                                                         & Ding ($2024$) \cite{ding2024}      \\ \hline
        \begin{tabular}[c]{@{}c@{}}Circular\\ holes\end{tabular}    &\raisebox{-0.38\totalheight}{\includegraphics[width=0.06\textwidth]{Ciruclar_unitcell_schematic_top.png}}            & $645$                                                        & \begin{tabular}[c]{@{}c@{}}$8000$\\ ($3.2\times10^{5*}$)\end{tabular}                                         & $0.35^*$                                                                      & $1.224$                 & Thin film                                                         & Jung ($2019$) \cite{jung2019}       \\ \hline
        \end{tabular}%
        }
        \begin{tablenotes}
            \item[1] \footnotesize{All values denoted with a "*" are simulated measures.}
        \end{tablenotes}
    \end{threeparttable}
    \label{tab:current_developments}
\end{table}

\newpage

%% file: methodology.tex
\section{Methodology}
\begin{figure}[h!]
        \centering
        \includegraphics[width=\linewidth]{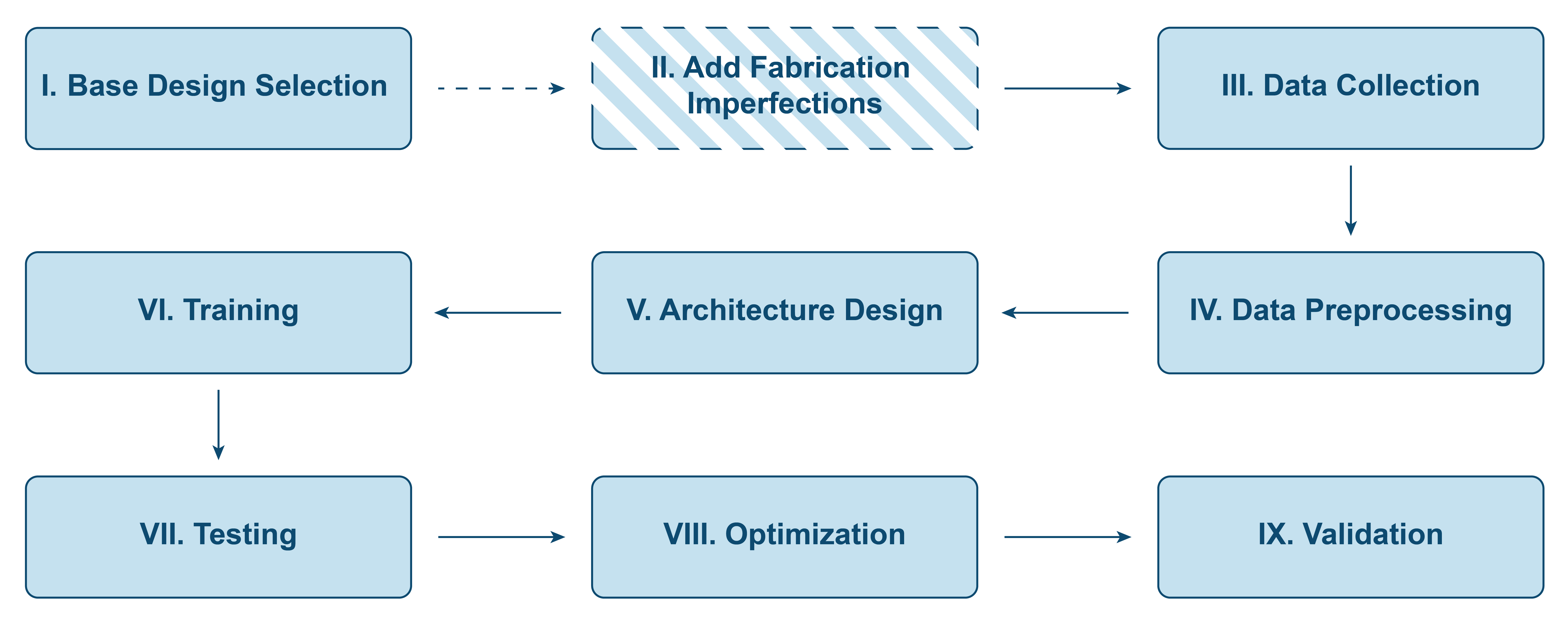}
        \caption{A flowchart illustrating the process for optimizing nanocavity designs using DL, highlighting the additional step introduced in this study.}
        \label{Flowchart_chapter6}
\end{figure}

The traditional process for optimizing the Q-factor of nanocavities with DL involves eight fundamental steps \cite{Takashi_CNN, Liu_genetic_algorithm, Baba_CMA-ES, Takashi_CNN_iter, Baba_ML_300}. This research extends and enhances this approach by integrating a comprehensive evaluation of fabrication tolerances, ensuring robust optimization that addresses real-world fabrication challenges. The nine optimization steps, in sequential order, are as follows (see Figure \ref{Flowchart_chapter6}):

\hypertarget{Flowchart_chapter6}{}

\paragraph{I. Base Design Selection:} The base nanocavity design is the design from which the final optimized structure will be derived. Therefore, it is important to consider which base nanocavity design has the right properties for the application in mind. For this research, the base nanocavity designs were selected for their inherent resilience to fabrication imperfections.

\paragraph{II. Add Fabrication Imperfections (additional step):} Perturbations like surface roughness or slanted sidewalls are introduced to the simulation to mimic the effect of fabrication imperfections (see Supplementary Information \hyperlink{supinfD1}{D1} and \hyperlink{supinfD2}{D2}). After adding the imperfections, the optimization process is proceeded as usual. The final optimized nanocavity will have the largest Q-factor in the presence of these fabrication errors. It is important to note that this does not imply a higher Q-factor under ideal conditions. Rather, it indicates that the Q-factor of the optimized structure is more resilient to fabrication imperfections. Hence, this step aims to identify nanocavity structures where the Q-factor is better preserved in the presence of fabrication errors compared to structures optimized without considering such imperfections.

\paragraph{III. Data Collection:} The optical properties of numerous random nanocavity designs are simulated using the FDTD method. These random nanocavity designs are derived from the base nanocavity by randomly altering specific design parameters. It is crucial to consider which parameters will be randomly varied, within what range, and according to which probability distribution, as this defines the optimization space. The defined optimization space dictates the amount of data required to effectively train a NN and the parameter space where it performs optimally.

\paragraph{IV. Data Preprocessing:} Data preprocessing plays a crucial role in the effectiveness and efficiency of NNs \cite{DL_general_book}. Firstly, the total dataset must be divided into a training dataset and a representative test dataset. Secondly, the input and output data are transformed to improve their quality. NNs demonstrate increased stability, compatibility, and faster convergence when trained on properly scaled data that follows a normal distribution.

\paragraph{V. Architecture design:} When designing a NN architecture, there are numerous options and techniques to consider. For instance, considerations include the type of NN architecture, the number of layers, the quantity of nodes per layer, activation functions, the choice of loss function, hyperparameters and the implementation of techniques such as dropout, L$^2$-regularization, batch normalization, and various other architectural design considerations. The specifications of the NN architecture depend on the complexity of the optimization problem at hand. A larger optimization space calls for a larger dataset, a more intricate NN, and the implementation of additional techniques to mitigate computational costs and unwanted effects like overfitting or vanishing gradients \cite{DL_general_book}.

\paragraph{VI. Training:} Throughout training, the weights and biases (internal parameters) of the NN are adjusted at the end of each loop (epoch) until the loss function is minimized. The loss function is computed by comparing the predicted optical properties, calculated with forward propagation, with the actual optical properties, computed with the FDTD method. This minimization process involves following the gradient, computed with backpropagation, in its reverse direction (gradient descent). This process continues, for a fixed amount of epochs, until the model converges to a satisfactory solution \cite{DL_general_book}.

\paragraph{VII. Testing:} Once training is complete, the final trained model is evaluated on a separate test dataset to assess its performance on unseen data. This step provides an unbiased estimate of the model's effectiveness. Afterwards, the relative prediction error and correlation coefficient from the training data and test data are compared to validate the generalizability of the NN on unseen nanocavity designs \cite{DL_general_book}.

\paragraph{VIII. Optimization:} The NN can now rapidly and accurately predict the optical properties of nanocavity designs, surpassing the computational speed of traditional methods such as the FDTD method. The goal of the optimization step is to use the trained NN to efficiently adjust design parameters, enhancing the optical properties of the base nanocavity design. This can be done with local optimization algorithms like gradient descent/ascent or global optimization algorithms like evolutionary strategies (ES). The NN, being significantly faster than the FDTD method, allows for more efficient exploration of the parameter space compared to these computationally intensive approaches.

\paragraph{IX. Validation:} A selection of nanocavities found with the NN, exhibiting potentially superior characteristics, is subjected to validation and confirmation of their optical properties through FDTD simulation. The nanocavity design with the most favorable performance is deemed as the optimized nanocavity design. \newline

\begin{figure}[h!]
    \centering
    \includegraphics[width=\linewidth]{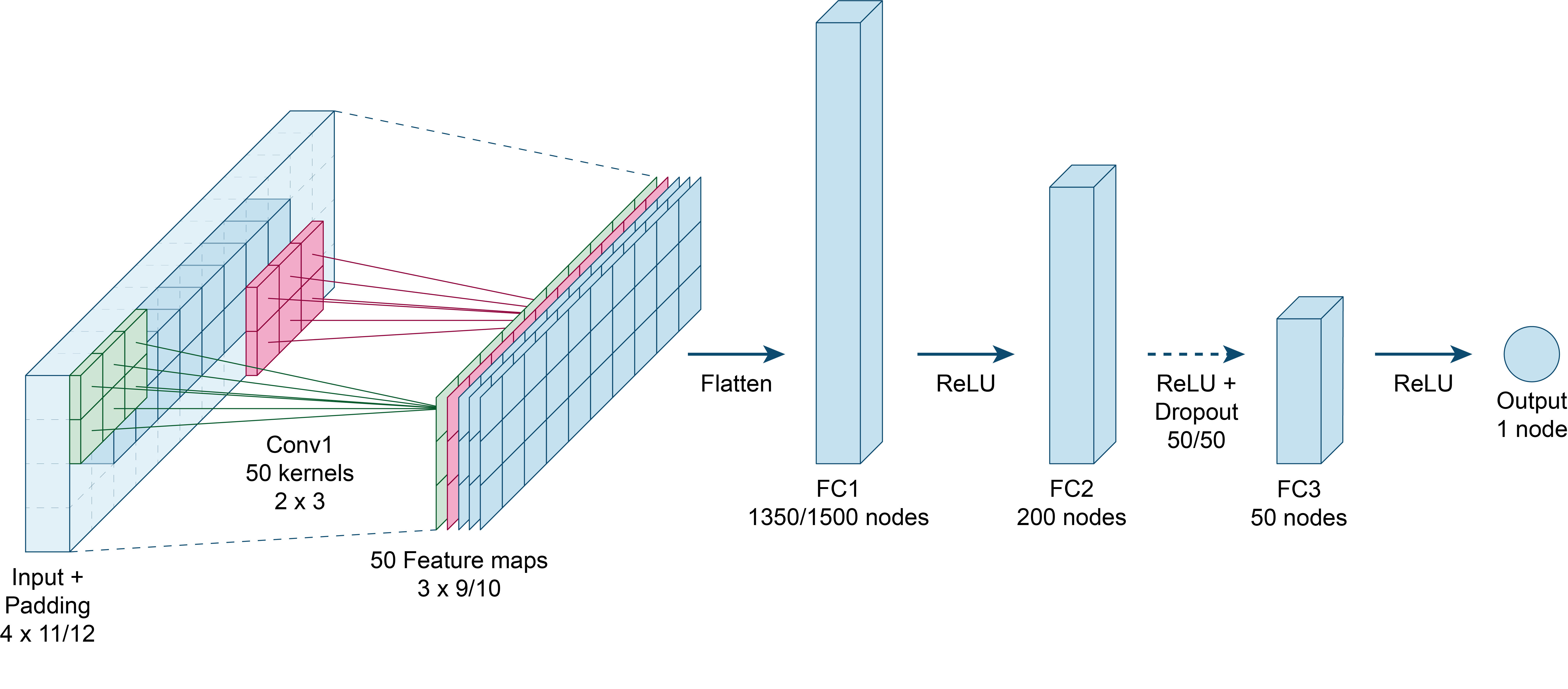}
    \caption{A schematic representation of the CNN architecture employed to optimize nanobeam cavities for fabrication error tolerance. The CNN consists of a convolutional layer followed by three FC layers, with dropout applied between the second and third FC layers to reduce overfitting. Zero-padding is applied to the input, and the ReLU activation function is utilized \protect\cite{Takashi_CNN, Takashi_CNN_iter}.}
    \label{fig:Conv_network}
\end{figure}

The NN architecture used to predict the Q-factor of the nanobeam cavities throughout this research is a CNN with one convolutional layer, followed by a series of fully connected layers (FC). Figure \ref{fig:Conv_network} shows a visual representation of this CNN architecture design (step \hyperlink{Flowchart_chapter6}{V}). This NN draws inspiration from the NN used in the research of T. Asano and S. Noda \cite{Takashi_CNN, Takashi_CNN_iter}, as they worked with a similar number of independent degrees of freedom (27 independent design parameters) and a comparable dataset size (1000 nanocavities). However, it is crucial to emphasize that their research primarily focuses on optimizing for maximum simulated Q-factors, whereas this research centers on fabrication error resilience.

The geometry of a nanobeam cavity is encoded as deviations in design parameters relative to the base nanobeam cavity. These encoded parameters are stored into a matrix that serves as input for the CNN (see Supplementary Information \hyperlink{supinfB}{B}). The design parameters are distributed over the input matrix in such a way that the kernels of the convolutional layer can cover the relations between the geometry of neighboring unit cells. Zero-padding is applied around this matrix before passing it to the convolutional layer to ensure that information near the edges of the input matrix is also adequately processed. The optimization process involves 13 independent degrees of freedom for the L2 nanobeam cavity and 16 for the fishbone nanobeam cavity. Consequently, the input matrix, including padding, differs between the two designs: 4$\times$11 for the L2 nanobeam and 4$\times$12 for the fishbone nanobeam.
\par
The convolutional layer makes use of $50$ different 2$\times$3 kernels with a stride of $1$ in both directions. This results in $50$ different feature maps, sized 3$\times$9 for the L2 nanobeam cavity and 3$\times$10 for the fishbone nanobeam cavity. These feature maps are flattened, before passed on to the first FC layer (FC1). FC1, consisting of $1350$ nodes for the L2 nanobeam cavity and $1500$ nodes for the fishbone nanobeam cavity, is connected to the second FC layer (FC2) through an affine transformation and a ReLU (Rectified linear unit) activation function \cite{activation_function} (see Supplementary Information \hyperlink{supinfE1}{E1} and \hyperlink{supinfE2}{E2}). FC2 consists of $200$ nodes and is the same way connected to the third FC layer (FC3), but with $50/50$ dropout \cite{DL_general_book} in between. FC3 has 50 nodes, and after the final affine transformations and ReLU activations, log$_{10}(Q_{\textrm{NN}})$ is obtained by summing the outputs from FC3. This logarithmic transformation is part of step \hyperlink{Flowchart_chapter6}{IV} and enhances the NN's learning capabilities \cite{Takashi_CNN, Renjie_CNN, Renjie_CNN_update, Renjie_transformer}.
\newline

The loss function of the NN is the function that will be minimized during training, by iteratively adjusting the weights and biases of the NN. The DL task addressed in this research is a regression task of a continuous variable. The default choice for such tasks is the mean-squared-error (MSE) loss function \cite{DL_general_book}. MSE is computed as the mean of the squared differences between predicted and actual values (see the first term of Equation \ref{Eq:Loss_function_NN}).

\begin{equation}
    \mathrm{loss} = \mathrm{MSE} + \mathrm{L2} = \left[\log_{10}(Q_{\mathrm{NN}}) - \log_{10}(Q_{\mathrm{FDTD}})\right]^2 + \frac{1}{2} \lambda \sum_{i} \theta_i^2
    \label{Eq:Loss_function_NN}
\end{equation}

The second term of this NN's loss function is an artificial loss term. It penalizes the use of large model parameters $\theta_i$ (weights and biases), resulting in the model learning smoother representations of the regression task and thus less overfitting. The hyperparameter $\lambda$ determines the strength of the so called L$^2$-regularization \cite{Takashi_CNN, Renjie_CNN_update}. L$^2$-regularization is a valuable technique for preventing overfitting in DL models \cite{L2_regularization, DL_general_book}. It encourages simpler models that generalize better to unseen data and prevents the memorization of noise or small fluctuations in the training data, leading to improved performance and robustness. \newline

Numerous minimization algorithms exist with which one can minimize the loss function. In the field of DL, this minimization algorithm is referred to as the optimizer. The goal of the optimizer is to find the set of model parameters that yield the best performance on the training data while generalizing well to unseen data. For this CNN, the stochastic gradient descent (SGD) optimizer was chosen for its simplicity, efficiency, effectiveness and compatibility with the MSE loss function \cite{DL_general_book}. SGD is a widely used optimization algorithm in DL. It iteratively updates model parameters by computing the gradient of the loss function with respect to each parameter of the model. 
\par
SGD can be used in combination with adaptive learning rates. Adaptive learning techniques are algorithms that dynamically adjust the learning rate during training based on various factors such as the magnitude of gradients, the history of parameter updates, or other characteristics of the optimization landscape. For this CNN, the SGD optimizer is extended with momentum-based adaptive learning rates. Momentum helps smooth out the update trajectory and accelerate convergence by considering the history of parameter updates \cite{momentum, DL_general_book}. When the gradients consistently point in the same direction, momentum accumulates and amplifies the effect, leading to faster progress along the gradient descent path. Conversely, when the gradients change direction or fluctuate, momentum helps dampen the effect, reducing the impact of noisy updates and preventing the algorithm from getting stuck in local minima or saddle points.
\par
The SGD algorithm with momentum based adaptive learning rates can be mathematically expressed as follows:

\begin{equation}
    \nu_{n} = \gamma\nu_{n-1} + \nabla_{\theta} L(\theta_{n-1})
\end{equation}
\begin{equation}
    \theta_{n} = \theta_{n-1} - \alpha \nu_{n}
\end{equation}

The model parameters from the previous iteration $\theta_{n-1}$ are updated by the momentum term $\nu_{n}$, scaled by the learning rate hyperparameter $\alpha$. This momentum term is determined by the gradient of the loss function $L(\theta_{n-1})$ with respect to the model parameters $\theta$ and the previous momentum term $\nu_{n-1}$, scaled by the momentum hyperparameter $\gamma$. Thus, this approach takes into account the history of the parameter updates when updating the model parameters. 
\par
Additionally, various strategies exist for initializing the weights and biases of the NNs before training. For this research, the He-initialization technique is used (see Supplementary Information \hyperlink{supinfE3}{E3}) \cite{Initialization_He}. After empirically studying the behavior of the NN with changes in hyperparameters, a learning rate of $\alpha = 0.001$ and a momentum of $\gamma = 0.9$ were determined to yield the optimal results. Additionally, the NNs were trained with a weight decay term of $\lambda = 0$ or $\lambda = 0.001$, allowing for the exploration of both scenarios with and without L$^2$-regularization. \newline

To train the NNs, a dataset containing $1250$ unique nanobeam cavities is created (step \hyperlink{Flowchart_chapter6}{III}) for each base design and fabrication error implementation (including under ideal, surface roughness, and sidewall slant conditions), resulting in a total of six datasets (see Supplementary Information \hyperlink{supinfF1}{F1}). Before training, these datasets must first be preprocessed (step \hyperlink{Flowchart_chapter6}{IV}) to facilitate the efficient learning of the underlying relationships by the NNs. This involves taking the logarithm of the Q-factors and scaling the design parameters (see Supplementary Information \hyperlink{supinfF2}{F2}).

\begin{figure}[h!]
    \centering
    \begin{subfigure}{0.325\textwidth}
        \centering
        \includegraphics[width=\linewidth]{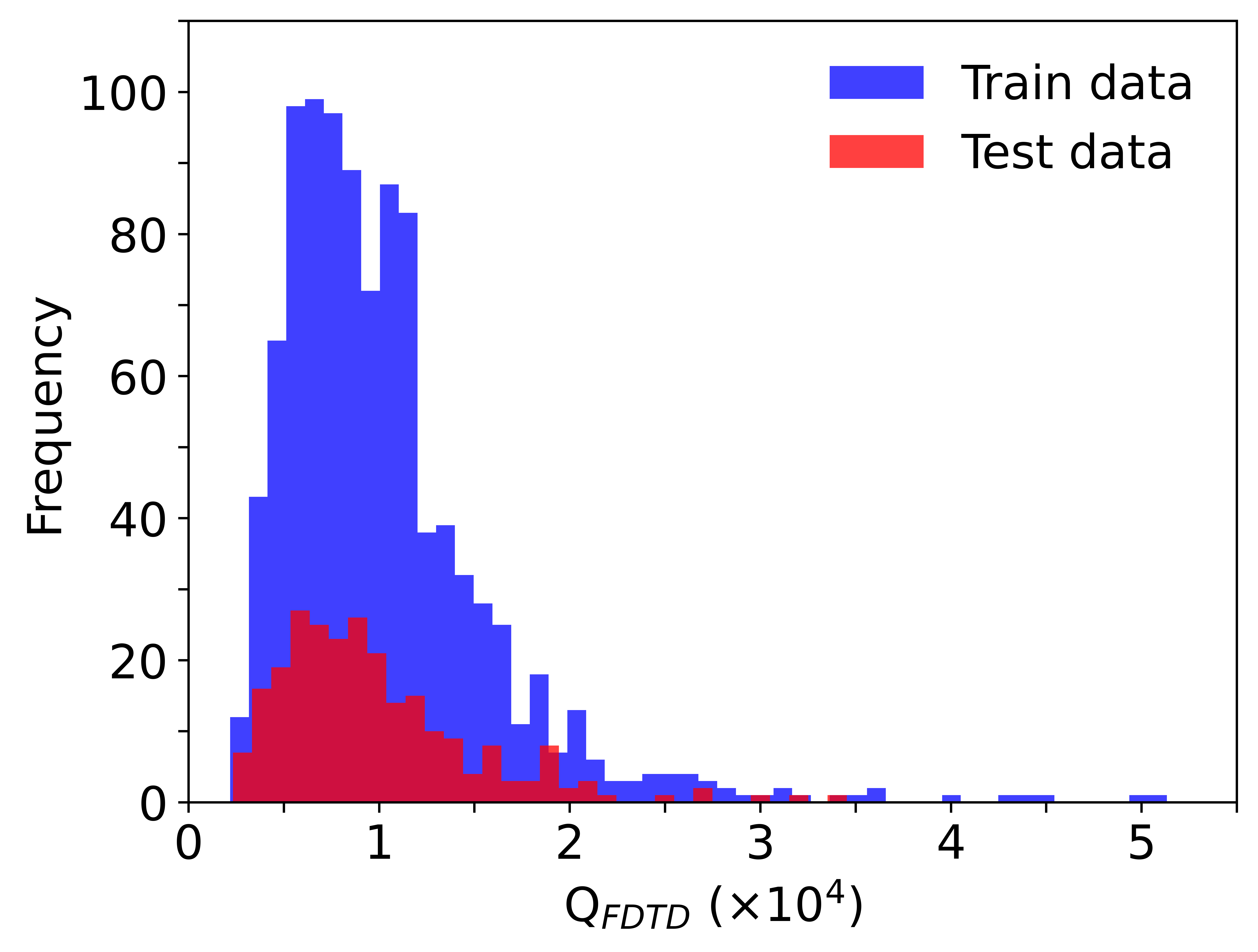}
        \caption{}
        \label{fig:dataset_NC123_notprocessed}
    \end{subfigure}
    \hspace{0.002\textwidth}
    \begin{subfigure}{0.32\textwidth}
        \centering
        \includegraphics[width=\linewidth]{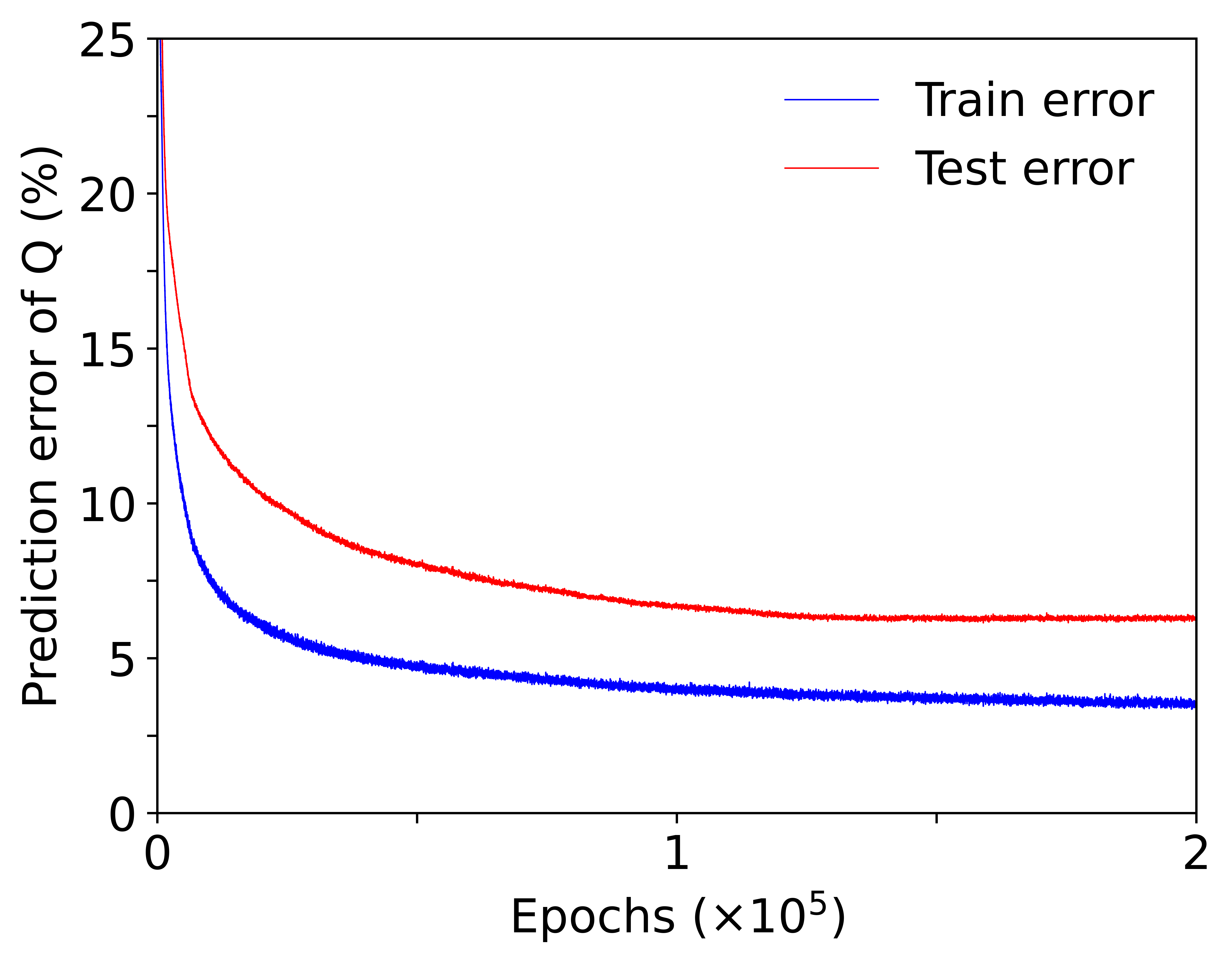}
        \caption{}
        \label{fig:Train_sequence_NC123_l0}
    \end{subfigure}
    \hspace{0.002\textwidth}
    \begin{subfigure}{0.315\textwidth}
        \centering
        \includegraphics[width=\linewidth]{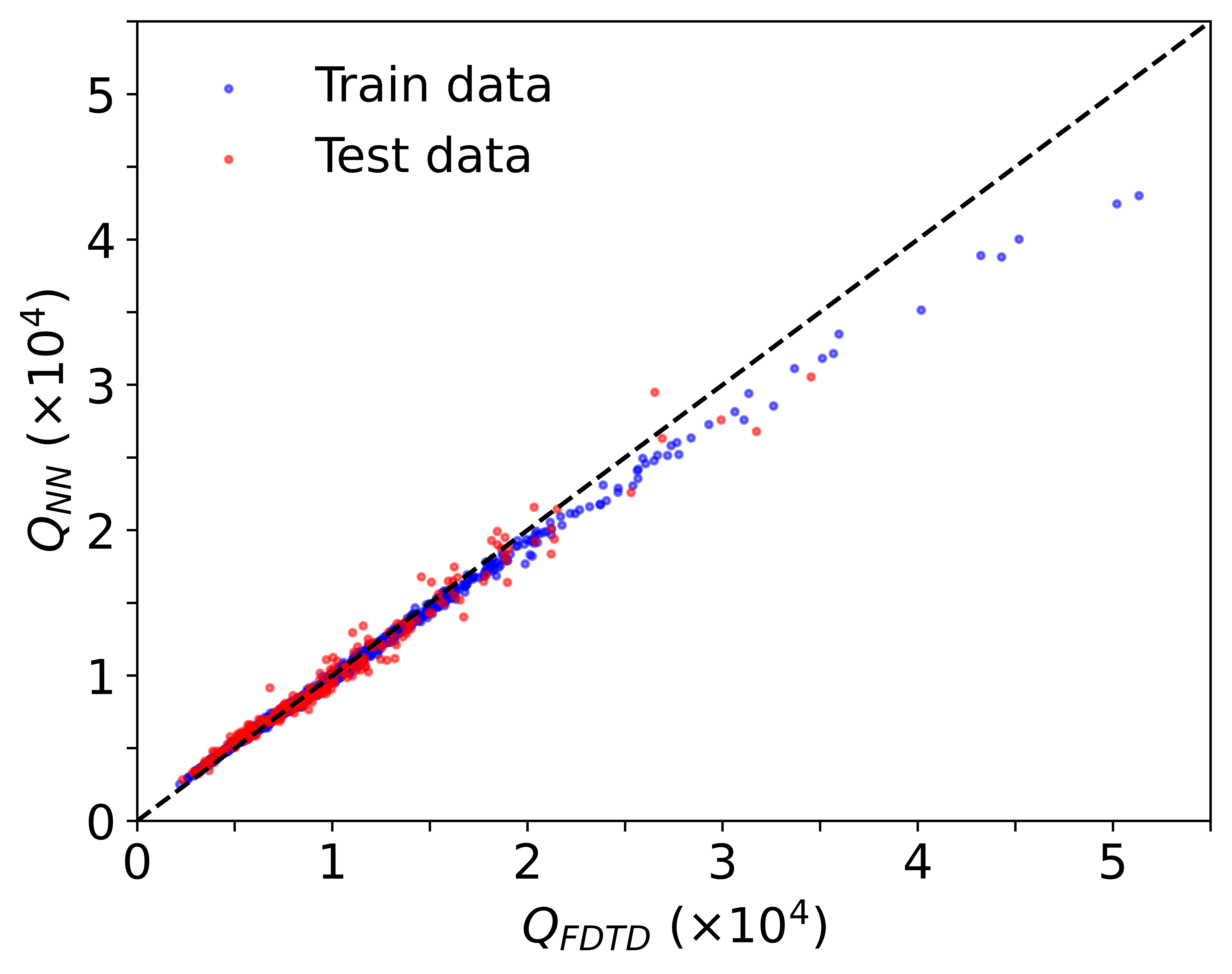}
        \caption{}
        \label{fig:Correlation_NC123_l0}
    \end{subfigure}

    \caption{An overview showcasing an example of the dataset, training, and correlation plots for L2 nanobeam cavities under ideal conditions, trained without L$^2$-regularization. Additional plots are provided in the \protect\hyperlink{supinfH}{Appendix}. \textbf{(a)} Dataset split before preprocessing. \textbf{(b)} Prediction error progression during training. \textbf{(c)} Correlation between predicted Q-factors ($Q_{\textrm{NN}}$) and simulated Q-factors ($Q_{\textrm{FDTD}}$).}
    \label{fig:combined_NC_plots}
\end{figure}

The total dataset collected is randomly divided into a training dataset and a test dataset with a ratio of 8:2, resulting in $1000$ training samples and $250$ test samples. The splitting is done randomly to ensure that all features from the training dataset are also captured in the test dataset (see Figure \ref{fig:dataset_NC123_notprocessed}). The training process is tracked by monitoring the relative prediction error of the Q-factor. The relative prediction error is defined as follows:

\begin{equation}
    \epsilon_{\textrm{pred}} = \frac{|Q_{\mathrm{NN}} - Q_{\mathrm{FDTD}}|}{Q_{\mathrm{FDTD}}} \times 100\%
    \label{Eq:pred}
\end{equation}

An example of the convergence of the training processes can be seen in Figure \ref{fig:Train_sequence_NC123_l0}. This graph demonstrates a rapid decrease in average prediction error during the initial $5\times10^4$ epochs. Thus, after $2\times10^5$ epochs the training process is deemed as converged. This process took approximately $150$ minutes. Furthermore, the difference between the training and testing prediction errors serves as a measure of the model's generalizability and potential overfitting. A large difference suggests that the model performs well only on the training data and struggles to generalize effectively to new, unseen data.
\par
Finally, the trained NNs are assessed on their ability to predict Q-factors from a given set of input design parameters. Their performance is quantified by analyzing the final relative prediction error (see Equation \ref{Eq:pred}) and the Pearson correlation coefficients \cite{Pearson_corr} (see Equation \ref{Eq:correlation}) between the predicted Q-factors ($Q_{\textrm{NN}}$) and the simulated Q-factors ($Q_{\textrm{FDTD}}$), for both the training and test datasets. The Pearson correlation coefficient quantifies how $Q_{\textrm{NN}}$ varies in relation to $Q_{\textrm{FDTD}}$. In the ideal scenario where $Q_{\textrm{NN}}$ equals $Q_{\textrm{FDTD}}$, the correlation coefficient would be 1. An example of a correlation graph is shown in Figure \ref{fig:Correlation_NC123_l0}. The covariance between $Q_{\textrm{NN}}$ and $Q_{\textrm{FDTD}}$ is denoted by $\textrm{cov}(Q_{\textrm{NN}}, Q_{\textrm{FDTD}})$, and their standard deviations are represented by $\sigma$. The Pearson correlation coefficient is defined as:

\begin{equation}
    R = \frac{\mathrm{cov}(Q_{\mathrm{NN}}, Q_{\mathrm{FDTD}})}{\sigma(Q_{\mathrm{NN}}) \sigma(Q_{\mathrm{FDTD}})}
    \label{Eq:correlation}
\end{equation}

After training and testing the NNs (step \hyperlink{Flowchart_chapter6}{VI} and \hyperlink{Flowchart_chapter6}{VII}) on their predictive performance (see Supplementary Information \hyperlink{supinfF3}{F3}), the NNs can be used to efficiently optimize the nanobeam cavities (step \hyperlink{Flowchart_chapter6}{VIII}). To achieve this, two different optimization algorithms are selected to interact with the trained NNs. One is based on a local optimization algorithm, while the other on a global optimization algorithm. Global optimization algorithms search broadly, exploring the entire solution space, while local optimization algorithms search intensively within a confined region around the initial condition. Both algorithms are deployed with a set of different hyperparameters to form a family of potentially high Q-factor nanobeam cavities (see Supplementary Information \hyperlink{supinfG1}{G1}). For this study, the best performing structure from this family is deemed as the optimized design.
\par
The local optimization algorithm is referred to as gradient ascent (GA) \cite{Takashi_CNN, Takashi_CNN_iter}, as the goal is to maximize the Q-factor by following a gradient in the design parameter landscape. Also, this way one can clearly distinguish between gradient descent, the optimizer that optimizes the weight and biases of the NN and GA, the optimization algorithm that maximizes the Q-factor by interacting with the trained NN.

\begin{equation}
    \textrm{loss} = \textrm{MSE} + \textrm{L2} = \left[\textrm{log}_{10}(Q_{\mathrm{NN}}) - \textrm{log}_{10}(Q_{\mathrm{target}})\right]^2 + \frac{1}{2} \lambda_{\mathrm{GA}} |\vec{x} - \vec{x}_0|^2
    \label{Eq:Artificial_loss_function_GA}
\end{equation}

To maximize the Q-factor using the local optimization algorithm, the artificial loss function described in Equation \ref{Eq:Artificial_loss_function_GA} is minimized. The target Q-factor is arbitrary and is set at a high value of $Q_{\textrm{target}} = 10^6$. To keep the parameter displacements $\vec{x}$ small and approximately within the initial design parameter space, an L$^2$-regularization term is added to the artificial loss function. This L$^2$-regularization is relative to the initial input structure defined by the parameter displacements $\vec{x}_0$. The gradient of the loss function with respect to $\vec{x}$ is calculated using backpropagation. The parameter displacements are then incrementally adjusted to minimize the loss, following the same principle as the SGD algorithm in optimizing the NN's loss function. Through empirical experimentation, a learning rate of $\alpha_{\textrm{GA}} = 10^{-5}$ and a momentum of $\gamma_{\textrm{GA}} = 0.9$ were selected. After $10^6$ iterations, the optimization process is considered converged (see Figure \ref{fig:GA_l0} and \ref{fig:GA_l0.001}). \newline

The global optimization algorithm used in this research belongs to the family of evolutionary strategies and is known as CMA-ES (covariance matrix adaptation evolution strategy) \cite{Baba_CMA-ES, CMA-ES}. As a population-based optimization approach, its population is determined through random sampling from a multivariate normal distribution. This multivariate normal distribution is updated every generation by adapting the covariance matrix and mean vector according to the elite solutions from the previous generation. These elite solutions are defined by the individuals with the highest fitness. Updating the covariance matrix in CMA-ES dynamically adjusts the algorithm's exploration strategy by learning from the performance of a subset of elite solutions. This approach enables CMA-ES to effectively explore the design parameter space and find a solution to the optimization problem.

\begin{equation}
    \textrm{fitness} = Q - \textrm{L2} = \textrm{log}_{10}(Q_{\mathrm{NN}}) - \frac{1}{2} \lambda_{\mathrm{ES}} |\vec{x} - \vec{x}_0|^2
    \label{Eq:fitness_CMA-ES}
\end{equation}

To ensure adequate exploration of the parameter space with the global optimization algorithm, a population size of $20$ is selected, with half of the population used for the adaptation of the covariance matrix. After $300$ iterations, the optimization is deemed as converged (see Figure \ref{fig:CMA_ES_l0} and \ref{fig:CMA_ES_l0.001}). The fitness function, described in Equation \ref{Eq:fitness_CMA-ES} and maximized by this optimization algorithm, consists of the predicted Q-factor, penalized by an L$^2$-regularization term. As before, this term penalizes large deviations of the parameter displacements $\vec{x}$ from the initial mean structure defined by $\vec{x}_0$. The algorithm is initialized with a multivariate normal distribution about the initial mean $\vec{x}_0$, with a initial standard deviation of $\sigma_0 = 1$.

\newpage

%% file: results.tex
\section{Results \& Discussion}
\begin{figure}[h!]
    \centering
    \begin{subfigure}{0.25\linewidth}
        \centering
        \includegraphics[width=\linewidth]{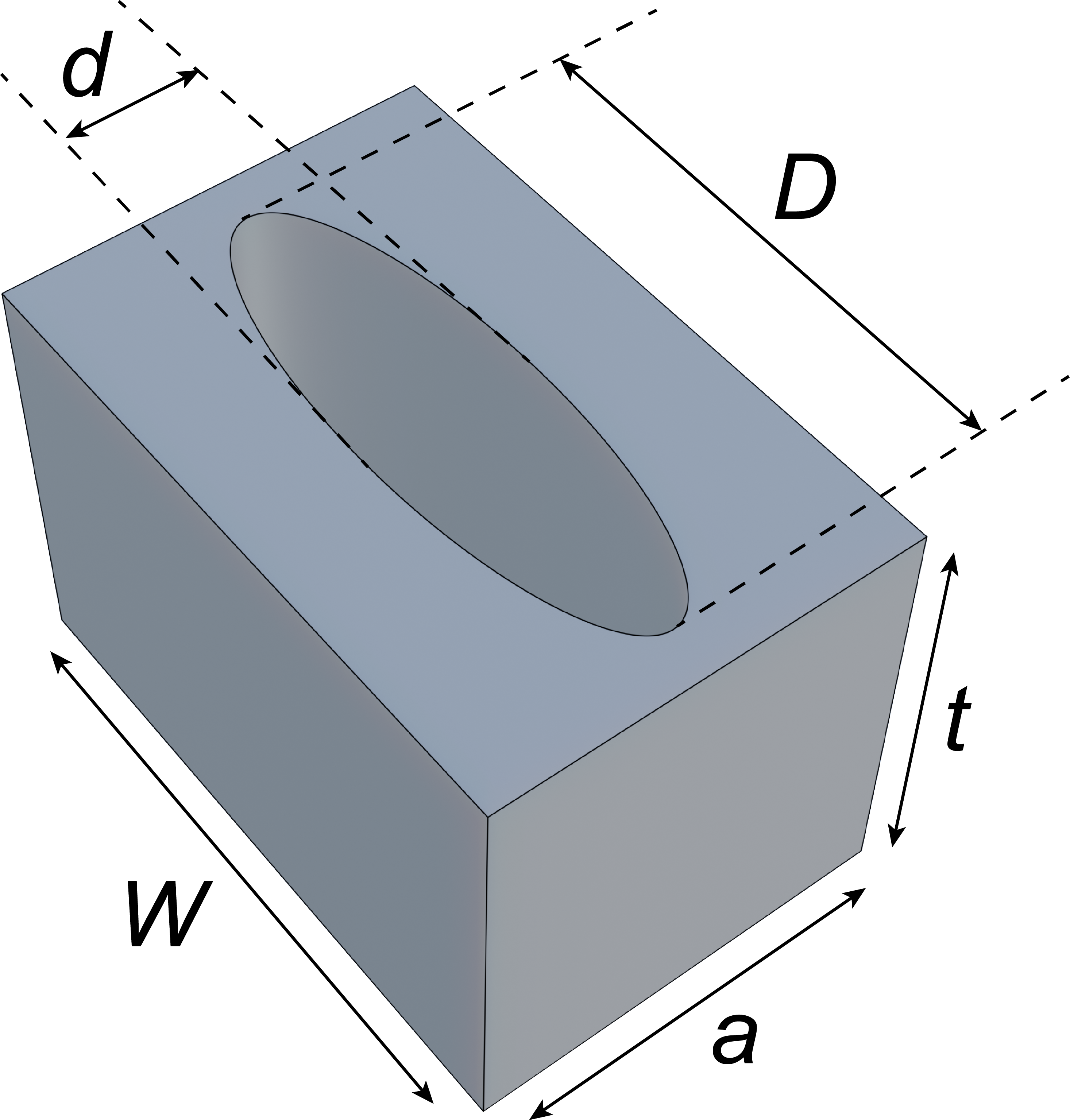}
        \caption{}
        \label{fig:NC_unitcell_3D}
    \end{subfigure}
    \hspace{0.05\linewidth}
    \begin{subfigure}{0.45\linewidth}
        \centering
        \includegraphics[width=\linewidth]{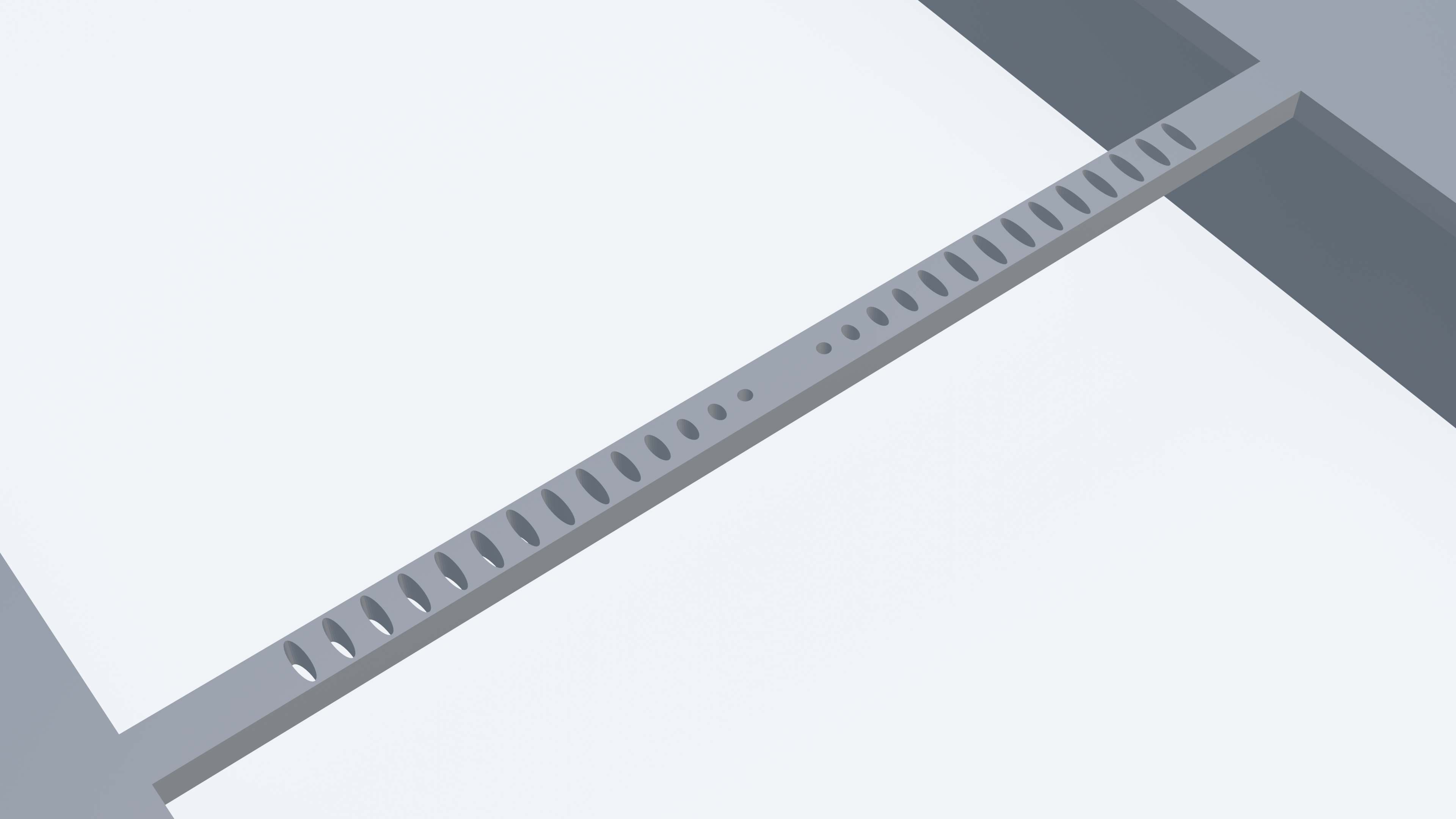}
        \caption{}
    \end{subfigure}

    \vspace{0.25cm}

    \begin{subfigure}{1\linewidth}
        \centering
        \includegraphics[width=\linewidth]{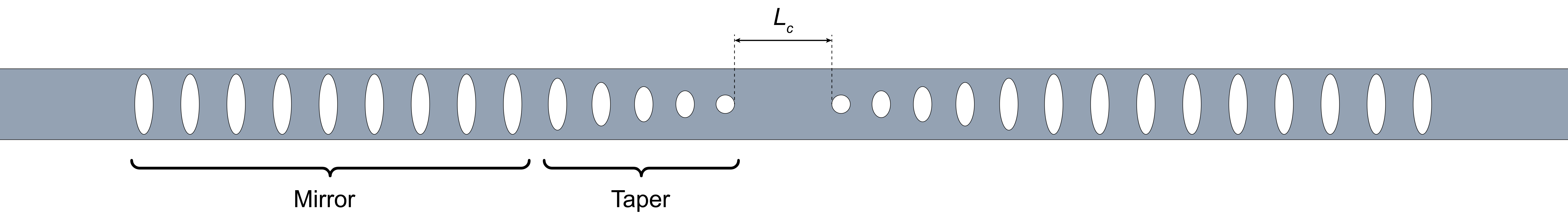}
        \caption{}
        \label{fig:NC_schematic_geom}
    \end{subfigure}
    
    \caption{A visual representation of the geometric properties of the base design for the L2 nanobeam cavity. \textbf{(a)} A 3D model of the elliptical airhole unit cell, highlighting its key design parameters. \textbf{(b)} A 3D model of the complete L2 nanobeam cavity. \textbf{(c)} A 2D schematic of the L2 nanobeam cavity, highlighting the mirror, taper and cavity region.}
    \label{fig:NC_geom}
\end{figure}

\begin{figure}[h!]
    \centering
    \begin{subfigure}{0.25\linewidth}
        \centering
        \includegraphics[width=\linewidth]{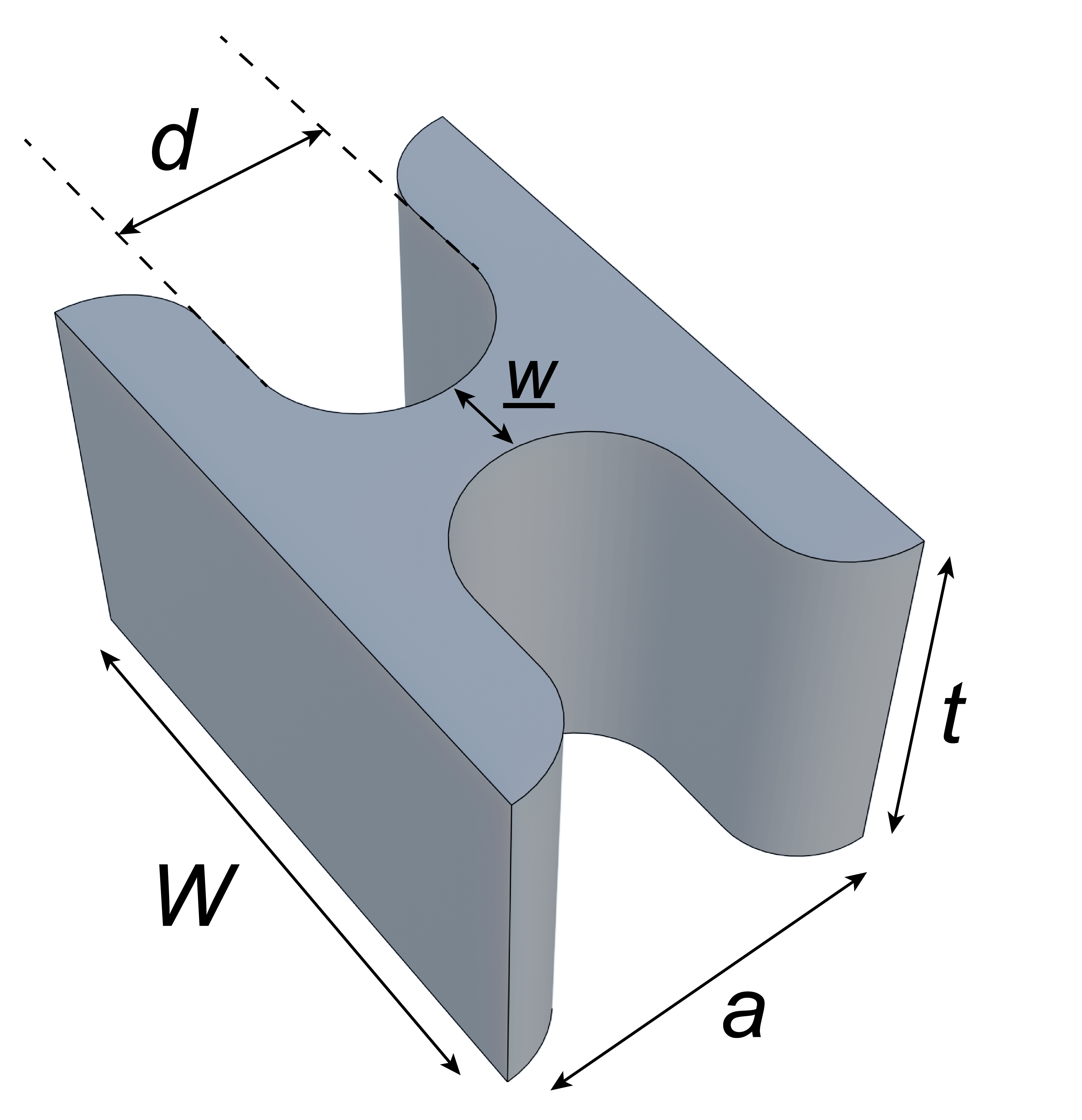}
        \caption{}
        \label{fig:FB_unitcell_3D}
    \end{subfigure}
    \hspace{0.05\linewidth}
    \begin{subfigure}{0.45\linewidth}
        \centering
        \includegraphics[width=\linewidth]{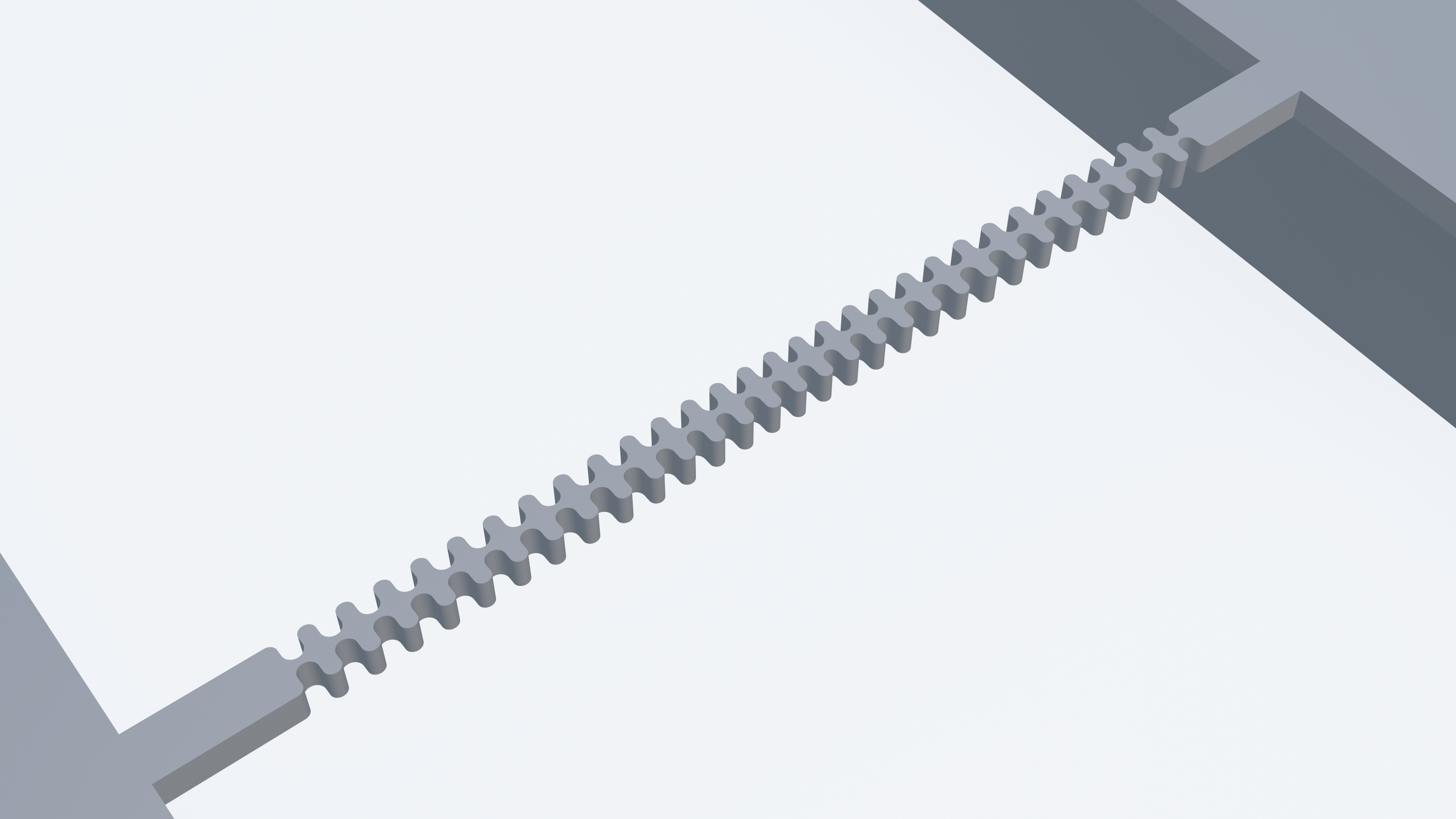}
        \caption{}
    \end{subfigure}

    \vspace{0.25cm}

    \begin{subfigure}{1\linewidth}
        \centering
        \includegraphics[width=\linewidth]{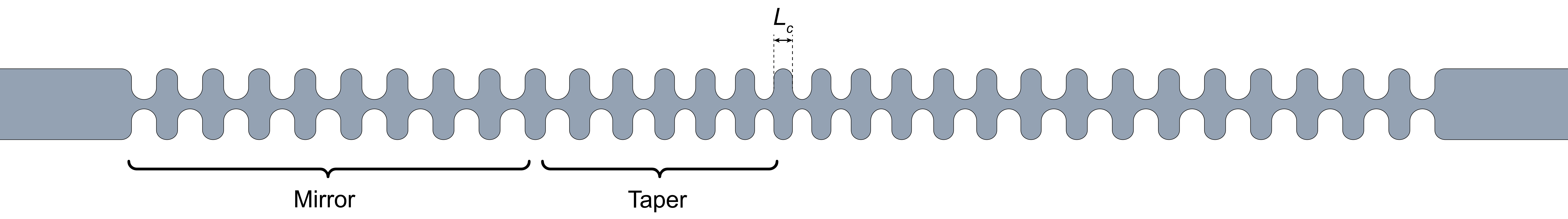}
        \caption{}
        \label{fig:FB_schematic_geom}
    \end{subfigure}
    
    \caption{A visual representation of the geometric properties of the base design for the fishbone nanobeam cavity. \textbf{(a)} A 3D model of the corrugated unit cell, highlighting its key design parameters. \textbf{(b)} A 3D model of the complete fishbone nanobeam cavity. \textbf{(c)} A 2D schematic of the fishbone nanobeam cavity, highlighting the mirror, taper and cavity region.}
    \label{fig:FB_geom}
\end{figure}

The two base nanobeam cavities are the elliptical-hole "L2" nanobeam cavity and the corrugated "fishbone" nanobeam cavity (see Figure \ref{fig:NC_geom} and \ref{fig:FB_geom}). Both cavity designs share the same primary structural features but exhibit distinct secondary or fine features. Each cavity is integrated within an in-air-suspended diamond waveguide ($n = 2.4$) with a width of $W = 300$ nm and a thickness of $t = 200$ nm. To form the optical cavity, the waveguide is perturbed by airholes arranged in a 1D lattice of $30$ unit cells. Both cavity geometries feature a mirror region (nine unit cells) and a taper region (six unit cells), symmetrically positioned on either side of the cavity. In the mirror region, the lattice constant is fixed at $a = 195$ nm. In the taper region, the spacing between the airholes, along with one other geometry are varied linearly. The spacing between the airholes is for both designs tapered from $\Delta x = 195 - 160$ nm with steps of $5$ nm.

The L2 nanobeam cavity is characterized by the absence of two airholes at its center, which is why it is referred to as the L2 nanobeam cavity (see Figure \ref{fig:NC_schematic_geom}). The mirror region includes nine elliptical airholes, each with a minor diameter of $d = 78$ nm and a major diameter of $D = 255$ nm. The taper region consists of the innermost five airholes on each side of the cavity (or six if considering the absent airhole). The major diameter and the spacing between these airholes taper linearly from $D = 255 - 78$ nm and from $\Delta x = 195 - 160$ nm, respectively (accounting for the missing airhole). This results in a cavity length of $L_c = 412$ nm.

The fishbone nanobeam cavity is characterized by its distinctive corrugated structure, which is why it is referred to as the fishbone nanobeam cavity (see Figure \ref{fig:FB_schematic_geom}). Although the airholes are shaped more like grooves or channels, they will be referred to as airholes for convenience. The pairs of airholes, facing each other transversely, are separated by a spacing of $\underline{w} = 40$ nm. Therefore, the fishbone nanobeam cavity consists of fins with a length of $l = 130$ nm, which are circularly rounded by arcs with a transverse diameter of $D_f = 80$ nm. The mirror region includes $18$ airholes (nine pairs), each with a longitudinal diameter of $d = 105.3$ nm and transverse diameter of $D_h = 100$ nm. The taper region consists of the innermost $12$ airholes on each side of the cavity (six pairs). The longitudinal diameter and the spacing between these airholes taper linearly from $d = 105.3 - 81.3$ nm and from $\Delta x = 195 - 160$ nm, respectively. This results in a cavity length of $L_c = 78.7$ nm. \newline

\begin{figure}[h!]
    \begin{subfigure}{0.515\textwidth}
        \centering
        \includegraphics[width=\linewidth]{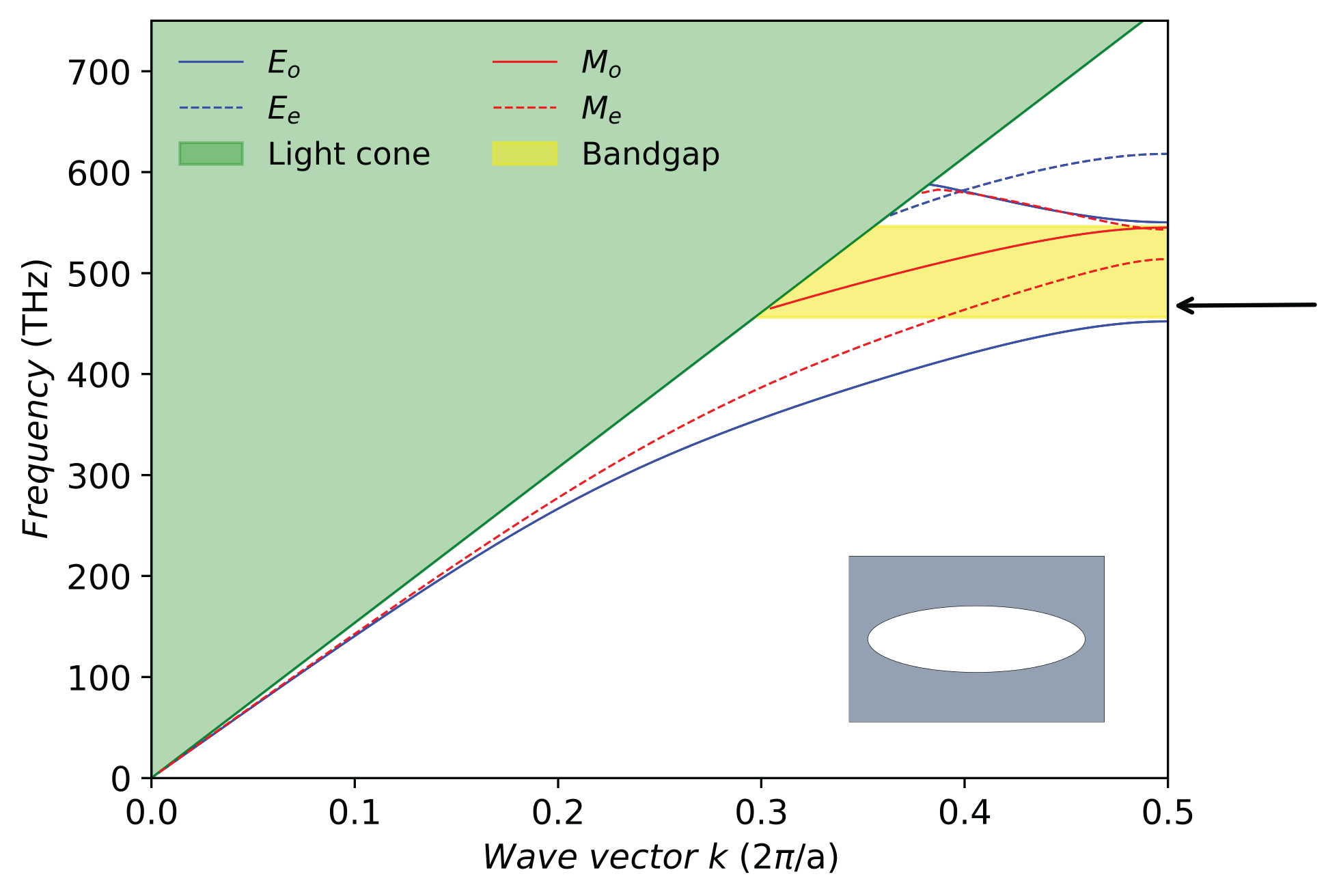}
        \caption{}
        \label{fig:NC_bandstructure_BandSolve}
    \end{subfigure}
    \hspace{0.005\textwidth}
    \begin{subfigure}{0.465\textwidth}
        \centering
        \includegraphics[width=\linewidth]{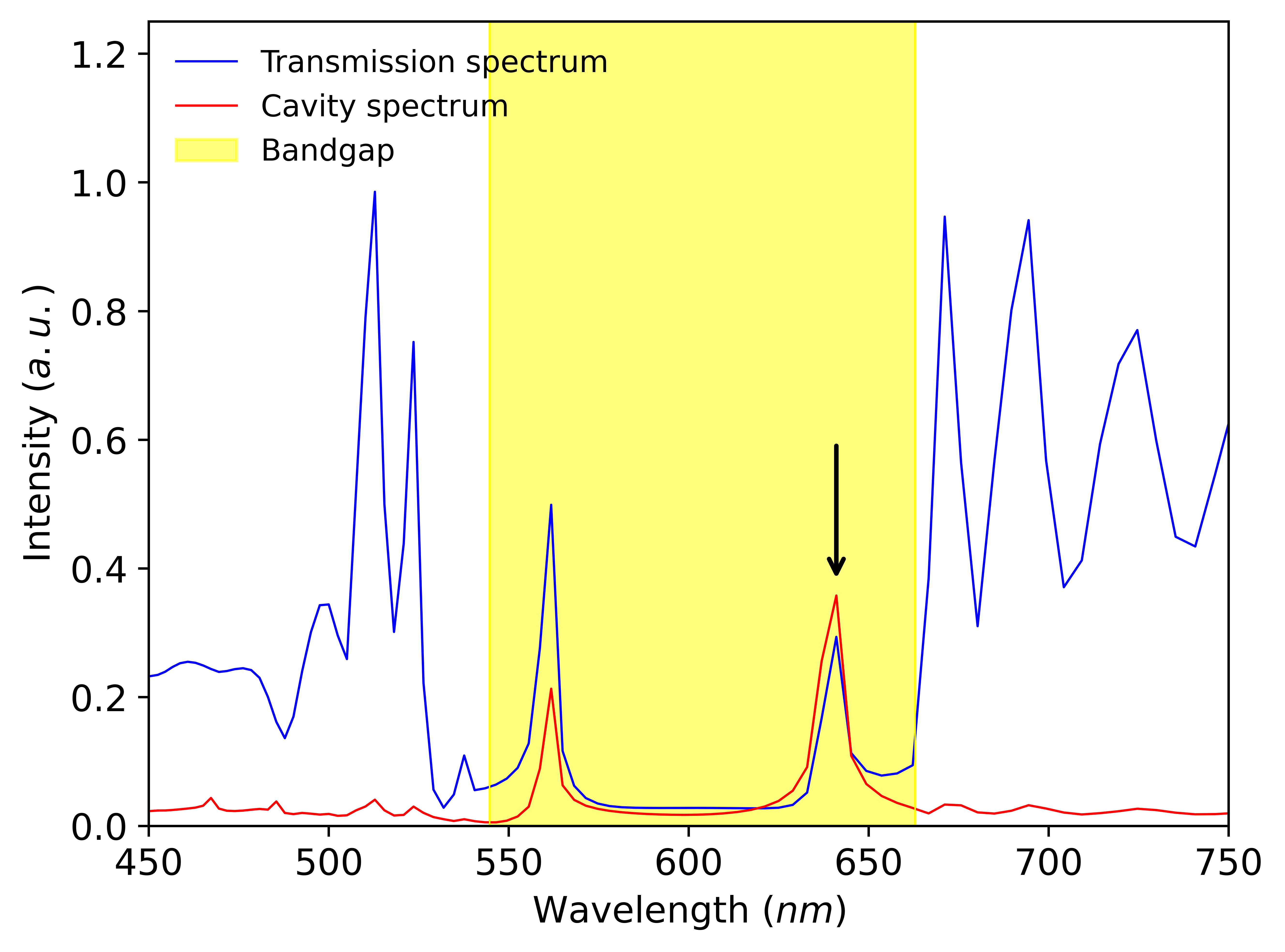}
        \caption{}
        \label{fig:NC_bandstructure_FullWave}
    \end{subfigure}
    
    \vspace{0.001\textwidth} 

    \begin{subfigure}{\linewidth}
        \centering
        \includegraphics[width=\linewidth]{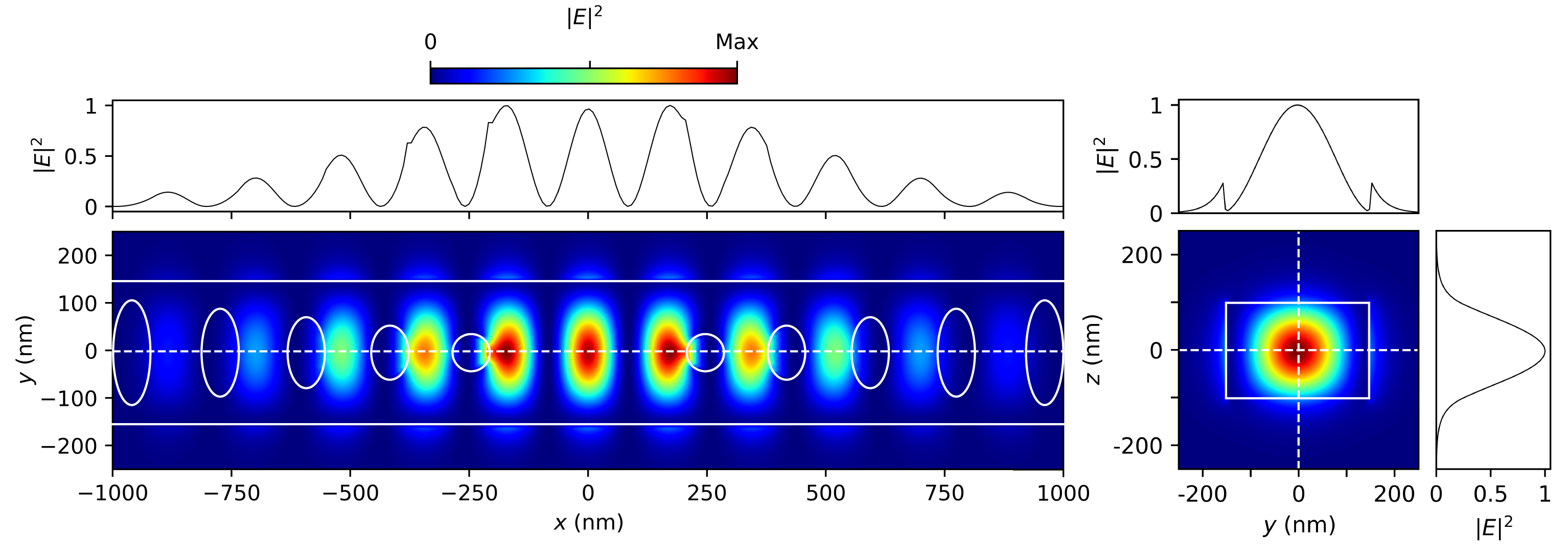}
        \caption{}
        \label{fig:NC_mode_profile}
    \end{subfigure}
    
    \caption{Optical properties of the L2 nanobeam cavity. \textbf{(a)} A visual representation of the band structure of the first six bands of the elliptical airhole unit cell (Rsoft BandSolve). \textbf{(b)} Transmission spectrum and cavity excitation spectrum plots of the L2 nanobeam cavity (Rsoft FullWave). The resonant peaks shown are not scaled to represent the actual peak intensities of the high Q-factors, ensuring clarity in the visualization. \textbf{(c)} 2D plots illustrating the mode distributions (electric field intensity $|E|^2$) of the L2 nanobeam cavity in the xy- and yz-planes, accompanied by 1D plots showing the mode distributions along the white dotted lines in the x-, y-, and z-directions.}
    \label{fig:NC_bandstructure}
\end{figure}

\begin{figure}[h!]
    \begin{subfigure}{0.515\textwidth}
        \centering
        \includegraphics[width=\linewidth]{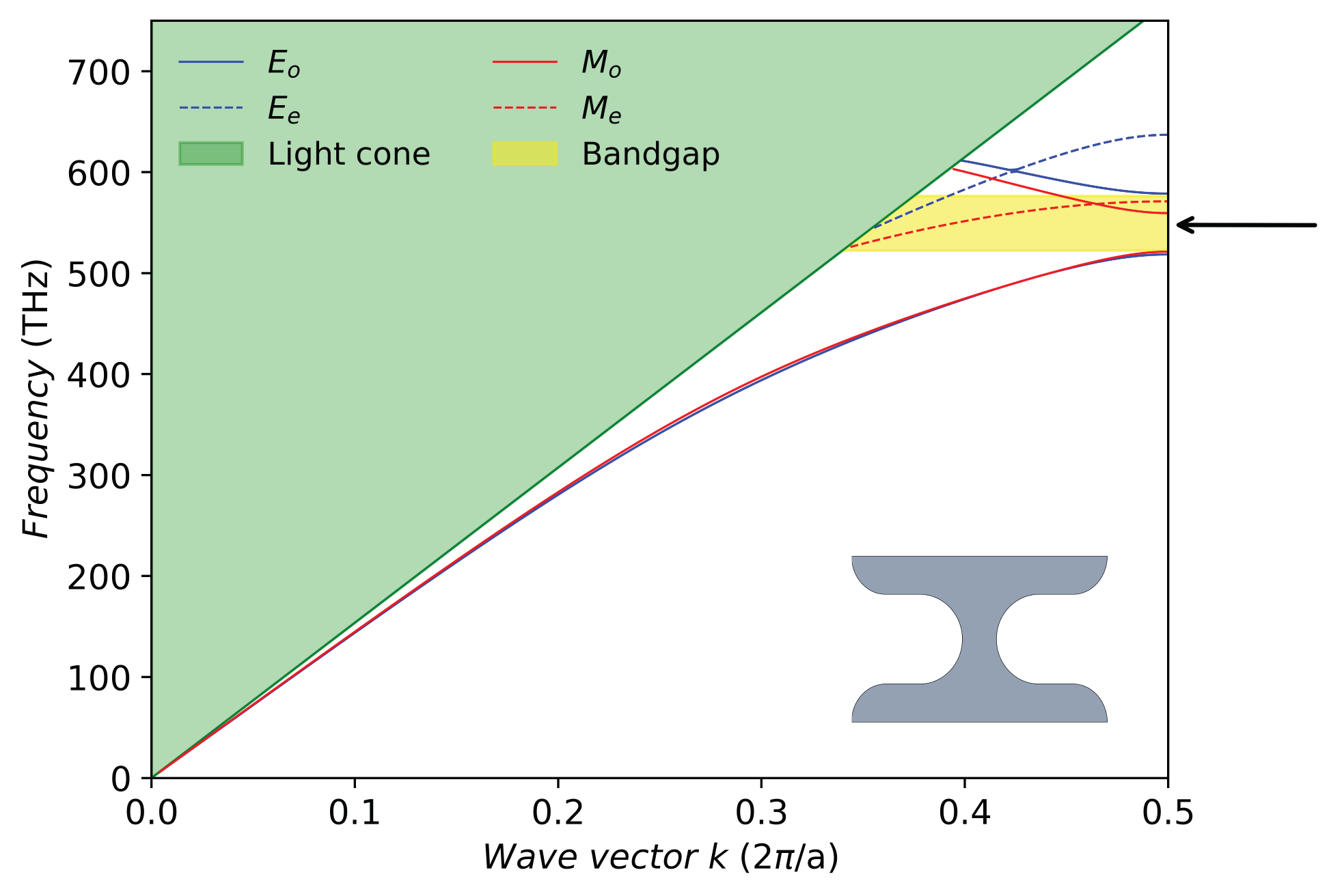}
        \caption{}
        \label{fig:FB_bandstructure_BandSolve}
    \end{subfigure}
    \hspace{0.005\textwidth}
    \begin{subfigure}{0.465\textwidth}
        \centering
        \includegraphics[width=\linewidth]{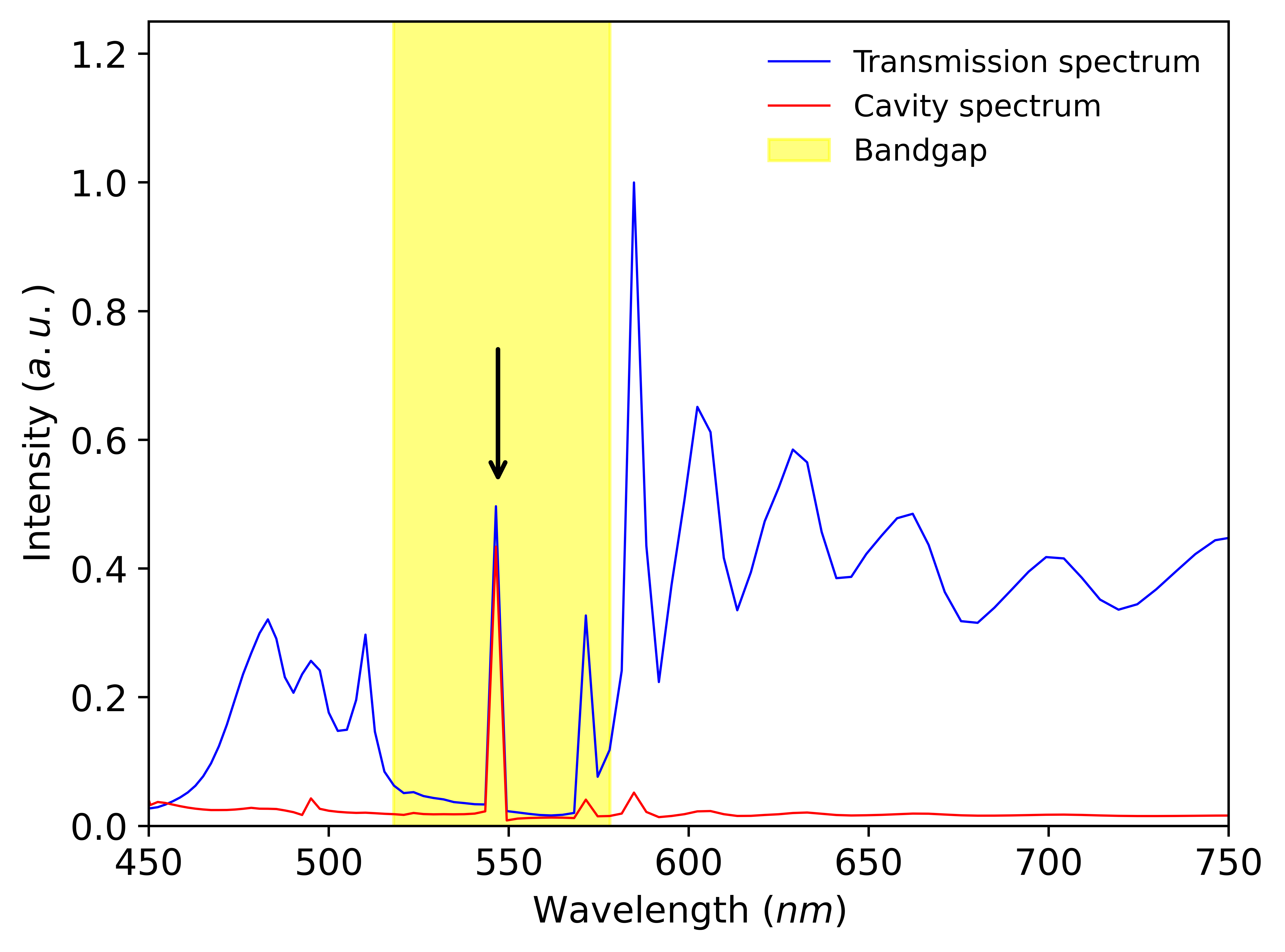}
        \caption{}
        \label{fig:FB_bandstructure_FullWave}
    \end{subfigure}
    
    \vspace{0.001\textwidth} 
    
    \begin{subfigure}{\linewidth}
        \centering
        \includegraphics[width=\linewidth]{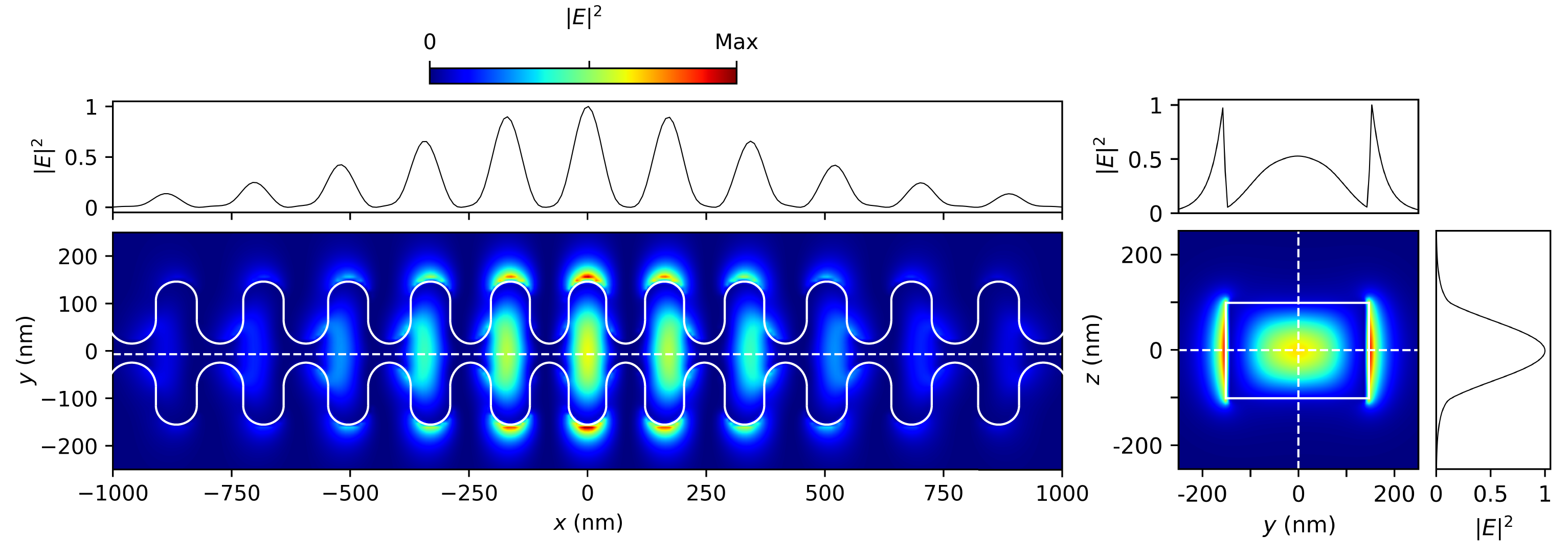}
        \caption{}
        \label{fig:FB_mode_profile}
    \end{subfigure}
    
    \caption{Optical properties of the fishbone nanobeam cavity.  \textbf{(a)} A visual representation of the band structure of the first six bands of the corrugated unit cell (Rsoft BandSolve). \textbf{(b)} Transmission spectrum and cavity excitation spectrum plots of the fishbone nanobeam cavity (Rsoft FullWave). The resonant peaks depicted do not represent the actual peak intensities of the high Q-factors, ensuring clarity in the visualization. \textbf{(c)} 2D plots illustrating the mode distributions (electric field intensity $|E|^2$) of the fishbone nanobeam cavity in the xy- and yz-planes, accompanied by 1D plots showing the mode distributions along the white dotted lines in the x-, y-, and z-directions.}
    \label{fig:FB_bandstructure}
\end{figure}

The two distinct unit cell designs (see Figure \ref{fig:NC_unitcell_3D} and \ref{fig:FB_unitcell_3D}) and tapers result in different optical properties, which can be quantitatively characterized by the resonant wavelength, Q-factor, mode volume, mode distribution, and the full width at half maximum (FWHM) of the mode distribution along all three spatial directions. Graphically, these properties can be represented by the mode distribution of the electric field in the xy- and yz-planes, the optical band gaps of the two unit cell designs, and the transmission spectra combined with the cavity excitation spectra of the two nanobeam cavities (see Figure \ref{fig:NC_bandstructure} and \ref{fig:FB_bandstructure}). For a detailed description of the numerical framework used to compute these optical properties, refer to Supplementary Information \hyperlink{supinfC}{C}.

The base L2 nanobeam cavity design has a resonant wavelength of $\lambda = 641$ nm, with a Q-factor of $Q_{\textrm{FDTD}} = 2.4 \times 10^4$ and a mode volume of $V = 0.7$ ($\lambda$/n)$^3$ (see Supplementary Information \hyperlink{supinfD3}{D3} for optical properties of the base models under fabrication imperfections). The mode distribution of this resonant mode is shown in Figure \ref{fig:NC_mode_profile}. The electric field distribution displays a fundamental mode profile with three peaks within the cavity, characteristic of this L2 cavity design. The FWHM$_{x,y,z}$ in all three directions are $84$ nm, $161$ nm, and $147$ nm, respectively. This results in a mode that is approximately twice as spread out in the x-direction compared to conventional state-of-the-art nanobeam cavity designs, which typically have an FWHM$_x$ around $40$ nm \cite{ding2024, rugar2021, lee2014, kuruma2021, mouradian2017, pregnolato2024, regan2021}. This broader mode profile indicates that the spatial accuracy required for ion implantation in this cavity design is significantly lower.
\par
The unit cell of the L2 nanobeam cavity (see Figure \ref{fig:NC_unitcell_3D}) gives rise to the band structure diagram shown in Figure \ref{fig:NC_bandstructure_BandSolve}. The band structure reveals a band gap for the $E_o$ modes between $452 - 550$ THz ($545 - 663$ nm), which are also referred to as the most TE-like modes \cite{molding_the_flow_of_light}.
\par
The transmission and cavity excitation spectra both show a resonant peak at $641$ nm, marked by a black arrow in Figure \ref{fig:NC_bandstructure_BandSolve} and \ref{fig:NC_bandstructure_FullWave}. Another resonant peak with a lower wavelength can be observed, which is typical for such airhole nanobeam cavities \cite{1D_PhC_cavitiy_PhD}. Finally, the band gap identified from the band structure diagram is again highlighted in yellow and approximately aligns with the band gap shown by the transmission spectrum shown in Figure \ref{fig:NC_bandstructure_FullWave}.

The base fishbone nanobeam cavity design has a resonant wavelength of $\lambda = 547$ nm, with a Q-factor of $Q_{\textrm{FDTD}} = 2.6 \times 10^4$ and a mode volume of $V = 1.4$ ($\lambda$/n)$^3$ (see Supplementary Information \hyperlink{supinfD3}{D3} for optical properties of the base models under fabrication imperfections). The mode distribution of this resonant mode is shown in Figure \ref{fig:FB_mode_profile}. The electric field distribution displays a fundamental mode profile with one peak within the cavity, characteristic of this fishbone cavity design. The FWHM$_{x,y,z}$ in all three directions are $68$ nm, $194$ nm, and $130$ nm, respectively. This indicates that the spatial accuracy required for ion implantation in these cavity designs is still lower compared to conventional state-of-the-art nanobeam cavity designs \cite{ding2024, rugar2021, lee2014, kuruma2021, mouradian2017, pregnolato2024, regan2021}.
\par
The unit cell of the fishbone nanobeam cavity (see Figure \ref{fig:FB_unitcell_3D}) gives rise to the band structure diagram shown in Figure \ref{fig:FB_bandstructure_BandSolve}. The band structure again reveals a band gap for the $E_o$ modes, this time between $579 - 519$ THz ($518 - 578$ nm) \cite{molding_the_flow_of_light}.
\par
The transmission and cavity excitation spectra both show a resonant peak at $547$ nm, marked by a black arrow in Figure \ref{fig:FB_bandstructure_BandSolve} and \ref{fig:FB_bandstructure_FullWave}. This is the only dominant resonant peak within the band gap. Noticeably, the resonant peak is situated more in the middle of the band gap, than with the L2 nanobeam cavity. Finally, the band gap identified from the band structure diagram is again highlighted in yellow and closely aligns with the band gap shown by the transmission spectrum shown in Figure \ref{fig:FB_bandstructure_FullWave}. 
\newline

For both the L2 and fishbone nanobeam cavities, three optimized designs are found, designated as cavity 1, cavity 2, and cavity 3 (see Supplementary Information \hyperlink{supinfG3}{G3}~-~\hyperlink{supinfG5}{G5}). Cavity 1 is optimized under ideal conditions without incorporating any fabrication imperfections, cavity 2 is optimized to account for surface roughness and cavity 3 is designed to reduce the effects of sidewall slant. The optical characteristics of each cavity are then evaluated and compared under ideal, rough surface, and slanted sidewall conditions. These characteristics are summarized in Table \ref{tab:results_L2} and \ref{tab:results_FB} for the L2 and fishbone nanobeam cavities, respectively.
\par
\begin{figure}[h!]
    \centering
    \includegraphics[width=0.75\textwidth]{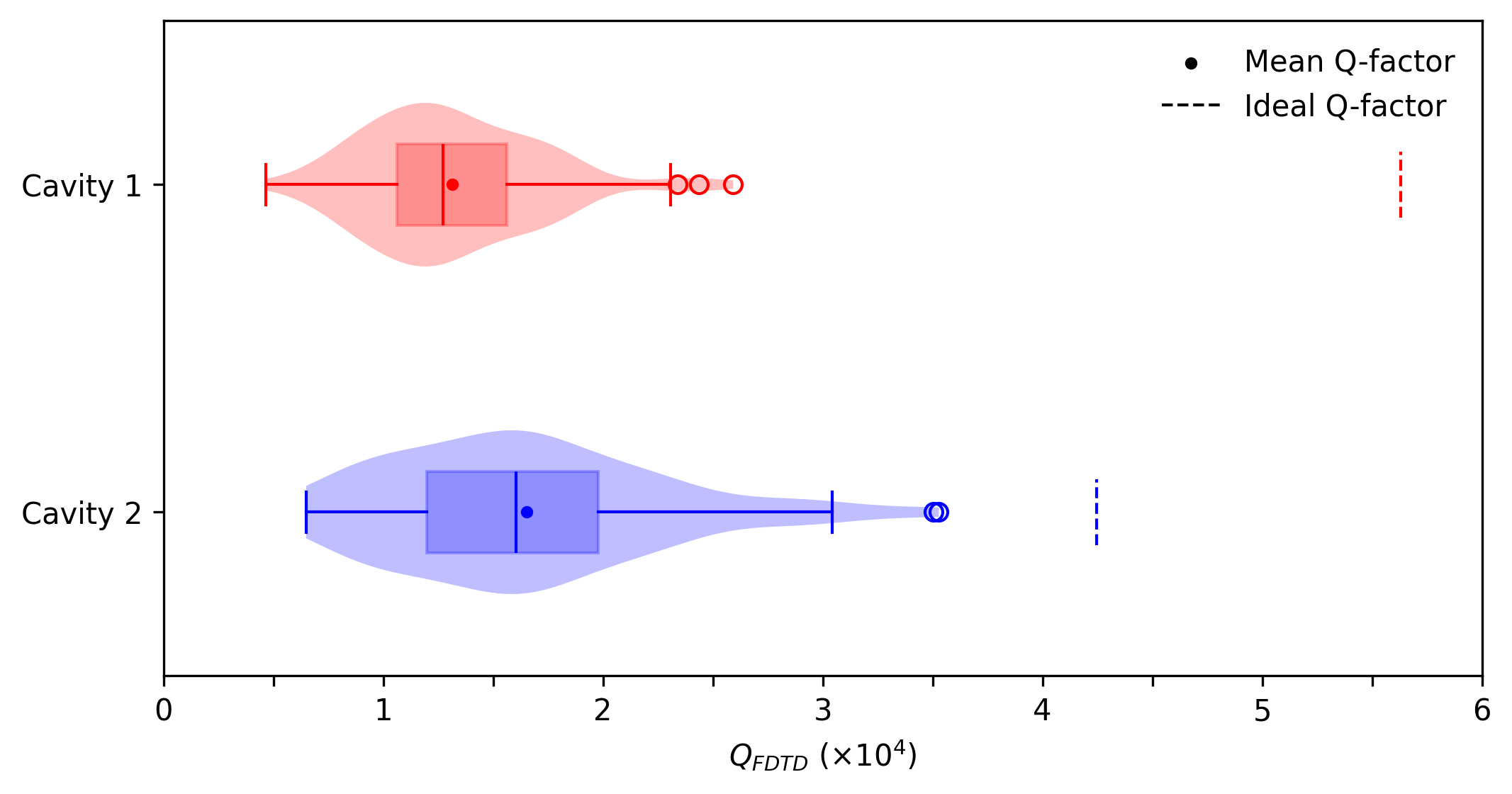}
    \caption{The distribution of Q-factors of fishbone cavity 1 and cavity 2 under different roughness configurations plotted as a box plot in combination with a violin plot. The Q-factor of both structures without surface roughness are denoted by a dotted line.}
    \label{fig:FB_boxplot_roughness}
\end{figure}
\begin{figure}[h!]
    \centering
    \includegraphics[width=0.75\textwidth]{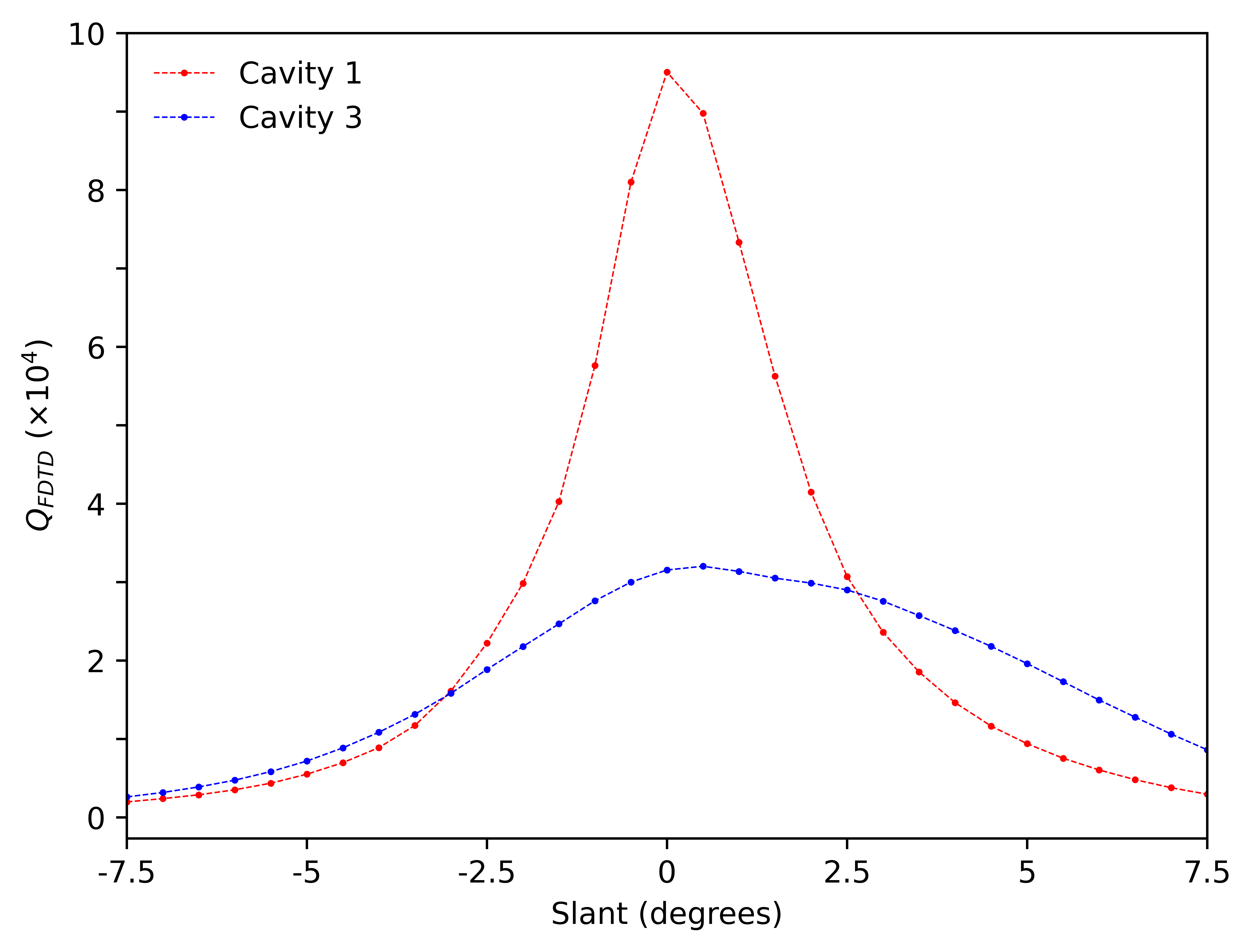}
    \caption{A graph presenting the Q-factor of L2 cavity 1 and cavity 3 under various slant angles.}
    \label{fig:NC_bestclean_vs_bestslant}
\end{figure}

Cavity 2 and cavity 3 are compared separately to cavity 1 to highlight the impact of optimizing against fabrication imperfections. Cavity 1 and cavity 2 are compared for their performance under surface roughness conditions. The results for $100$ simulated structures with varying roughness seeds are depicted using a combination of box and violin plots (see Figures \ref{fig:FB_boxplot_roughness} and \ref{fig:NC_boxplot_roughness}). Cavity 1 and cavity 3 are compared based on their performance under different angles of sidewall slant (see Figures \ref{fig:NC_bestclean_vs_bestslant} and \ref{fig:FB_bestclean_vs_bestslant}). Structures are deemed more robust to fabrication imperfections if their Q-factor remains higher despite these imperfections and exhibits relatively less degradation from the ideal Q-factor. Two of the four cases are described here, with the remaining cases detailed in Supplementary Information \hyperlink{supinfG4}{G4} and \hyperlink{supinfG5}{G5}.
\par
The first optimized cavity evaluated is fishbone cavity 2, which is characterized by a significant reduction in the distance between opposing holes, $\underline{w}$, compared to its base model (see Table \ref{tab:FB_cavities_optim}). The average Q-factor of this structure under surface roughness, predicted by the NN, is $Q_{\textrm{NN}} = 1.090 \times 10^4$, while the FDTD simulation yields $Q_{\textrm{FDTD}} = 1.652 \times 10^4$. This results in a relative prediction error of approximately $\epsilon_{\textrm{pred}} = 34.02\%$ (see Table \ref{tab:CMA-ES_or_GA}). This optimized design was obtained using a NN with L$^2$-regularization on its weights and biases ($\lambda = 0.001$) and optimized with the GA algorithm, initialized at the base design with L$^2$-regularization on the design parameters ($\lambda_{\textrm{GA}} = 0.15$).
\par
In Figure \ref{fig:FB_boxplot_roughness}, the distribution of Q-factors for fishbone cavities 1 and 2 under different surface roughness configurations (random seeds) is visualized. Under ideal conditions (denoted by the dotted line), cavity 2 has a lower Q-factor ($Q_{\textrm{ideal}} = 4.247 \times 10^4$) compared to cavity 1 ($Q_{\textrm{ideal}} = 5.629 \times 10^4$). However, when surface roughness is applied, cavity 2 ($Q_{\textrm{rough}} = 1.652 \times 10^4$) outperforms cavity 1 ($Q_{\textrm{rough}} = 1.314 \times 10^4$) by $25.72\%$ on average. As a result, under surface roughness, cavity 2 demonstrates improved performance and experiences a $15.55\%$ smaller average degradation compared to cavity 1, making it more robust against the fabrication imperfection of surface roughness.

The second optimized cavity evaluated is L2 cavity 3, which is characterized by a shift of all holes towards the cavity (see Table \ref{tab:NC_cavities_optim}). The average Q-factor of this structure under sidewall slant, predicted by the NN, is $Q_{\textrm{NN}} = 1.607 \times 10^4$, while the FDTD simulation yields $Q_{\textrm{FDTD}} = 1.962 \times 10^4$. This results in a relative prediction error of approximately $\epsilon_{\textrm{pred}} = 18.09\%$ (see Table \ref{tab:CMA-ES_or_GA}). This optimized design was obtained using a NN without L$^2$-regularization on its weights and biases ($\lambda = 0$) and optimized with the CMA-ES algorithm, initialized at the best structure of the training data with L$^2$-regularization on the design parameters ($\lambda_{\textrm{ES}} = 0.15$).
\par
In Figure \ref{fig:NC_bestclean_vs_bestslant}, the Q-factors of L2 cavities 1 and 3 are compared under different angles of sidewall slant. Despite cavity 3 initially having a lower Q-factor ($Q_{\textrm{ideal}} = 3.208 \times 10^4$) compared to cavity 1 ($Q_{\textrm{ideal}} = 1.076 \times 10^5$) under $0^{\circ}$ slant, it begins to outperform cavity 1 at slant angles exceeding $2.5^{\circ}$ and performs comparably for slant angles below $-2.5^{\circ}$. Cavity 3 was specifically optimized against a $5^{\circ}$ slant. At this angle, the Q-factor of cavity 1 degrades to $Q_{\textrm{slant}} = 9403$, while cavity 3 maintains a higher Q-factor of $Q_{\textrm{slant}} = 1.962 \times 10^4$. As a result, under sidewall slant, cavity 3 achieves more than twice the performance of cavity 1 and experiences $52.42\%$ less degradation, making it more robust against the fabrication imperfection of sidewall slant. \newline

The NN-driven optimization consistently produced structures with significantly higher Q-factors than the original base models, achieving up to a 5$\times$ improvement over the initial Q-factor. Furthermore, incorporating fabrication imperfections into the datasets proved effective, resulting in more robust device designs.
\par
However, some patterns and complications emerged. For instance, the fishbone nanobeam cavity, despite its initially higher Q-factor, showed less improvement under ideal fabrication conditions compared to the L2 nanobeam cavity. This difference may stem from the NNs performing less effectively on the fishbone cavity due to differences in the parameters and the size of the optimization space (see Table \hyperlink{supinfTabS1}{S1} and \ref{tab:FB_parameter_space}). Additionally, it is plausible that the fishbone cavity's optimization space offers lower maxima, indicating inherently less potential for improvement compared to the L2 cavity.
\par
The design parameters of the optimized structures largely remained within the initial dataset's range (see Table \ref{tab:NC_cavities_optim} and \ref{tab:FB_cavities_optim}). This limitation arises from the NNs' degrading performance outside their training parameter space, leaving a large portion of the entire parameter space unexplored.
\par
In addition, the NNs struggled to predict high Q-factors, with the lowest prediction error for the optimized designs being $\epsilon_{\textrm{pred}} = 18.09\%$ (see Table \ref{tab:CMA-ES_or_GA}). Notably, NNs with L$^2$-regularization faced the greatest difficulty in accurately predicting high Q-factors, consistently underestimating them (see Figure \ref{fig:combined_correlation_plots}). This underprediction is attributed to the lack of high-Q training data. Despite this limitation, the NNs were successful in identifying peaks in the optimization landscape, accurately evaluating the gradient direction of Q-factors with respect to the design parameters.
\par
The iterative optimization approach \cite{Takashi_CNN_iter, Baba_ML_300, Baba_CMA-ES, Renji_RL} addresses the challenge of limited high-Q data and inefficient parameter space exploration. By training NNs on data beyond the initial parameter space, this method provides high-Q training data after the first iteration, enabling the exploration of higher Q-factors and more fabrication error-tolerant devices.

\newpage

%% file: conslusions.tex
\section{Conclusions}

Diamond’s exceptional material properties cause unavoidable fabrication
imperfections that degrade photonic device performances. This research focuses on counteracting the affects of such structural imperfections through design optimizations and advances scalable quantum photonic hardware by introducing an effective design methodology for fabrication-tolerant PhC nanobeam cavities, enabling on-chip diamond quantum systems.
\par
The considered fabrication imperfections include surface roughness, sidewall slant, and non-optimal emitter positioning. Two nanobeam cavity designs were optimized: one with elliptical holes and one with a corrugated structure. These designs were selected for their potential to enhance fabrication tolerance by providing large cavity regions that improve spatial-spectral coupling between the emitter and resonant mode and reduce noise from nearby surface charge variations, thereby maintaining high performance even under non-ideal conditions. \newline

The optimization of the two nanobeam cavity designs utilized DL, replacing FDTD simulations with CNNs trained on both ideal and realistic fabrication conditions. By incorporating imperfections into the training data, the NNs learned to produce designs that perform well even under real-world conditions. These models predict Q-factors for nanobeam cavities based on the two base designs with prediction errors as low as 3.99\% and correlation coefficients up to 0.988, while operating a million times faster than FDTD simulations. This substantial speed-up allows for efficient exploration and optimization across large parameter spaces, enabling the discovery of robust, high-performing designs that would otherwise be computationally unattainable.
\par
The two nanobeam cavity designs were optimized under three conditions: ideal, surface roughness, and sidewall slant. This process resulted in six distinct optimized designs. When tested against fabrication imperfections, nearly all four real-world-optimized designs demonstrated better performance and significantly less degradation compared to the two ideal-world-optimized designs. These results confirm that optimizing for fabrication imperfections using this DL-based approach produces more robust devices. Furthermore, this study demonstrates that fabrication challenges can be addressed through design optimization rather than relying solely on fabrication process optimization and most importantly that optimizing for extremely high Q-factors may not always yield the best overall device performance, especially when dealing with such fabrication imperfections. \newline

While effective, the current DL approach faces limitations in handling extremely high Q-factors. Future work could address these challenges by iteratively expanding the dataset with high-Q candidates validated through FDTD simulations and by improving NN architectures for better performance, such as through enhanced regularization, hyperparameter optimization, or exploring alternative architectures like transformers or RL-based models.
\par
However, the most critical next step is realizing these optimized structures through nanofabrication. Testing the performance of both the ideal-world-optimized and real-world-optimized designs will provide valuable insights into the actual improvements in device robustness achieved by this approach. Ultimately, integrating this optimization strategy with active fabrication processes could further enhance device reliability by directly addressing the fabrication imperfections dealt with during manufacturing. \newline

In conclusion, this research provides an important stepping stone for the scalable integration of color centers in diamond for quantum information processing and highlights the potential of design-based optimization to overcome fabrication challenges in nanophotonic devices. It shows that robust, high-performance devices can be achieved even under less-than-ideal manufacturing conditions. Looking ahead, this design optimization strategy can be expanded to a wider range of quantum systems, paving the way for more resilient and scalable quantum technologies that can withstand the inherent imperfections of real-world fabrication processes.

%% file: acknowledgements.tex
\begin{acknowledgement}
The authors thank Ir. M. van Beusekom (Q\&CE) for essential IT support and Dr. E. Akiki (RSoft) for assistance with the simulation software. They also appreciate insightful discussions with Asst. Prof. E. Greplová.
\end{acknowledgement}

%% file: A.tex
\newpage
\hypertarget{supinfA}{}
\section{\fontsize{24}{24}\selectfont Supplementary Information}

\section{A. Enhancing Optical Properties: Color Centers in Diamond}
The color centers emit light across a broad spectrum, consisting of two components: one from the coherent zero-phonon line (ZPL) and the other from the phonon sideband (PSB). Only the coherent photons from the ZPL can be entangled with the spin of the qubit. Therefore, to improve spin-spin entanglement efficiency, it is crucial to enhance spontaneous emission into the ZPL. This can be achieved through Purcell enhancement, which can be realized using resonant optical trapping \cite{QCED2, Review_groupIV}. 

\subsection{A1. Resonant optical trapping}
Resonant optical trapping refers to the confinement of specific wavelengths of light within an optical cavity. Optical resonators come in various shapes and forms, such as ring resonators, racetrack resonators, Fabry-Perot resonators, and photonic crystal (PhC) micro- and nano-resonators \cite{Optical_microcavities_review, QCED2}. These optical cavities have a broad range of applications, including laser optics, telecommunications, interferometry, bio-sensing, optical switches, and quantum photonics, emphasizing their versatility and importance across different fields \cite{Optical_cavity_sensing, Cavity_sensing2, Optical_cavity_switch, Fabry_perot_applications}.
\par
These cavities can be characterized by their resonant modes, which are the electromagnetic field distributions that determine the wavelength, Q-factor, and mode volume at which the cavity efficiently traps and stores light. They are not true modes of the system, as light cannot be trapped indefinitely within the cavity due to imperfect confinement or unavoidable scattering losses. Therefore, these resonant modes are often referred to as the leaky modes of the cavity. \newline

\newpage
The energy inside a cavity dissipates exponentially over time. The Q-factor of a resonant mode is a dimensionless quantity that measures how effectively the cavity can confine light at that specific mode, quantifying temporal confinement. It characterizes the rate at which energy is lost in the cavity and can be defined and calculated in various ways \cite{the_story_of_Q}.
\par
The most straightforward way would be to divide the total stored energy by the energy lost per cycle, as shown in Equation \ref{Eq:Q_factor_energy}. In this equation, $f_r$ represents the resonant frequency. An interpretation of this definition in the context of optical cavities is that the Q-factor represents the number of times a photon can traverse the cavity path before being lost from the system due to scattering or absorption

\begin{equation}
    Q \overset{\textrm{def}}{=} 2\pi\times\frac{\textrm{energy stored}}{\textrm{energy dissipated per cycle}} = 2\pi f_r \times \frac{\textrm{energy stored}}{\textrm{power loss}}
    \label{Eq:Q_factor_energy}
\end{equation}

Another nearly equivalent definition of the Q-factor is related to the frequency-to-bandwidth ratio. In this context, the Q-factor represents the ratio of the resonant frequency of the cavity to the bandwidth of the resonator. A higher Q-factor indicates a narrower resonance linewidth. This definition is expressed in Equation \ref{Eq:Q_factor_bw}, where $\Delta f$ and $\Delta\omega$ represent the full width at half maximum (FWHM) of the resonant frequency power.

\begin{equation}
    Q \overset{\textrm{def}}{=} \frac{f_r}{\Delta f} = \frac{\omega_r}{\Delta\omega}
    \label{Eq:Q_factor_bw}
\end{equation}

Another important figure of merit for the optical cavity is its mode volume, $V_{\textrm{mode}}$, expressed in units of $(\lambda_{\textrm{free}}/n)^3$. The mode volume refers to the region within the optical cavity where the electromagnetic field associated with the resonant mode is predominantly confined, quantifying spatial confinement. Over time, the definition of mode volume has evolved, but the most widely accepted form relates to the energy of the electric field, $U_E(\mathbf{r})$, at its maximum, distributed over a spatial volume $V$. This relationship is described in Equations \ref{Eq:electric_field_energy} and \ref{Eq:mode_volume} \cite{the_story_of_V}. 

\begin{equation}
    U_E(\mathbf{r}) = \frac{1}{2}\epsilon(\mathbf{r})|\mathbf{E}(\mathbf{r})|^2
    \label{Eq:electric_field_energy}
\end{equation}

\begin{equation}
    V_{\textrm{mode} }=  \int_{V}^{} \frac{U_E(\mathbf{r})}{\textrm{max}[U_E(\mathbf{r})]} \,dV \
    \label{Eq:mode_volume}
\end{equation}

\subsection{A2. Purcell enhancement}
The color centers in diamond are quantum emitters that can be modeled as a two-level system. A photon is emitted when the system transitions from the excited state to the lower-energy ground state. This process occurs without external stimulation and is known as spontaneous emission. Fermi's golden rule, a fundamental principle in quantum mechanics, describes the transition rate between an initial eigenstate $|i\rangle$ and a final eigenstate $|f\rangle$ of the quantum system, subjected to a perturbation Hamiltonian $\hat{H}'$ \cite{Advanced_QM_book}. Fermi's golden rule is expressed by Equation \ref{Eq:Fermi}, where $\rho(E_f)$ denotes the density of states at the energy of the final state.

\begin{equation}
    \Gamma_{i\rightarrow f} = \frac{2\pi}{\hbar}|\langle f|\hat{H}'|i \rangle|^2\rho(E_f)
    \label{Eq:Fermi}
\end{equation}

Purcell enhancement refers to the increase in the spontaneous emission rate of a quantum emitter when placed inside an optical cavity at its resonant wavelengths. In an optical cavity, only specific discrete wavelengths of light resonate, forming cavity modes. These modes have a much higher density of states compared to free space at those resonant wavelengths. When a quantum emitter interacts with these cavity modes, the probability of spontaneous emission into these modes increases significantly. This is because the higher density of states at the resonant wavelengths enhances the likelihood of the emitter coupling to these modes, resulting in a higher photon emission rate at those wavelengths \cite{Purcell_enhancement, Advanced_QM_book}.
\par
\newpage
In the context of optical cavities, the perturbation Hamiltonian $\hat{H}'$ describes the interaction between the quantum emitter and the cavity's electromagnetic modes. According to Fermi's golden rule, the transition rate to a final state increases when its density of states increases, which is exactly what an optical cavity provides. The cavity modes define the electromagnetic field distribution within the cavity, which affects the spatial overlap between the emitter and these modes. The transition matrix element $|\langle f|\hat{H}'|i \rangle|^2$ is maximized when the spatial overlap between the emitter and cavity modes is maximized. Therefore, optimizing the coupling between the emitter and the cavity mode is essential for enhancing spontaneous emission at a specific wavelength.
\par
The Purcell factor, denoted by $F_P$ (as shown in Equation \ref{Eq:purcell_factor}, which represents a simplified version), quantifies the enhancement of spontaneous emission at the vacuum wavelength $\lambda_{\textrm{free}}$. It is defined as the ratio of the spontaneous emission rate of the emitter inside the cavity $\Gamma_{\textrm{cavity}}$ to the emission rate in free space $\Gamma_{\textrm{free}}$. Intuitively, the Purcell factor’s dependence on the Q-factor and mode volume, expressed as $F_P \propto Q/V$, makes sense: higher Q-factors (which indicate sharper resonance) and smaller mode volumes (which imply stronger spatial confinement) both lead to an increase in the density of states, thereby enhancing the emission rate.

\begin{equation}
    F_P = \frac{\Gamma_{\textrm{cavity}}}{\Gamma_{\textrm{free}}} \approx \frac{3}{4\pi^2}\left(\frac{\lambda_{\textrm{free}}}{n}\right)^3\frac{Q}{V}
    \label{Eq:purcell_factor}
\end{equation}

Therefore, an optical cavity can enhance the spontaneous emission of a quantum emitter at its resonant wavelength through Purcell enhancement. In the case of color centers as quantum emitters, optical cavities can be used to boost the spontaneous emission at the ZPL, as long as the cavity is resonant with this wavelength. This amplification increases the coherent emission of photons, thereby facilitating more efficient spin-photon entanglement.
\newpage

%% file: B.tex
\hypertarget{supinfB}{}
\section{B. Structural Design Parameters to be Optimized}
A set of structural design parameters is selected for optimization for each base design. Once the parameters are chosen, a set of nanobeam structures are generated by randomly varying the selected design parameters (step III in the process flowchart of Figure \ref{Flowchart_chapter6}). These fluctuations are made within a specified range, following a uniform distribution. The nanobeam designs are kept symmetric along all three axes. Consequently, modifying one side of the cavity simultaneously affects the opposing side.
\par
The complexity of the regression problem and the required amount of data depend on the number of independent parameters and their ranges. Therefore, it is essential to carefully balance the size of the optimization space and the amount of available data.

\subsection{B1. L2 nanobeam cavity}
\begin{figure}[h!]
    \centering
    \includegraphics[width=1\textwidth]{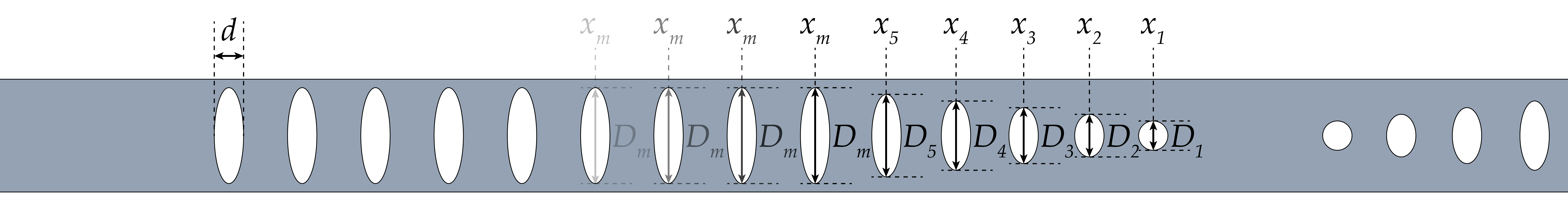}
    \caption{A schematic representation of the L2 nanobeam cavity, highlighting the structural design parameters used for optimization.}
    \label{fig:NC_parameter_space}
\end{figure}

For the L2 nanobeam cavity design, 13 independent design parameters were selected for optimization. These parameters, detailed in Figure \ref{fig:NC_parameter_space}, include the position of the first five airholes within a range of $x_{1-5} \in [-10,\,10]$ nm, the major diameter of these airholes within $D_{1-5} \in [-7.5,\,7.5]$ nm, the collective position of all airholes in the mirror region within $x_m \in [-15,\,15]$ nm, the major diameter of all airholes in the mirror region within $D_m \in [-7.5,\,2.5]$ nm, and the minor diameter of all airholes within $d \in [-7.5,\,7.5]$ nm (see Table \hyperlink{supinfTabS1}{S1}).

\begin{table}[h!]
\centering
\caption{An overview of the parameter space of the nanobeam cavities constituting the large dataset for the L2 design.}
\label{tab:NC_parameter_space}
\begin{tabular}{cccccccc}
\hline
            & $\mathit{x_1}$ (nm) & $\mathit{x_2}$ (nm) & $\mathit{x_3}$ (nm) & $\mathit{x_4}$ (nm) & $\mathit{x_5}$ (nm) & $\mathit{x_m}$ (nm) & $\mathit{d}$ (nm) \\ \hline
\textbf{max} & $10$               & $10$               & $10$               & $10$               & $10$               & $15$               & $7.5$            \\ \hline
\textbf{min} & $-10$              & $-10$              & $-10$              & $-10$              & $-10$              & $-15$              & $-7.5$           \\ \hline
            & $\mathit{D_1}$ (nm) & $\mathit{D_2}$ (nm) & $\mathit{D_3}$ (nm) & $\mathit{D_4}$ (nm) & $\mathit{D_5}$ (nm) & $\mathit{D_m}$ (nm) &                  \\ \hline
\textbf{max} & $7.5$              & $7.5$              & $7.5$              & $7.5$              & $7.5$              & $2.5$              &                  \\ \hline
\textbf{min} & $-7.5$             & $-7.5$             & $-7.5$             & $-7.5$             & $-7.5$             & $-7.5$             &                  \\ \hline
\end{tabular}
\end{table}
\hypertarget{supinfTabS1}{}

\subsection{B2. Fishbone nanobeam cavity}
\begin{figure}[h!]
    \centering
    \includegraphics[width=1\textwidth]{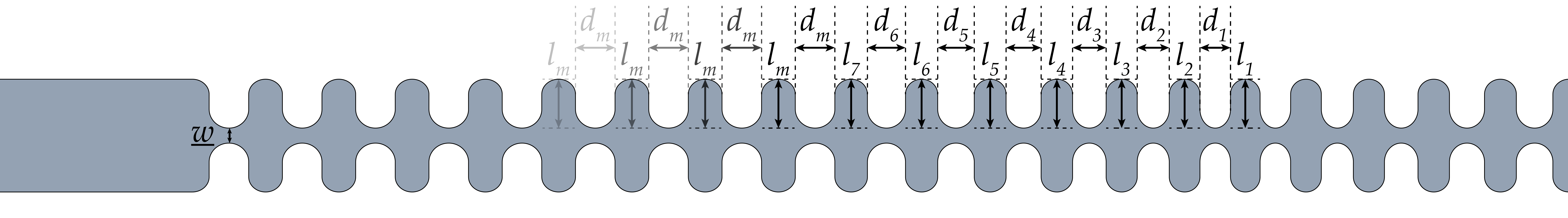}
    \caption{A schematic representation of the fishbone nanobeam cavity, highlighting the structural design parameters used for optimization.}
    \label{fig:FB_parameter_space}
\end{figure}

For the fishbone nanobeam cavity design, 16 independent structural parameters were selected for optimization. These parameters, detailed in Figure \ref{fig:FB_parameter_space}, include the minor diameter of the first six airholes within a range of $d_{1-6} \in [-5,\,5]$ nm, the length of the first seven fins within $l_{1-7} \in [-10,\,10]$ nm, the minor diameter of all airholes in the mirror region within $d_m \in [-7.5,\,7.5]$ nm, the length of all fins in the mirror region within $l_m \in [-10,\,10]$ nm, and the distance between all opposing airholes within $w \in [-5,\,15]$ nm (see Table \ref{tab:FB_parameter_space}).

\begin{table}[h!]
\centering
\caption{An overview of the parameter space of the nanobeam cavities constituting the large dataset for the fishbone design.}
\label{tab:FB_parameter_space}
\begin{tabular}{ccccccccc}
\hline
         & $\mathit{l_1}$ (nm) & $\mathit{l_2}$ (nm) & $\mathit{l_3}$ (nm) & $\mathit{l_4}$ (nm) & $\mathit{l_5}$ (nm) & $\mathit{l_6}$ (nm) & $\mathit{l_7}$ (nm) & $\mathit{l_m}$ (nm) \\ \hline
\textbf{max} & $10$             & $10$             & $10$             & $10$             & $10$             & $10$             & $10$             & $10$             \\ \hline
\textbf{min} & $-10$            & $-10$            & $-10$            & $-10$            & $-10$            & $-10$            & $-10$            & $10$             \\ \hline
         & $\mathit{d_1}$ (nm) & $\mathit{d_2}$ (nm) & $\mathit{d_3}$ (nm) & $\mathit{d_4}$ (nm) & $\mathit{d_5}$ (nm) & $\mathit{d_6}$ (nm) & $\mathit{d_m}$ (nm) & $\mathit{\underline{w}}$ (nm)   \\ \hline
\textbf{max} & $5$              & $5$              & $5$              & $5$              & $5$              & $5$              & $7.5$            & $15$             \\ \hline
\textbf{min} & $-5$             & $-5$             & $-5$             & $-5$             & $-5$             & $-5$             & $-7.5$           & $-5$             \\ \hline
\end{tabular}
\end{table}

\subsection{B3. Input matrix neural network}
\begin{figure}[h!]
    \centering
    \begin{subfigure}{0.45833333\linewidth}
        \centering
        \includegraphics[width=\linewidth]{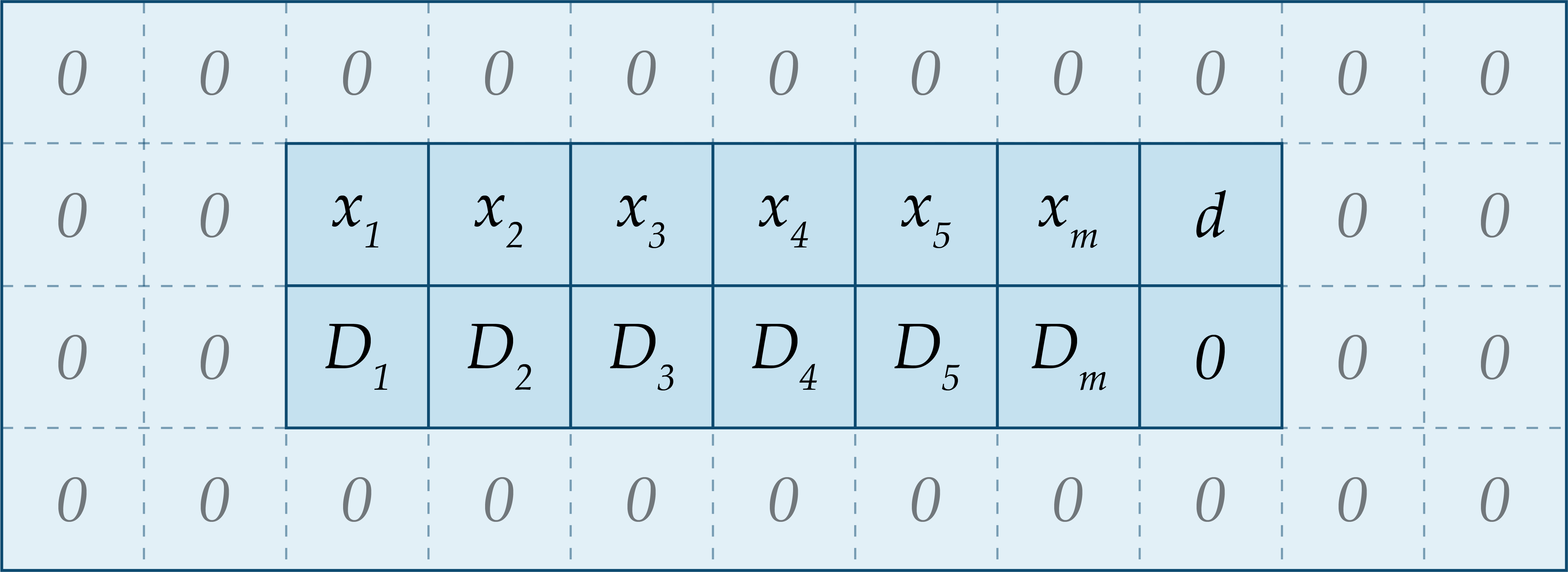}
        \caption{}
        \label{fig:NC_variables_matrix}
    \end{subfigure}
    \hfill
    \begin{subfigure}{0.5\linewidth}
        \centering
        \includegraphics[width=\linewidth]{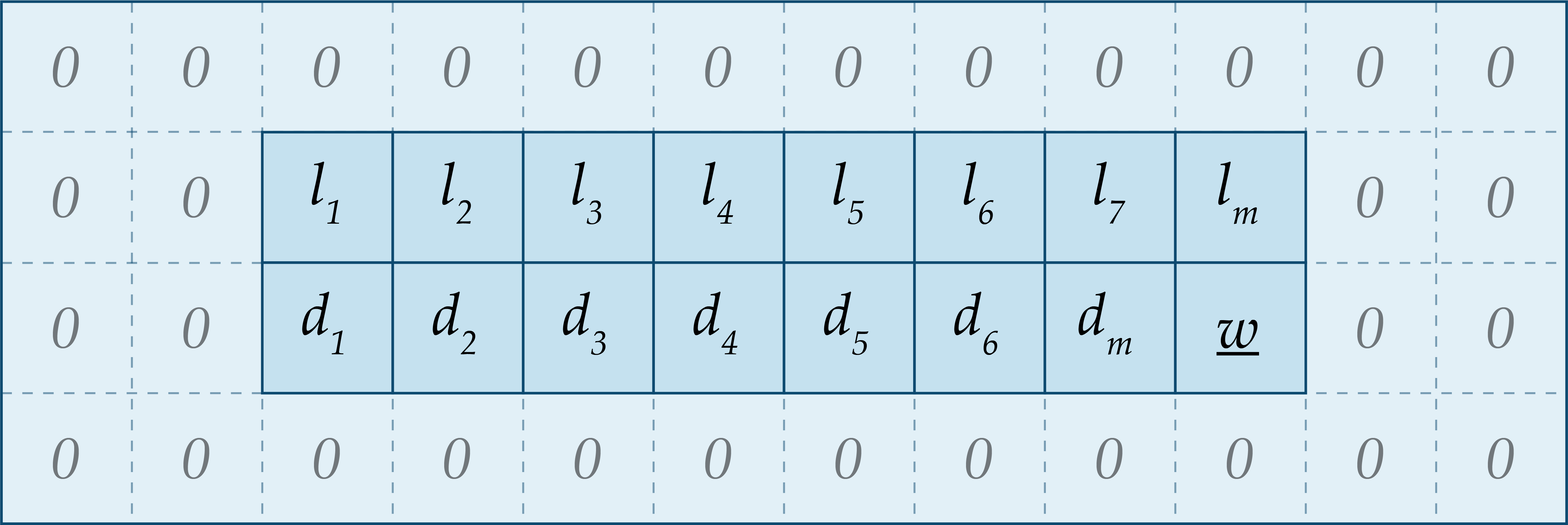}
        \caption{}
        \label{fig:FB_variables_matrix}
    \end{subfigure}
    \caption{A visual representation of the distribution of input parameters within the input matrix, with zero-padding applied. \textbf{(a)} Input matrix showing the parameters of the L2 nanobeam cavity after zero-padding. \textbf{(b)} Input matrix showing the parameters of the fishbone nanobeam cavity after zero-padding.}
    \label{fig:both_variables_matrix}
\end{figure}

The number of independent degrees of freedom for the L2 and fishbone nanobeam cavities to optimize differ, which leads to variations in the shape of the input matrices (see Figure \ref{fig:both_variables_matrix}). The L2 nanobeam cavities have 13 independent variables, resulting in an input matrix of size $2 \times 7$. In contrast, the fishbone nanobeam cavities have 16 independent variables, leading to an input matrix of size $2 \times 8$. This difference in input matrix dimensions results in differently shaped feature maps, which in turn requires a different number of nodes in the first fully connected layer (FC1). Specifically, the L2 nanobeam cavities require a FC1 with 1350 nodes, while the fishbone nanobeam cavities require 1500 nodes in FC1 (for CNN architecture see Figure \ref{fig:Conv_network}).

\newpage

%% file: C.tex
\hypertarget{supinfC}{}
\section{C. Numerical Framework for Cavity Analysis}
All simulations are conducted using Rsoft's FullWAVE and BandSOLVE in 3D. Optical properties such as the resonant wavelength, Q-factor, and mode volume are extracted from FullWAVE simulations, while the band structures shown in Figures \ref{fig:NC_bandstructure_BandSolve} and \ref{fig:FB_bandstructure_BandSolve} are computed using BandSOLVE. This section provides an overview of the numerical framework used in the simulation environment, including key simulation parameters like grid size, domain size, and simulation duration.

\subsection{C1. Rsoft FullWAVE}
The FullWAVE numerical framework for both nanobeam cavity designs is kept identical, as both waveguides share similar dimensions and support resonant wavelengths within approximately the same range. The only difference in simulation criteria arises from the structural differences between the two unit cells.
\par
The grid is set to be uniform, with spatial grid sizes of $dx = a/24$, $dy = a/24$, and $dz = a/8$. The grid size in the z-direction is coarser than in the x- and y-directions, as the nanobeams exhibit less detail in the z-direction. The time step is then automatically determined by the Courant–Friedrichs–Lewy condition, which ensures numerical stability during the simulation \cite{Principles_of_photonic}.
\par
The domain size is chosen to ensure that the resonant wavelength in free space extends twice the distance beyond the periodic structure in all directions. Symmetric boundary conditions (SBCs) are applied along all geometric axes of symmetry, with exceptions for cases such as slanted sidewalls or transmission measurements. This setup—featuring odd SBCs in the x-direction, even SBCs in the y-direction, and even SBCs in the z-direction—ensures that the simulations support modes symmetric across all dimensions (i.e. the fundamental mode). The Q-finder is configured to use fast harmonic analysis as its spectral analysis method, with a sequence of two calculations of $2^{12}$ and $2^{14}$ time steps, respectively.
\par
The FullWAVE simulations are executed on the Q\&CE cluster server \cite{QCEInfrastructure} provide by Delft University of Technology and distributed across eight CPU threads. These threads are spread over eight CPU cores, each operating at a clock speed of 2.8 GHz. The workload is parallelized to fully utilize the available cores, with computational efficiency depending on the system's parallelization capabilities. Consequently, the approximate simulation time to determine the resonant wavelength, Q-factor, and mode volume of a cavity design, with SBCs applied in all three directions, is about 5 minutes.

\subsection{C2. Rsoft BandSOLVE}
The BandSOLVE numerical framework for both unit cell designs is chosen to be identical because both unit cells have the same dimensions. BandSOLVE applies periodic boundary conditions (PBCs) along all axes. Therefore, careful attention is given to defining the domain size of the unit cell. The domain size is selected to cover $X = a$, $Y = 3W$, and $Z = 3t$. Coupling between distant nanobeam cavities is minimized by selecting a sufficiently large domain size in the y- and z-directions. This ensures that BandSOLVE can apply PBCs in all three directions and accurately simulate the band structure of the 1D nanocavities.
\par
Three bands are simulated for both parities along the $z = 0$ plane using the plane-wave expansion method. The bands are measured for the wave vector in the direction of propagation, with the wave vector divided into $128$ points between $k = 0$ and $k = \pi/a$. The light cone is defined by the semi-infinite surroundings of air. To ensure the accuracy of the results, the eigenvalue tolerance ($10^{-10}$) and lattice domain steps ($128$) are set to higher values than the minimum required. This conservative approach guaranteed the precision of the simulations, resulting in a simulation duration of approximately $5$ hours per band structure.

\newpage

%% file: D.tex
\section{D. Incorporating Fabrication Imperfections}
To identify nanobeam cavities that are more resilient to fabrication errors, imperfections are intentionally introduced into the nanobeam cavities that form the training and testing datasets (step II in the process flowchart of Figure \ref{Flowchart_chapter6}). The NNs are then trained on these imperfect structures, allowing them to learn the relations between the input design parameters and the affected Q-factors. The imperfections considered in this research include surface roughness and sidewall slant, as these are prevalent fabrication errors encountered in thin-film fabrication processes.

\hypertarget{supinfD1}{}
\subsection{D1. Surface roughness}
Surface roughness introduces randomized peaks and valleys to the otherwise smooth surfaces of the nanobeam cavity, affecting the top, bottom, and sidewall surfaces. This imperfection is inherently stochastic, causing variability in the computed Q-factor for each nanobeam cavity. To address this randomness, the Q-factors are determined by averaging the computed Q-factors of the same nanobeam cavity across multiple simulations with different roughness random seeds. Consequently, the Q-factors for structures affected by roughness are reported as averages.
\par
\begin{figure}[h!]
    \centering
    \includegraphics[width=1\textwidth]{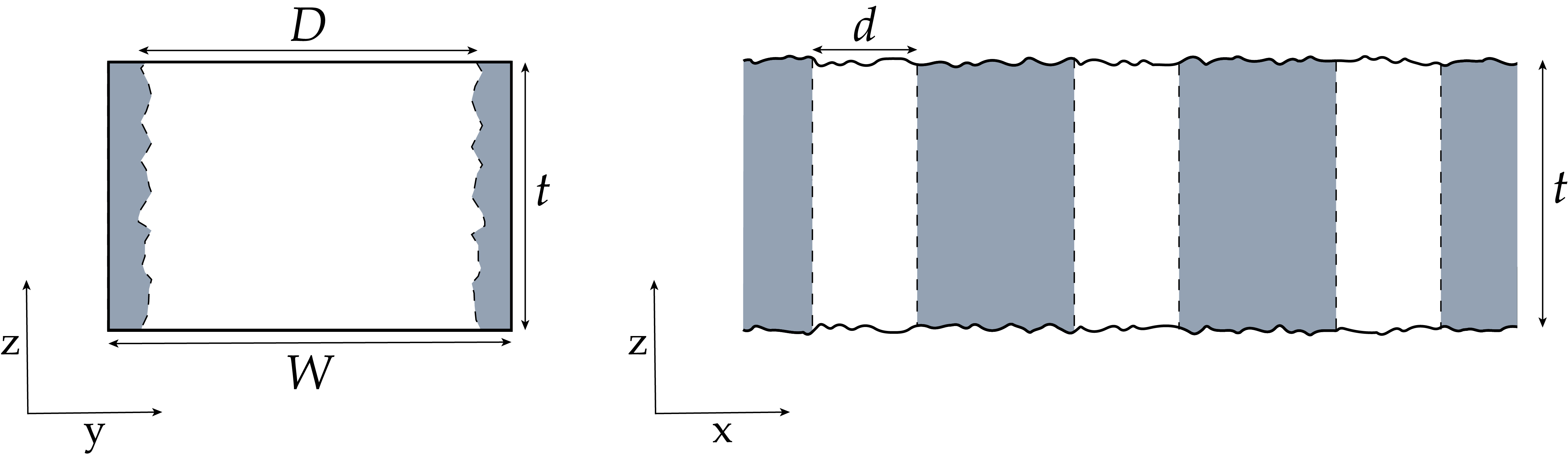}
    \caption{A visualization of an example of a roughness configuration for the L2 nanobeam cavity. The schematic on the \textbf{left} emphasizes the roughness of the curved surfaces (i.e. sidewalls of airholes, indicated by lighter areas), while the schematic on the \textbf{right} highlights the roughness of the flat surfaces (i.e. top and bottom of the waveguide).}
    \label{fig:NC_schematic_roughness}
\end{figure}
\begin{figure}[h!]
    \centering
    \includegraphics[width=1\textwidth]{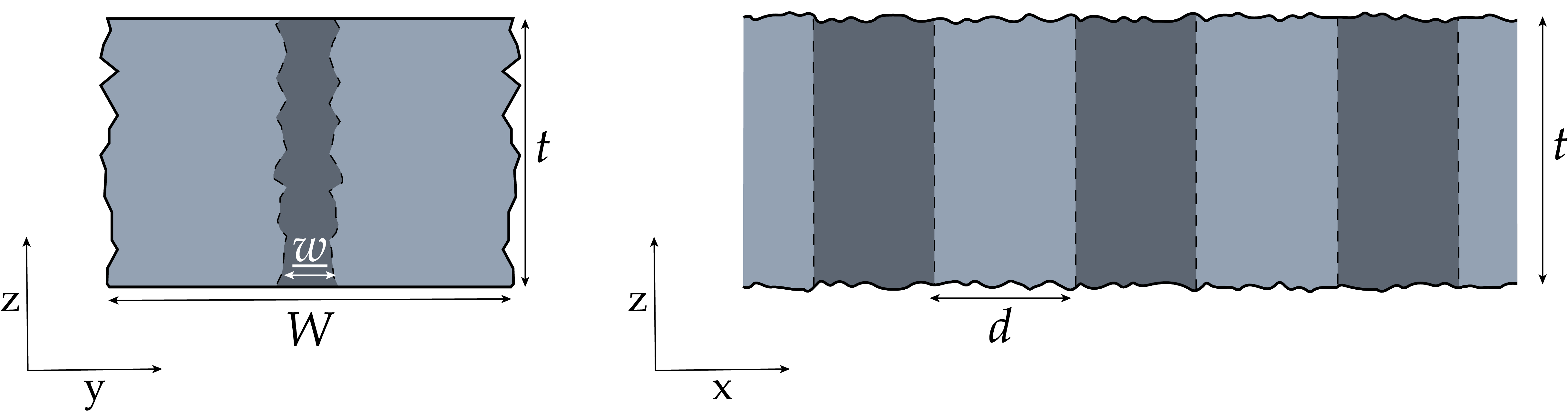}
    \caption{A visualization of an example of a roughness configuration for the fishbone nanobeam cavity. The schematic on the \textbf{left} emphasizes the roughness of the curved surfaces (i.e. sidewalls of the narrow ridge between opposing holes and fin ends, indicated by darker areas), while the schematic on the \textbf{right} highlights the roughness of the flat surfaces (i.e. top and bottom of the waveguide).}
    \label{fig:FB_schematic_roughness}
\end{figure}

Although roughness disrupts the symmetry of the structures, SBCs are applied in all three directions to ensure the simulations remain computationally manageable. The roughness introduced to both base designs is conceptually identical: flat surfaces experience surface roughness with an amplitude of $\pm 2.5$ nm, while curved surfaces have a roughness amplitude of $\pm 4.0$ nm. Examples of surface roughness configurations for each base design, along with their effects on performance, are detailed below (see Figure \ref{fig:NC_schematic_roughness} and \ref{fig:FB_schematic_roughness}) and summarized in Table \ref{tab:base_model_imperfections}.
\newpage

\paragraph{L2 nanobeam cavity:} The impact of surface roughness on the L2 nanobeam base design is relatively minor, with the average Q-factor decreasing by approximately $16.59\%$ and the mode volume increasing by $14.50\%$. The resonant wavelength remains almost entirely unaffected. This assessment is based on $100$ random roughness seeds, yielding a standard deviation of approximately $0.7$ nm for the wavelength, $3 \times 10^3$ for the Q-factor, and negligible standard deviation for the mode volume. An example of an imperfect structure is shown in Figure \ref{fig:NC_schematic_roughness}, highlighting the increased roughness amplitude on the curved sidewalls of the airholes compared to other surfaces.

\paragraph{Fishbone nanobeam cavity:}
The impact of surface roughness on the fishbone nanobeam base design is more pronounced compared to the L2 nanobeam cavity, with an approximate $73.91\%$ reduction in the Q-factor. However, the resonant wavelength and mode volume remain almost entirely unaffected. This evaluation is based on $100$ random roughness seeds, resulting in a standard deviation of approximately $0.7$ nm for the wavelength, $1 \times 10^2$ for the Q-factor, and negligible standard deviation for the mode volume. An example of an imperfect structure is shown in Figure \ref{fig:FB_schematic_roughness}, where the sidewalls of the curved airhole sections exhibit a larger roughness amplitude compared to other surfaces.

\hypertarget{supinfD2}{}
\subsection{D2. Sidewall slant}
\begin{figure}[h!]
    \centering
    \includegraphics[width=1\textwidth]{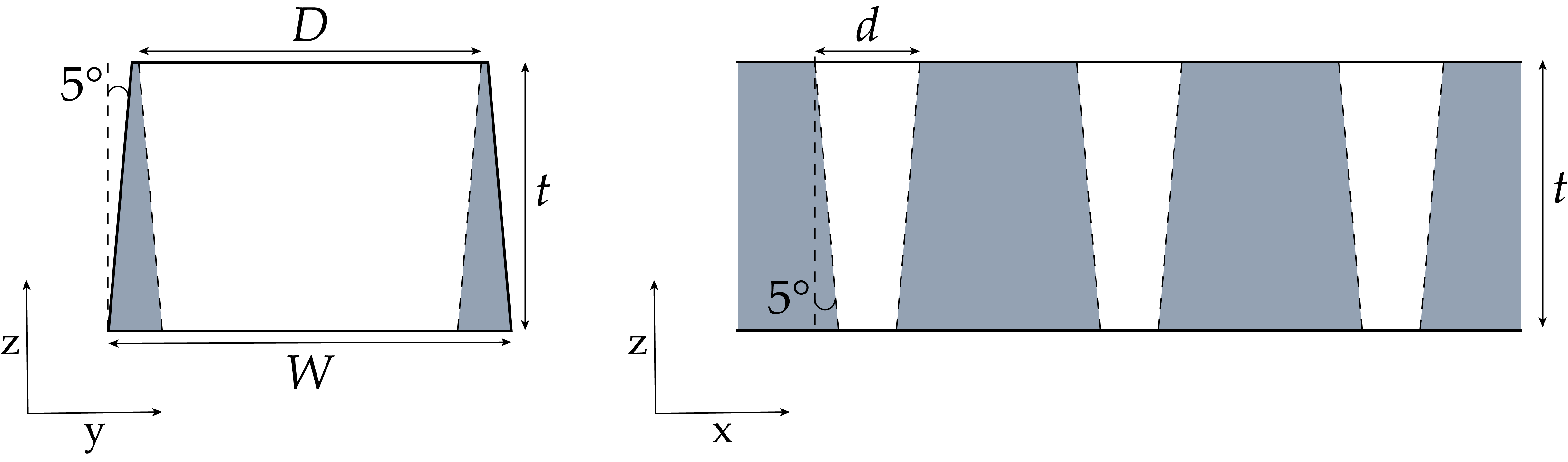}
    \caption{A visualization of the configuration of sidewall slant for the L2 nanobeam cavity. The airholes are indicated by lighter areas.}
    \label{fig:NC_schematic_slant}
\end{figure}

\begin{figure}[h!]
    \centering
    \includegraphics[width=1\textwidth]{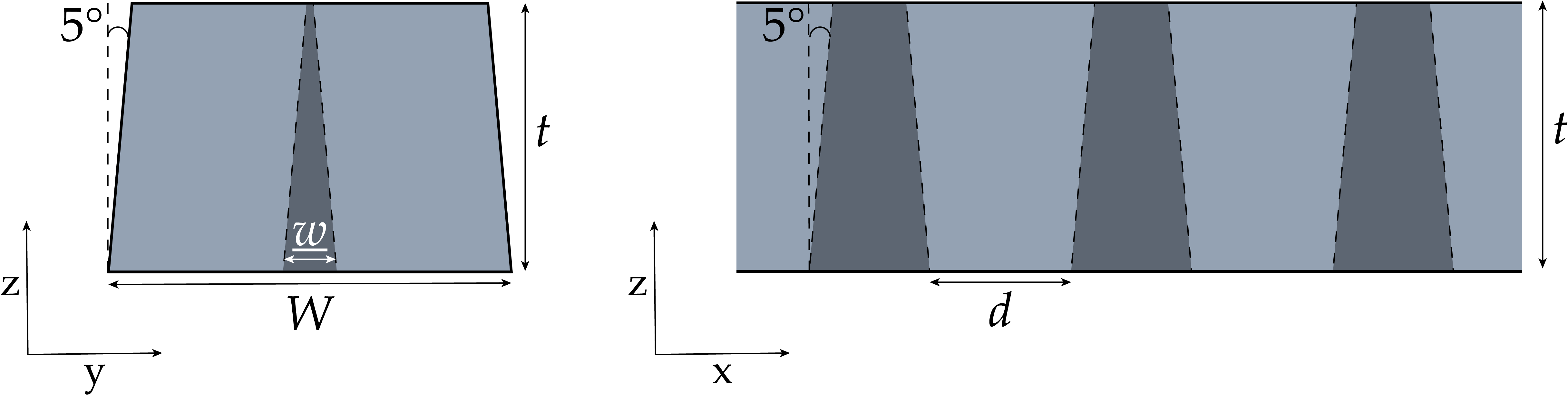}
    \caption{A visualization of the configuration of sidewall slant for the fishbone nanobeam cavity. The ridge between opposing holes \textbf{(left)} and the fins \textbf{(right)} are indicated by darker areas, respectively.}
    \label{fig:FB_schematic_slant}
\end{figure}

Sidewall slant introduces angled sidewalls to the nanobeam structures, resulting in a deterministic imperfection where each Q-factor computation yields identical results. This imperfection disrupts the symmetry of the structures in the z-direction. Therefore, SBCs are not implemented in this direction. The sidewall slant added to both models is conceptually the same, with all vertical surfaces (i.e., sidewalls) slanted with a $5^\circ$ slope. The exact configurations of the sidewall slant for each base design, along with its impact on the designs' performance, are illustrated in Figures \ref{fig:NC_schematic_slant} and \ref{fig:FB_schematic_slant} and detailed in Table \ref{tab:base_model_imperfections}.

\paragraph{L2 nanobeam cavity:}
The sidewall slant significantly impacts the L2 nanobeam base design, causing a $54.87\%$ reduction in the Q-factor and a $29.54\%$ increase in the mode volume. Unlike surface roughness, the sidewall slant also alters the resonant wavelength, introducing a red shift of $1.43\%$. An example of this imperfect structure is shown in Figure \ref{fig:NC_schematic_slant}. While the dimensions of the waveguide's bottom and the airholes' top surface remain unchanged from the original base model, the dimensions of other surfaces are reduced due to the angled sidewalls.

\paragraph{Fishbone nanobeam cavity:}
The sidewall slant has an unexpectedly positive effect on the fishbone nanobeam base design, leading to a $50.25\%$ increase in the Q-factor and an $8.42\%$ decrease in the mode volume. Additionally, the resonant wavelength experiences a blue shift of $5.29\%$. An example of this imperfect structure is shown in Figure \ref{fig:FB_schematic_slant}. Note that the geometry of the bottom surface of the entire structure remains unchanged from the original base model, while the top surface geometry is reduced due to the sidewall slant.
\newpage

\hypertarget{supinfD3}{}
\subsection{D3. Optical Properties under fabrication imperfections}

\begin{table}[h!]
\centering
\caption{The impact of fabrication imperfections on the performance of the base designs.}
\label{tab:my-table}
\begin{tabular}{c|cccc}
\hline
   &       & \begin{tabular}[c]{@{}c@{}}\textbf{Wavelength} (nm)\end{tabular} & \begin{tabular}[c]{@{}c@{}}\textbf{Q-factor} ($\times10^3$)\end{tabular} & \begin{tabular}[c]{@{}c@{}}\textbf{Mode volume} $((\lambda / n)^3)$\end{tabular} \\ \hline
\textbf{L2} & \textbf{Ideal} & 640.6                                                     & 24                                                               & $0.7$                                         \\ \cline{2-5} 
   & \textbf{Rough} & $640.2$                                           & $20$                                                     & 0.8                                                           \\ \cline{2-5} 
   & \textbf{Slant} & 649.7                                                     & 11                                                               & $0.9$                                                           \\ \hline
\textbf{FB} & \textbf{Ideal} & 546.6                                                     & 26                                                               & $1.4$                                         \\ \cline{2-5} 
   & \textbf{Rough} & 546.3                                           & 7                                                     & 1.4                                                           \\ \cline{2-5} 
   & \textbf{Slant} & 517.6                                                        & 39                                                                & $1.3$                                                           \\ \hline
\end{tabular}
\label{tab:base_model_imperfections}
\end{table}

As shown in Table \ref{tab:base_model_imperfections}, the L2 nanobeam cavity is generally more resilient to surface roughness than the fishbone nanobeam cavity. This can be attributed to the fact that the fishbone design has more curved surfaces, which give rise to larger amplitudes of roughness. However, the situation reverses when considering sidewall slant. Surprisingly, the optical properties of the fishbone structure improve with sidewall slant. This could be due to the fact that during the manual optimization of the base model, the distance between opposing airholes, $\underline{w}$, was intentionally kept larger to account for fabrication error tolerance. The sidewall slant effectively reduces this design parameter, potentially enhancing the optical performance of the fishbone structure \cite{pregnolato2024, sawfish2023}. In addition to their effects on the Q-factor, sidewall slant also significantly influences the resonant wavelength, especially the fishbone design.
\newpage

%% file: E.tex
\section{E. Affine Transformation, Activation, and Initialization}
Beyond the details of the loss function, optimizer, and node configuration, several additional concepts are crucial to fully characterize the NN architecture. These include the affine transformation, activation function, and model initialization. Together, these components define how the NN processes input data and learns from it, offering a complete overview of the architecture.

\hypertarget{supinfE1}{}
\subsection{E1. Affine transformation}
Each artificial neuron in the NN holds a real number referred to as its "activation." The activations from one layer influence those in the next layer through interconnected edges. Each neuron is associated with a bias, and each edge has a weight; both are adjusted during the learning process. When transitioning from one layer to the next, the activations of all input neurons $x_i$ are multiplied by the weights $w_i$ of their corresponding edges. Optionally, a bias $b$ associated with the target neuron is added \cite{DL_general_book}. Following this affine transformation, the sum of the weighted activations are passed through a non-linear function $f$, which determines the activation of the output neuron $y$ (see Equation \ref{Eq:NN_func}). This function is known as the activation function.

\begin{equation}
    y = f\left(\sum_{i=1}^{n} (x_i \cdot w_i) + b\right)
    \label{Eq:NN_func}
\end{equation}

\hypertarget{supinfE2}{}
\subsection{E2. Activation function}
Every node in a NN applies both an affine transformation and a nonlinear transformation to its input. The affine transformation involves scaling the input with a weight factor and adding a bias term. These parameters define the model and enable it to perform various transformations while learning iteratively during training. The activation function introduces the nonlinearity into the model, which allows the NN to capture complex, non-linear relationships. Common activation functions include ReLU, sigmoid, and hyperbolic tangent functions \cite{activation_function, DL_general_book}. 
\par
ReLU, short for rectified linear unit, is one of the most widely used activation functions in NNs. It is simple yet highly effective, making it a popular choice in DL architectures. The ReLU function is mathematically defined as:

\begin{equation}
    f(x) = max(0,x)
\end{equation}

\begin{figure}[h!]
    \centering
    \includegraphics[width=0.5\linewidth]{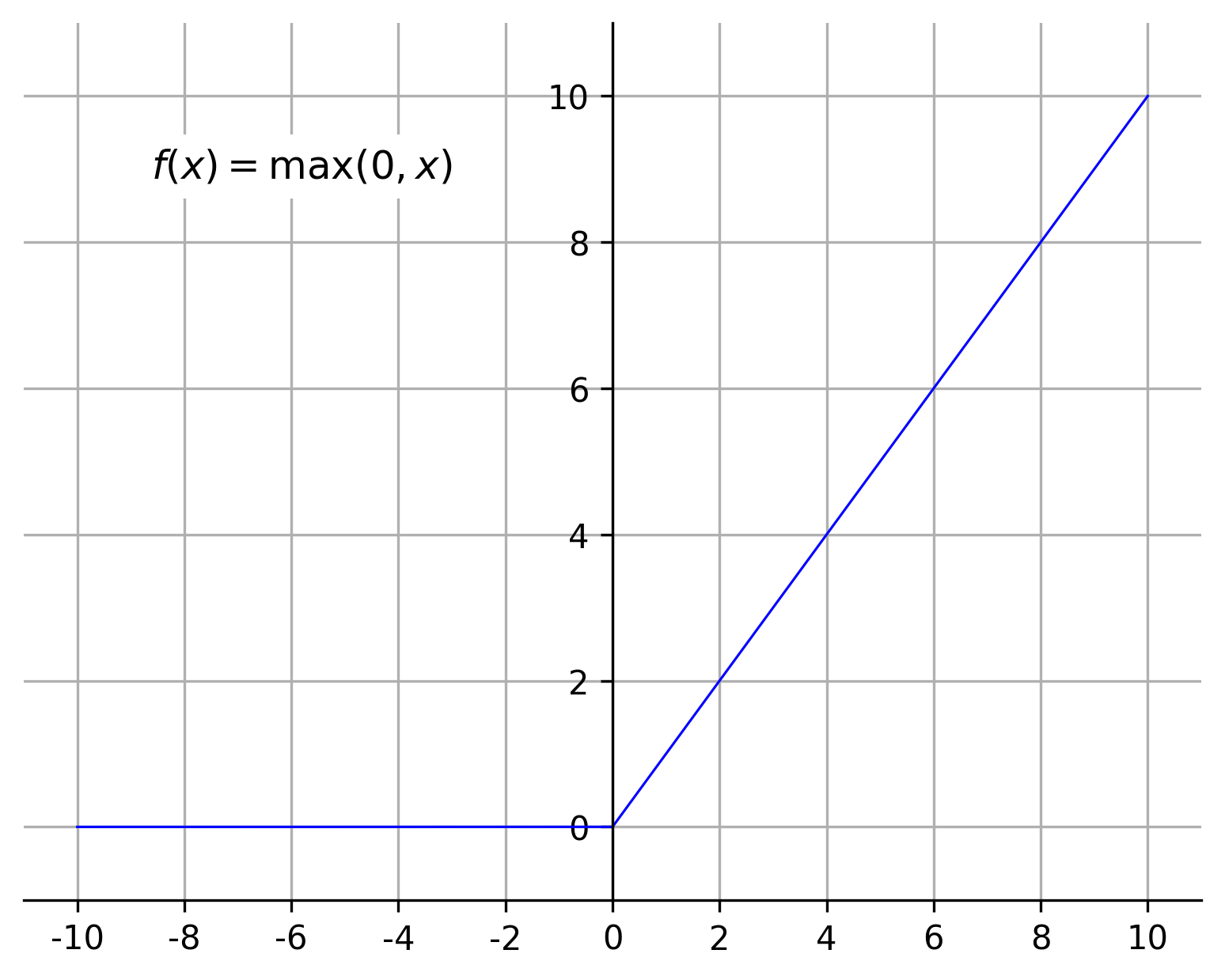}
    \caption{A plotted representation of the behavior of the ReLU activation function.}
    \label{fig:ReLU}
\end{figure}

In other words, the ReLU function outputs the input $x$ if $x > 0$, and outputs 0 otherwise. Visually, the ReLU function resembles a ramp, with a flat segment for negative inputs (slope of 0) and a linear segment for positive inputs (slope of 1), as illustrated in Figure \ref{fig:ReLU}. ReLU addresses common challenges such as the vanishing gradient problem. Unlike activation functions such as the sigmoid, where gradients saturate and diminish in magnitude during backpropagation, the gradients in ReLU neurons remain stable, enabling more effective learning \cite{activation_function, DL_general_book}. 
\par
Additionally, ReLU is computationally efficient due to its simplicity and promotion towards sparse activation. Neurons that do not activate have zero gradients, effectively bypassing those neurons during backpropagation. This reduces unnecessary computations and helps streamline the training process.

A common issue with ReLU is the occurrence of the "dying ReLU" problem, where neurons become inactive for almost all inputs and effectively stop contributing to the model. This reduces the model's capacity to learn. Activation functions such as leaky ReLU and GELU have been specifically designed to address this problem \cite{activation_function, DL_general_book, GeLU}. Nonetheless, ReLU remains a strong default choice for shallow and simple CNNs due to its effectiveness and computational efficiency. Therefore, ReLU is selected as the activation function for this work.

\hypertarget{supinfE3}{}
\subsection{E3. Model initialization}
Proper initialization of model parameters is crucial for avoiding issues like exploding and vanishing gradients \cite{DL_general_book}. The weights and biases of the NN should be set within a specific range, with suitable variance, to align with the activation function used and the size of the network. One popular technique for initializing these weights and biases is He-initialization, which is designed specifically to be compatible with the ReLU activation function. Hence, He-initialization was employed for this NN \cite{Initialization_He}. It initializes the weights of each layer using a Gaussian distribution with zero mean and a variance that depends on the number of input connections to the neuron.
\par
He-initialization of the model’s weights and biases is based on a Gaussian distribution, which introduces a random component, meaning it is dependent on a random seed. Similarly, techniques like dropout, which deactivate a random subset of neurons during each iteration of training, also rely on a random seed. As a result, the random seed influences the initialization process, affecting the learning dynamics and, ultimately, the final performance of the NN. Different seeds can lead to variations in how the NN behaves. Therefore, it is crucial to select random seeds that promote convergence to a useful configuration.

\newpage

%% file: F.tex
\section{F. Dataset Generation, Preprocessing, and Performance}
This section outlines the procedures for dataset generation, preprocessing, and evaluation of NN performance. It covers the creation of datasets for the L2 and fishbone nanobeam cavities under ideal, rough surface, and slanted sidewall fabrication conditions, with particular attention to the variability introduced by surface roughness. Preprocessing steps are detailed, including the normalization of input parameters and the transformation of Q-factors to ensure stable training. The performance of the NNs is then evaluated, with an emphasis on the impact of L2-regularization on prediction accuracy and generalization.

\hypertarget{supinfF1}{}
\subsection{F1. Dataset generation}
The dataset generation for the nanobeam cavities under ideal and sidewall slant conditions involves creating a single dataset of $1250$ distinct devices. Since the Q-factor output remains constant for a given structure under these conditions, one dataset is sufficient. However, this is not the case for the surface roughness condition, where each roughness random seed produces a different Q-factor. To account for this variability, the datasets for the surface roughness case consist of data points that represent the average Q-factor of the same structure, computed across three roughness seeds. This averaging approach helps reduce the variability introduced by the stochastic nature of surface roughness.
\par
The datasets for the different versions of the L2 and fishbone nanobeam cavities share identical parameter displacements as inputs for each respective design. Consequently, all datasets for the L2 and fishbone devices consist of the same cavity designs. This consistency ensures that comparisons between different fabrication error implementations are not affected by random variations in dataset quality. 
\par
An example of one of the three datasets consisting of L2 nanobeam cavities is presented in Figures \ref{fig:dataset_NC123_notprocessed_suppinf} and \ref{fig:dataset_NC123_processed}. Similarly, an example of one of the three datasets consisting of fishbone nanobeam cavities is shown in Figures \ref{fig:dataset_FB32_notprocessed} and \ref{fig:dataset_FB32_processed}. 

\hypertarget{supinfF2}{}
\subsection{F2. Data preprocessing}
Before training the NNs on the datasets, preprocessing is required. The input for the NNs consists of the design parameter shifts of the base nanobeam cavities, which approximately range from $-10$ nm to $10$ nm. To ensure stable gradients during backpropagation, it is standard practice to normalize these input values to lie within the range $[-1, 1]$. Consequently, the input values are rescaled and expressed in tens of nanometers.

\begin{figure}[h!]
    \begin{subfigure}{0.5\textwidth}
        \centering
        \includegraphics[width=\linewidth]{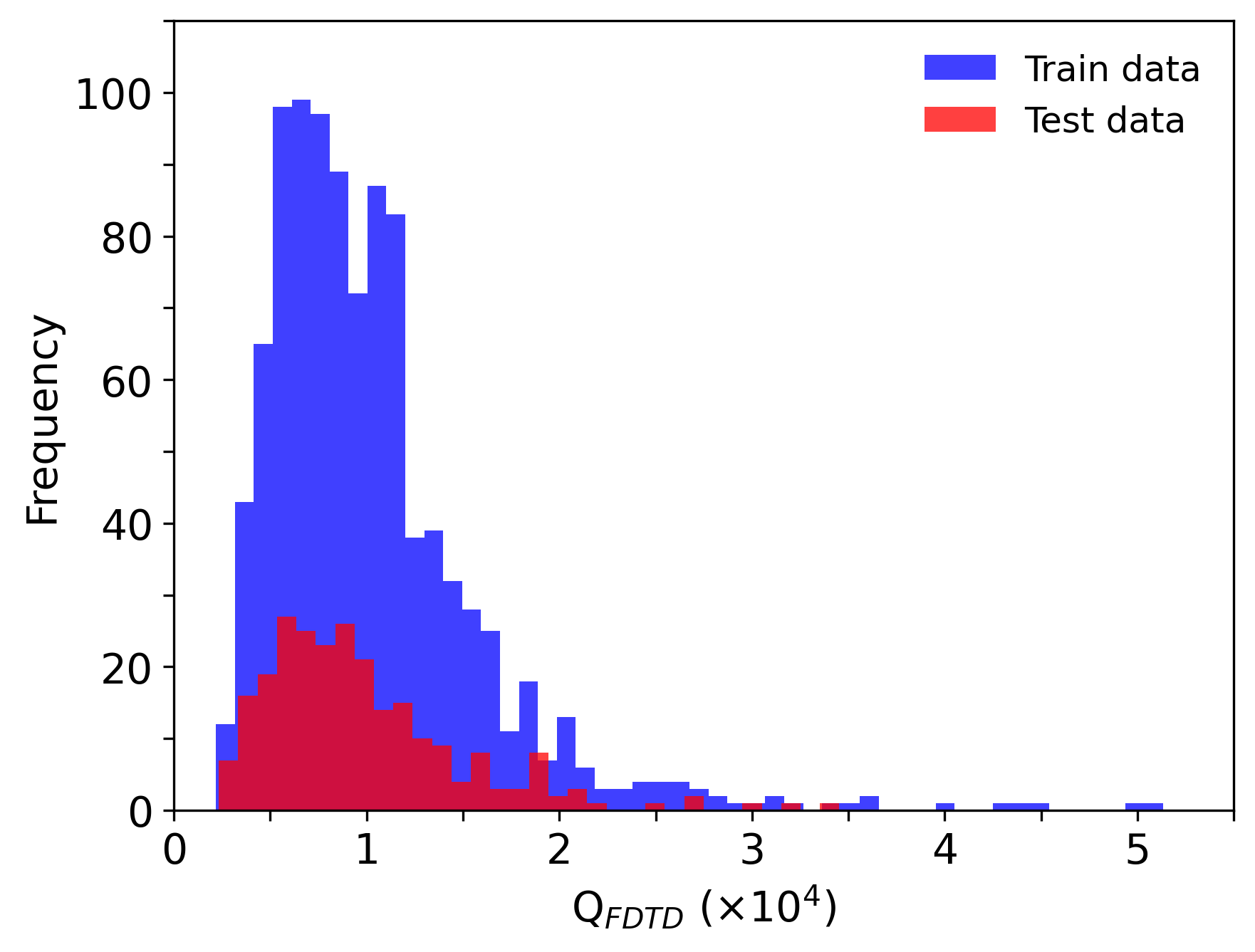}
        \caption{}
        \label{fig:dataset_NC123_notprocessed_suppinf}
    \end{subfigure}%
    \begin{subfigure}{0.5\textwidth}
        \centering
        \includegraphics[width=\linewidth]{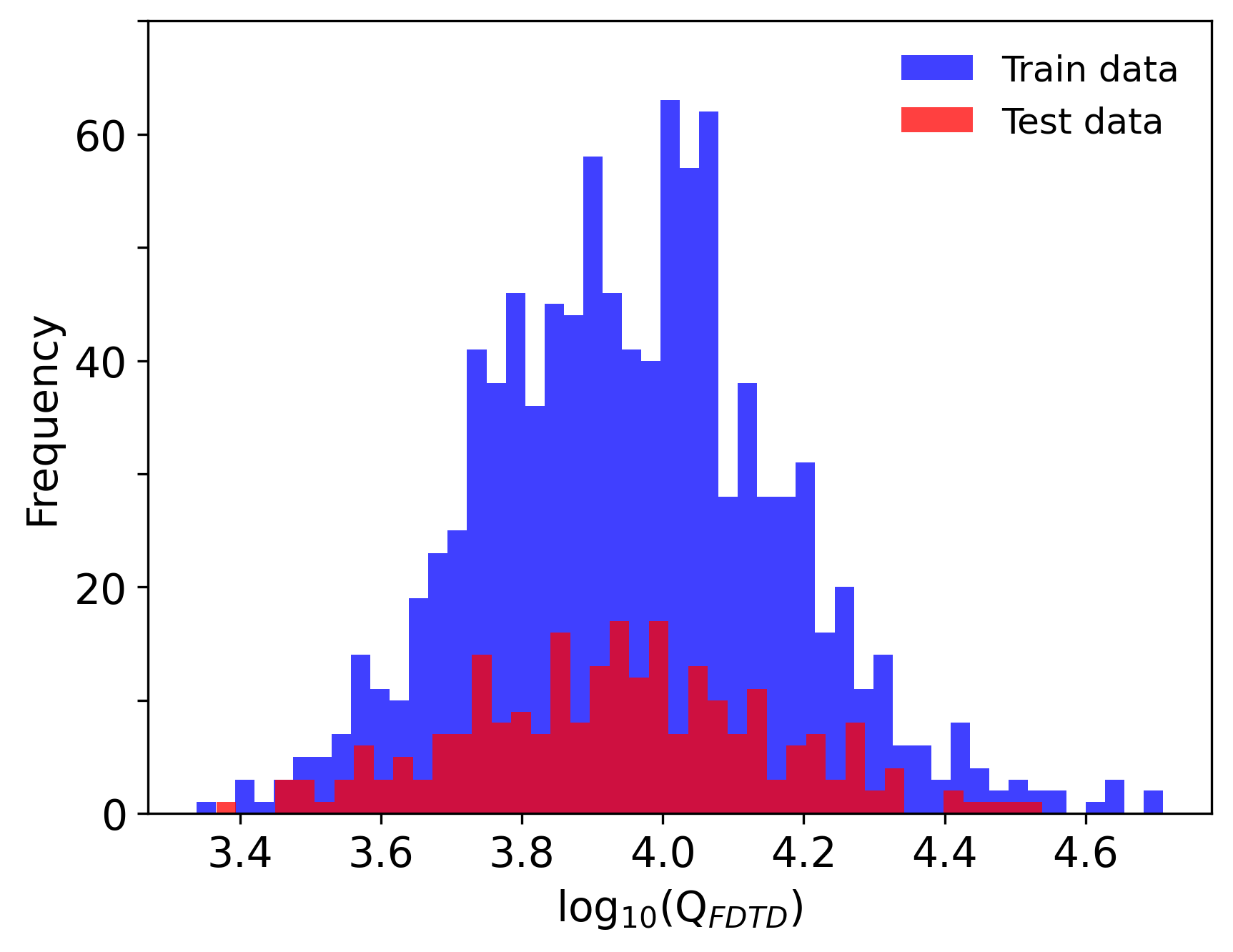}
        \caption{}
        \label{fig:dataset_NC123_processed}
    \end{subfigure}
    
    \begin{subfigure}{0.5\textwidth}
        \centering
        \includegraphics[width=\linewidth]{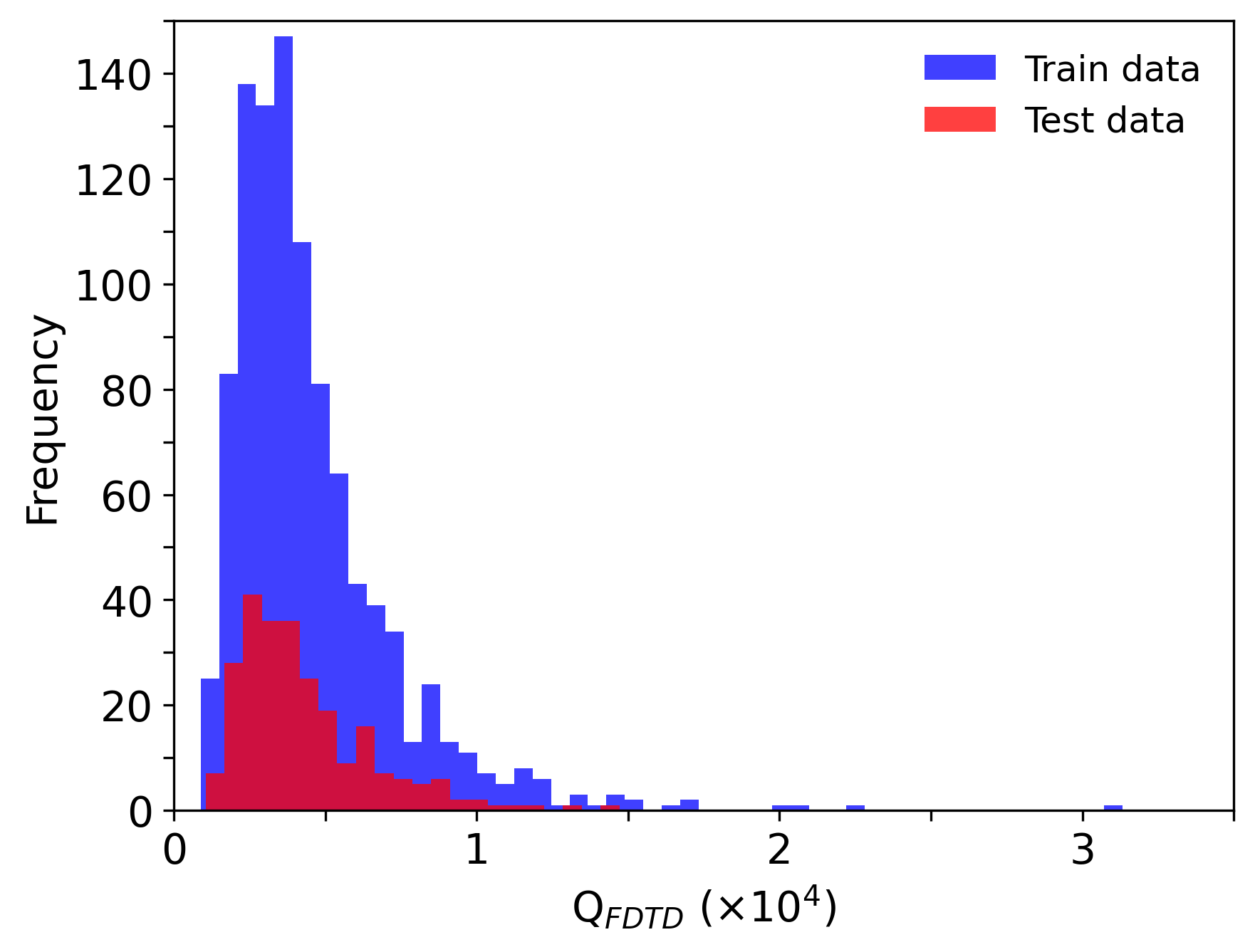}
        \caption{}
        \label{fig:dataset_FB32_notprocessed}
    \end{subfigure}%
    \begin{subfigure}{0.5\textwidth}
        \centering
        \includegraphics[width=\linewidth]{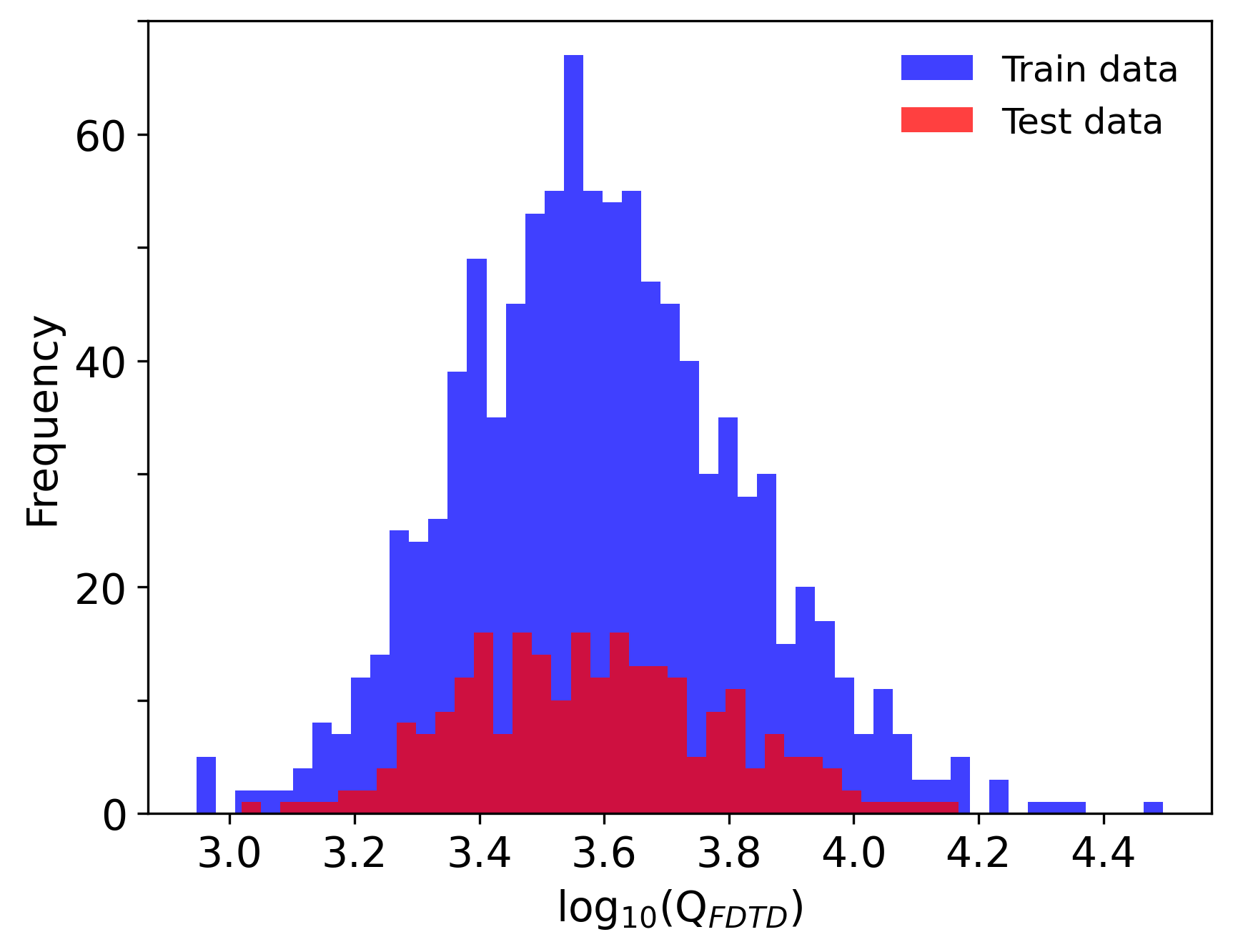}
        \caption{}
        \label{fig:dataset_FB32_processed}
    \end{subfigure}
    
    \caption{Examples of histograms illustrating dataset splits from different nanobeam cavities without fabrication imperfections. \textbf{(a)} and \textbf{(b)} show the dataset split from the L2 nanobeam cavity before and after preprocessing, respectively. \textbf{(c)} and \textbf{(d)} show the dataset split from the fishbone nanobeam cavity before and after preprocessing, respectively.}
    \label{fig:combined_datasets}
\end{figure}

The output of the NNs corresponds to the Q-factor of the nanobeam cavity design, $Q_{\textrm{FDTD}}$. The distribution of $Q_{\textrm{FDTD}}$ values is typically highly skewed and spans a wide range. To reduce the variance in the output data, the logarithm of the Q-factors is taken. Transforming the output data to $\log_{10}(Q_{\textrm{FDTD}})$ reshapes the data into a distribution closer to a normal distribution, as illustrated in Figure \ref{fig:combined_datasets} (additional data split graphs are provided in the \hyperlink{supinfH}{Appendix}). This transformation enhances the suitability of the data for modeling by the NN, facilitating more effective learning and prediction.

\hypertarget{supinfF3}{}
\subsection{F3. Model Performance}
The relative prediction errors of the Q-factor and the Pearson correlation coefficients between $Q_{\textrm{NN}}$ and $Q_{\textrm{FDTD}}$ are presented in Table \ref{tab:prediction_errors_NN} and Table \ref{tab:correlation_coeff_NN}, respectively (corresponding training process and correlation graphs are provided in the \hyperlink{supinfH}{Appendix}). A comparison of the performance of the NNs trained on the L2 and fishbone nanobeam cavities reveals that the NNs trained on the fishbone structures generally perform worse. This difference is likely due to the variation in the number of design parameters: both NNs are trained on datasets of the same size, but the fishbone cavities feature a larger optimization space. Consequently, the NNs trained on the L2 nanobeam cavities benefit from a denser sampling of their optimization space, resulting in more effective training and better predictive accuracy.

\begin{table}[h!]
\centering
\caption{An overview of the relative prediction errors of the Q-factors under ideal, surface roughness, and sidewall slant conditions is provided, comparing different NNs with and without L$^2$-regularization. The test prediction errors are shown, with the train prediction errors in brackets. The prediction errors in bold indicate the NNs responsible for identifying the optimized cavity.}
\label{tab:prediction_errors_NN}
\begin{tabular}{c|ccccc}
\hline
& & \textbf{NN1:} $\lambda = 0$ & \textbf{NN1:} $\lambda = 0.001$ & \textbf{NN2}: $\lambda = 0$ & \textbf{NN2:} $\lambda = 0.001$ \\ \hline
\textbf{L2} & \textbf{Ideal} & 6.33\% (3.54\%) & 5.62\% (3.65\%) & \textbf{5.76\% (3.08\%)} & 5.39\% (3.45\%) \\ \hline
& \textbf{Rough} & 13.41\% (4.73\%) & \textbf{11.50\% (6.62\%)} & 12.17\% (4.50\%) & 11.52\% (6.39\%) \\ \hline
& \textbf{Slant} & 4.41\% (2.67\%) & 3.99\% (3.13\%) & \textbf{4.22\% (2.43\%)} & 4.55\% (2.98\%) \\ \hline
\textbf{FB} & \textbf{Ideal} & 10.25\% (3.82\%) & 8.29\%(4.42\%) & 9.42\% (4.09\%) & \textbf{8.21\% (4.39\%)}  \\ \hline
& \textbf{Rough} & 14.05\% (4.54\%) & 13.00\% (6.49\%) & 15.17\% (4.94\%) & \textbf{12.58\% (6.72\%)} \\ \hline
& \textbf{Slant} & 8.62\% (3.97\%) &  \textbf{8.25\% (4.52\%)} & 9.24\% (4.32\%) & 7.95\% (4.67\%) \\ \hline
\end{tabular}
\end{table}

\begin{table}[h!]
\centering
\caption{An overview of the correlation coefficients between the predicted Q-factors $Q_{\textrm{NN}}$ and the simulated Q-factors $Q_{\textrm{FDTD}}$ under ideal, surface roughness, and sidewall slant conditions is provided, comparing different NNs with and without L$^2$-regularization. The test correlation coefficients are shown, with the train correlation coefficients in brackets. The correlation coefficients in bold indicate the NNs responsible for identifying the optimized cavity.}
\label{tab:correlation_coeff_NN}
\begin{tabular}{c|ccccc}
\hline
& & \textbf{NN1:} $\lambda = 0$ & \textbf{NN1:} $\lambda = 0.001$ & \textbf{NN2}: $\lambda = 0$ & \textbf{NN2:} $\lambda = 0.001$ \\ \hline
\textbf{L2} & \textbf{Ideal} & 0.988 (0.998) & 0.987 (0.991) & \textbf{0.979 (0.998)} & 0.965 (0.991) \\ \hline
& \textbf{Rough} & 0.924 (0.995) & \textbf{0.940 (0.981)} & 0.927 (0.995) & 0.935 (0.983) \\ \hline
& \textbf{Slant} & 0.985 (0.997) & 0.983 (0.990) & \textbf{0.988 (0.998)} & 0.984 (0.992) \\ \hline
\textbf{FB} & \textbf{Ideal} & 0.948 (0.998) & 0.956 (0.988) & 0.955 (0.998) & \textbf{0.957 (0.990)} \\ \hline
& \textbf{Rough} & 0.917 (0.997) & 0.938 (0.988) & 0.938 (0.997) & \textbf{0.952 (0.986)} \\ \hline
& \textbf{Slant} & 0.965 (0.989) & \textbf{0.962 (0.993)} & 0.966 (0.998) & 0.969 (0.992) \\ \hline
\end{tabular}
\end{table}

As discussed earlier, roughness introduces randomness, making it a stochastic imperfection. NNs are particularly sensitive to this added variability, which is evident in their performance. Incorporating L$^2$-regularization addresses these challenges by preventing overfitting and enhancing performance on noisy training datasets. This is demonstrated by the significant improvement in performance for NNs trained with L$^2$-regularization, as well as the reduced difference between test and training results. Consequently, L$^2$-regularization proves indispensable when training NNs on datasets influenced by stochastic imperfections, ensuring better generalization and more reliable performance.
\par
\begin{figure}[h!]
    \begin{subfigure}{0.5\textwidth}
        \centering
        \includegraphics[width=\linewidth]{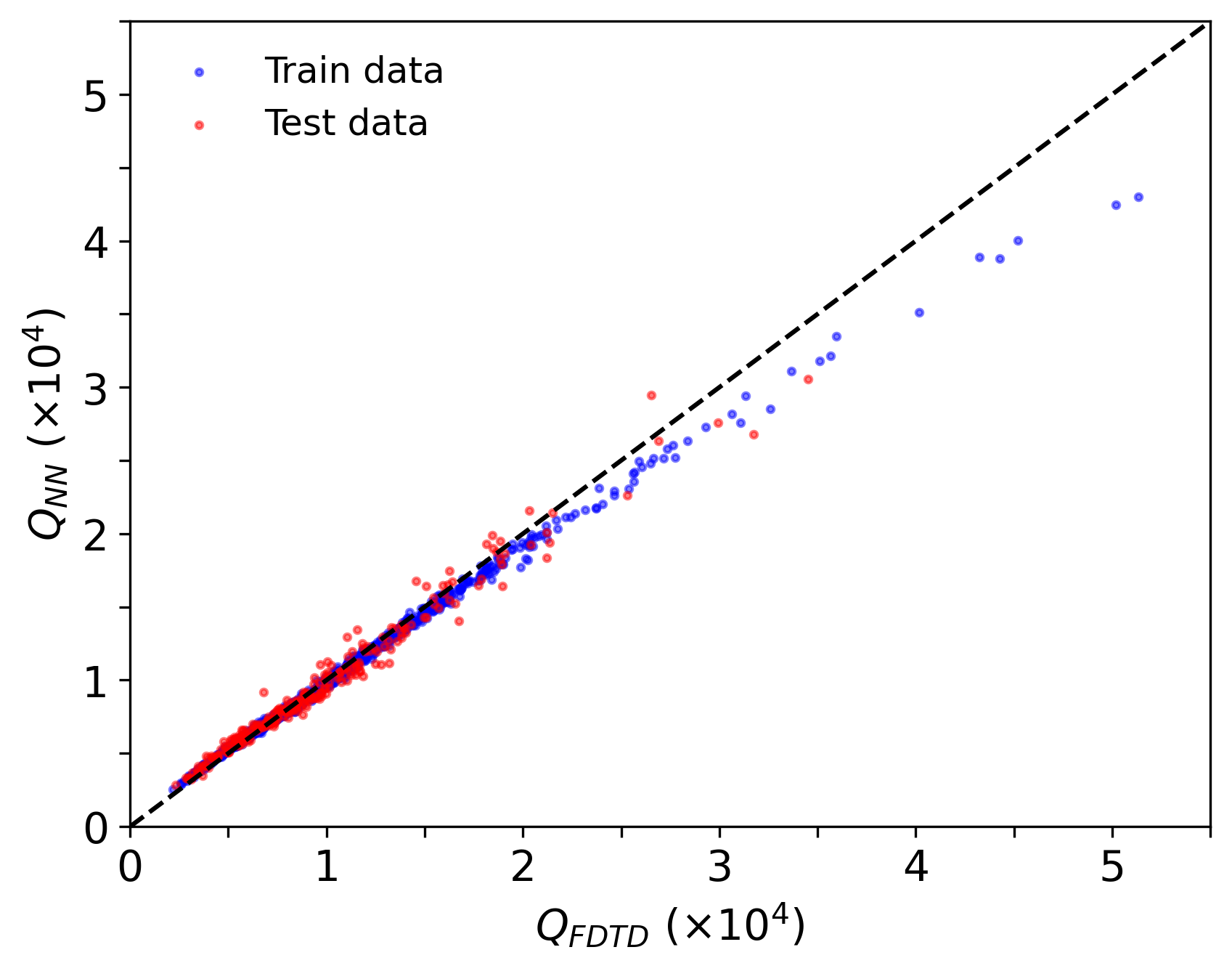}
        \caption{}
        \label{fig:Correlation_NC123_l0_suppinf}
    \end{subfigure}%
    \begin{subfigure}{0.5\textwidth}
        \centering
        \includegraphics[width=\linewidth]{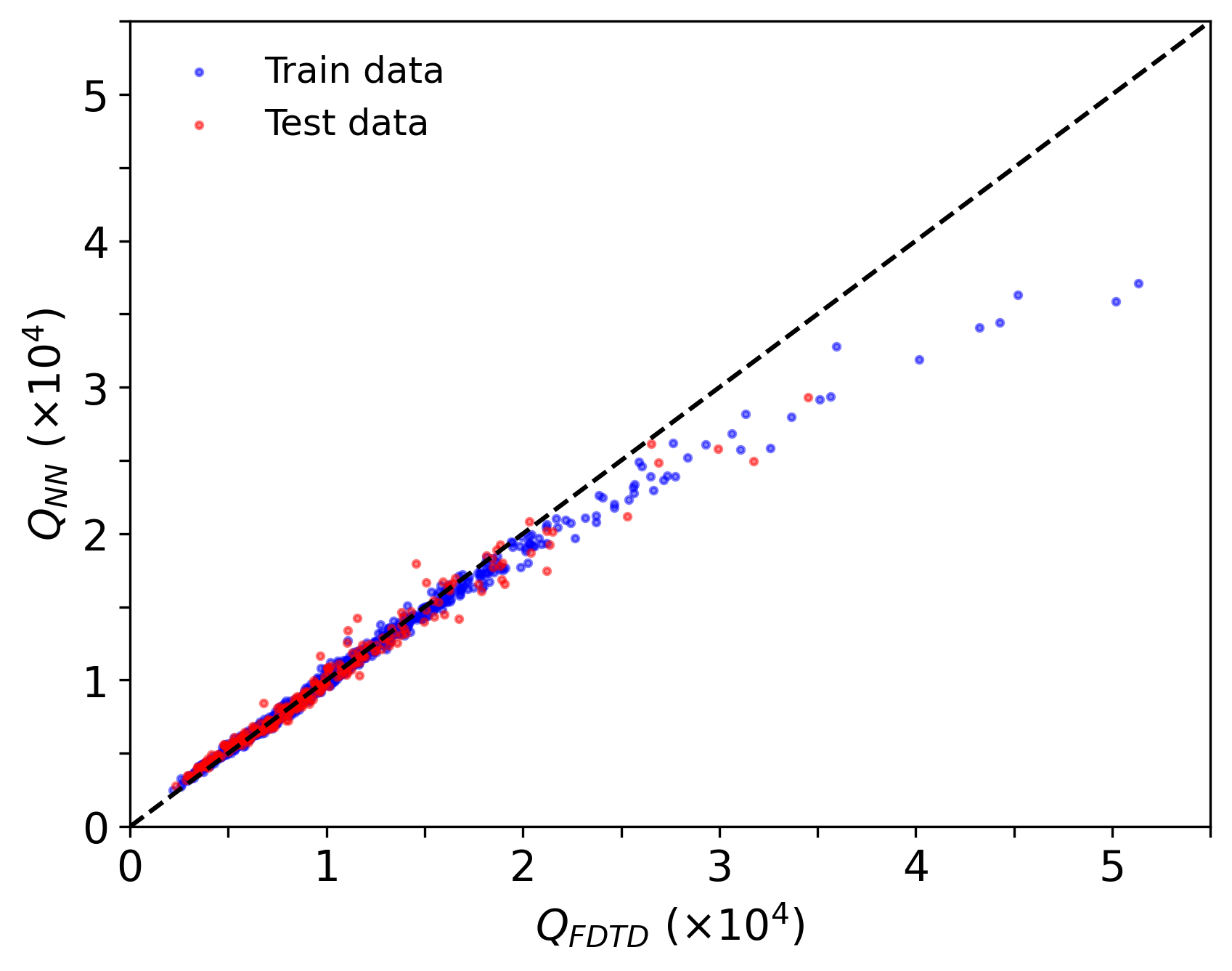}
        \caption{}
        \label{fig:Correlation_NC123_l0.001}
    \end{subfigure}
    
    \begin{subfigure}{0.5\textwidth}
        \centering
        \includegraphics[width=\linewidth]{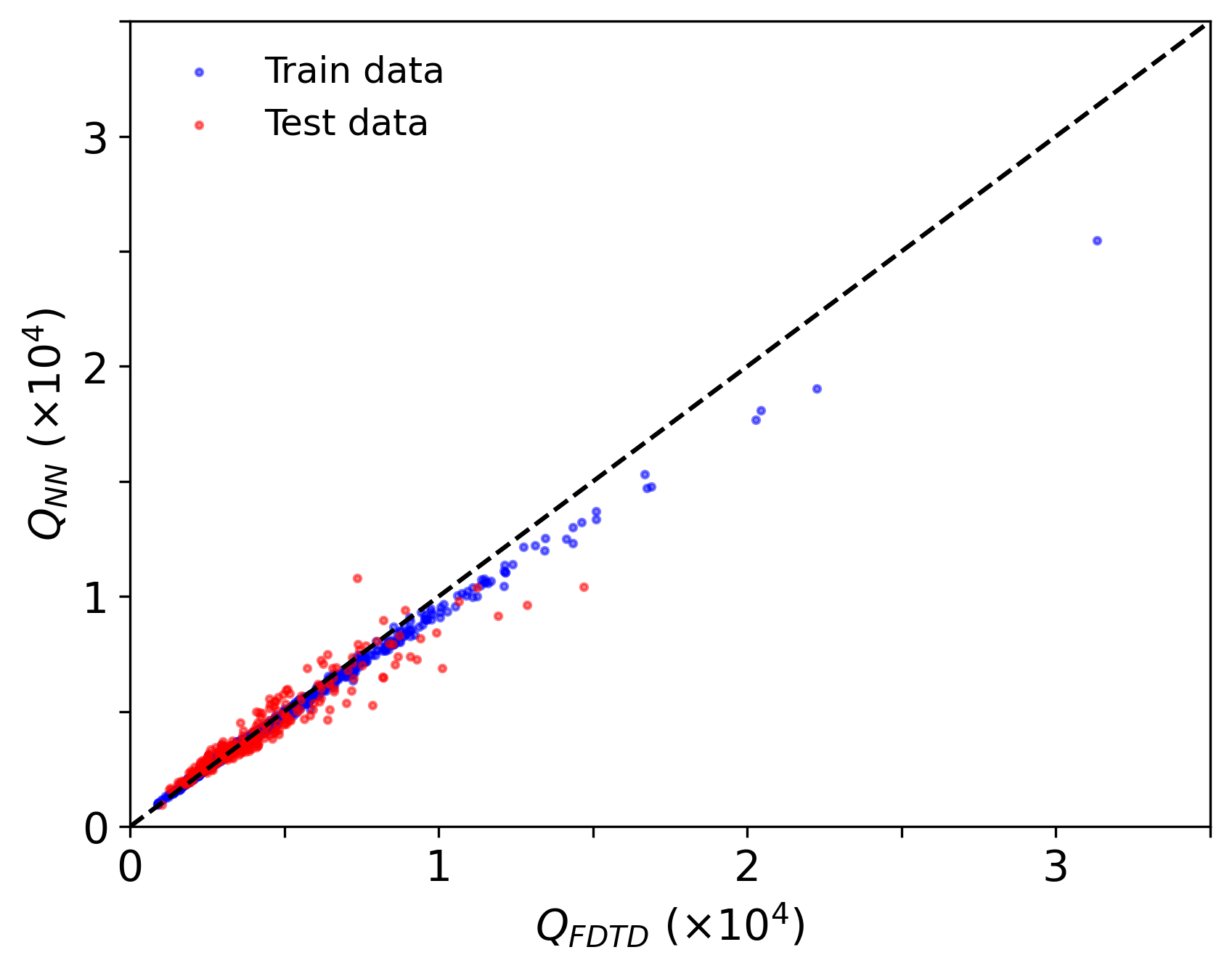}
        \caption{}
        \label{fig:Correlation_FB32_l0}
    \end{subfigure}%
    \begin{subfigure}{0.5\textwidth}
        \centering
        \includegraphics[width=\linewidth]{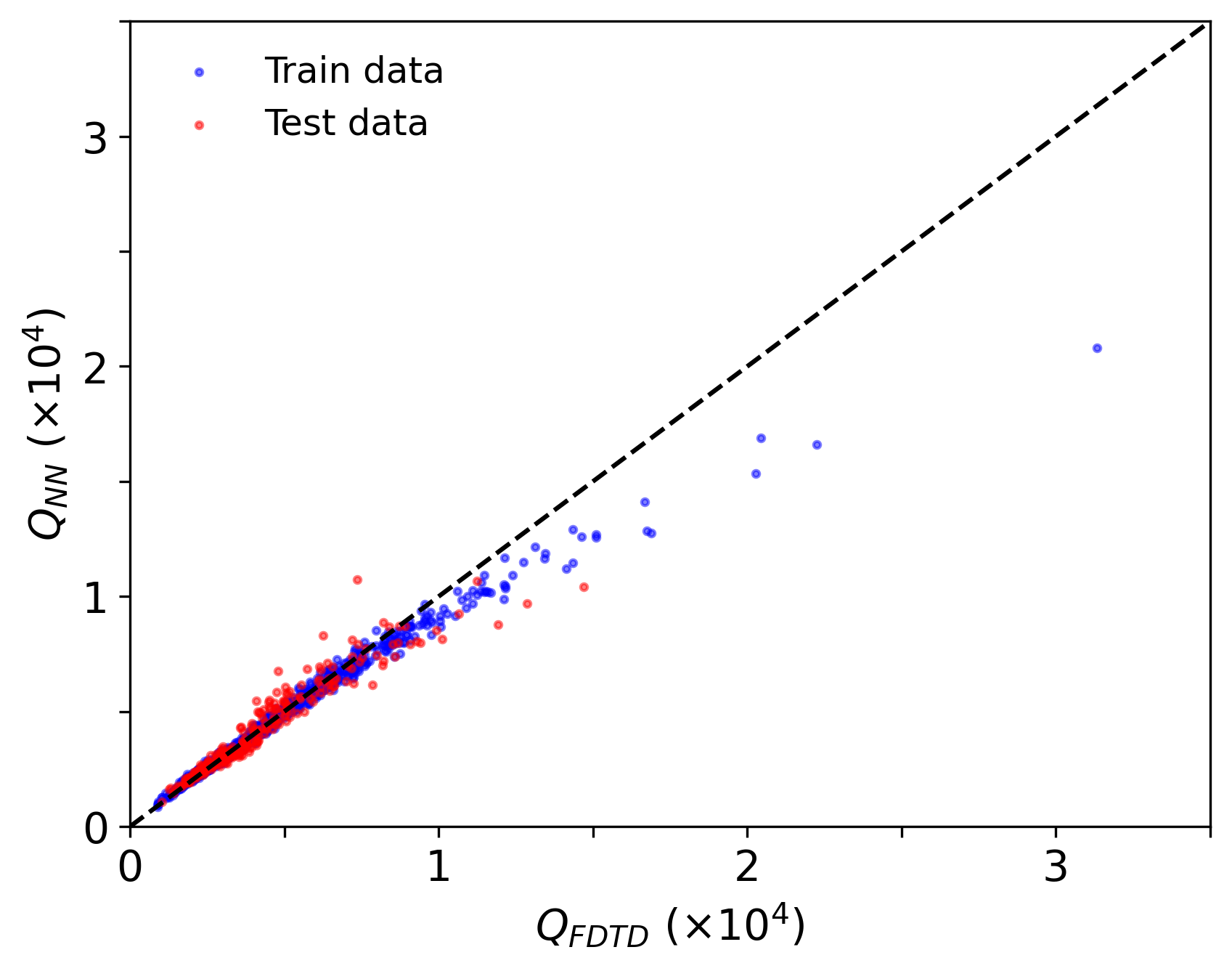}
        \caption{}
        \label{fig:Correlation_FB32_l0.001}
    \end{subfigure}
    
    \caption{Examples of scatter plots illustrating the predictive performance through correlation of two NNs trained by datasets consisting of different nanobeam cavities without fabrication imperfections. The dashed line represents the line of equality. \textbf{(a)} and \textbf{(b)} show the correlation plots of NNs trained without and with L$^2$-regularization on their weights and biases, respectively, using data from L2 nanobeam cavities. \textbf{(c)} and \textbf{(d)} show the correlation plots of NNs trained without and with L$^2$-regularization on their weights and biases, respectively, using data from fishbone nanobeam cavities.}
    \label{fig:combined_correlation_plots}
\end{figure}

The performance of the NN within a specific range strongly depends on the density of training data in that range. This is evident from Figures \ref{fig:combined_datasets} and \ref{fig:combined_correlation_plots}, where the data becomes more scattered at larger Q-factors, correlating with sparser training data density. High Q-factor data is particularly valuable, as it represents the range where optimal performance of the models is desired during optimization tasks using trained NNs. 
\par
In general, L$^2$-regularization effectively reduces relative prediction errors. However, as seen in the correlation graphs, models trained with L$^2$-regularization show reduced performance at higher Q-factors (see Figure \ref{fig:combined_correlation_plots}). This appears to result from the models saturating and failing to reach the higher output values. L$^2$-regularization restricts the magnitude of the weights and biases in the models, which, while beneficial for generalization and preventing overfitting, seemingly also limits the range of the output. Consequently, this restriction leads to degraded performance in predicting higher Q-factor values. However, this does not rule out the model's ability to detect peaks within the optimization space. Despite the saturation at higher Q-factors, the models can still identify regions with local maxima. 

Overall, the NNs are effectively trained and successfully capture the dependencies between input design parameters and output Q-factors within each dataset. They achieve relative test prediction errors ranging from $3.99\%$ to $15.17\%$, with test correlation coefficients spanning from $0.917$ to $0.988$.

\newpage

%% file: G.tex
\section{G. Optimization with Trained Neural Network}
This section discusses the optimization of nanobeam cavities using the trained NNs in combination with local and global optimization algorithms (i.e. GA and CMA-ES, respectively). The process generates a family of 80 distinct nanobeam cavity designs, each validated through FDTD simulations. The optimization results, including the origin of the optimized device, prediction errors, and the structure's performance, are summarized. Finally, the design parameter displacements of all six optimized nanobeam cavities are presented, along with their performance under ideal, rough surface, and slanted sidewall fabrication conditions.

\hypertarget{supinfG1}{}
\subsection{G1. Family of candidate structures}

\begin{figure}[h!]
        \centering
        \includegraphics[width=\linewidth]{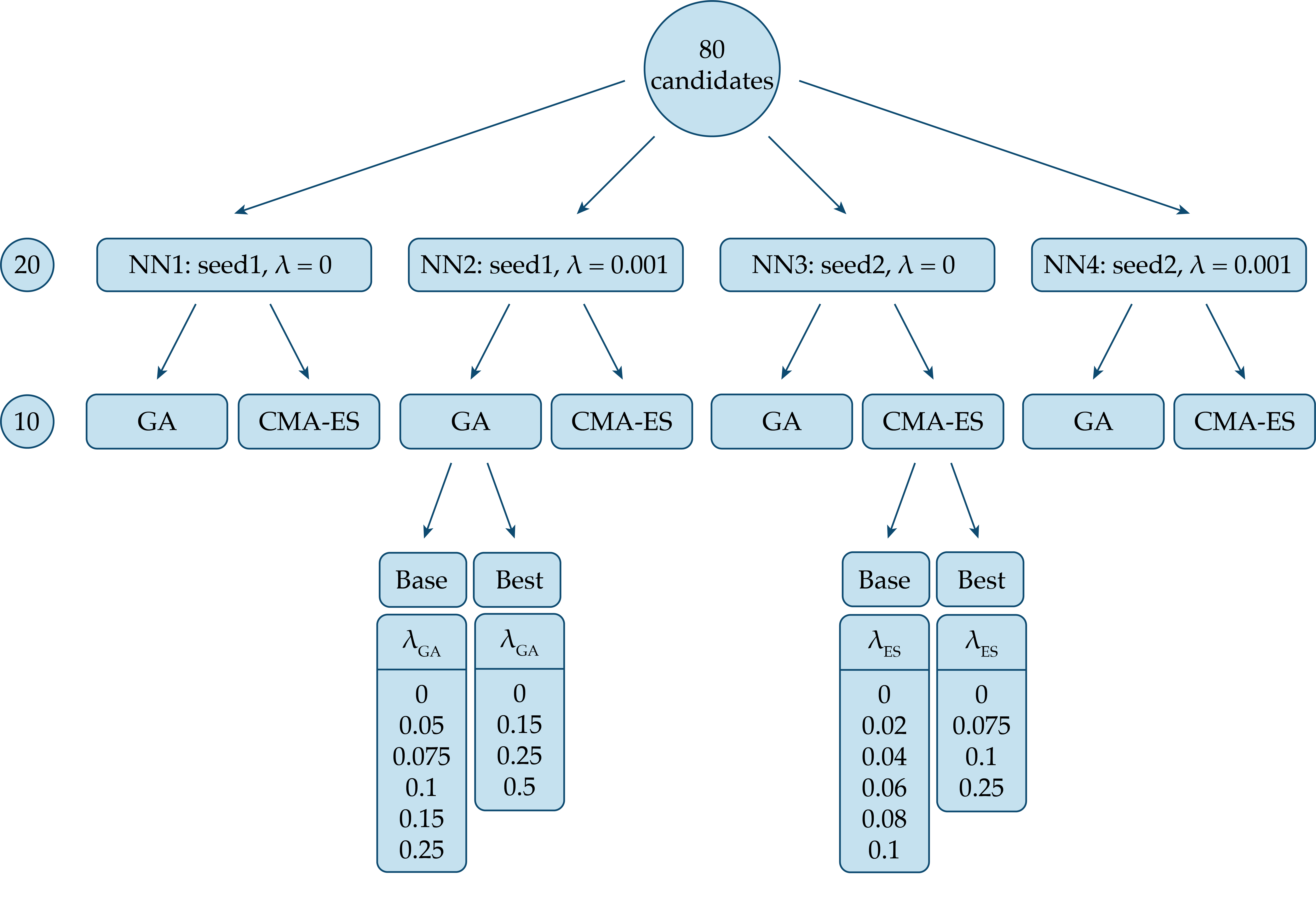}
        \caption{An overview of the family of candidate structures.}
        \label{Fig:Family_tree}
\end{figure}

The local and global optimization algorithms work collaboratively to generate a family of high-Q-factor candidate structures (see Figure \ref{Fig:Family_tree}), which are subsequently validated using FDTD simulations. This process results in a total of $80$ distinct nanobeam cavity designs. For each combination of base design and fabrication imperfection implementation, four NNs are trained. Among these NNs, two share the same random seed, and two share the same training L$^2$-regularization parameter $\lambda$. 
\par
From each NN, $20$ candidates are extracted, $10$ from each optimization algorithm. Of these 10 candidates, six are derived from an algorithm initialized with the base design $\vec{x_0} = \vec{0}$, while the remaining four originate from an algorithm initialized with the structure exhibiting the highest Q-factor in the dataset, referred to as the best structure.
\par
For the GA optimization algorithm, the six candidates derived from the "base" initialization were constrained by the following set of optimization L$^2$-regularization parameters: $\lambda_{\textrm{GA}} \in [0, 0.05, 0.075, 0.1, 0.15, 0.25]$. The other four candidates derived from the "best" initialization were constrained by a different set of L$^2$-regularization values: $\lambda_{\textrm{GA}} \in [0, 0.15, 0.25, 0.5]$. When $\lambda_{\textrm{GA}} = 0$, the optimization domain is limited to the parameter space defined by the datasets (see Tables \ref{tab:NC_parameter_space} and \ref{tab:FB_parameter_space}). 
\par
For the CMA-ES optimization algorithm, the six candidates derived from the "base" initialization were constrained by the following set of L$^2$-regularization values: $\lambda_{\textrm{ES}} \in [0, 0.02, 0.04, 0.06, 0.08, 0.1]$. Similarly, the four candidates derived from the "best" initialization were constrained by another set of L$^2$-regularization values: $\lambda_{\textrm{ES}} \in [0, 0.075, 0.1, 0.25]$. As with GA, when $\lambda_{\textrm{ES}} = 0$, the optimization domain is restricted to the parameter space of the datasets (see Tables \ref{tab:NC_parameter_space} and \ref{tab:FB_parameter_space}).

\hypertarget{supinfG2}{}
\subsection{G2. Optimization curves}
\begin{figure}[h!]
    \begin{subfigure}{0.49\textwidth}
        \centering
        \includegraphics[width=\linewidth]{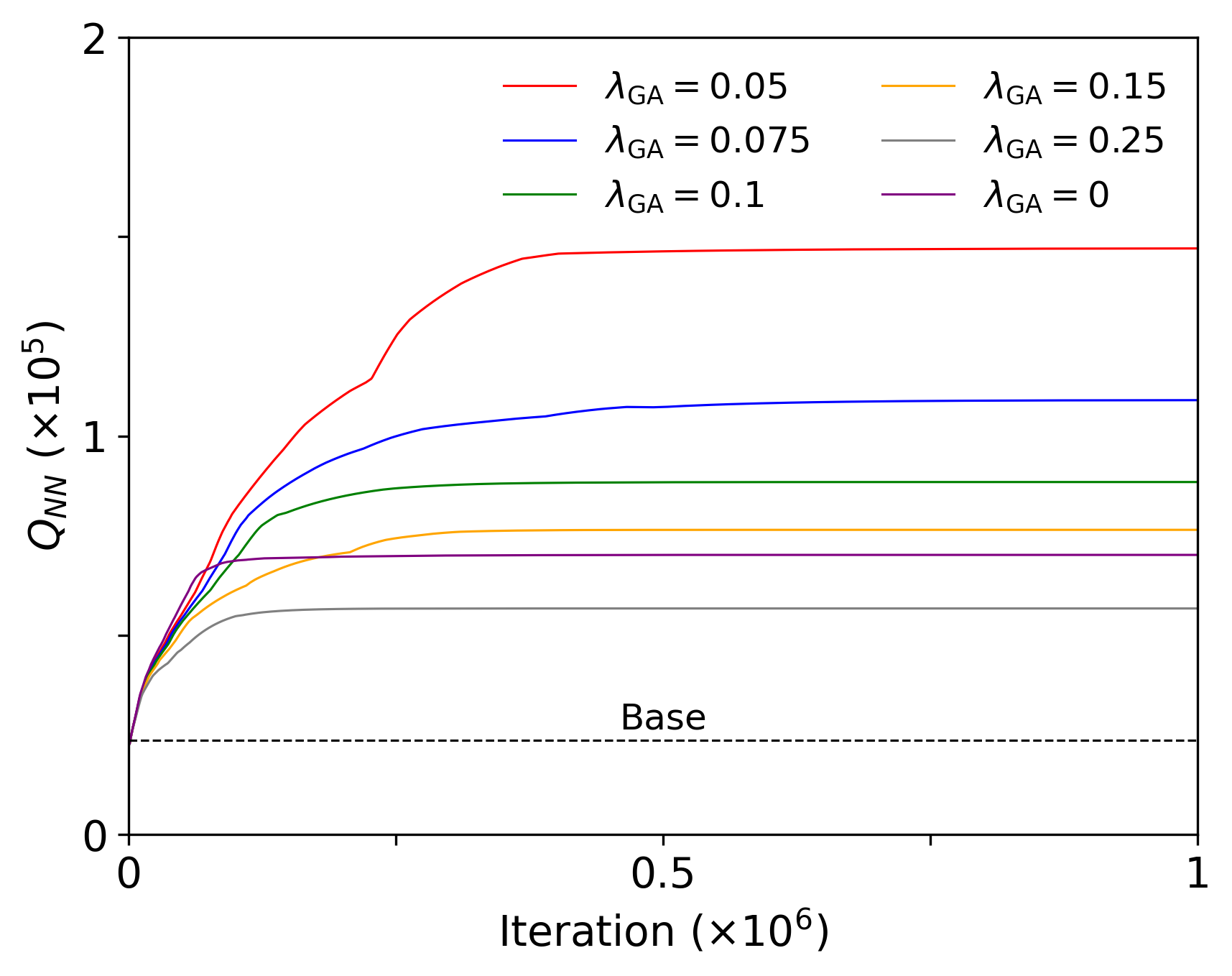}
        \caption{}
        \label{fig:GA_l0}
    \end{subfigure}%
    \begin{subfigure}{0.51\textwidth}
        \centering
        \includegraphics[width=\linewidth]{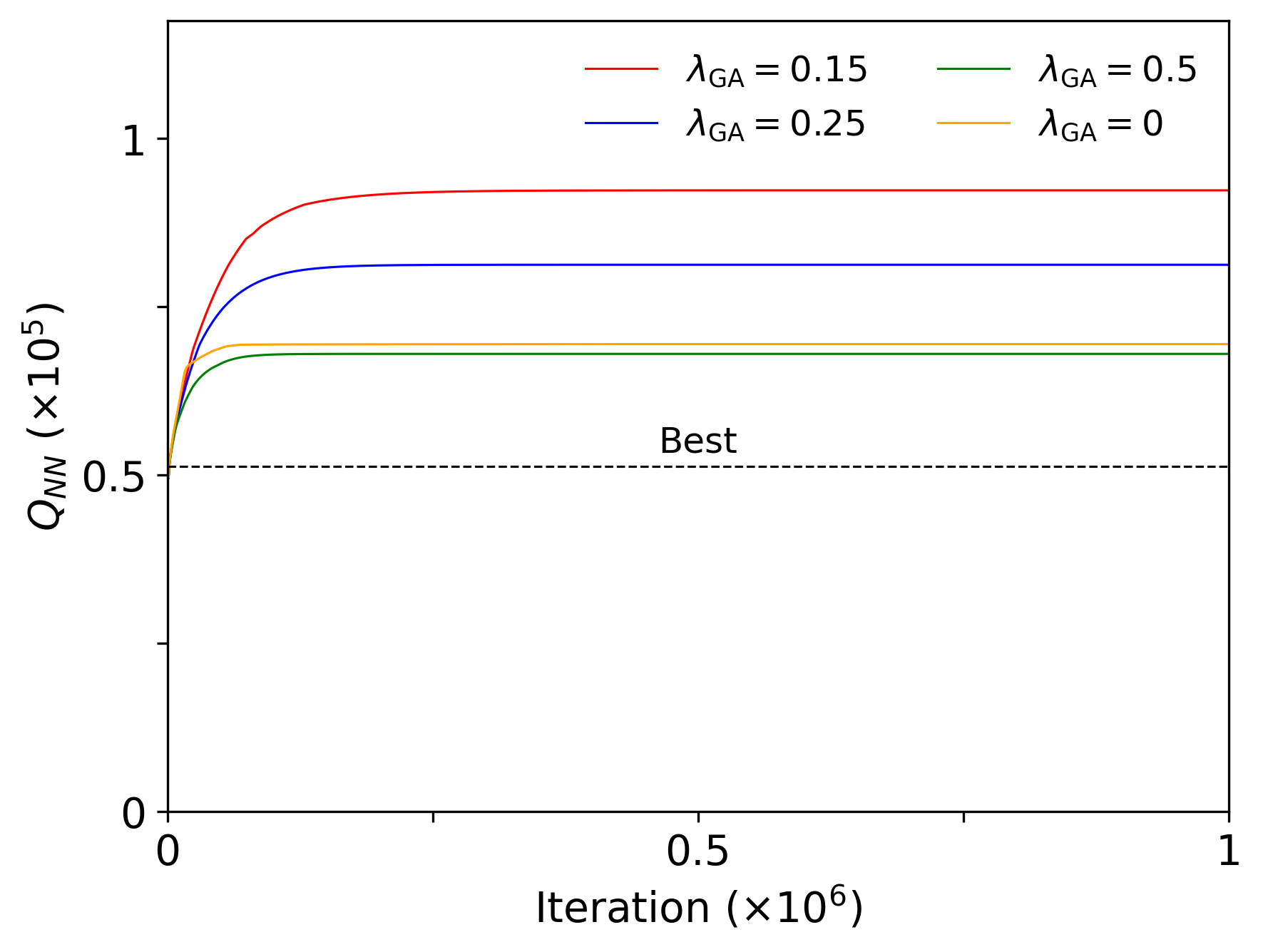}
        \caption{}
        \label{fig:GA_l0.001}
    \end{subfigure}
    
    \begin{subfigure}{0.495\textwidth}
        \centering
        \includegraphics[width=\linewidth]{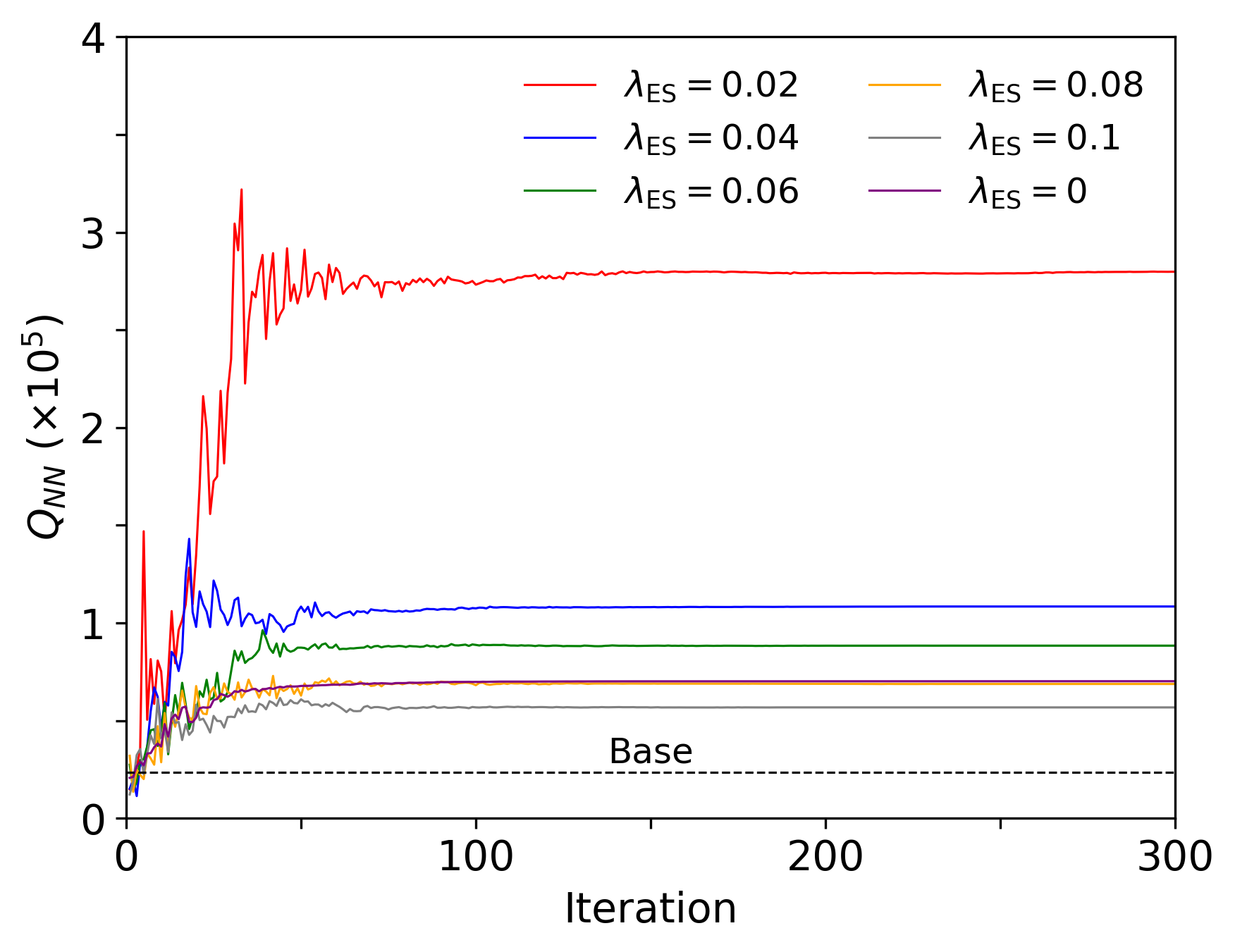}
        \caption{}
        \label{fig:CMA_ES_l0}
    \end{subfigure}%
    \begin{subfigure}{0.505\textwidth}
        \centering
        \includegraphics[width=\linewidth]{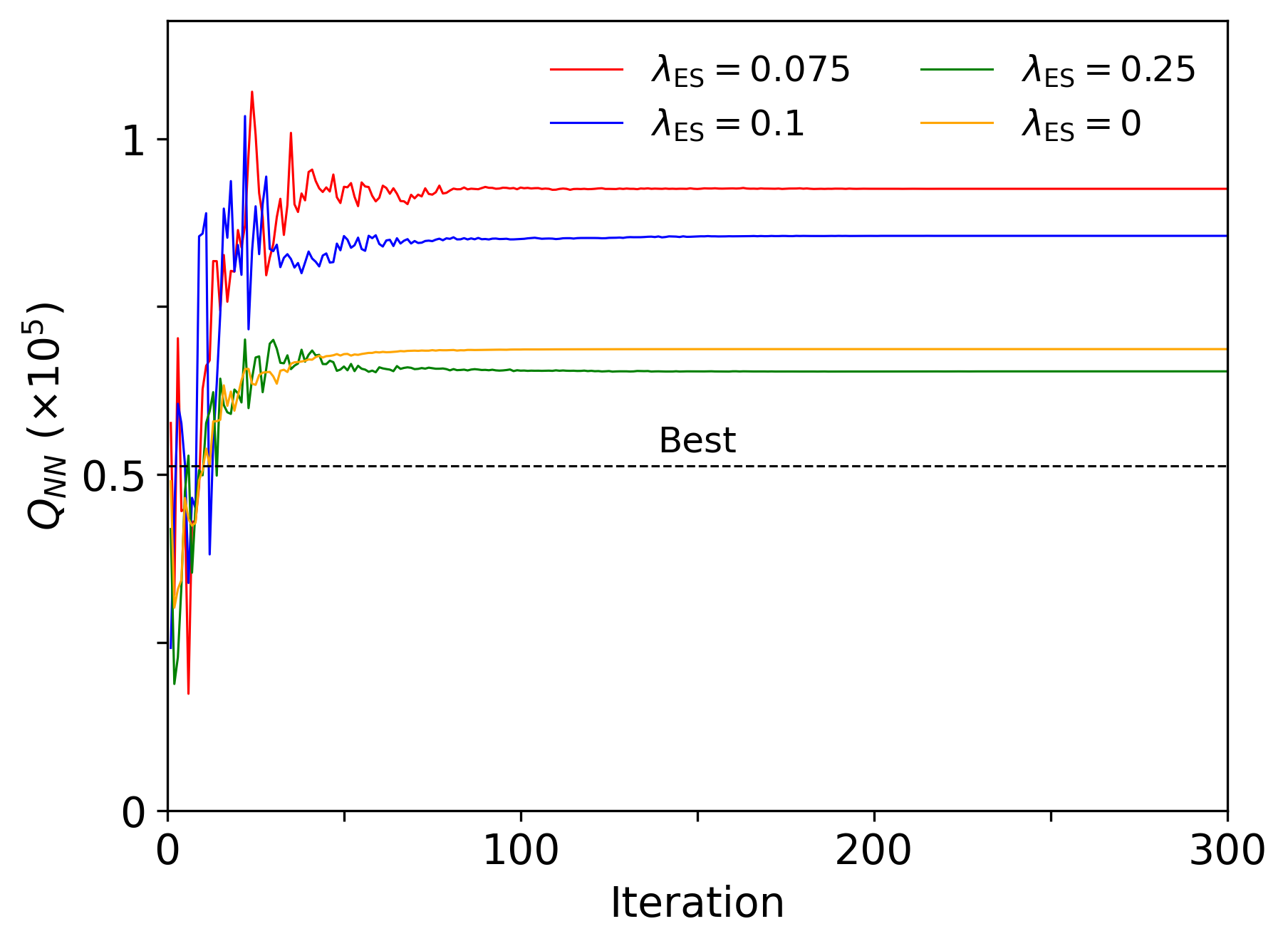}
        \caption{}
        \label{fig:CMA_ES_l0.001}
    \end{subfigure}
    
    \caption{The evolution of $Q_{\textrm{NN}}$ during optimization is illustrated through examples using various optimization algorithms, each with different parameter displacement constraints (i.e., L$^2$-regularization denoted by $\lambda_{\textrm{GA}}$ or $\lambda_{\textrm{ES}}$). \textbf{(a)} and \textbf{(b)} show the optimization curves using the local optimization algorithm (GA) with initial structures of "base" and "best," respectively. \textbf{(c)} and \textbf{(d)} show the optimization curves using the global optimization algorithm (CMA-ES) with initial structures of "base" and "best," respectively.}
    \label{fig:combined_optimization_curves}
\end{figure}

During the optimization process, the Q-factor converges to an optimized value. This convergence is demonstrated in Figure \ref{fig:combined_optimization_curves}, which presents examples of the GA and CMA-ES algorithms. Both graphs reveal a clear relationship between the degree of L$^2$-regularization, determined by $\lambda_{\textrm{GA}}$ and $\lambda_{\textrm{ES}}$, and the optimization result. 
\par
A higher L$^2$-regularization imposes stronger constraints on the design parameters, effectively limiting the parameter space available for exploration. As a result, the optimization process produces a lower predicted Q-factor under these more restrictive conditions. This indicates that while L$^2$-regularization enhances control over the optimization domain, it can also hinder the ability to achieve higher Q-factor predictions.

However, a higher predicted Q-factor, $Q_{\textrm{NN}}$, does not necessarily correlate with a higher simulated Q-factor, $Q_{\textrm{FDTD}}$, due to increasing prediction inaccuracies as the design parameters move further away from the training parameter space. To reduce this limitation, it is essential to incrementally adjust the degree of parameter restriction during optimization.
\par
The optimization sequences illustrated in Figures \ref{fig:GA_l0} and \ref{fig:CMA_ES_l0} are initialized from the base design, while those in Figures \ref{fig:GA_l0.001} and \ref{fig:CMA_ES_l0.001} start from the best design in the generated training datasets. The application of L$^2$-regularization with varying reference points alters the optimization landscape, resulting in different optimized Q-factors.
\par
The GA optimization algorithm typically requires significantly more iterations to achieve convergence compared to the CMA-ES algorithm, resulting in a longer total computation time. Although each GA iteration is computationally faster than CMA-ES, the total duration is longer, with CMA-ES completing in approximately $1.5$ seconds compared to $450$ seconds for GA. Despite its slower convergence, the GA algorithm remains practical and effective for this application.

\hypertarget{supinfG3}{}
\subsection{G3. Results of optimization}
The Q-factors predicted by the NNs for the $80$ candidate structures are validated using FDTD simulations. Based on these validated results, the best-performing device is selected for each fabrication error implementation. The best performing device is determined as the one with the highest Q-factor. An overview of the combinations of training L$^2$-regularization, optimization L$^2$-regularization, optimization algorithm, and initial structure that resulted in the optimized predicted Q-factors ($Q_{\textrm{NN}}$), simulated Q-factors ($Q_{\textrm{FDTD}}$), and their corresponding prediction errors is presented in Table \ref{tab:CMA-ES_or_GA}.

\begin{table}[h!]
\centering
\caption{An overview indicating which NN, in combination with which optimization algorithm, led to the discovery of the optimized device, along with the corresponding prediction error.}
\label{tab:CMA-ES_or_GA}
\resizebox{\textwidth}{!}{%
\begin{tabular}{c|cccccccc}
\hline
   &          & $\boldsymbol{\lambda}$ & \begin{tabular}[c]{@{}c@{}}\textbf{Optimization}\\ \textbf{algorithm}\end{tabular} & \begin{tabular}[c]{@{}c@{}}\textbf{Initial}\\ \textbf{structure}\end{tabular} & \begin{tabular}[c]{@{}c@{}}$\boldsymbol{\lambda_{\textrm{GA}}}$ \textbf{or} $\boldsymbol{\lambda_{\textrm{ES}}}$\end{tabular} & $\mathbf{Q_{\textrm{NN}}}\, (\times10^3)$ & $\mathbf{Q_{\textrm{FDTD}}}\, (\times10^3)$ & \begin{tabular}[c]{@{}c@{}}\textbf{Prediction}\\ \textbf{error} $(\%)$\end{tabular} \\ \hline
L2 & \textbf{Cavity 1} & 0         & GA                                                               & base                                                        & 0.1                                                                        & 65                    & 108                     & 39.35                                                             \\ \cline{2-9} 
   & \textbf{Cavity 2} & 0.001     & CMA-ES                                                           & best                                                        & 0                                                                          & 36                    & $50$              & 28.11                                                             \\ \cline{2-9} 
   & \textbf{Cavity 3} & 0         & GA                                                               & best                                                        & 0.15                                                                       & 16                    & 20                      & 18.09                                                             \\ \hline
FB & \textbf{Cavity 1} & 0.001     & CMA-ES                                                           & base                                                        & 0.04                                                                       & 24                    & 56                      & 57.41                                                             \\ \cline{2-9} 
   & \textbf{Cavity 2} & 0.001     & GA                                                               & base                                                        & 0.15                                                                       & 11                    & 17                      & 34.02                                                             \\ \cline{2-9} 
   & \textbf{Cavity 3} & 0.001     & GA                                                               & base                                                        & 0.1                                                                        & 34                    & 108                     & 68.19                                                             \\ \hline
\end{tabular}%
}
\end{table}

For ideal devices and those with sidewall slant, this selection process is relatively straightforward due to their deterministic nature, requiring only a single simulation per candidate. However, for devices affected by surface roughness, the selection process is more complex and requires additional simulations. 
\par
Initially, the Q-factors of the $80$ candidates are simulated under three distinct random roughness configurations. From these results, $10$ candidates are selected based on their average Q-factors and standard deviations. The Q-factors of these $10$ structures are then simulated with a total of $30$ different random roughness seeds to ensure a more reliable comparison. Finally, the top-performing structure under surface roughness conditions is determined. This structure is compared to the best-performing ideal structure by simulating $100$ different roughness configurations for each, providing a comprehensive evaluation of their robustness and performance while dealing with surface roughness.

\hypertarget{supinfG4}{}
\subsection{G4. Optimized L2 cavities}
The three identified L2 nanobeam cavities are characterized by the design parameters listed in Table \ref{tab:NC_cavities_optim}. Each cavity demonstrates distinct design parameter values, underlining their uniqueness. Notable trends include the shifts in positions $x_3$, $x_4$, and $x_5$ towards the cavity region, as well as a preference for increasing $D_3$, $D_4$, and $D_5$. The most prominent observation, however, is the significant reduction in the major diameter of the holes $D_m$ within the mirror region. The optical properties of the three optimized L2 cavities are compared under ideal, surface roughness, and sidewall slant conditions in Table \ref{tab:results_L2}.

\begin{table}[h!]
\centering
\caption{An overview of the design parameter displacements of the L2 nanobeam cavity optimized under ideal (cavity 1), surface roughness (cavity 2) and sidewall slant (cavity 3) conditions. All values are in nanometers (nm).}
\label{tab:NC_cavities_optim}
\begin{tabular}{c|c|c|c|}
      & \textbf{Cavity 1} & \textbf{Cavity 2} & \textbf{Cavity 3} \\ \hline
$\mathit{x_1}$ & 8.4      & 5.2      & -5.0     \\
$\mathit{x_2}$ & 4.0      & 2.0      & -3.8     \\
$\mathit{x_3}$ & -3.6     & -7.5     & -3.6     \\
$\mathit{x_4}$ & -8.9     & -10.0    & -6.3     \\
$\mathit{x_5}$ & -3.9     & -4.5     & -7.0     \\
$\mathit{x_m}$ & 4.0      & 8.0      & -2.9     \\
$\mathit{D_1}$ & -1.5     & 1.6      & -6.3     \\
$\mathit{D_2}$ & -5.6     & 5.9     & -4.5     \\
$\mathit{D_3}$ & 3.5      & 4.9      & 6.9      \\
$\mathit{D_4}$ & 7.4      & 7.5      & 2.8      \\
$\mathit{D_5}$ & 5.5      & 7.2      & 2.6      \\
$\mathit{D_m}$ & -14.8    & -7.0     & -12.8    \\
$\mathit{d}$   & -3.9     & 1.2     & 5.7      \\ 
\end{tabular}
\end{table}

L2 cavity 2 is optimized for surface roughness and is characterized by an increase in the major diameters $D_i$, with the exception of $D_m$. The average Q-factor of this structure under surface roughness, predicted by the NN, is $Q_{\textrm{NN}} = 3.610 \times 10^4$, while the FDTD simulation yields $Q_{\textrm{FDTD}} = 5.025 \times 10^4$. This results in a relative prediction error of approximately $\epsilon_{\textrm{pred}} = 28.11\%$ (see Table \ref{tab:CMA-ES_or_GA}). This optimized design was obtained using a NN with L$^2$-regularization on its weights and biases ($\lambda = 0.001$) and optimized with the CMA-ES algorithm, initialized at the best structure of the training data without L$^2$-regularization on the design parameters ($\lambda_{\textrm{ES}} = 0$).

\begin{figure}[h!]
    \centering
    \includegraphics[width=0.75\textwidth]{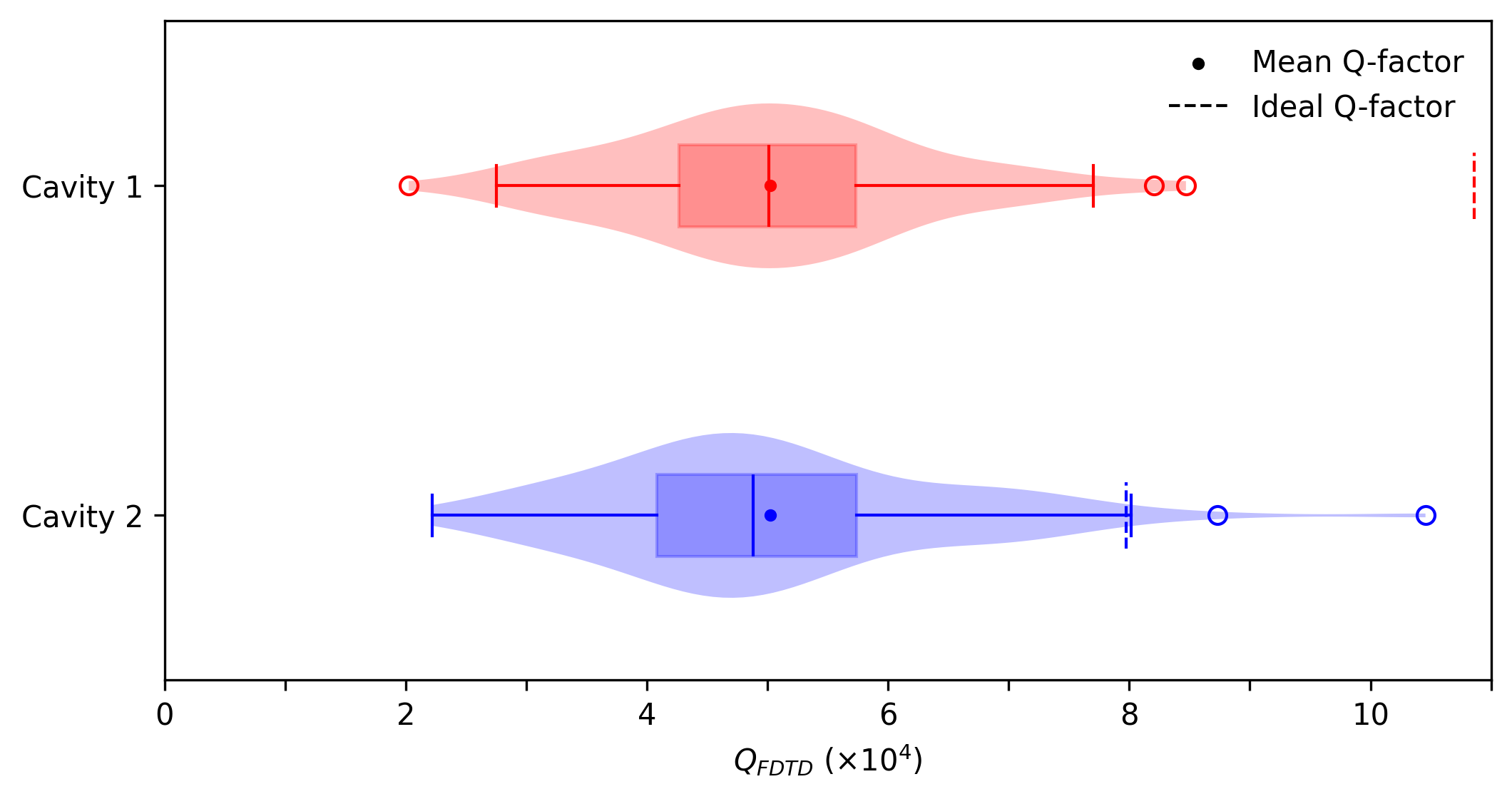}
    \caption{The distribution of Q-factors of L2 cavity 1 and cavity 2 under different roughness configurations plotted as a box plot in combination with a violin plot. The Q-factor of both structures without surface roughness are denoted by a dotted line.}
    \label{fig:NC_boxplot_roughness}
\end{figure}

In Figure \ref{fig:NC_boxplot_roughness}, the distribution of Q-factors for L2 cavities 1 and 2 under different surface roughness configurations (random seeds) is visualized. Under ideal conditions (denoted by the dotted line), cavity 2 has a lower Q-factor ($Q_{\textrm{ideal}} = 7.977 \times 10^4$) compared to cavity 1 ($Q_{\textrm{ideal}} = 1.076 \times 10^5$). However, when surface roughness is applied, cavity 2 ($Q_{\textrm{rough}} = 5.021 \times 10^4$) performs similarly to cavity 1 ($Q_{\textrm{rough}} = 5.025 \times 10^4$). As a result, under surface roughness, cavity 2 experiences a $16.24\%$ smaller average degradation compared to cavity 1, making it more robust to the fabrication imperfection of surface roughness.

\begin{table}[h!]
\centering
\caption{An overview of the optical properties of L2 cavity 1, cavity 2 and cavity 3 under ideal, surface roughness and sidewall slant conditions.}
\label{tab:results_L2}
\begin{tabular}{c|cccc}
\hline
      &          & \textbf{Wavelength} (nm) & \textbf{Q-factor} ($\times 10^3$) & \textbf{Mode volume} $((\lambda / n)^3)$ \\ \hline
\textbf{Ideal} & \textbf{Cavity 1} & 640.4           & 108                     & $1.0$                 \\ \cline{2-5} 
      & \textbf{Cavity 2} &     638.1       & 80                      &   1.0  \\ \cline{2-5} 
      & \textbf{Cavity 3} & 637.0           & 32                      & $0.8$                 \\ \hline
\textbf{Rough} & \textbf{Cavity 1} & 639.9 & $50$            & 1.0                 \\ \cline{2-5} 
      & \textbf{Cavity 2} & 637.9 & $50$            & 0.9                 \\ \cline{2-5} 
      & \textbf{Cavity 3} & 636.8  & 29                     &   0.8               \\ \hline
\textbf{Slant} & \textbf{Cavity 1} & 649.7           & 9                      & $1.0$                 \\ \cline{2-5} 
      & \textbf{Cavity 2} &      648.0         &     10                   &    1.0              \\ \cline{2-5} 
      & \textbf{Cavity 3} & 646.7           & 20                      & $0.9$                 \\ \hline
\end{tabular}
\end{table}

\hypertarget{supinfG5}{}
\subsection{G5. Optimized fishbone cavities}
The three identified fishbone nanobeam cavities are characterized by the design parameters listed in Table \ref{tab:FB_cavities_optim}. Each cavity demonstrates distinct design parameter values, underlining their uniqueness. Notable trends include a preference for an increase in most of the hole diameters $d_i$, with the exception of cavity 2. Additionally, the fin lengths $l_i$ tend to decrease for cavity 1, whereas an opposite trend is observed for cavity 3. The most prominent observation, however, is the reduction in the distance between the holes $\underline{w}$ on transversely opposing sides. The optical properties of the three optimized L2 cavities are compared under ideal, surface roughness, and sidewall slant conditions in Table \ref{tab:results_FB}.

\begin{table}[h!]
\centering
\caption{An overview of the design parameter displacements of the fishbone nanobeam cavity optimized under ideal (cavity 1), surface roughness (cavity 2) and sidewall slant (cavity 3) conditions. All values are in nanometers (nm).}
\label{tab:FB_cavities_optim}
\begin{tabular}{c|c|c|c|}
      & \textbf{Cavity 1} & \textbf{Cavity 2} & \textbf{Cavity 3} \\ \hline
$\mathit{d_1}$ & 1.2      & 1.0      & 1.0      \\
$\mathit{d_2}$ & 0.5      & -1.7      & 1.4      \\
$\mathit{d_3}$ & 3.0      & -3.1      & 3.5      \\
$\mathit{d_4}$ & 4.2      & -1.4      & 2.1      \\
$\mathit{d_5}$ & 3.6      & 4.2      & 1.4      \\
$\mathit{d_6}$ & 7.0      & 2.8      & -1.6     \\
$\mathit{d_m}$ & 6.7      & 4.1      & -0.3     \\
$\mathit{l_1}$ & -3.1     & 0.7     & 6.6      \\
$\mathit{l_2}$ & -5.3     & -3.4     & 4.9      \\
$\mathit{l_3}$ & -3.6     & -3.9     & 6.0      \\
$\mathit{l_4}$ & -3.5     & -2.3     & 5.5      \\
$\mathit{l_5}$ & -1.5     & 1.5     & 4.6      \\
$\mathit{l_6}$ & -2.0     & 1.6      & 1.7      \\
$\mathit{l_7}$ & 2.9      & 3.1      & 1.1      \\
$\mathit{l_m}$ & 4.7      & 1.9      & -1.0     \\
$\mathit{\underline{w}}$   & -0.1     & -4.4     & -2.6     \\
\end{tabular}
\end{table}

Fishbone cavity 3 is optimized against sidewall slant and is characterized by an increase in most of the fin lengths $l_i$. The average Q-factor of this structure under sidewall slant, predicted by the NN, is $Q_{\textrm{NN}} = 3.430 \times 10^4$, while the FDTD simulation yields $Q_{\textrm{FDTD}} = 1.078 \times 10^5$. This results in a relative prediction error of approximately $\epsilon_{\textrm{pred}} = 68.19\%$ (see Table \ref{tab:CMA-ES_or_GA}). This optimized design was obtained using a NN with L$^2$-regularization on its weights and biases ($\lambda = 0.001$) and optimized with the GA algorithm, initialized at the base design with L$^2$-regularization on the design parameters ($\lambda_{\textrm{GA}} = 0.10$).

\begin{figure}[h!]
    \centering
    \includegraphics[width=0.75\textwidth]{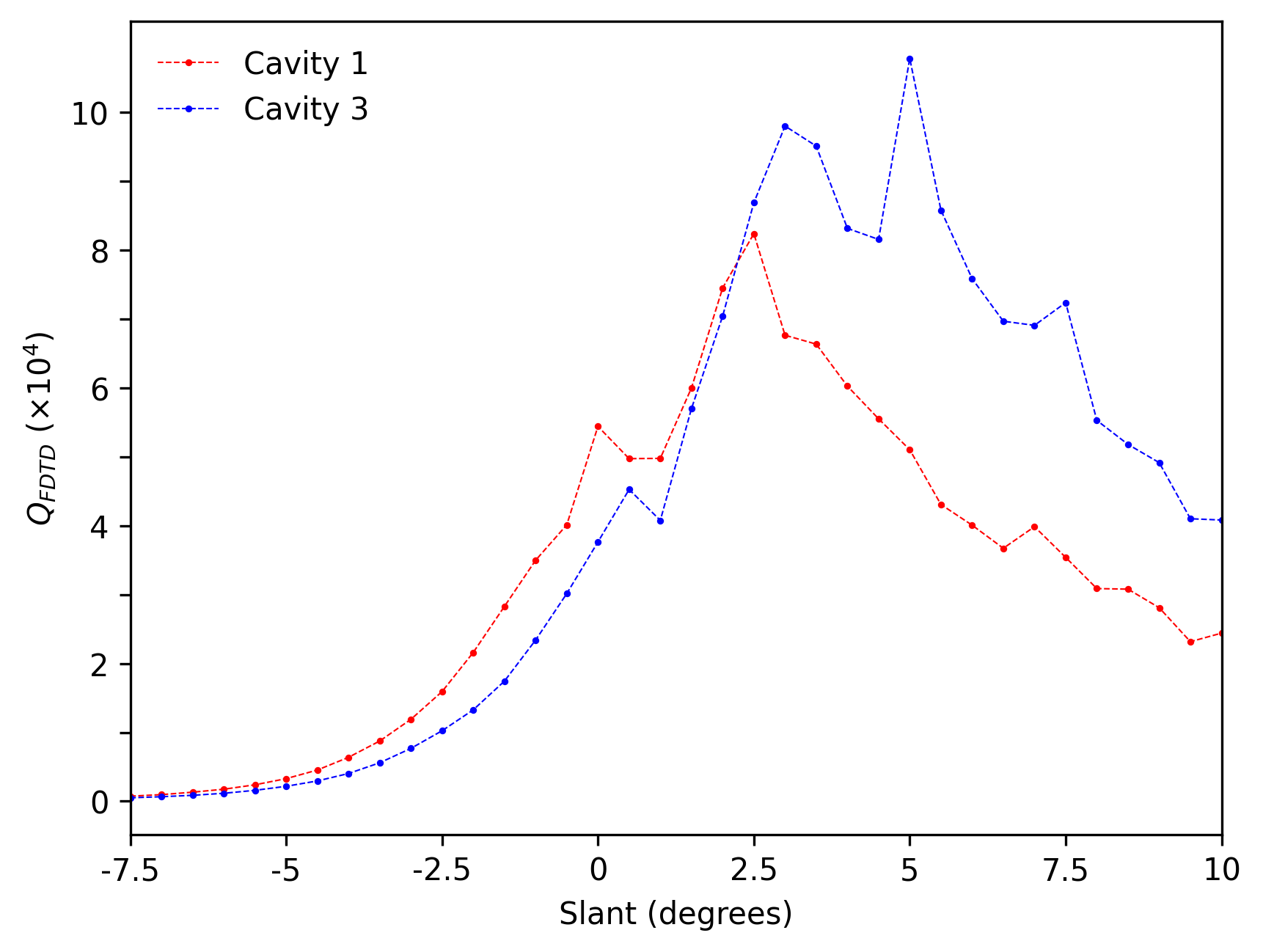}
    \caption{A graph presenting the Q-factor of fishbone cavity 1 and cavity 3 under various slant angles.}
    \label{fig:FB_bestclean_vs_bestslant}
\end{figure}

In Figure \ref{fig:FB_bestclean_vs_bestslant}, the Q-factors of fishbone cavities 1 and 3 are compared under different angles of sidewall slant. Despite cavity 3 initially having a lower Q-factor ($Q_{\textrm{ideal}} = 3.765 \times 10^4$) compared to cavity 1 ($Q_{\textrm{ideal}} = 5.629 \times 10^4$) under $0^{\circ}$ slant, it begins to outperform cavity 1 at slant angles exceeding $2.5^{\circ}$. Cavity 3 was specifically optimized against a $5^{\circ}$ slant. At this angle, the Q-factor of cavity 1 degrades to $Q_{\textrm{slant}} = 5.104 \times 10^4$, while cavity 3 shows a higher Q-factor of $Q_{\textrm{slant}} = 1.078 \times 10^5$. As a result, under sidewall slant, cavity 3 achieves more than twice the performance of cavity 1. 
\par
However, does this imply that cavity 3 is inherently more robust against fabrication errors, or has its ideal configuration shifted to a model with a $5^\circ$ slant? Figure \ref{fig:FB_bestclean_vs_bestslant} indicates the latter: the ideal scenario for cavity 3 occurs around a $5^\circ$ sidewall slant, whereas for cavity 1, it is closer to a $2.5^\circ$ slant. To assess fabrication error tolerance, it is essential to evaluate how the Q-factor of each device degrades relative to its ideal scenario. As shown in Figure \ref{fig:FB_bestclean_vs_bestslant}, the Q-factor of cavity 3 decreases slightly faster than that of cavity 1 as deviations from their respective ideal scenarios increase. Nonetheless, this does not exclude cavity 3 as a viable solution for reducing fabrication imperfections, particularly in cases where unavoidable sidewall slants exceed $2.5^\circ$.

\newpage

\begin{table}[ht!]
\centering
\caption{An overview of the optical properties of fishbone cavity 1, cavity 2 and cavity 3 under ideal, surface roughness and sidewall slant conditions.}
\label{tab:results_FB}
\begin{tabular}{c|cccc}
\hline
      &          & \textbf{Wavelength} (nm) & \textbf{Q-factor} ($\times 10^3$) & \textbf{Mode volume} $((\lambda / n)^3)$ \\ \hline
\textbf{Ideal} & \textbf{Cavity 1} & 541.0           & 56                     & $1.4$                 \\ \cline{2-5} 
      & \textbf{Cavity 2} & 542.5           & 42                      & 1.4                 \\ \cline{2-5} 
      & \textbf{Cavity 3} & 548.7           & 39                      & $1.3$                 \\ \hline
\textbf{Rough} & \textbf{Cavity 1} &  539.7            & 13   & 1.4              \\ \cline{2-5} 
      & \textbf{Cavity 2} & 541.6            & 17        & 1.4        \\ \cline{2-5} 
      & \textbf{Cavity 3} & 547.1              & 11                       & 1.4                 \\ \hline
\textbf{Slant} & \textbf{Cavity 1} & 511.7           & 51                      & $1.3$                 \\ \cline{2-5} 
      & \textbf{Cavity 2} & 514.1              & 19                       & 1.3                 \\ \cline{2-5} 
      & \textbf{Cavity 3} & 519.0           & 108                      & $1.3$                 \\ \hline
\end{tabular}
\end{table}

\newpage

%% file: H.tex
\hypertarget{supinfH}{}
\section{\fontsize{24}{24}\selectfont Appendix}

\begin{figure}[h!]
    \centering
    \begin{subfigure}{0.495\textwidth}
        \centering
        \includegraphics[width=\linewidth]{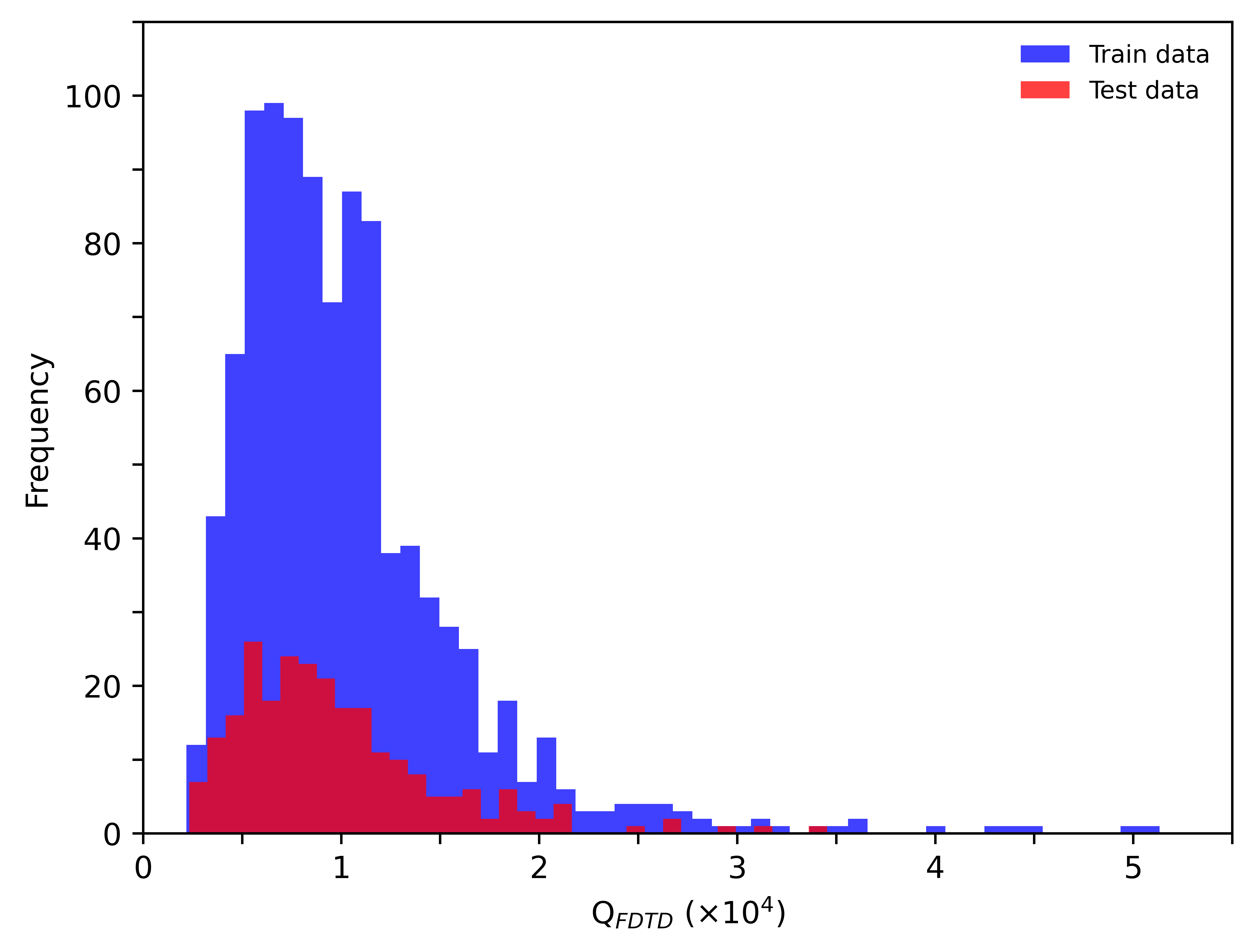}
        \caption{NN1: raw}
        \label{fig:NC123_dataset_clean_split_pross}
    \end{subfigure}
    \begin{subfigure}{0.495\textwidth}
        \centering
        \includegraphics[width=\linewidth]{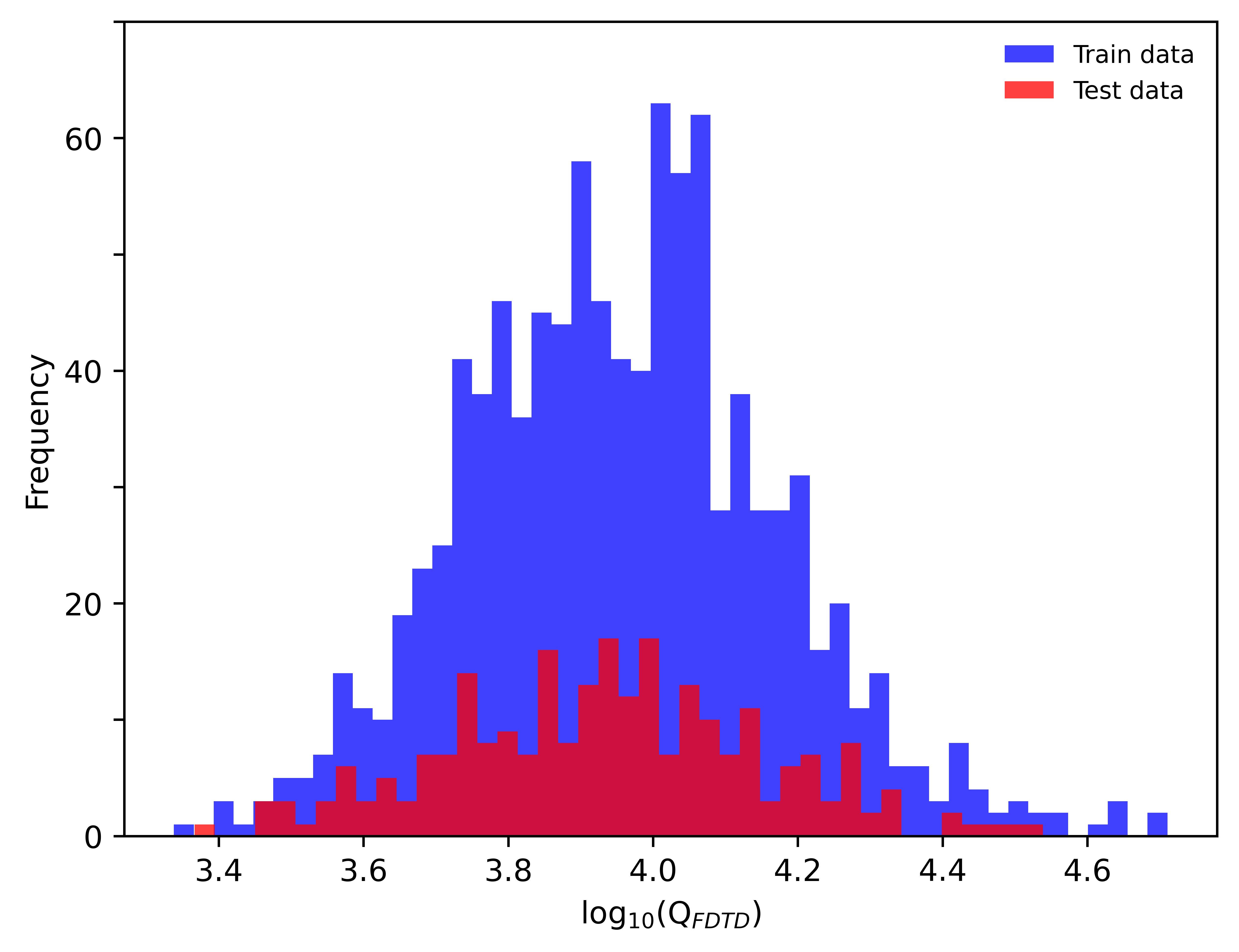}
        \caption{NN1: processed}
        \label{fig:NC123_dataset_clean_split}
    \end{subfigure}
    
    \begin{subfigure}{0.495\textwidth}
        \centering
        \includegraphics[width=\linewidth]{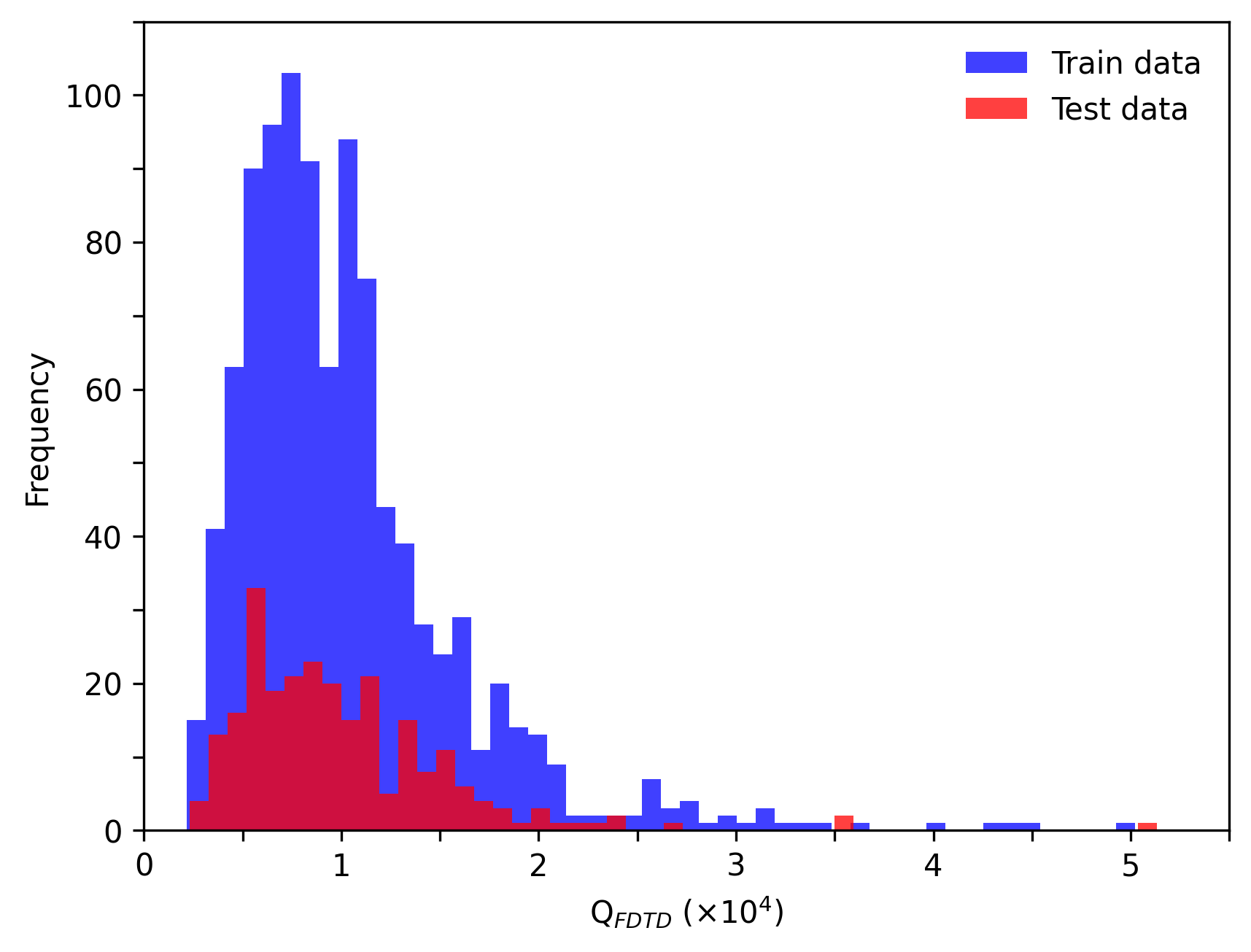}
        \caption{\textbf{NN2}: raw}
        \label{fig:NC21_dataset_clean_split}
    \end{subfigure}
    \begin{subfigure}{0.495\textwidth}
        \centering
        \includegraphics[width=\linewidth]{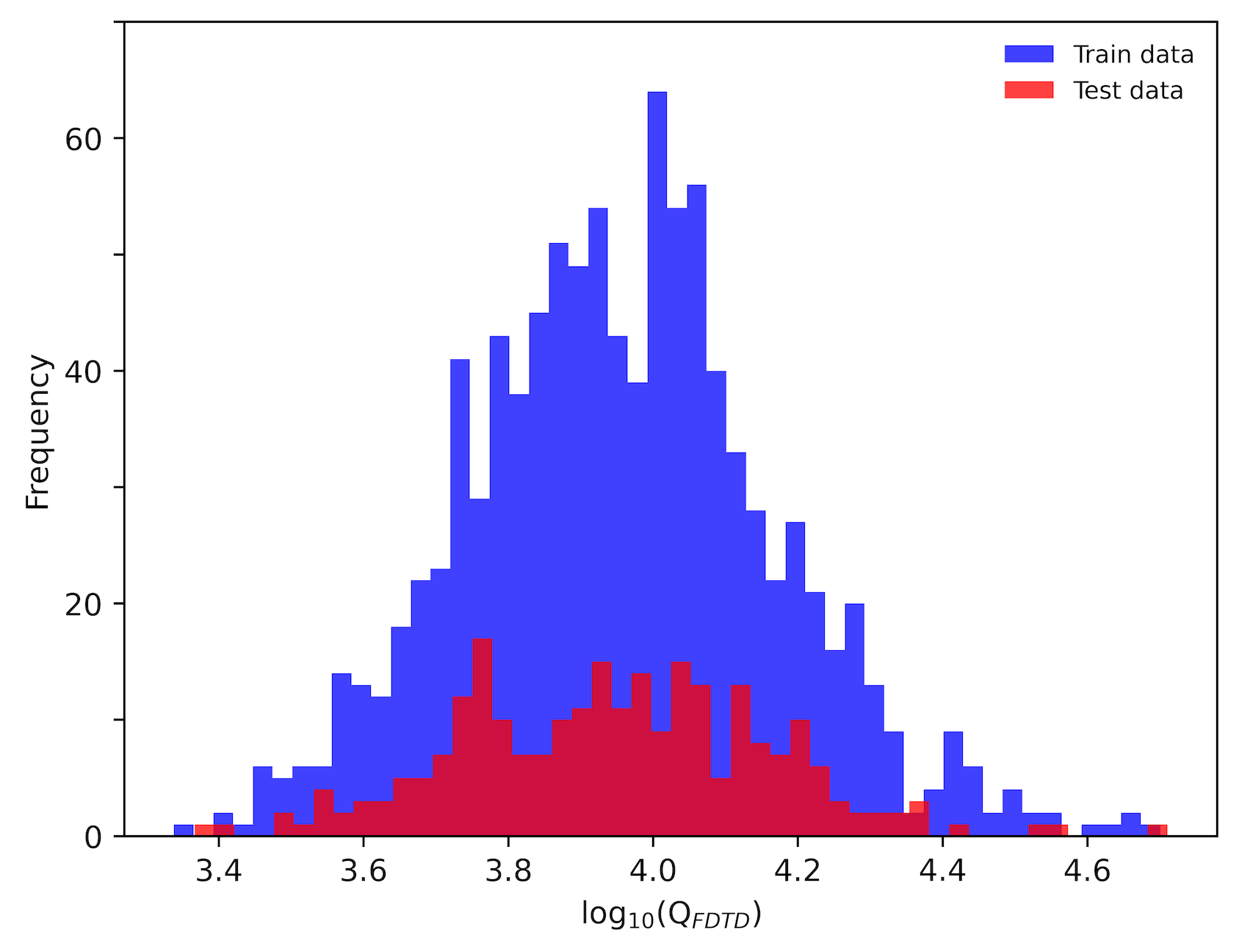}
        \caption{\textbf{NN2}: processed}
        \label{fig:NC21_dataset_clean_split_pross}
    \end{subfigure}
    \caption{data split, L2 without imperfections (NN (bold) responsible for optimizing L2 cavity 1)}
    \label{fig:NC_dataset_clean_split}
\end{figure}
\newpage

\begin{figure}[h!]
    \centering
    \begin{subfigure}{0.495\textwidth}
        \centering
        \includegraphics[width=\linewidth]{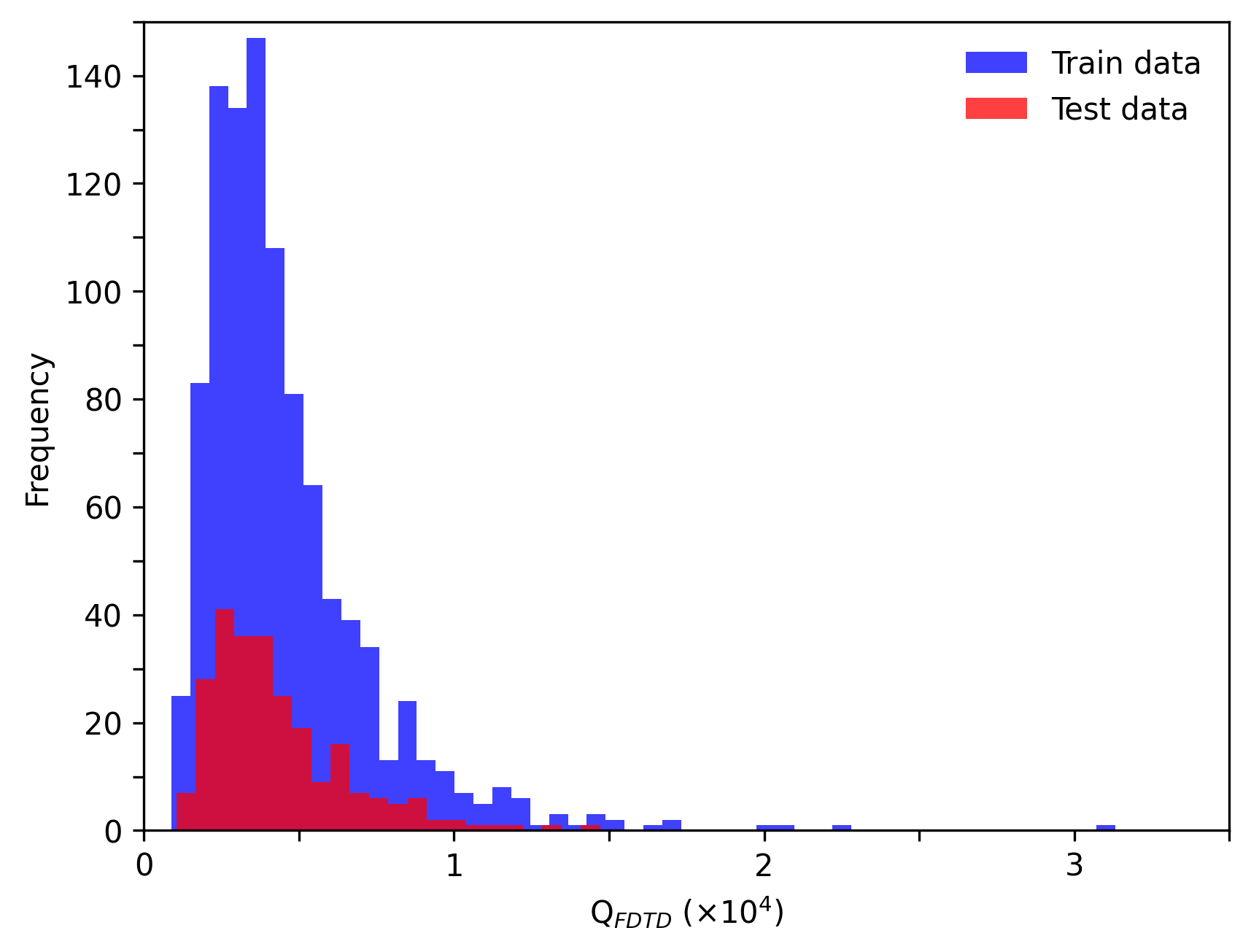}
        \caption{NN1: raw}
        \label{fig:FB32_dataset_clean_split_pross}
    \end{subfigure}
    \begin{subfigure}{0.495\textwidth}
        \centering
        \includegraphics[width=\linewidth]{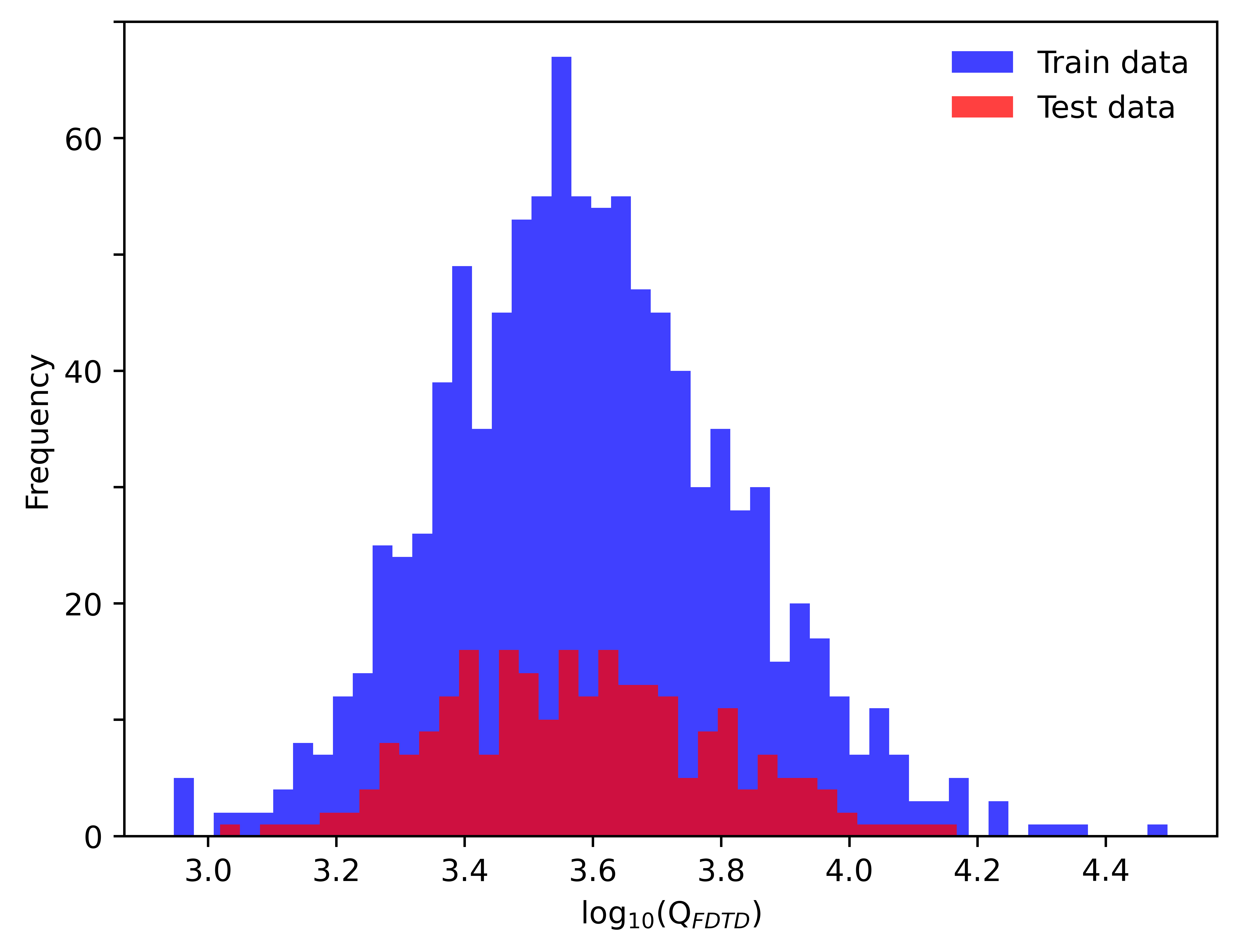}
        \caption{NN1: processed}
        \label{fig:FB32_dataset_clean_split}
    \end{subfigure}
    
    \begin{subfigure}{0.495\textwidth}
        \centering
        \includegraphics[width=\linewidth]{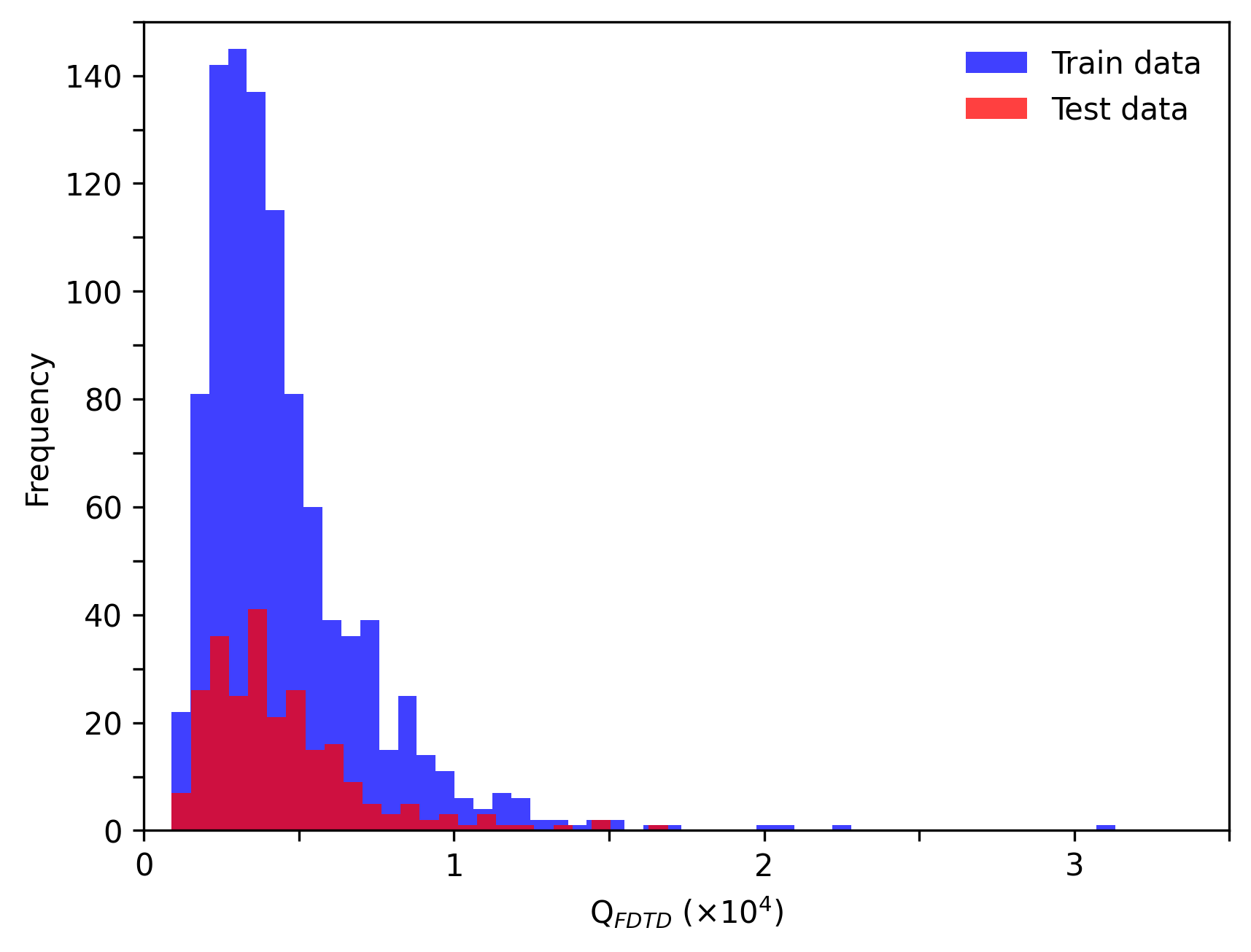}
        \caption{\textbf{NN2}: raw}
        \label{fig:FB212_dataset_clean_split}
    \end{subfigure}
    \begin{subfigure}{0.495\textwidth}
        \centering
        \includegraphics[width=\linewidth]{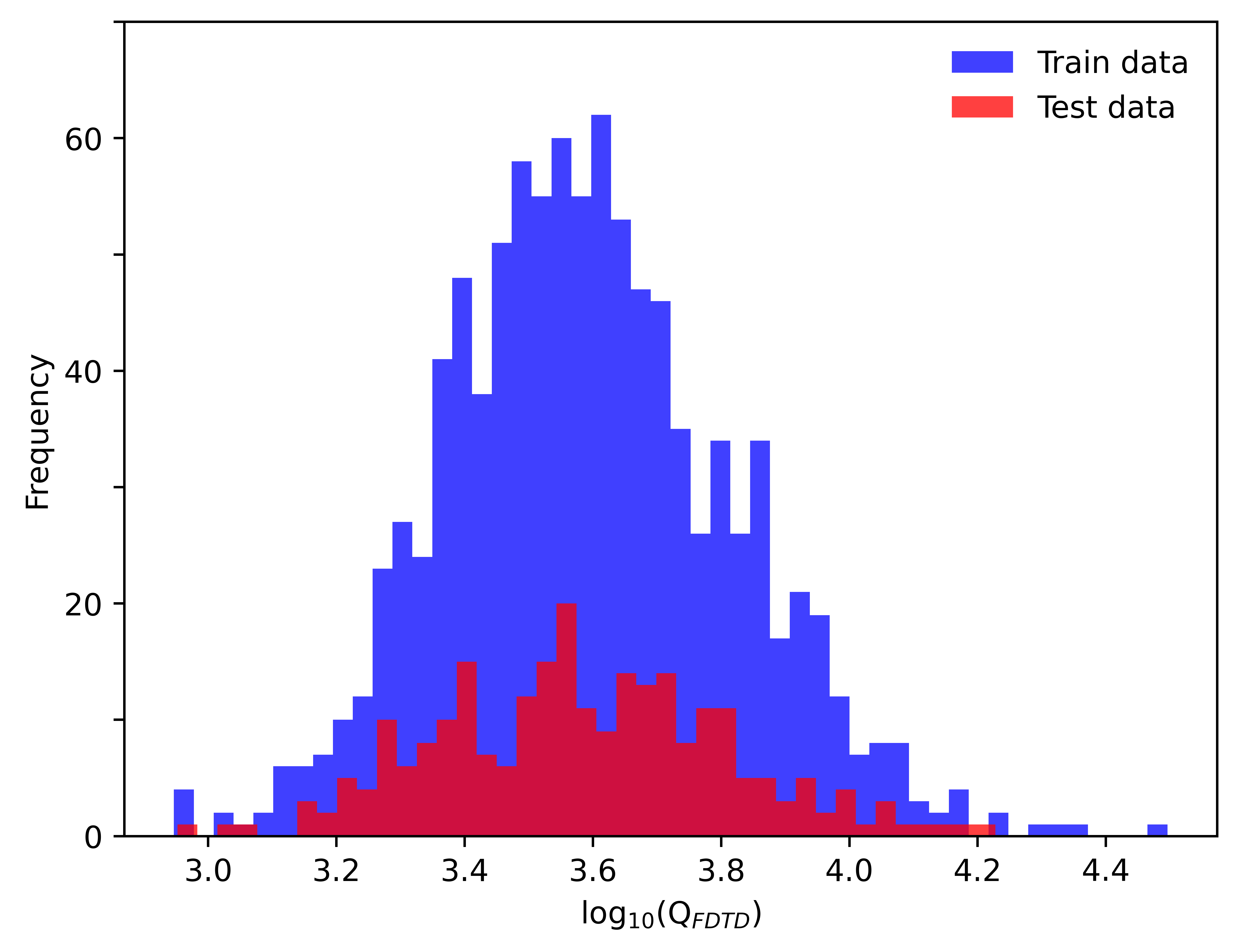}
        \caption{\textbf{NN2}: processed}
        \label{fig:FB212_dataset_clean_split_pross}
    \end{subfigure}
    \caption{data split, fishbone without imperfections (NN (bold) responsible for optimizing fishbone cavity 1)}
    \label{fig:FB_dataset_clean_split}
\end{figure}
\newpage

\begin{figure}[h!]
    \centering
    \begin{subfigure}{0.495\textwidth}
        \centering
        \includegraphics[width=\linewidth]{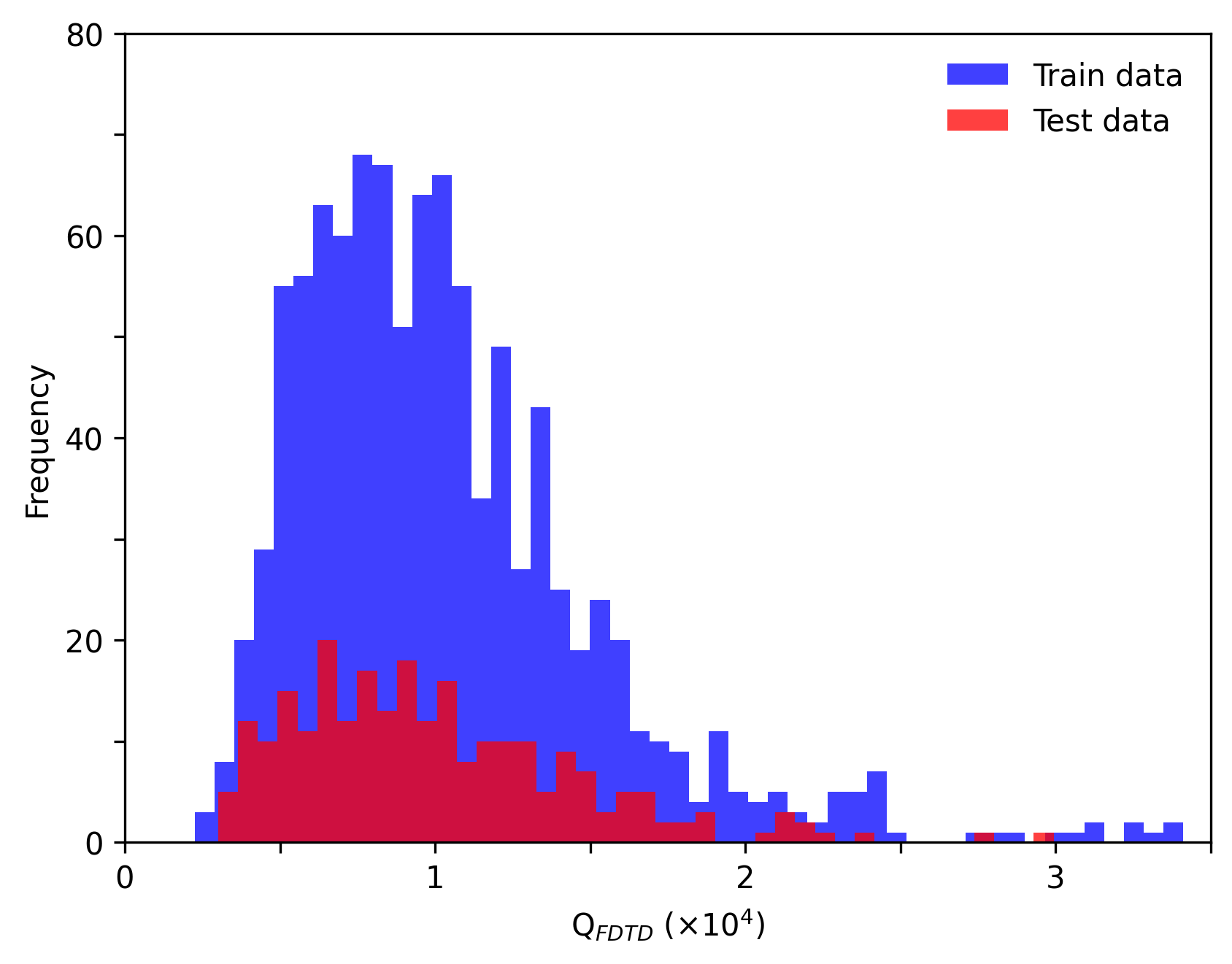}
        \caption{\textbf{NN1}: raw}
        \label{fig:NC123_dataset_rough_split}
    \end{subfigure}
    \begin{subfigure}{0.495\textwidth}
        \centering
        \includegraphics[width=\linewidth]{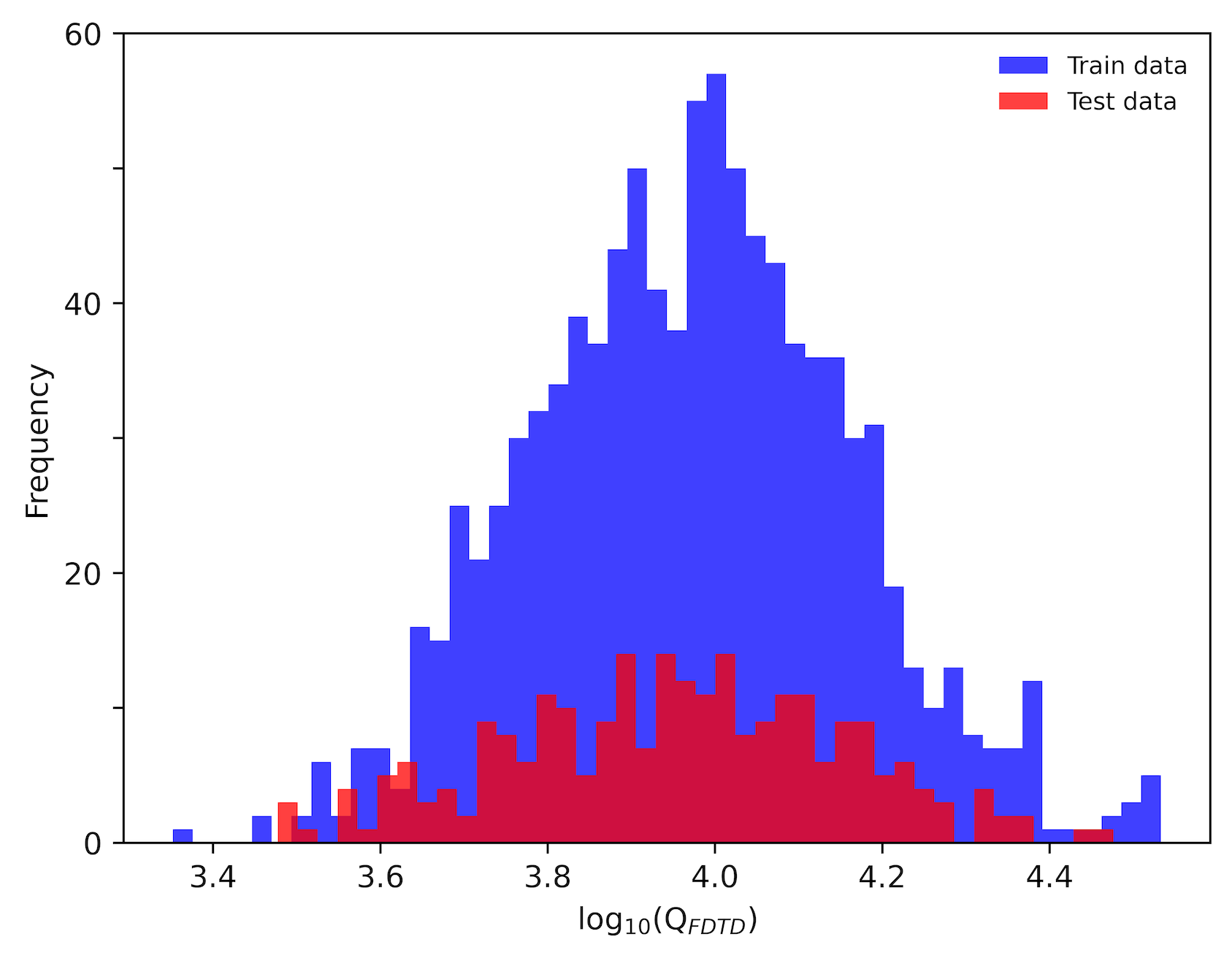}
        \caption{\textbf{NN1}: processed}
        \label{fig:NC123_dataset_rough_split_pross}
    \end{subfigure}
    
    \begin{subfigure}{0.495\textwidth}
        \centering
        \includegraphics[width=\linewidth]{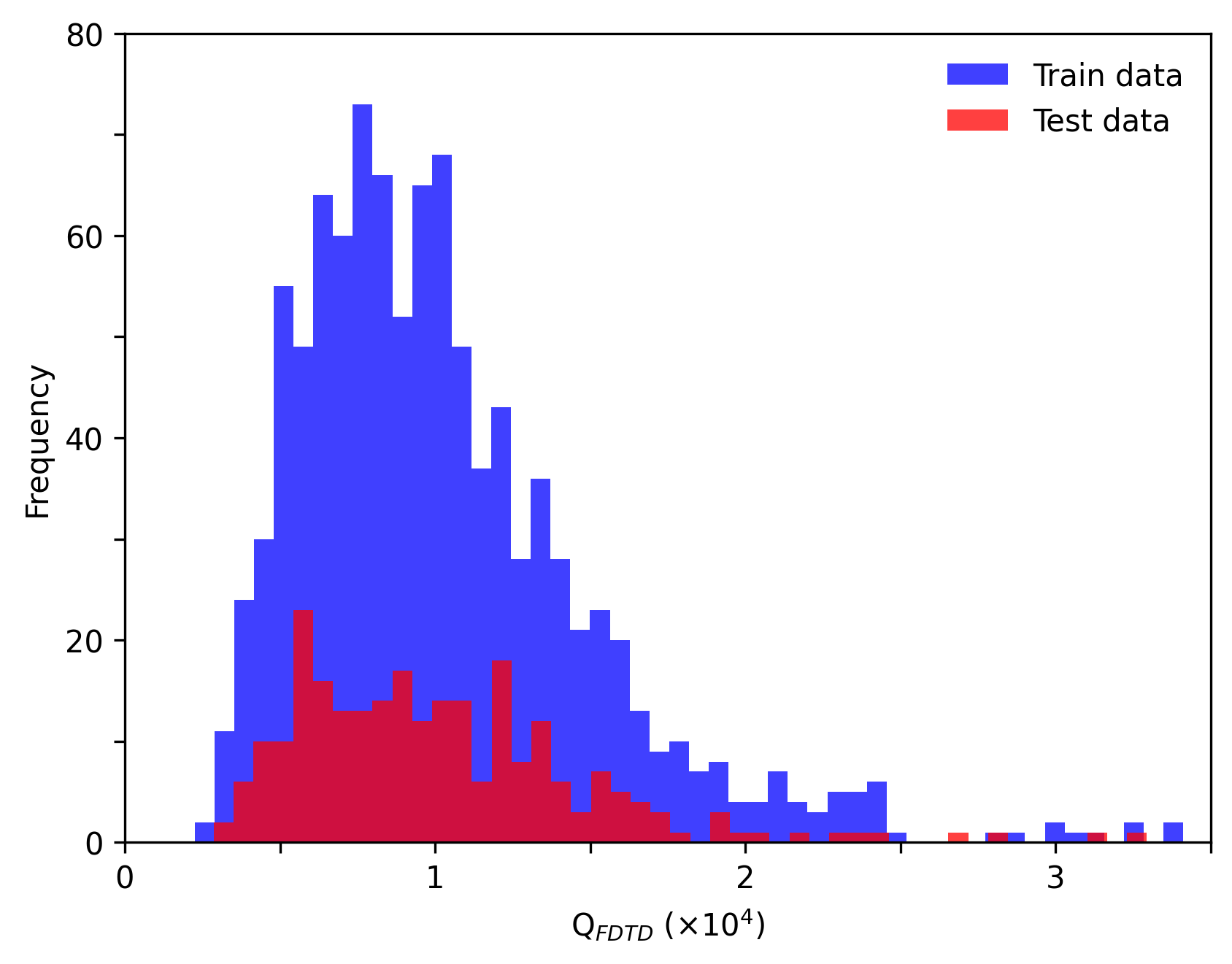}
        \caption{NN2: raw}
        \label{fig:NC21_dataset_rough_split_split}
    \end{subfigure}
    \begin{subfigure}{0.495\textwidth}
        \centering
        \includegraphics[width=\linewidth]{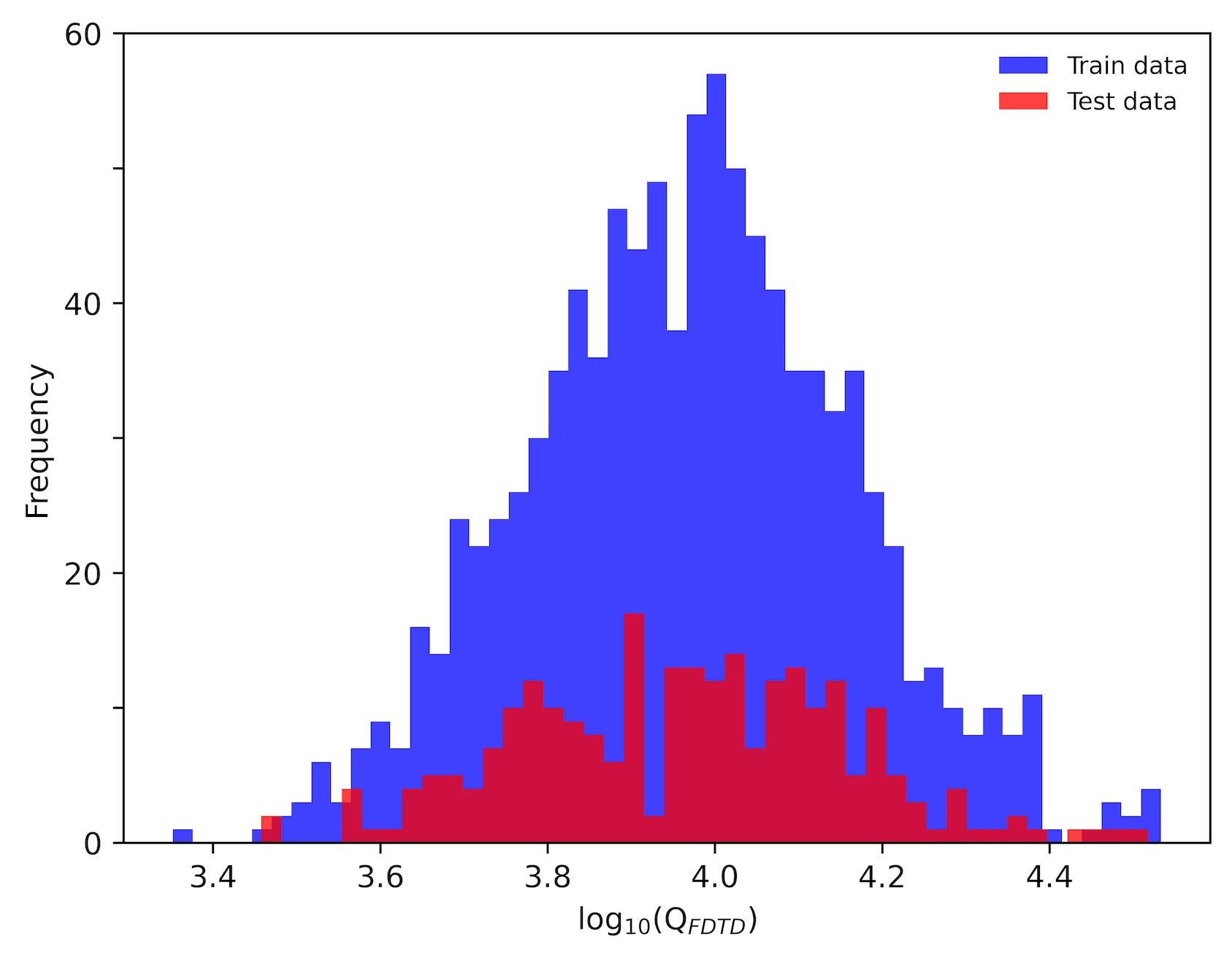}
        \caption{NN2: processed}
        \label{fig:NC21_dataset_rough_split_split_pross}
    \end{subfigure}
    \caption{data split, L2 with surface roughness (NN (bold) responsible for optimizing L2 cavity 2)}
    \label{fig:NC_dataset_rough_split}
\end{figure}
\newpage 

\begin{figure}[h!]
    \centering
    \begin{subfigure}{0.495\textwidth}
        \centering
        \includegraphics[width=\linewidth]{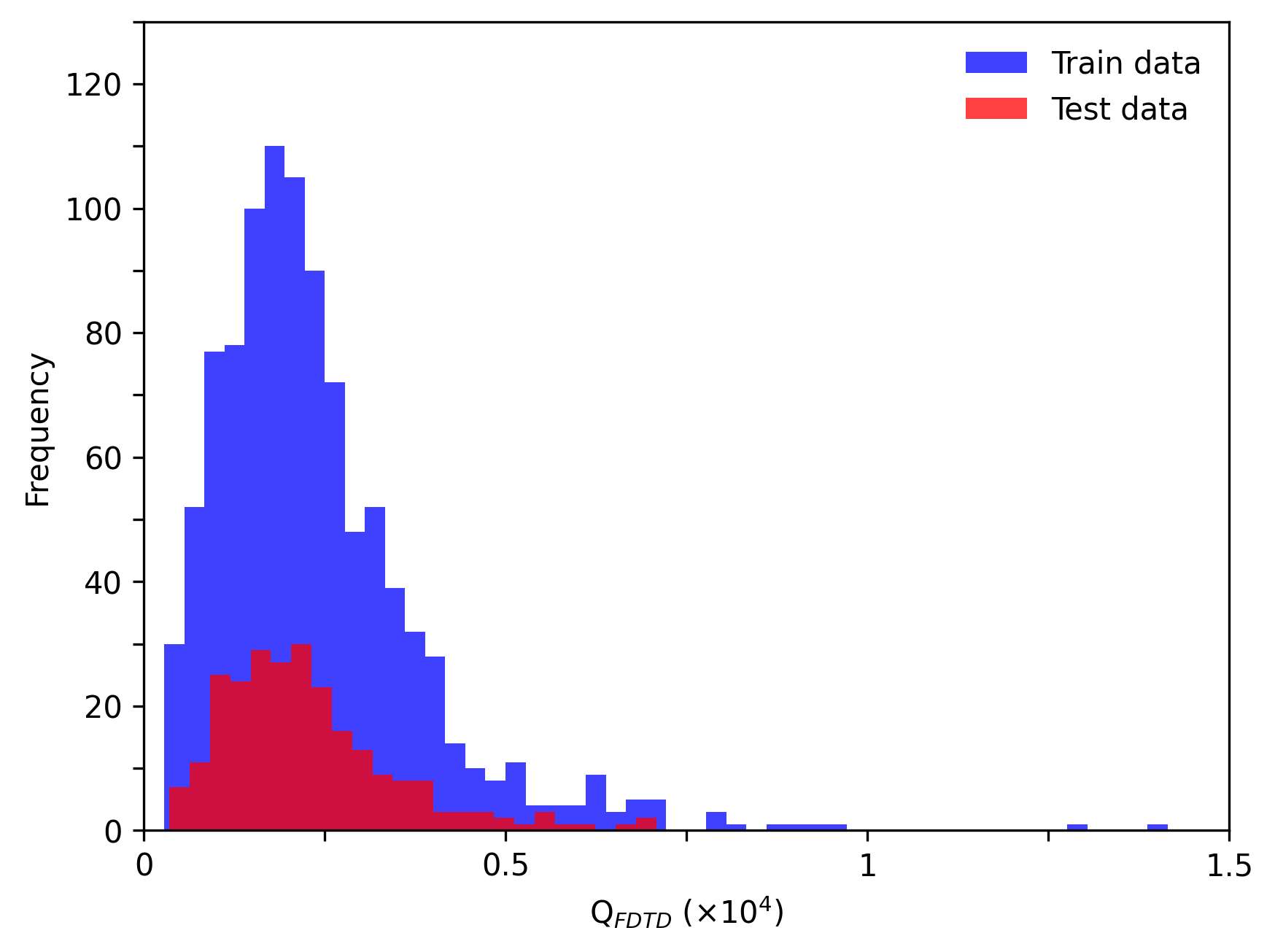}
        \caption{NN1: raw}
        \label{fig:FB32_dataset_rough_split}
    \end{subfigure}
    \begin{subfigure}{0.495\textwidth}
        \centering
        \includegraphics[width=\linewidth]{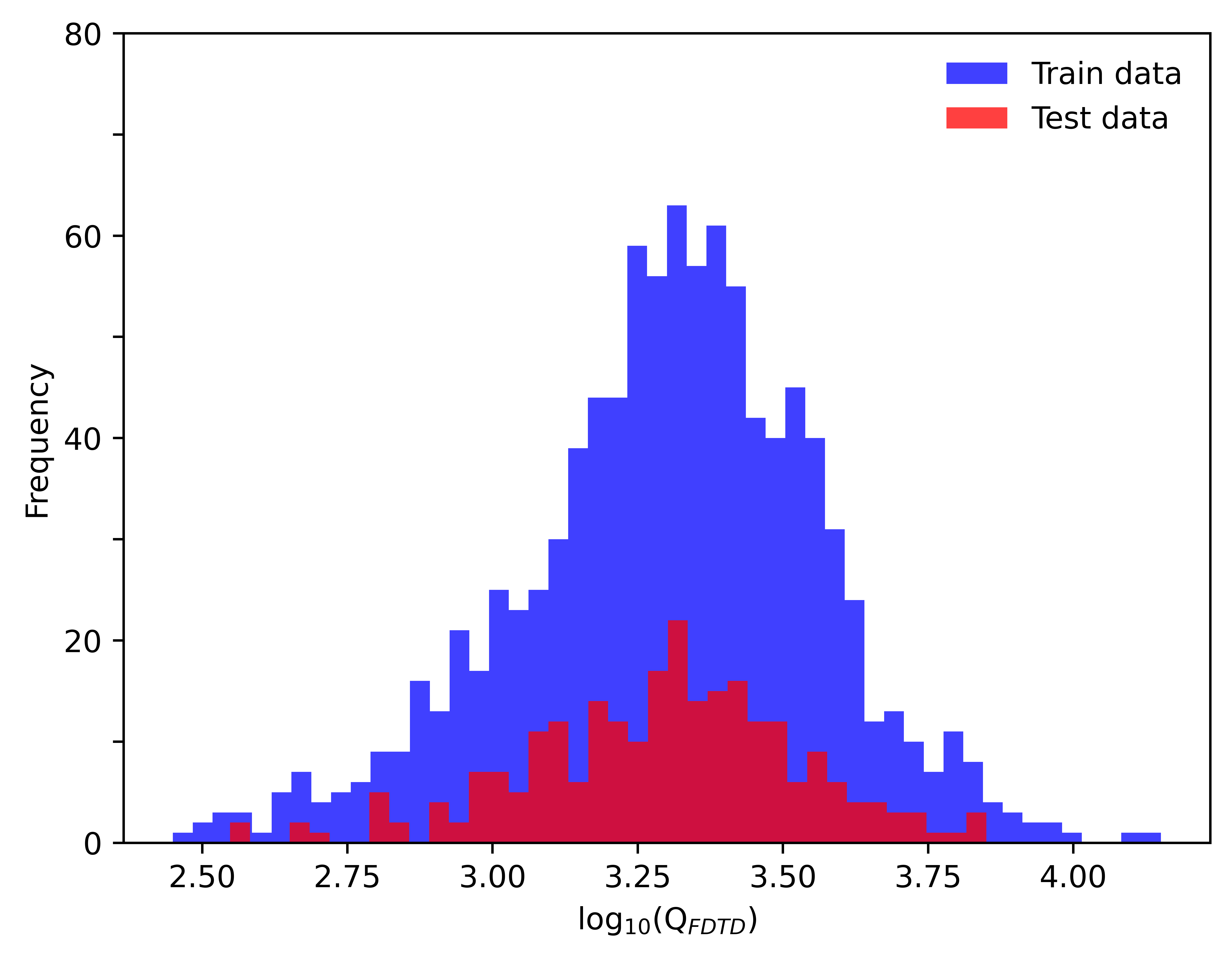}
        \caption{NN1: processed}
        \label{fig:FB32_dataset_rough_split_pross}
    \end{subfigure}
    
    \begin{subfigure}{0.495\textwidth}
        \centering
        \includegraphics[width=\linewidth]{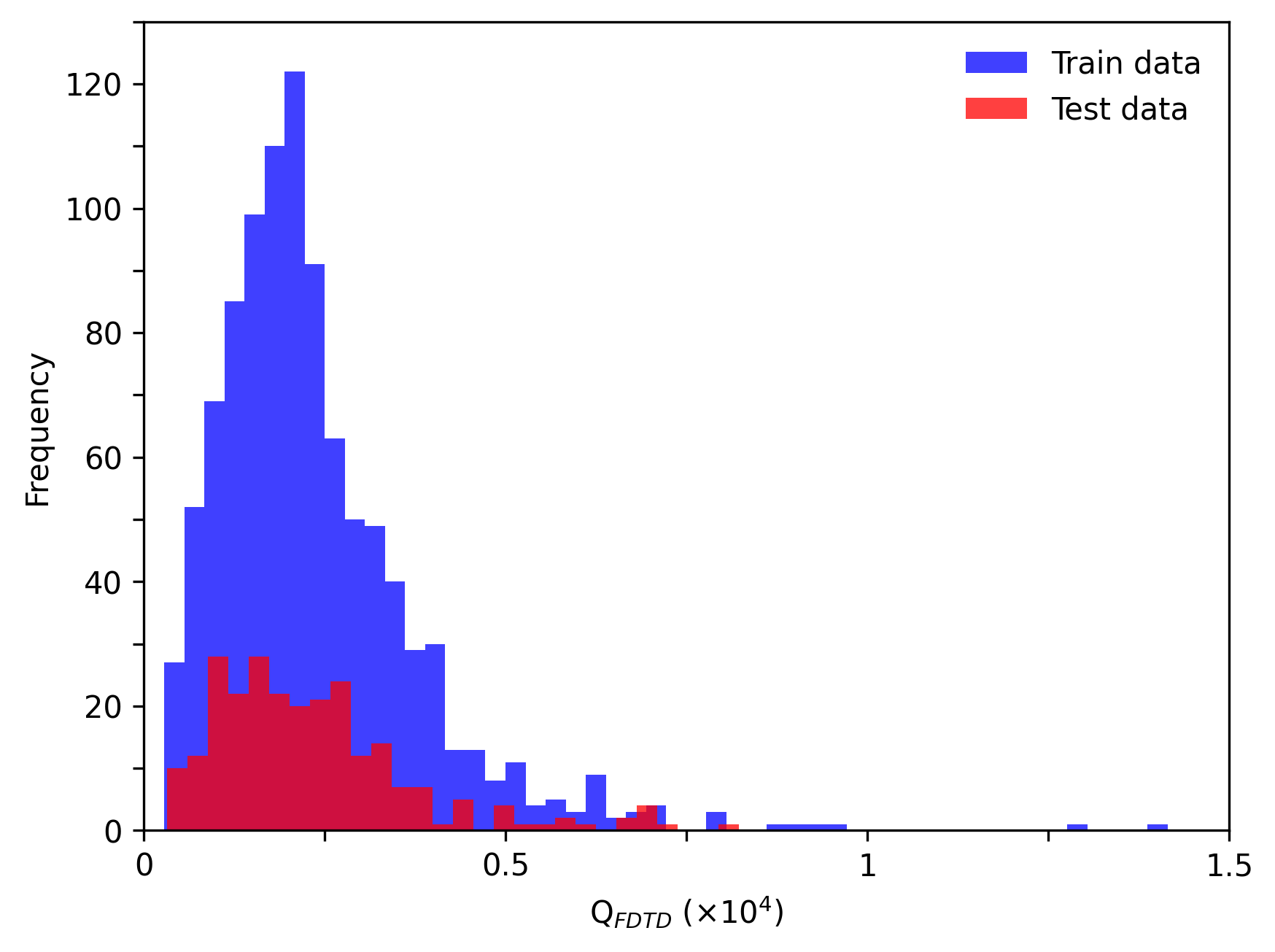}
        \caption{\textbf{NN2}: raw}
        \label{fig:FB212_dataset_rough_split}
    \end{subfigure}
    \begin{subfigure}{0.495\textwidth}
        \centering
        \includegraphics[width=\linewidth]{FB_dataset_rough32_split_processed.png}
        \caption{\textbf{NN2}: processed}
        \label{fig:FB212_dataset_rough_split_pross}
    \end{subfigure}
    \caption{data split, fishbone with surface roughness (NN (bold) responsible for optimizing fishbone cavity 2)}
    \label{fig:FB_dataset_rough_split}
\end{figure}
\newpage

\begin{figure}[h!]
    \centering
    \begin{subfigure}{0.495\textwidth}
        \centering
        \includegraphics[width=\linewidth]{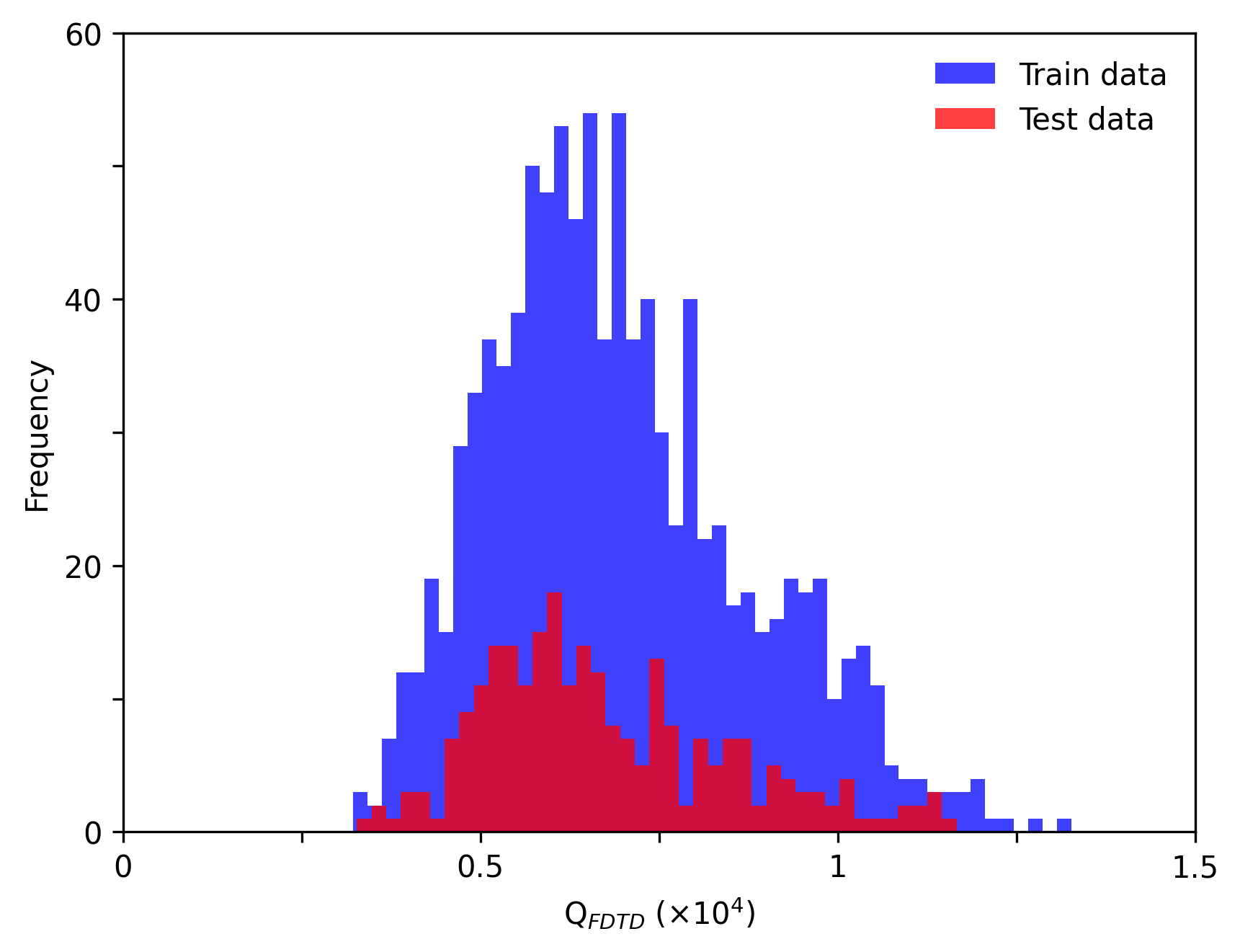}
        \caption{NN1: raw}
        \label{fig:NC123_dataset_slant_split}
    \end{subfigure}
    \begin{subfigure}{0.495\textwidth}
        \centering
        \includegraphics[width=\linewidth]{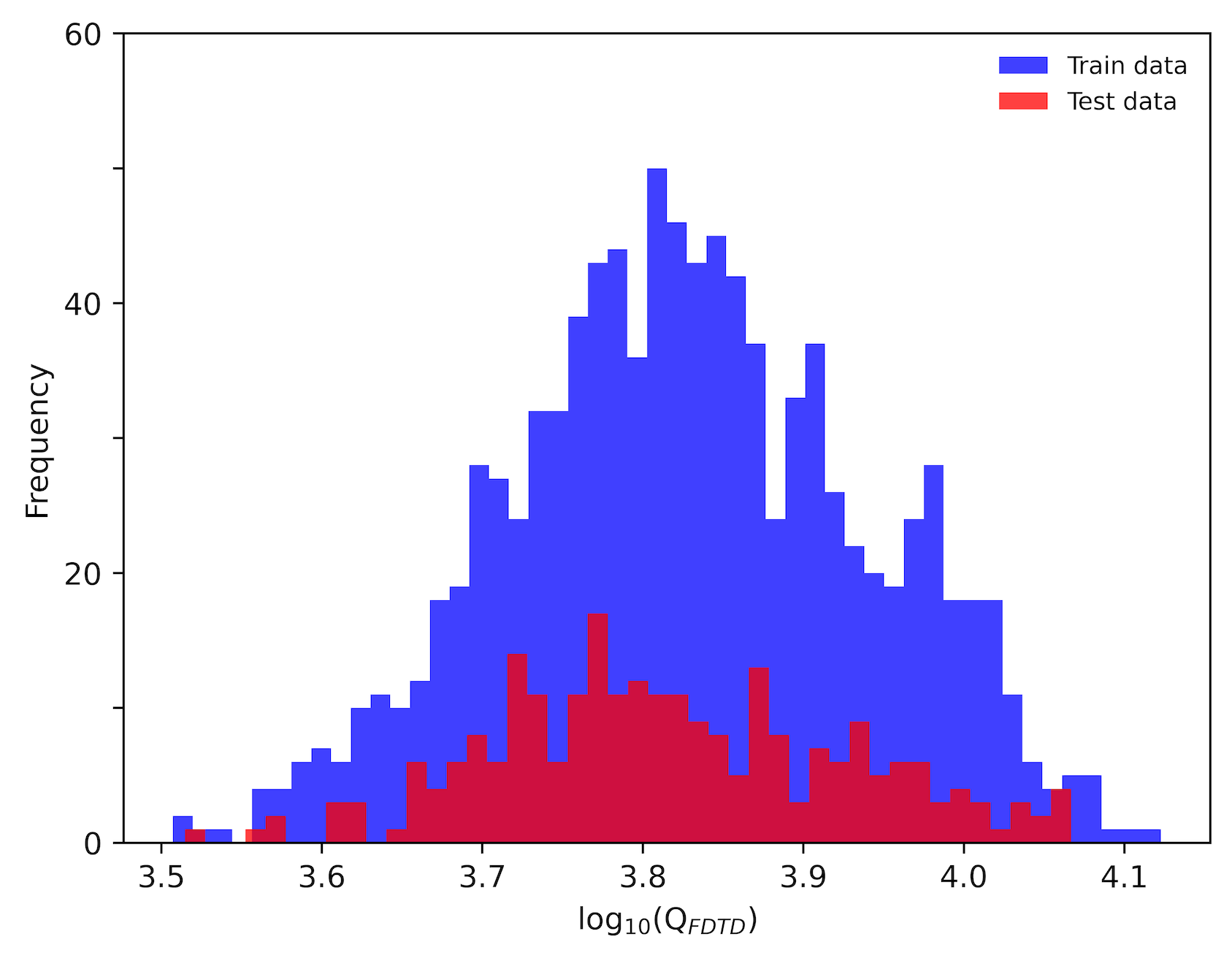}
        \caption{NN1: processed}
        \label{fig:NC123_dataset_slant_split_pross}
    \end{subfigure}
    
    \begin{subfigure}{0.495\textwidth}
        \centering
        \includegraphics[width=\linewidth]{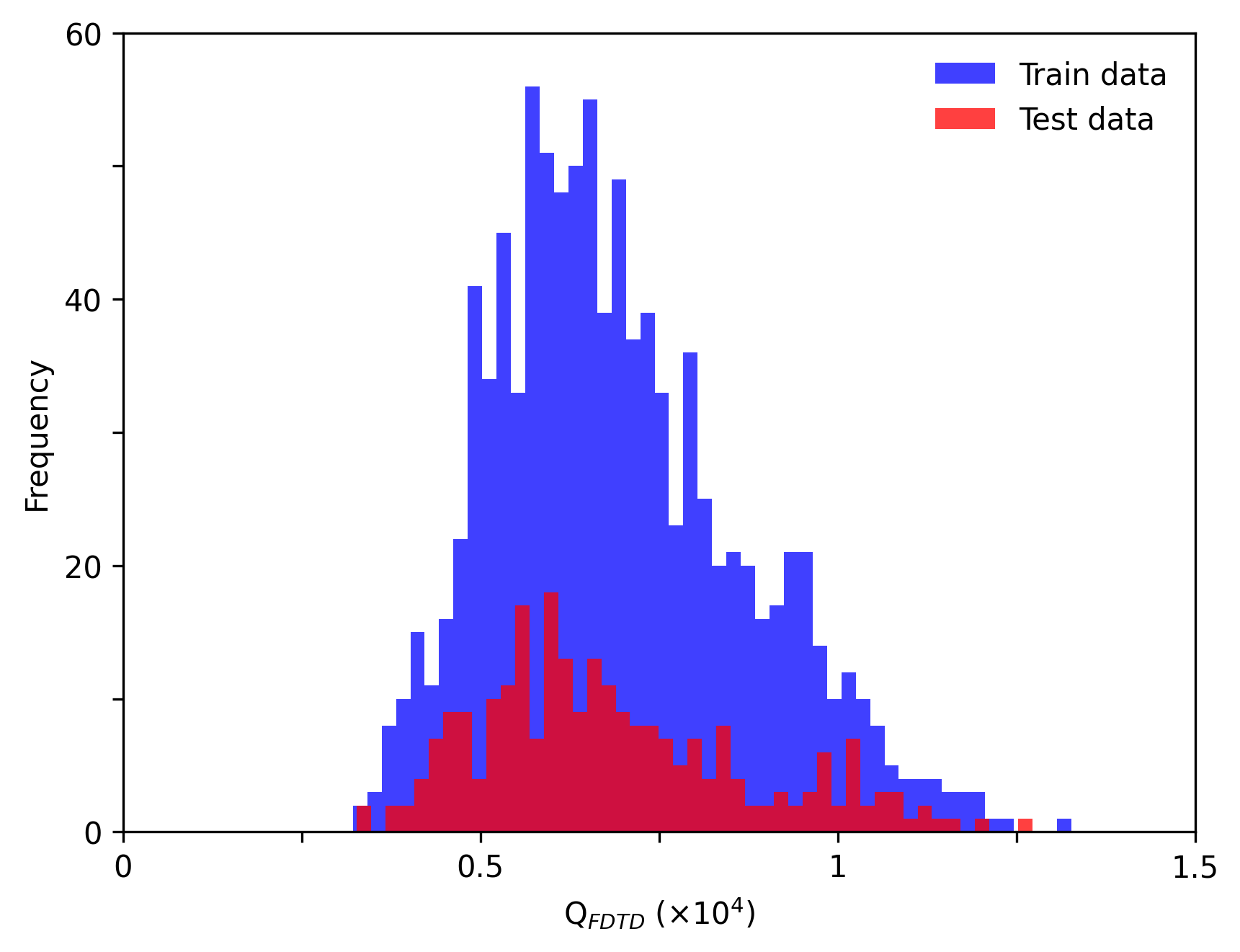}
        \caption{\textbf{NN2}: raw}
        \label{fig:NC21_dataset_slant_split_split}
    \end{subfigure}
    \begin{subfigure}{0.495\textwidth}
        \centering
        \includegraphics[width=\linewidth]{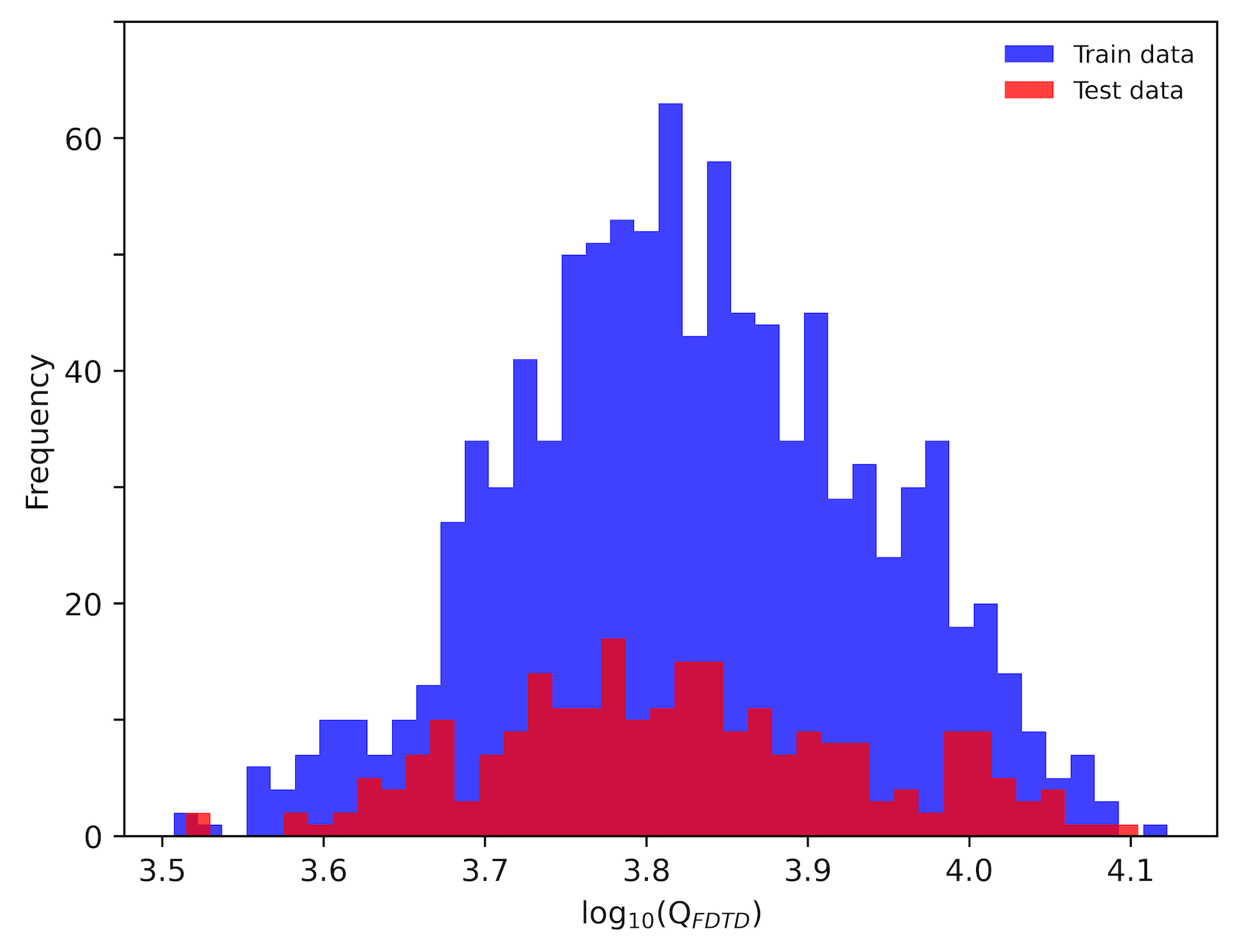}
        \caption{\textbf{NN2}: processed}
        \label{fig:NC21_dataset_slant_split_split_pross}
    \end{subfigure}
    \caption{data split, L2 with sidewall slant (NN (bold) responsible for optimizing L2 cavity 3)}
    \label{fig:NC_dataset_slant_split}
\end{figure}
\newpage

\begin{figure}[h!]
    \centering
    \begin{subfigure}{0.495\textwidth}
        \centering
        \includegraphics[width=\linewidth]{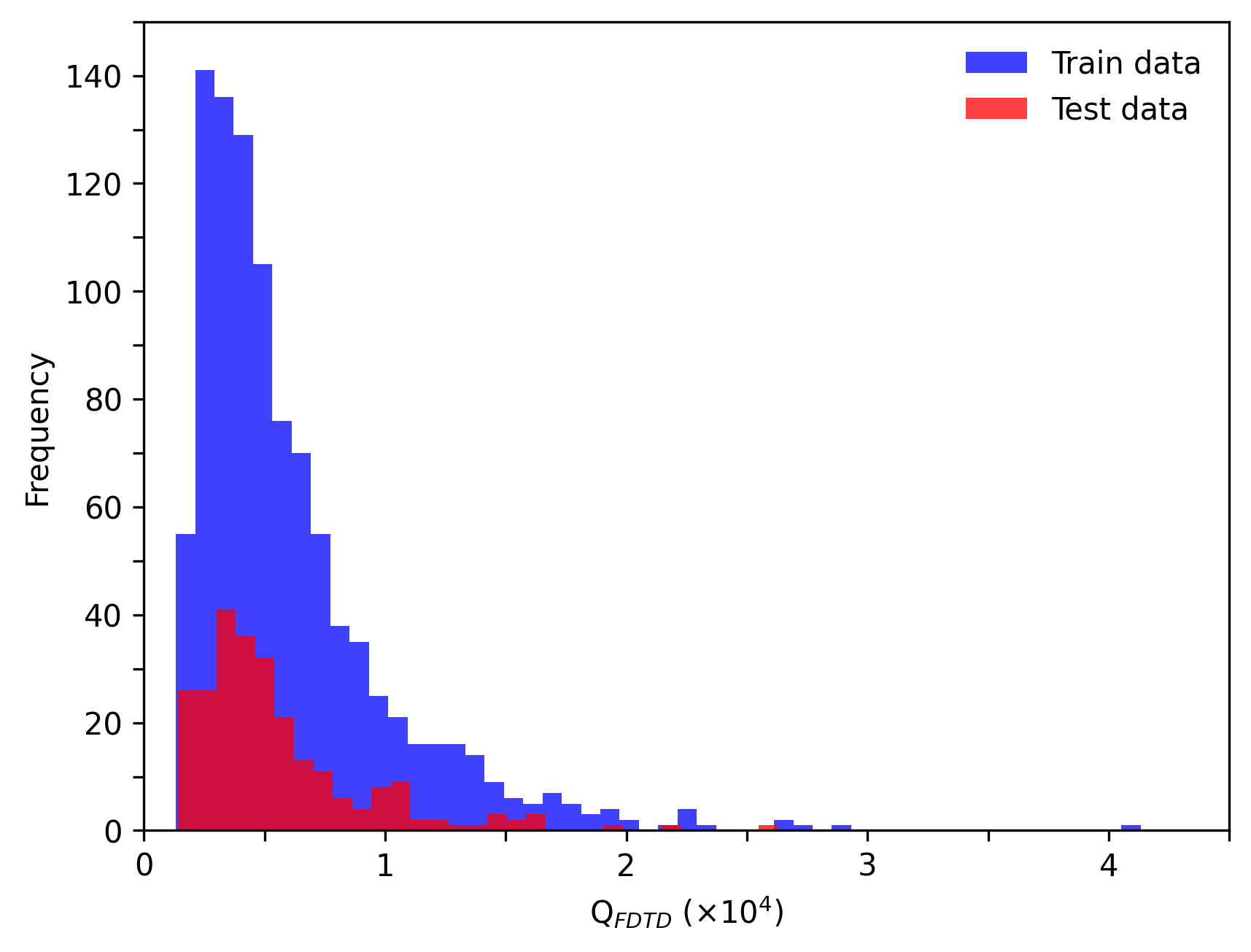}
        \caption{\textbf{NN1}: raw}
        \label{fig:FB41_dataset_slant_split}
    \end{subfigure}
    \begin{subfigure}{0.495\textwidth}
        \centering
        \includegraphics[width=\linewidth]{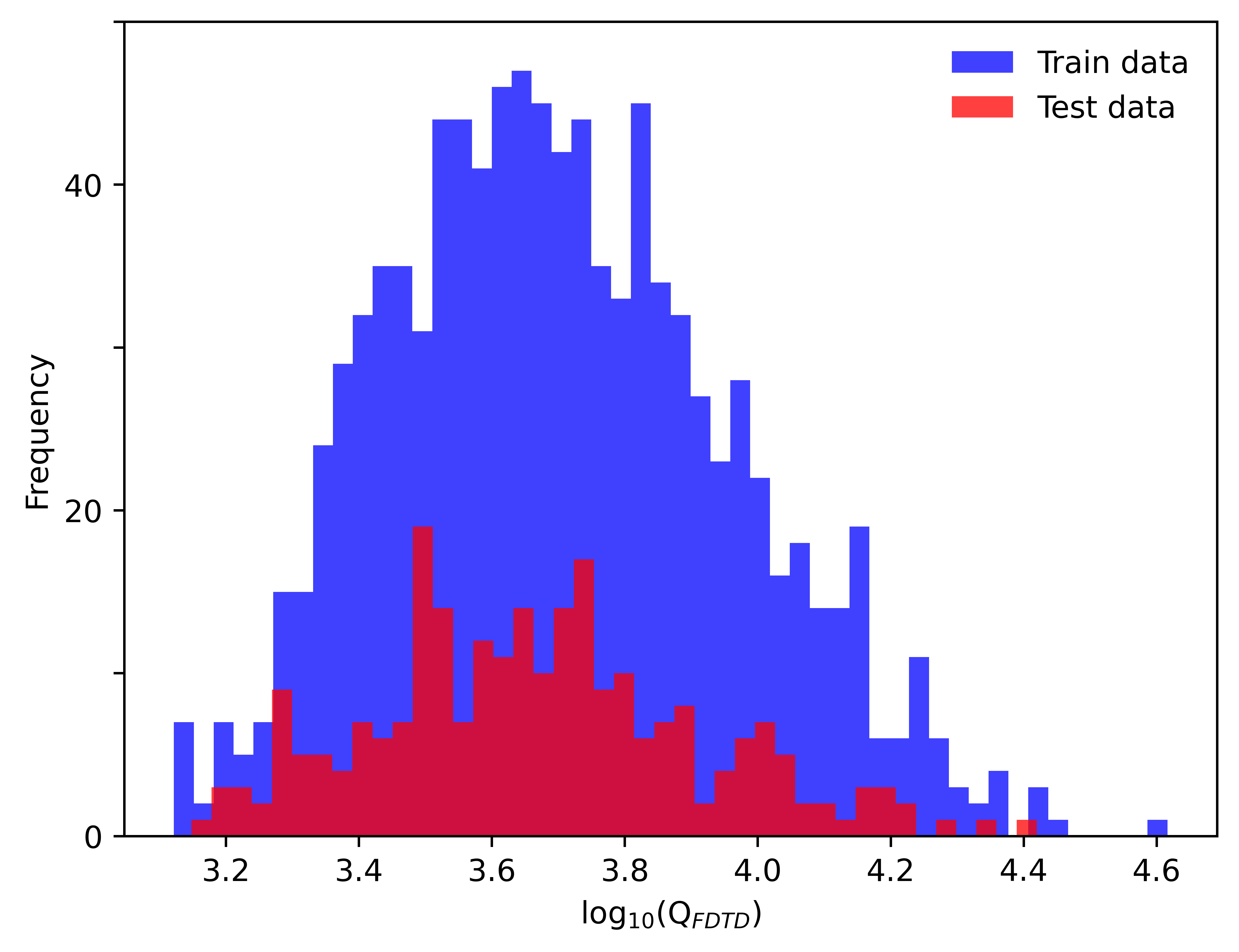}
        \caption{\textbf{NN1}: processed}
        \label{fig:FB41_dataset_slant_split_pross}
    \end{subfigure}
    
    \begin{subfigure}{0.495\textwidth}
        \centering
        \includegraphics[width=\linewidth]{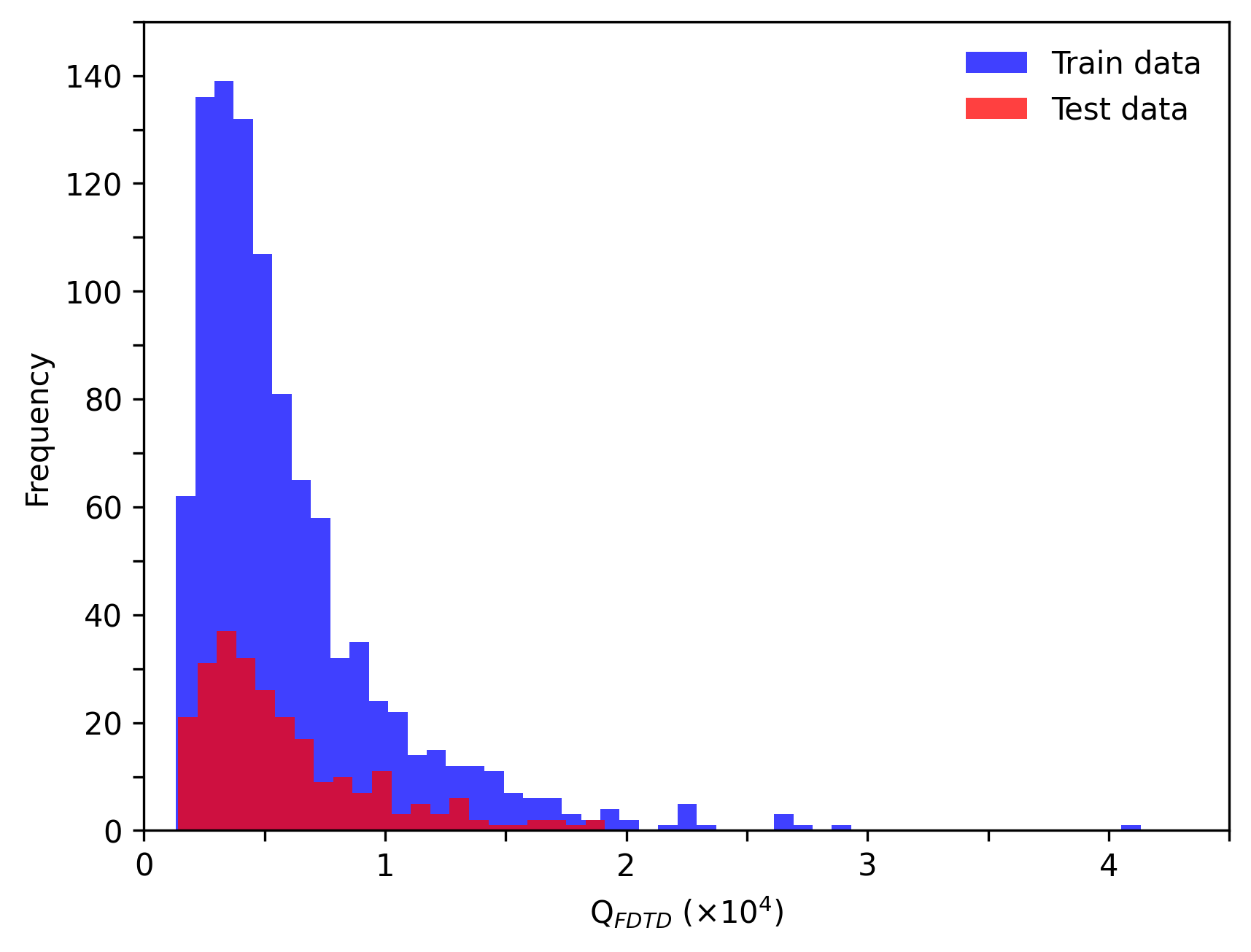}
        \caption{NN2: raw}
        \label{fig:FB44_dataset_slant_split}
    \end{subfigure}
    \begin{subfigure}{0.495\textwidth}
        \centering
        \includegraphics[width=\linewidth]{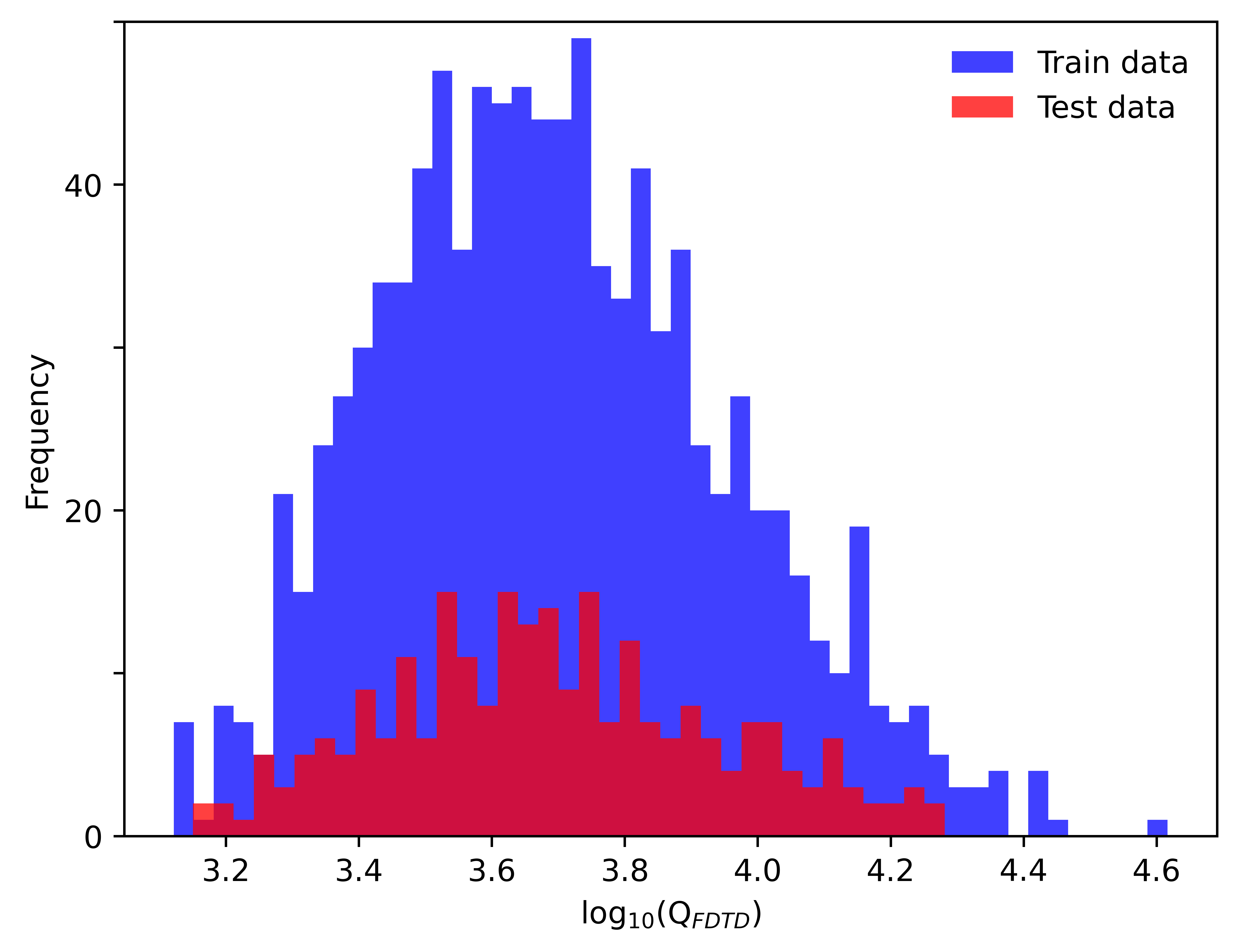}
        \caption{NN2: processed}
        \label{fig:FB44_dataset_slant_split_pross}
    \end{subfigure}
    \caption{data split, fishbone with sidewall slant (NN (bold) responsible for optimizing fishbone cavity 3)}
    \label{fig:FB_dataset_slant_split}
\end{figure}
\newpage

\begin{figure}[h!]
    \centering
    \begin{subfigure}{0.495\textwidth}
        \centering
        \includegraphics[width=\linewidth]{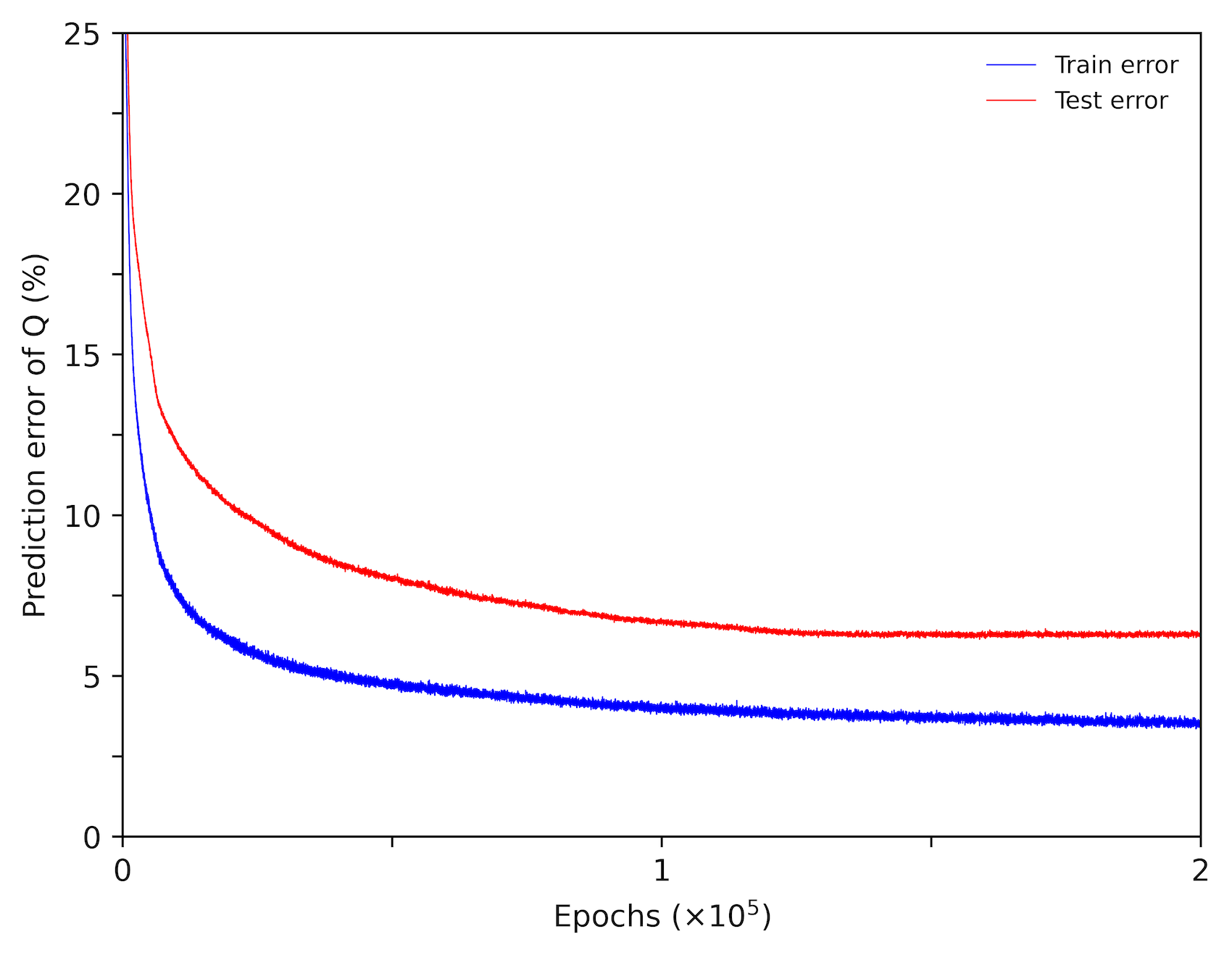}
        \caption{NN1: $\lambda = 0$, $\epsilon_{\mathrm{pred}} = 6.33\%\,(3.54\%)$}
        \label{fig:NC123_training_l0_clean}
    \end{subfigure}
    \begin{subfigure}{0.495\textwidth}
        \centering
        \includegraphics[width=\linewidth]{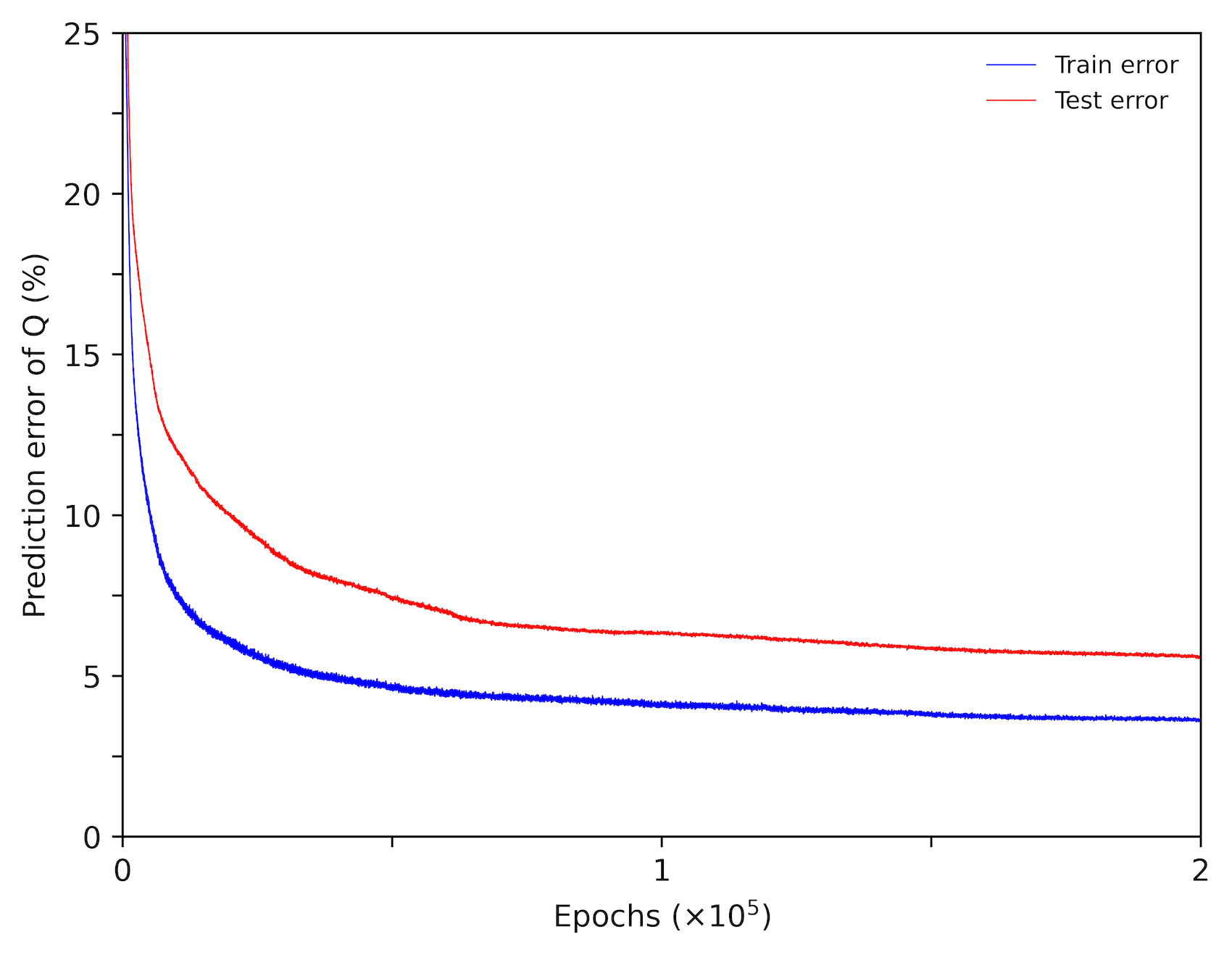}
        \caption{NN1: $\lambda = 0.001$, $\epsilon_{\mathrm{pred}} = 5.62\%\,(3.65\%)$}
        \label{fig:NC123_training_l0.001_clean}
    \end{subfigure}
    
    \begin{subfigure}{0.495\textwidth}
        \centering
        \includegraphics[width=\linewidth]{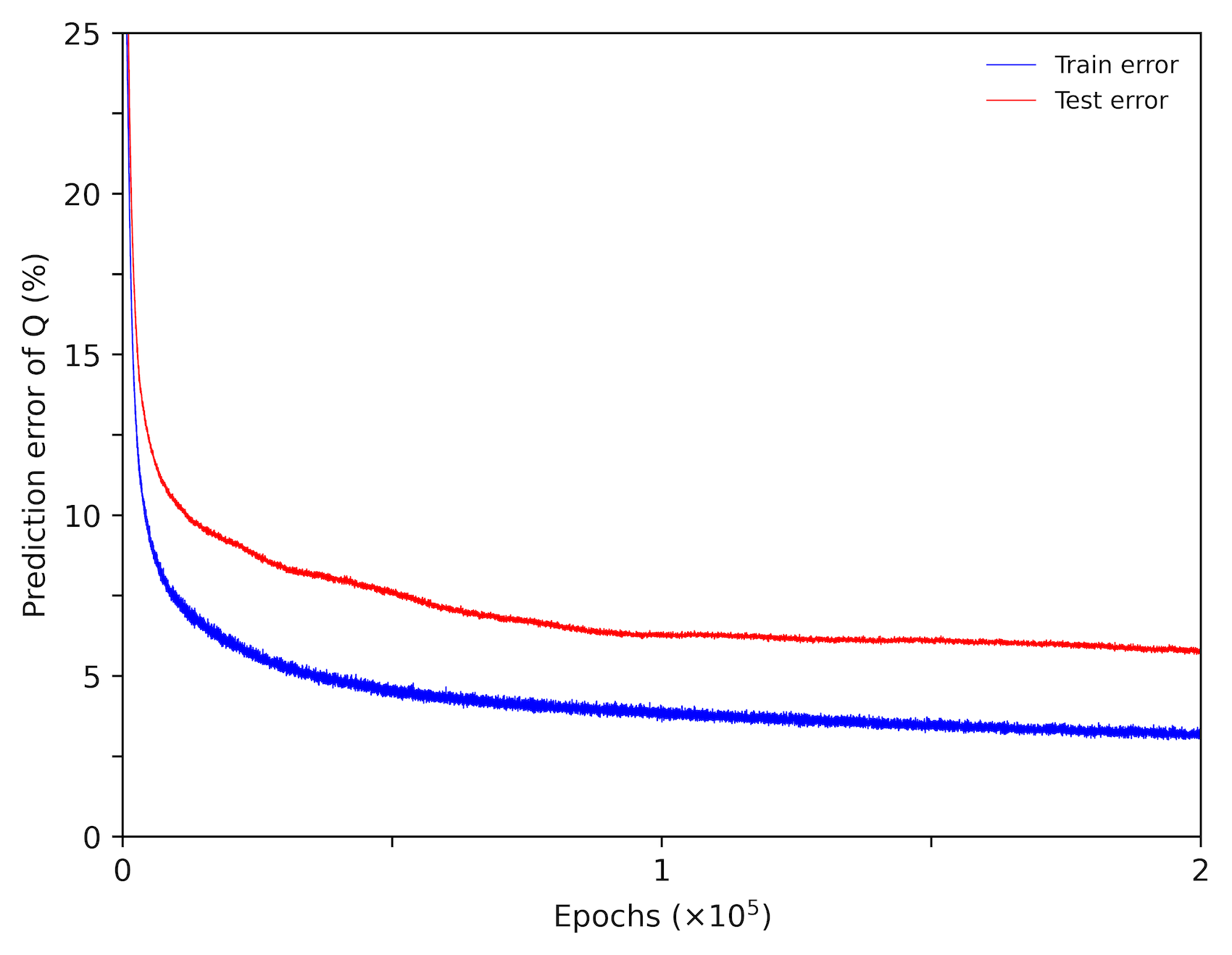}
        \caption{\textbf{NN2}: $\lambda = 0$, $\epsilon_{\mathrm{pred}} = 5.76\%\,(3.08\%)$}
        \label{fig:NC21_training_l0_clean}
    \end{subfigure}
    \begin{subfigure}{0.495\textwidth}
        \centering
        \includegraphics[width=\linewidth]{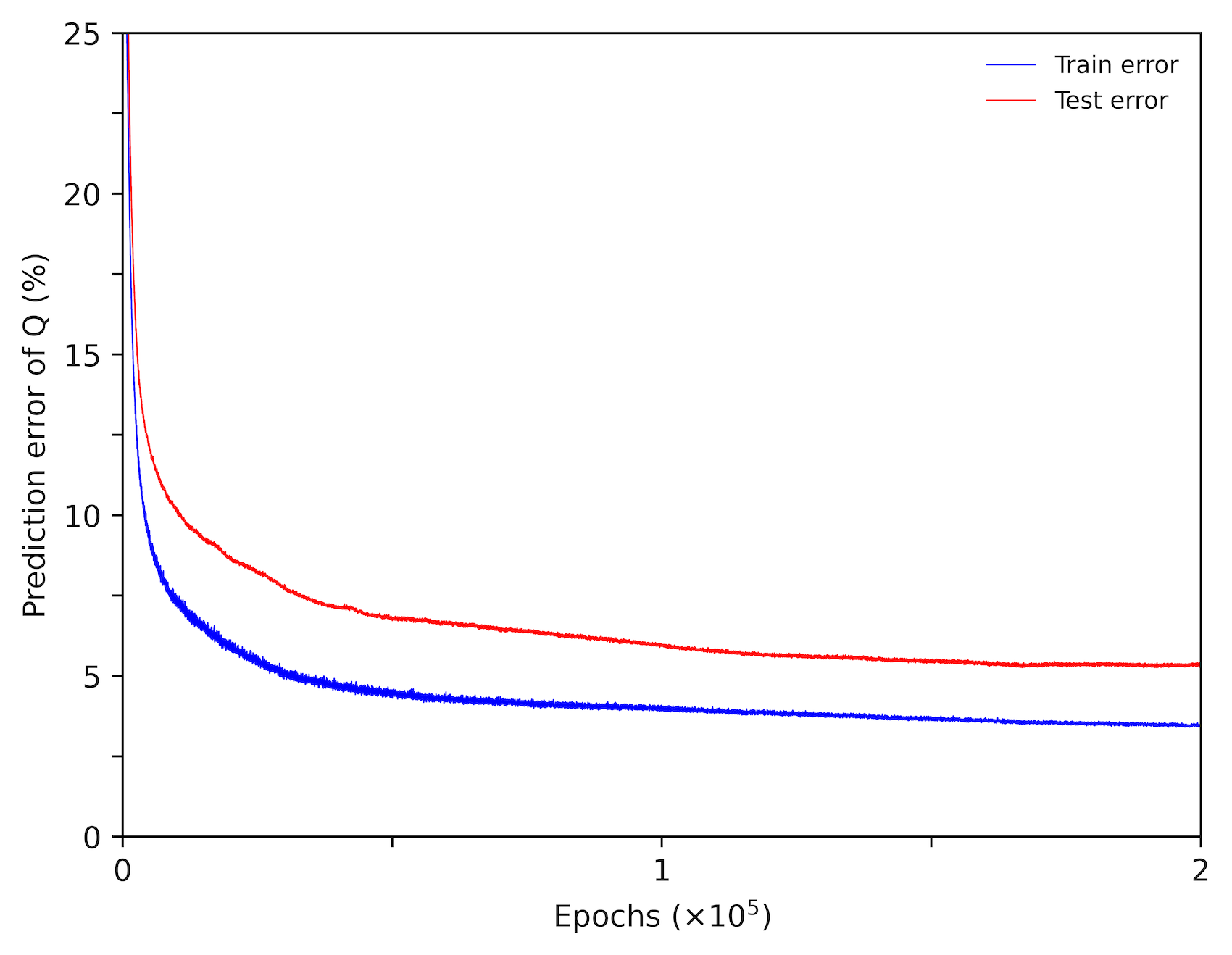}
        \caption{NN2: $\lambda = 0.001$, $\epsilon_{\mathrm{pred}} = 5.39\%\,(3.45\%)$}
        \label{fig:NC21_training_l0.001_clean}
    \end{subfigure}
    \caption{Training process, L2 without imperfections (NN (bold) responsible for optimizing L2 cavity 1)}
    \label{fig:NC_training_clean}
\end{figure}
\newpage

\begin{figure}[h!]
    \centering
    \begin{subfigure}{0.495\textwidth}
        \centering
        \includegraphics[width=\linewidth]{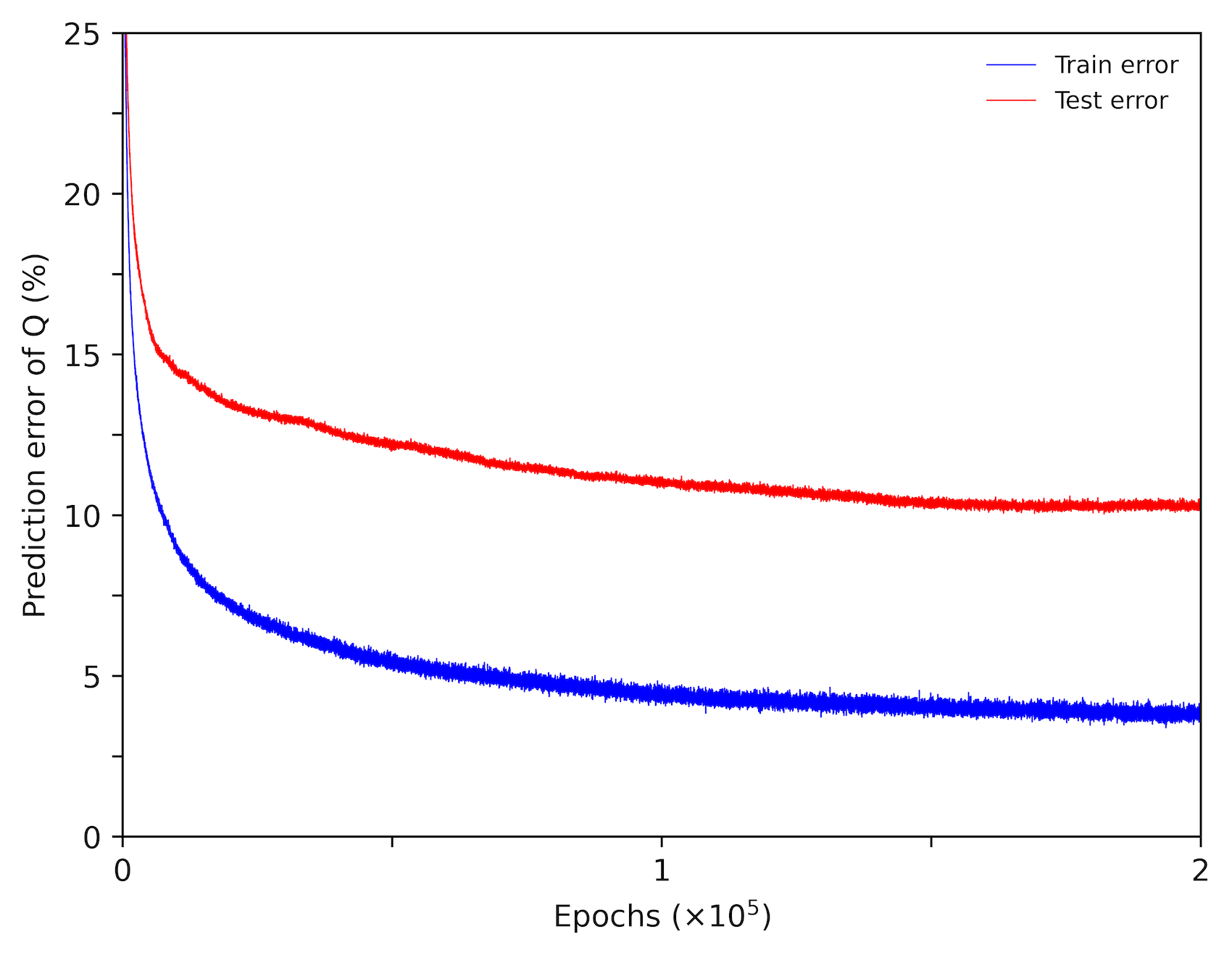}
        \caption{NN1: $\lambda = 0$, $\epsilon_{\mathrm{pred}} = 10.25\%\,(3.82\%)$}
        \label{fig:FB32_training_l0_clean}
    \end{subfigure}
    \begin{subfigure}{0.495\textwidth}
        \centering
        \includegraphics[width=\linewidth]{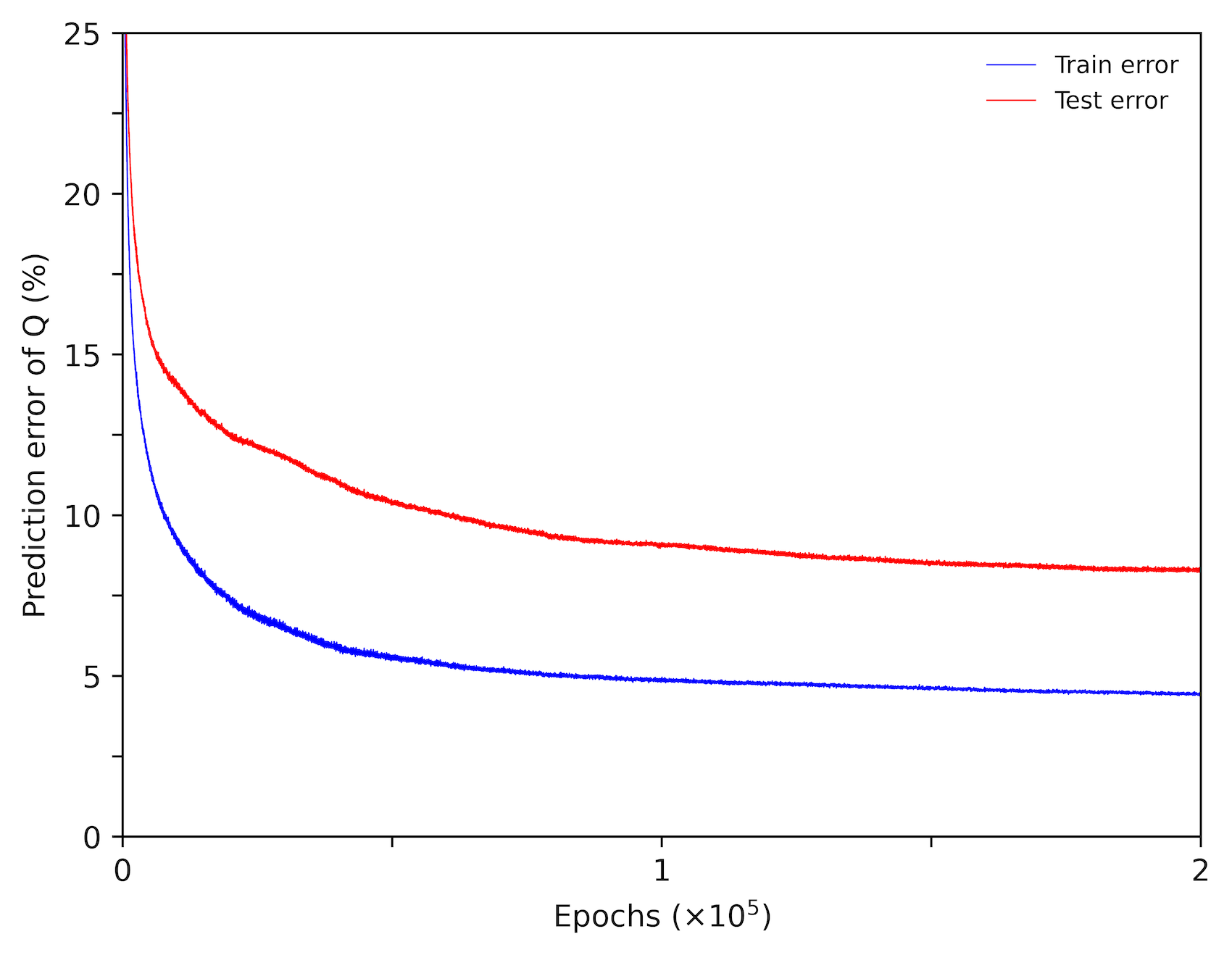}
        \caption{NN1: $\lambda = 0.001$, $\epsilon_{\mathrm{pred}} = 8.29\%\,(4.42\%)$}
        \label{fig:FB32_training_l0.001_clean}
    \end{subfigure}
    
    \begin{subfigure}{0.495\textwidth}
        \centering
        \includegraphics[width=\linewidth]{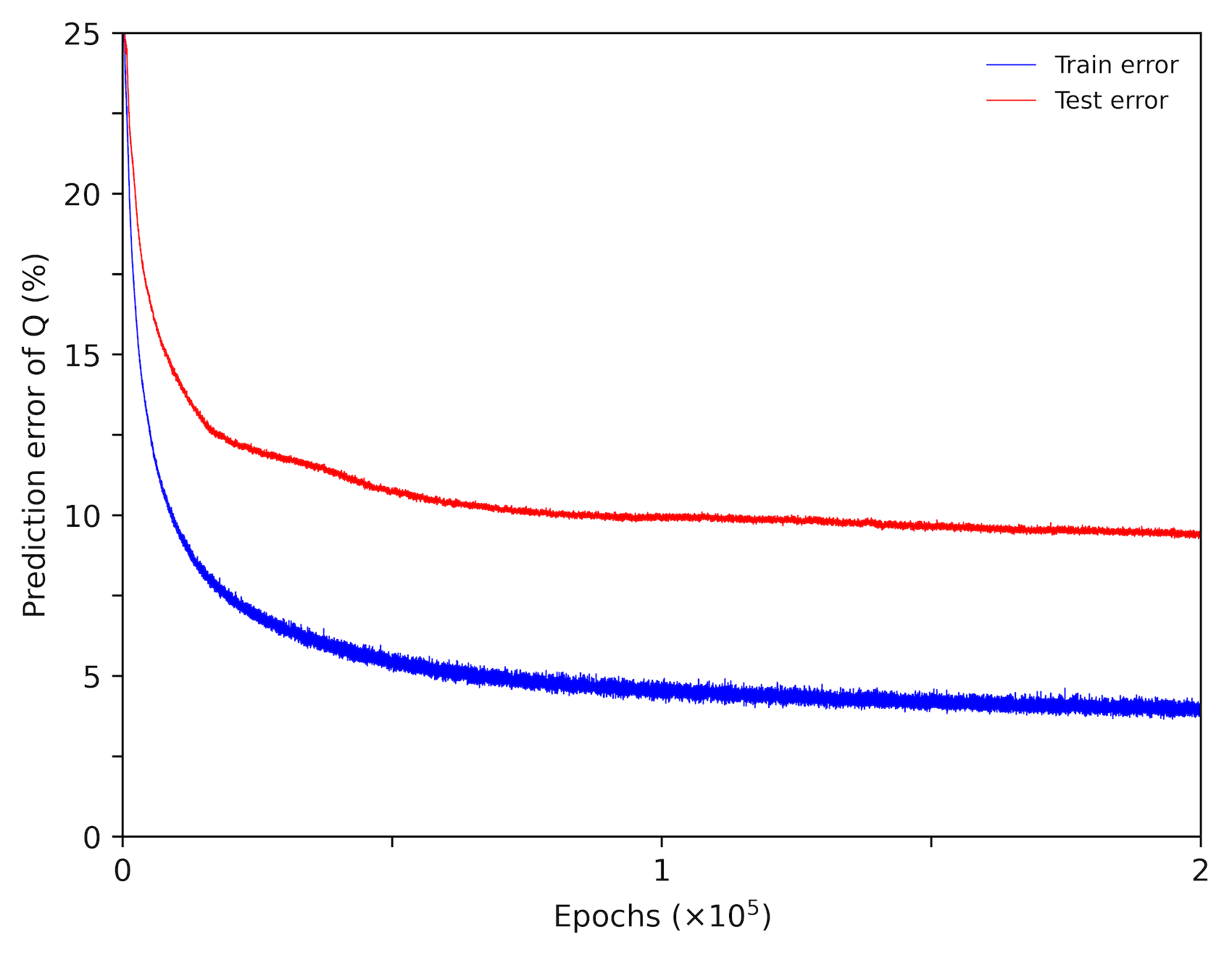}
        \caption{NN2: $\lambda = 0$, $\epsilon_{\mathrm{pred}} = 9.42\%\, (4.09\%)$}
        \label{fig:FB212_training_l0_clean}
    \end{subfigure}
    \begin{subfigure}{0.495\textwidth}
        \centering
        \includegraphics[width=\linewidth]{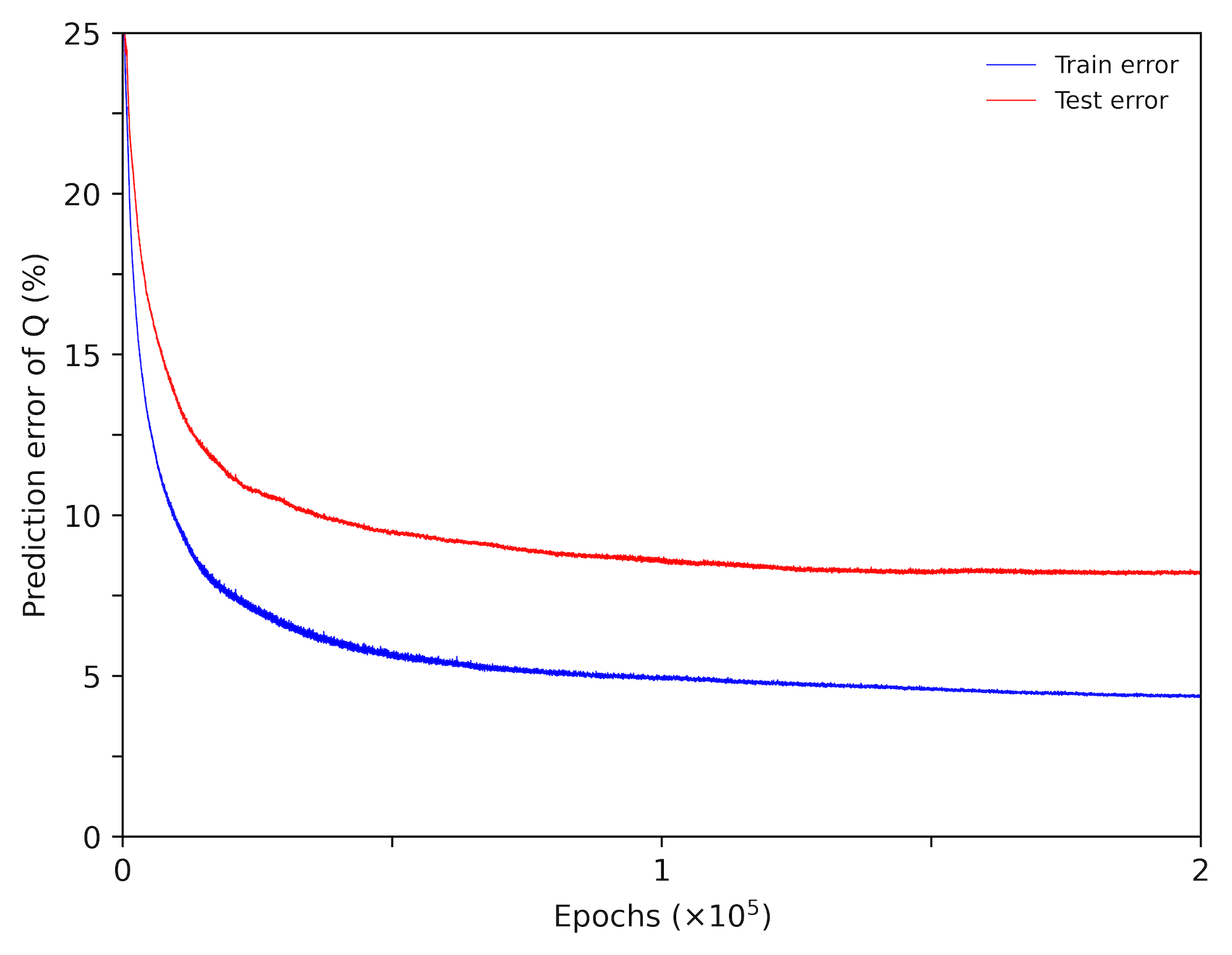}
        \caption{\textbf{NN2}: $\lambda = 0.001$, $\epsilon_{\mathrm{pred}} = 8.21\%\,(4.39\%)$}
        \label{fig:FB212_training_l0.001_clean}
    \end{subfigure}
    \caption{Training process, fishbone without imperfections (NN (bold) responsible for optimizing fishbone cavity 1)}
    \label{fig:FB_training_clean}
\end{figure}
\newpage

\begin{figure}[h!]
    \centering
    \begin{subfigure}{0.495\textwidth}
        \centering
        \includegraphics[width=\linewidth]{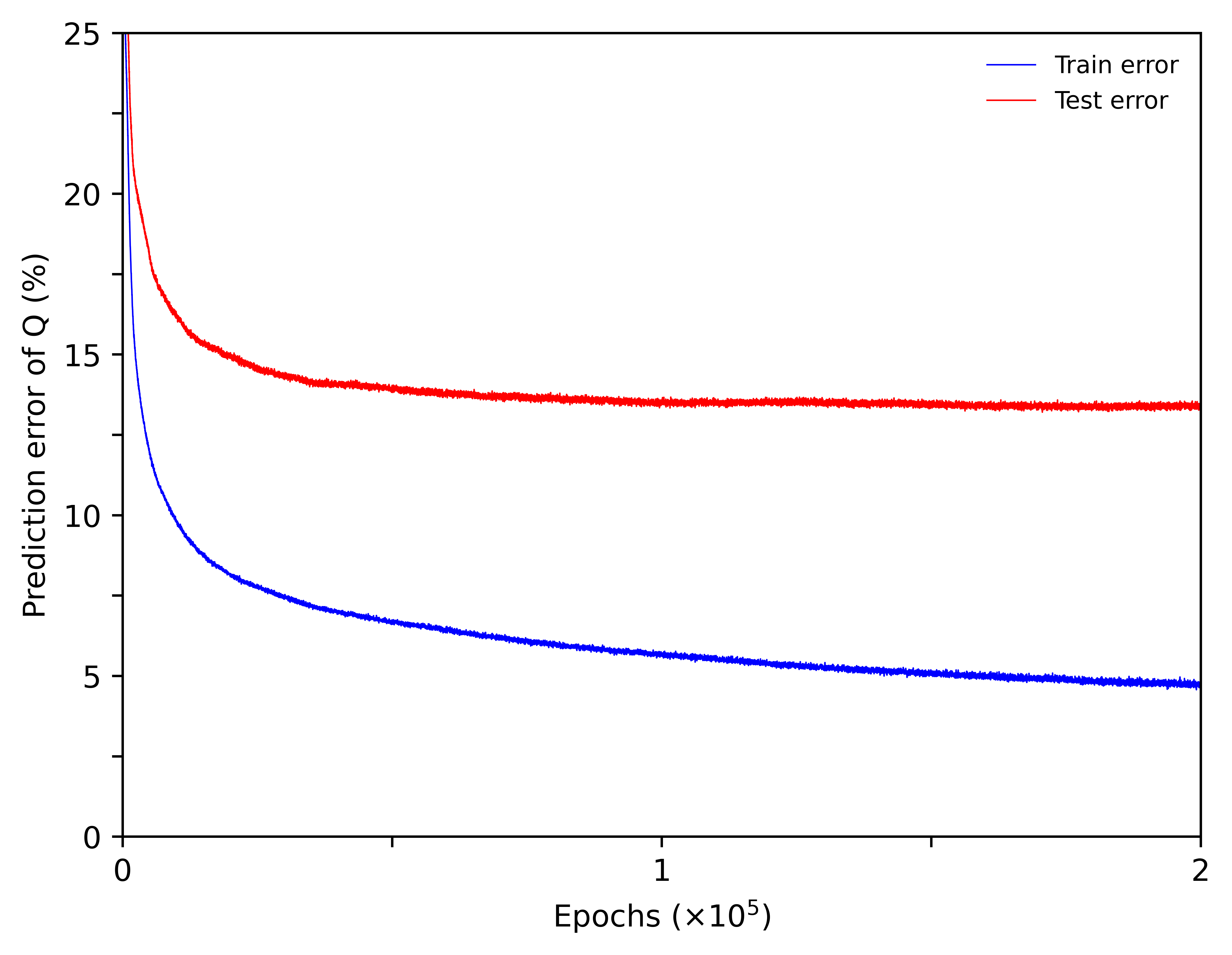}
        \caption{NN1: $\lambda = 0$, $\epsilon_{\mathrm{pred}} = 13.41\%\,(4.73\%)$}
        \label{fig:NC123_training_l0_rough}
    \end{subfigure}
    \begin{subfigure}{0.495\textwidth}
        \centering
        \includegraphics[width=\linewidth]{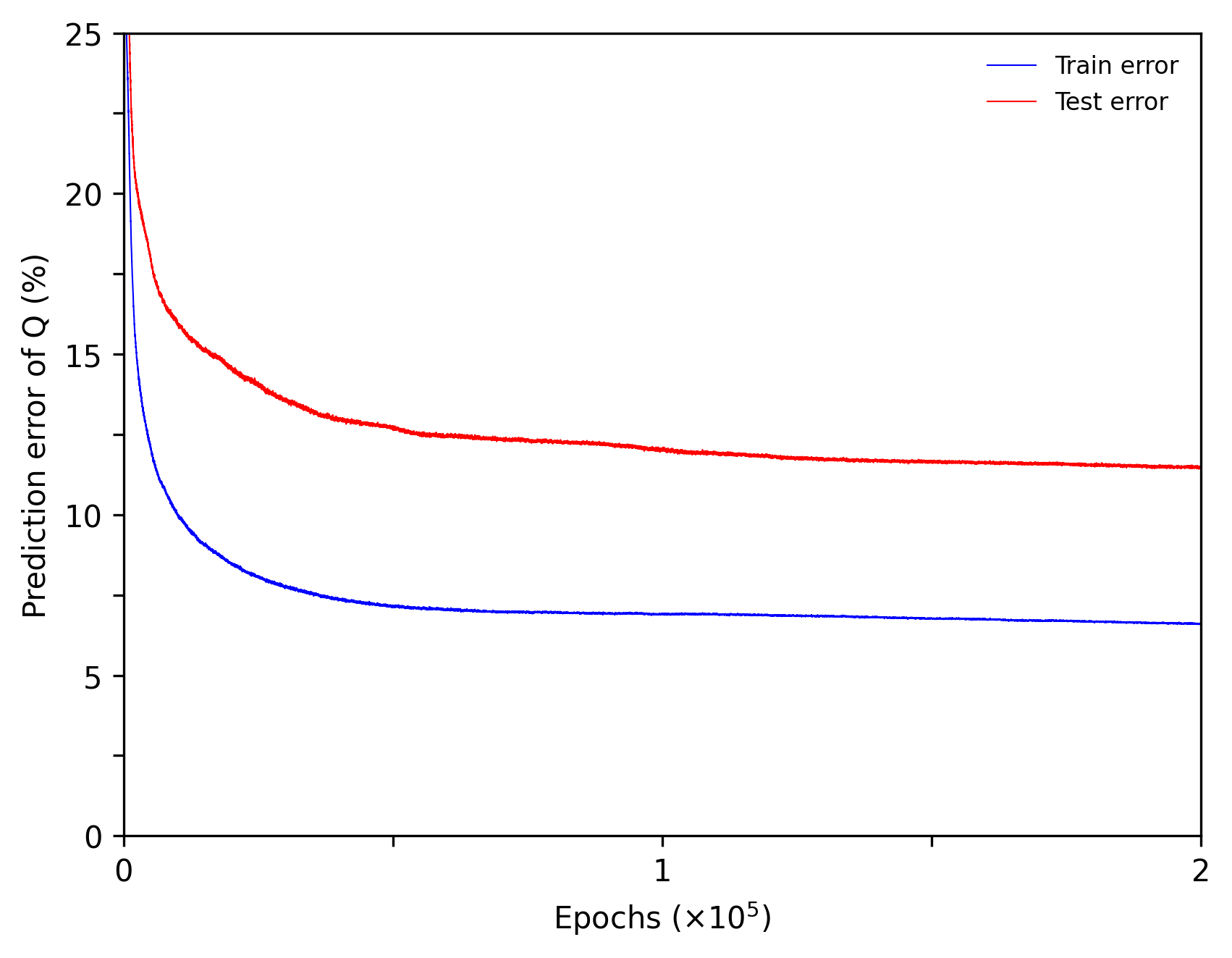}
        \caption{\textbf{NN1}: $\lambda = 0.001$, $\epsilon_{\mathrm{pred}} = 11.50\%\,(6.62\%)$}
        \label{fig:NC123_training_l0.001_rough}
    \end{subfigure}
    
    \begin{subfigure}{0.495\textwidth}
        \centering
        \includegraphics[width=\linewidth]{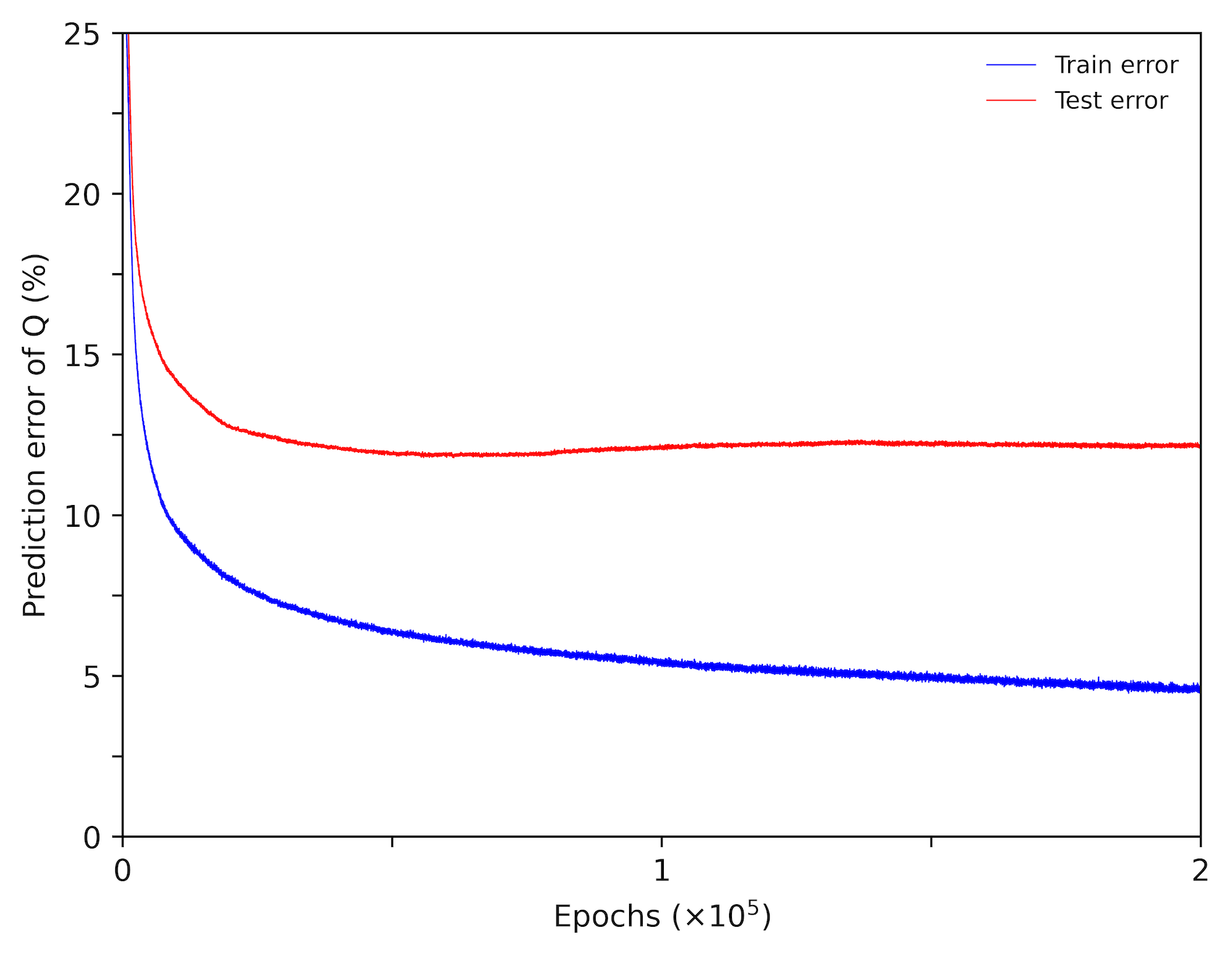}
        \caption{NN2: $\lambda = 0$, $\epsilon_{\mathrm{pred}} = 12.17\%\,(4.50\%)$}
        \label{fig:NC21_training_l0_rough}
    \end{subfigure}
    \begin{subfigure}{0.495\textwidth}
        \centering
        \includegraphics[width=\linewidth]{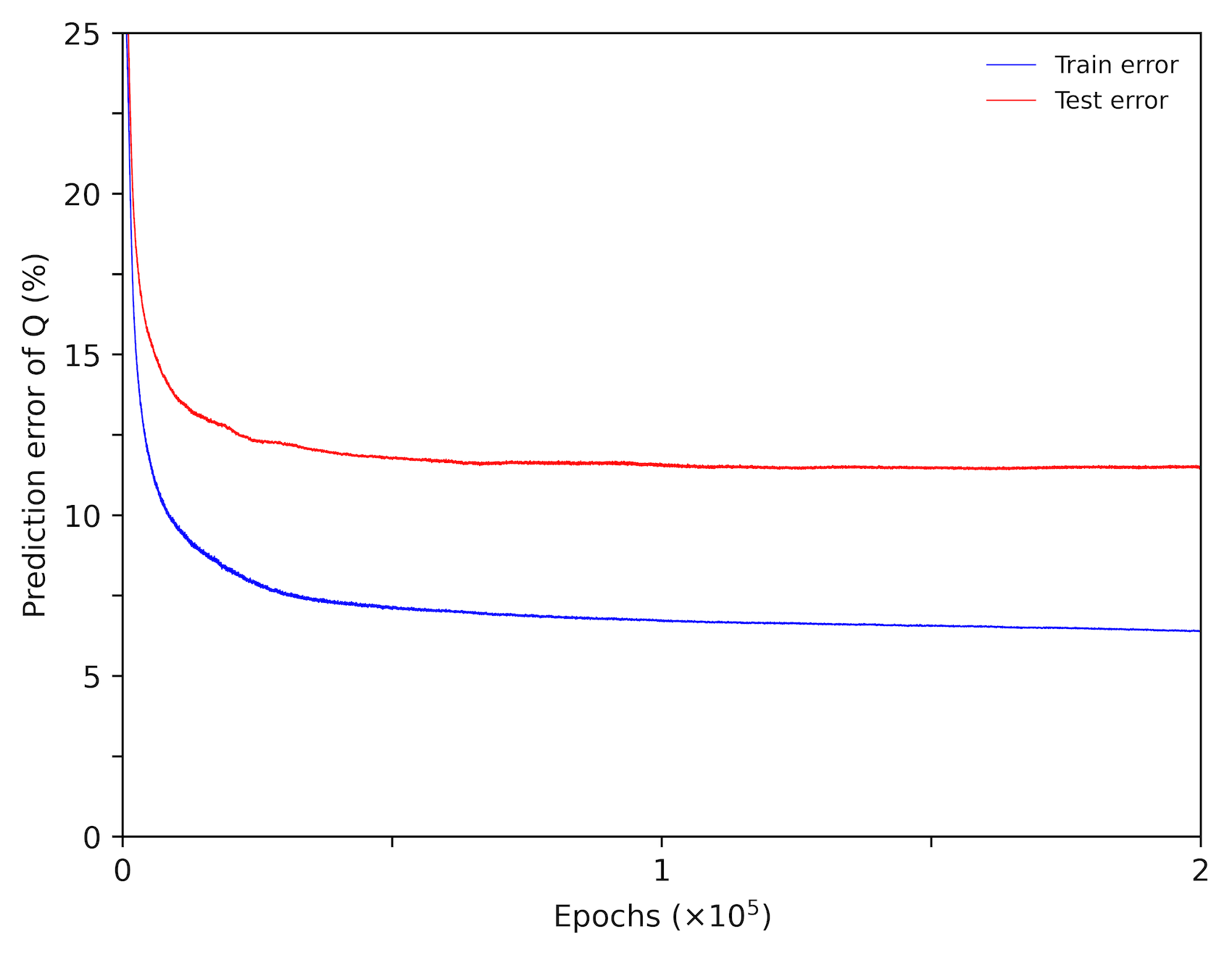}
        \caption{NN2: $\lambda = 0.001$, $\epsilon_{\mathrm{pred}} = 11.52\%\,(6.39\%)$}
        \label{fig:NC21_training_l0.001_rough}
    \end{subfigure}
    \caption{Training process, L2 with surface roughness (NN (bold) responsible for optimizing L2 cavity 2)}
    \label{fig:NC_training_rough}
\end{figure}
\newpage

\begin{figure}[h!]
    \centering
    \begin{subfigure}{0.495\textwidth}
        \centering
        \includegraphics[width=\linewidth]{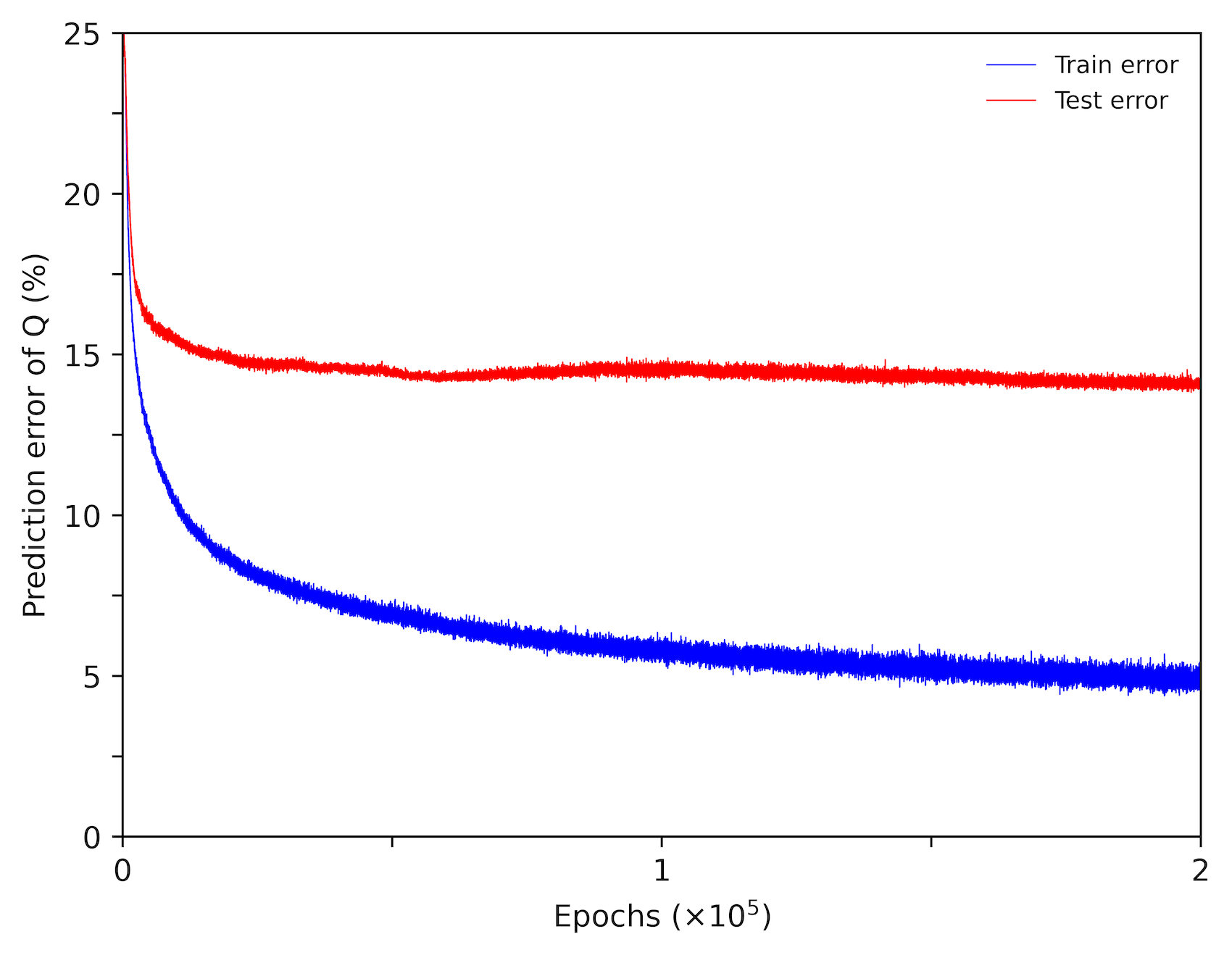}
        \caption{NN1: $\lambda = 0$, $\epsilon_{\mathrm{pred}} = 14.05\% (4.54\%)$}
        \label{fig:FB32_training_l0_rough}
    \end{subfigure}
    \begin{subfigure}{0.495\textwidth}
        \centering
        \includegraphics[width=\linewidth]{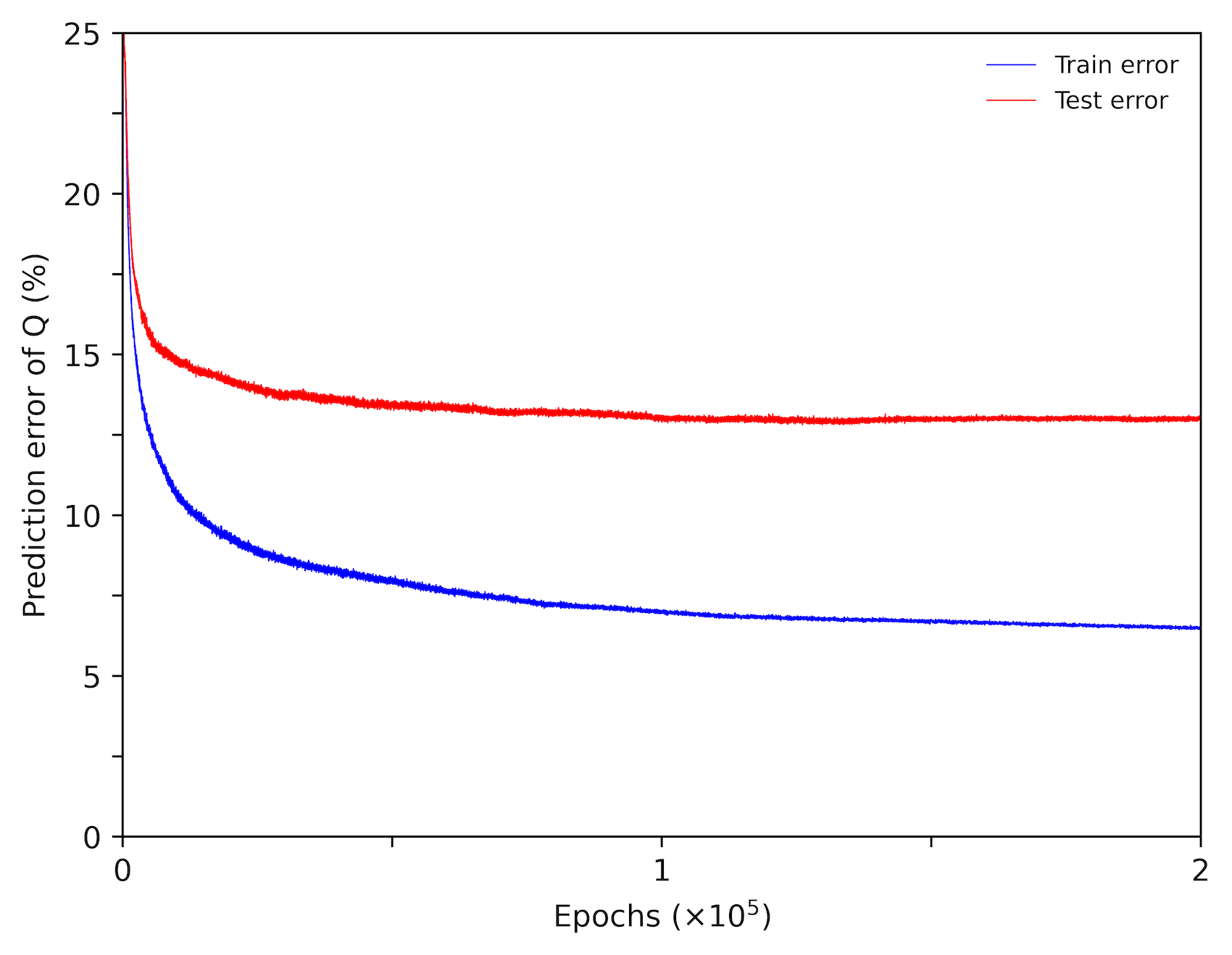}
        \caption{NN1: $\lambda = 0.001$, $\epsilon_{\mathrm{pred}} =  13.00\%\,(6.49\%)$}
        \label{fig:FB32_training_l0.001_rough}
    \end{subfigure}
    
    \begin{subfigure}{0.495\textwidth}
        \centering
        \includegraphics[width=\linewidth]{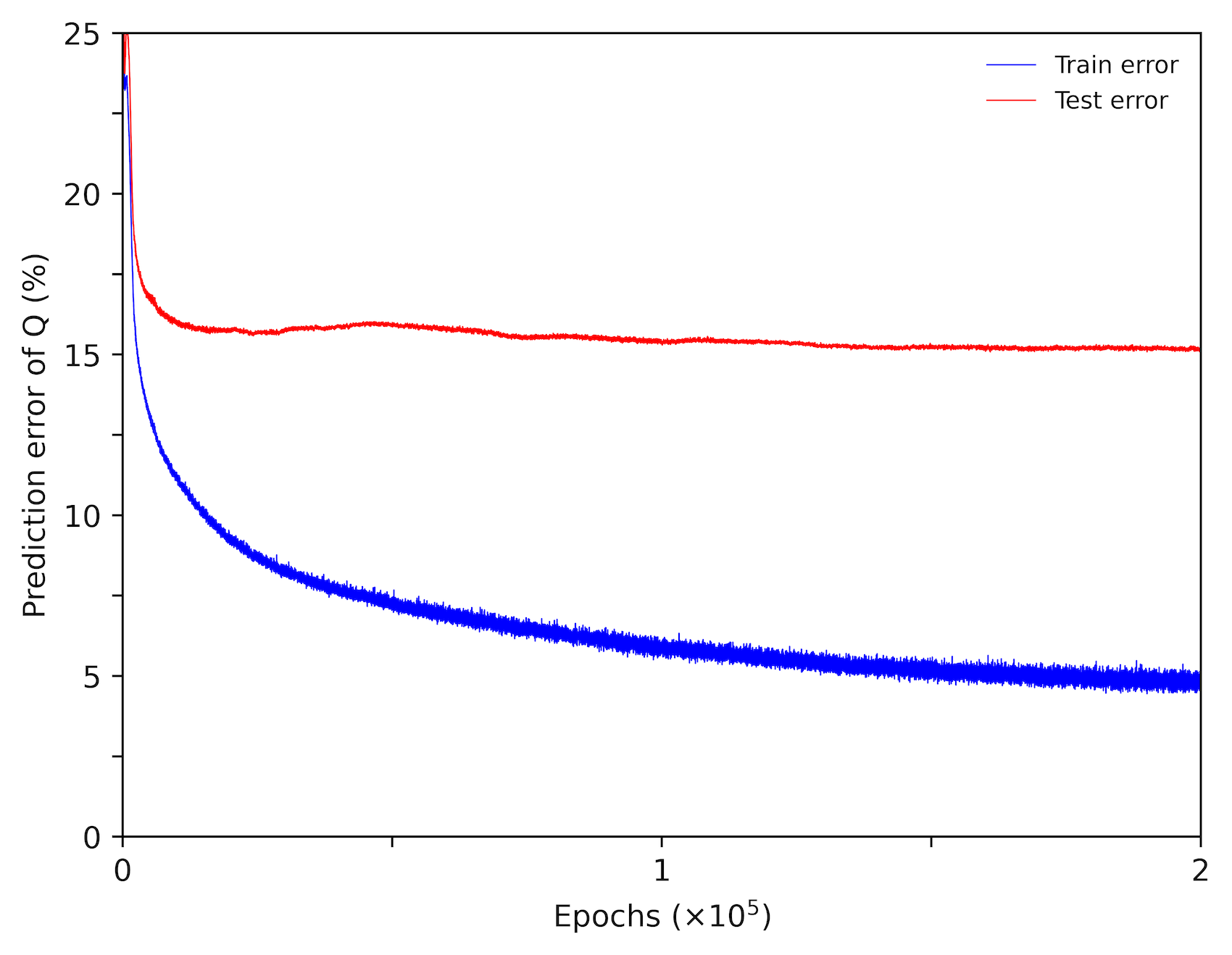}
        \caption{NN2: $\lambda = 0$, $\epsilon_{\mathrm{pred}} = 15.17\%\,(4.94\%)$}
        \label{fig:FB212_training_l0_rough}
    \end{subfigure}
    \begin{subfigure}{0.495\textwidth}
        \centering
        \includegraphics[width=\linewidth]{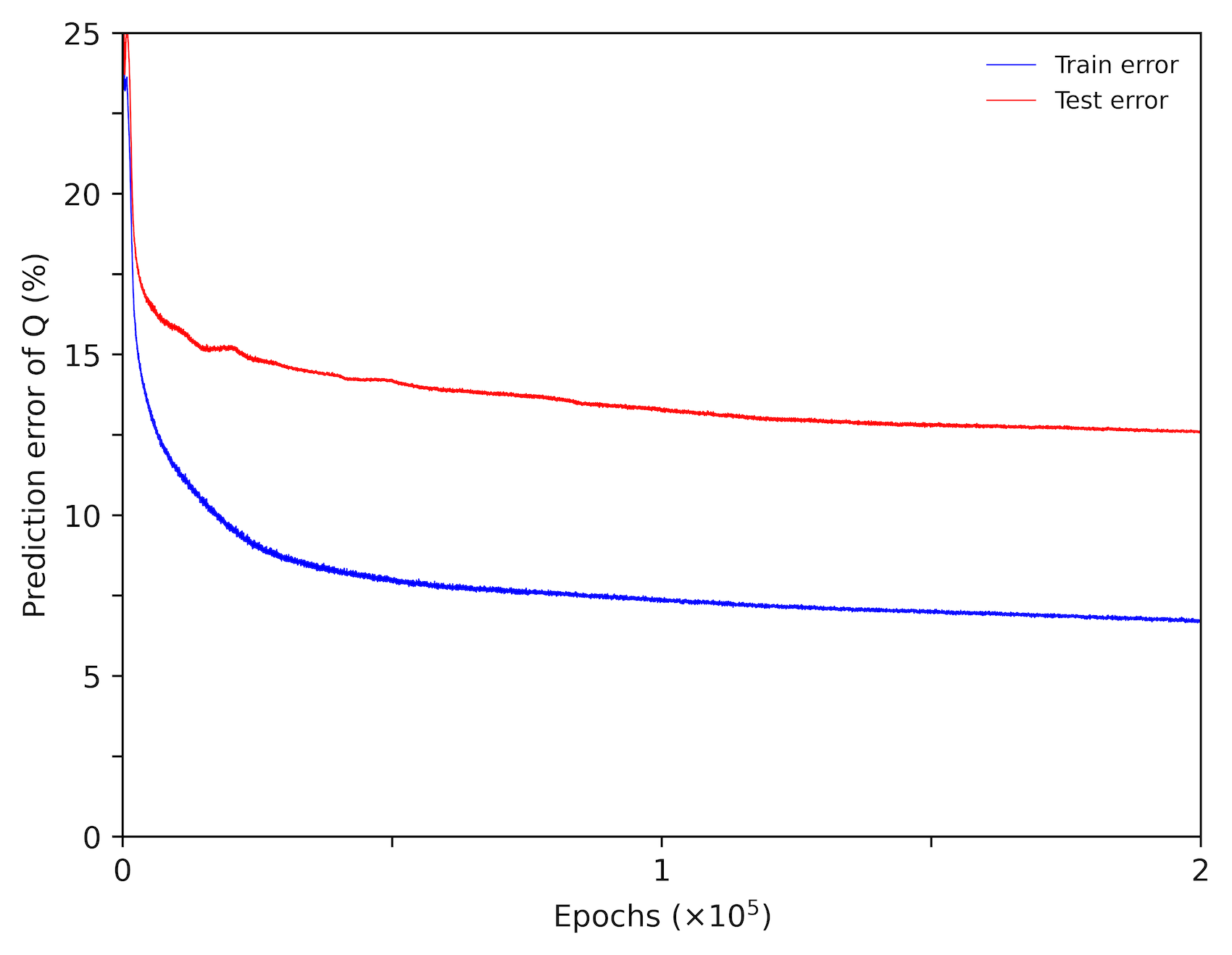}
        \caption{\textbf{NN2}: $\lambda = 0.001$, $\epsilon_{\mathrm{pred}} = 12.58\% (6.72\%)$}
        \label{fig:FB212_training_l0.001_rough}
    \end{subfigure}
    \caption{Training process, fishbone with surface roughness (NN (bold) responsible for optimizing fishbone cavity 2)}
    \label{fig:FB_training_rough}
\end{figure}
\newpage

\begin{figure}[h!]
    \centering
    \begin{subfigure}{0.495\textwidth}
        \centering
        \includegraphics[width=\linewidth]{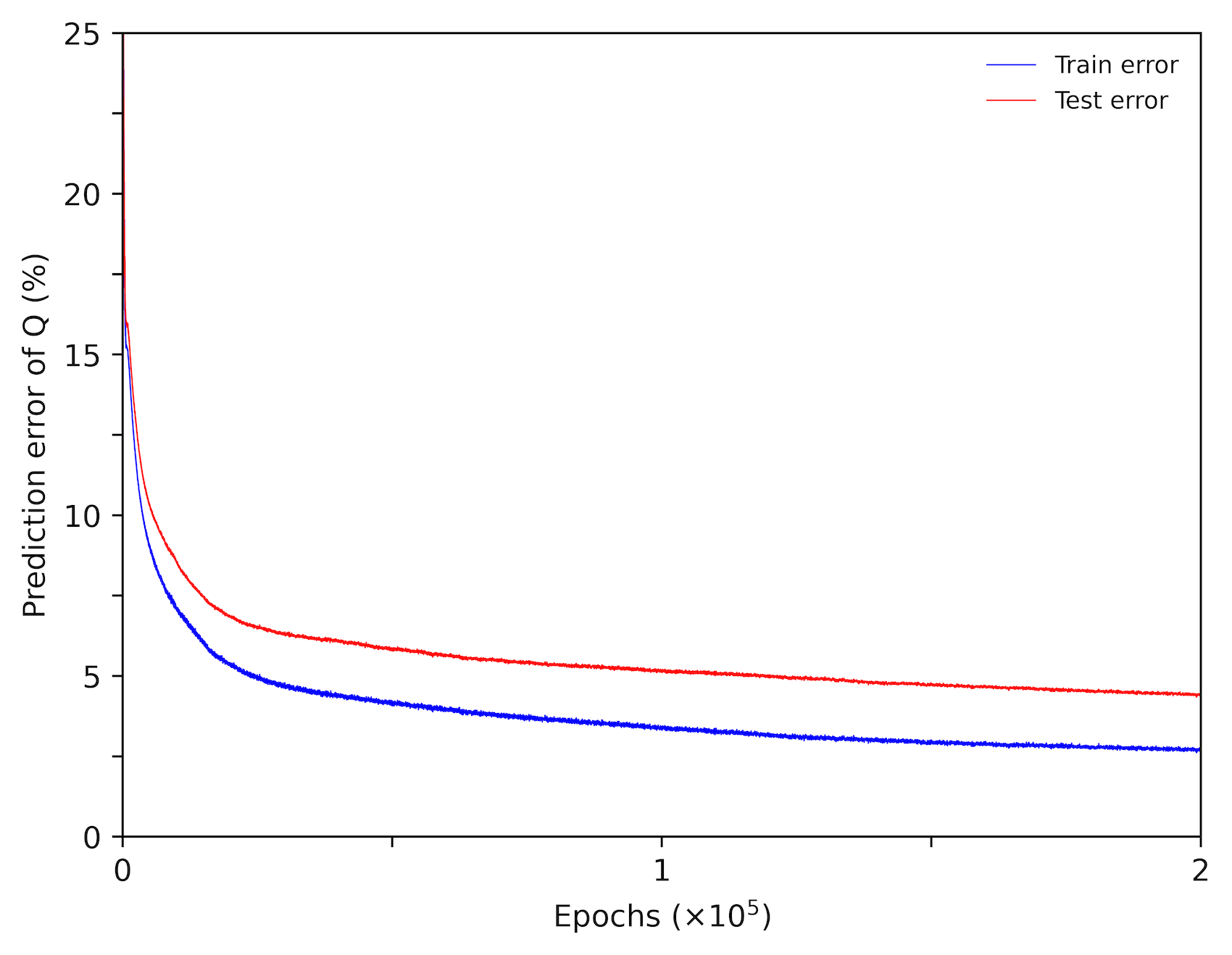}
        \caption{NN1: $\lambda = 0$, $\epsilon_{\mathrm{pred}} = 4.41\%\, (2.67\%)$}
        \label{fig:NC123_training_l0_slant}
    \end{subfigure}
    \begin{subfigure}{0.495\textwidth}
        \centering
        \includegraphics[width=\linewidth]{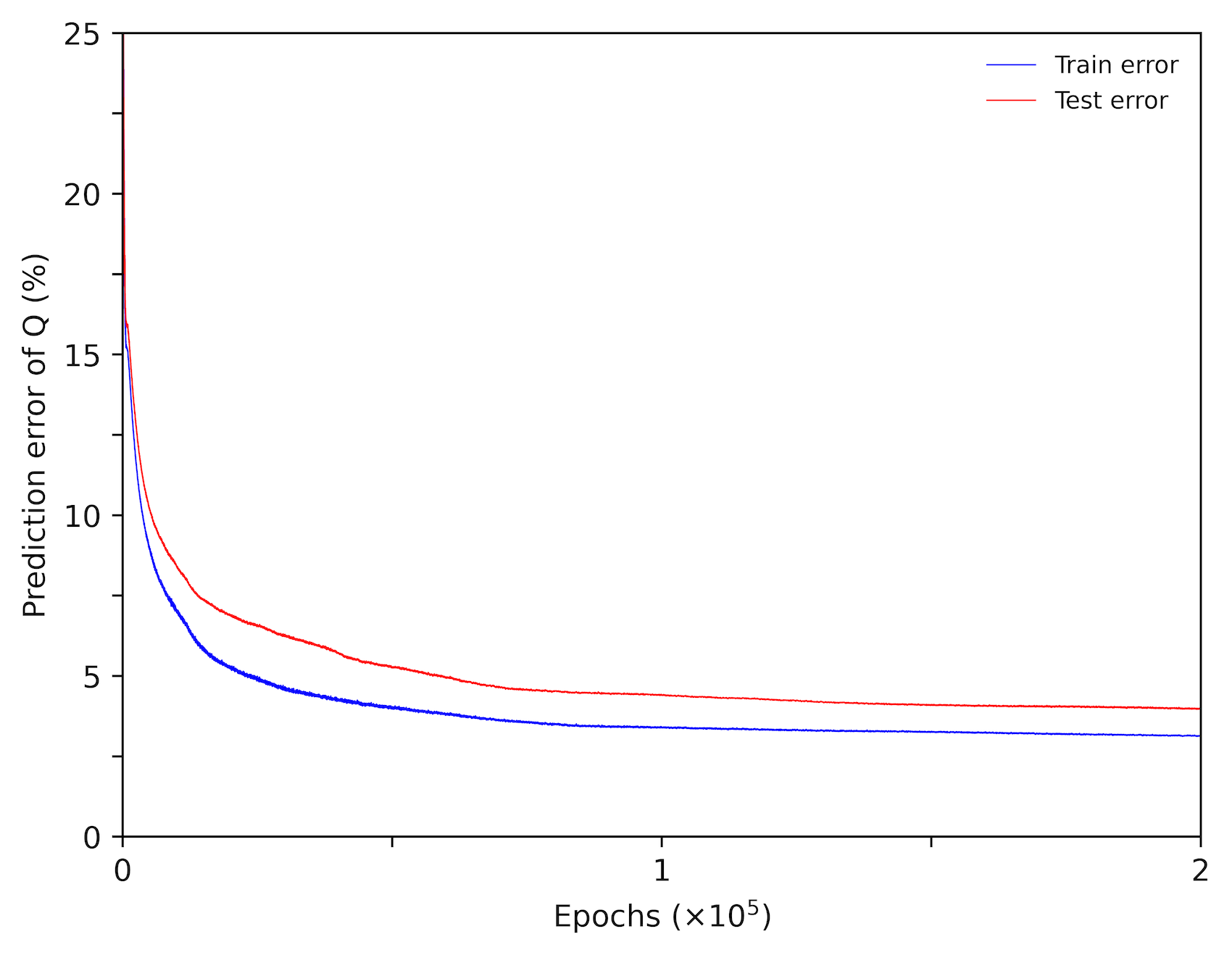}
        \caption{NN1: $\lambda = 0.001$, $\epsilon_{\mathrm{pred}} = 3.99\%\,(3.13\%)$}
        \label{fig:NC123_training_l0.001_slant}
    \end{subfigure}
    
    \begin{subfigure}{0.495\textwidth}
        \centering
        \includegraphics[width=\linewidth]{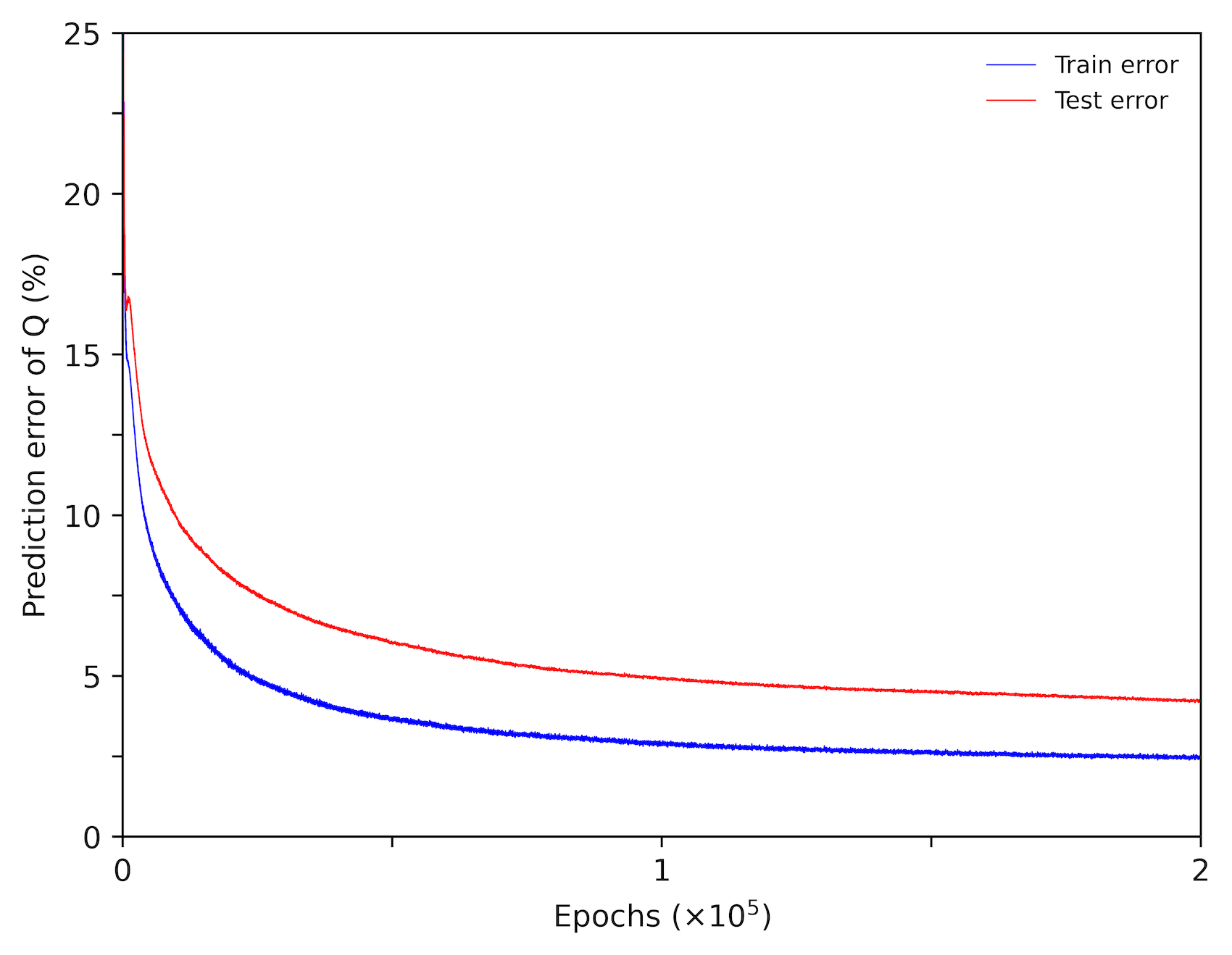}
        \caption{\textbf{NN2}: $\lambda = 0$, $\epsilon_{\mathrm{pred}} = 4.22\%\,(2.43\%)$}
        \label{fig:NC21_training_l0_slant}
    \end{subfigure}
    \begin{subfigure}{0.495\textwidth}
        \centering
        \includegraphics[width=\linewidth]{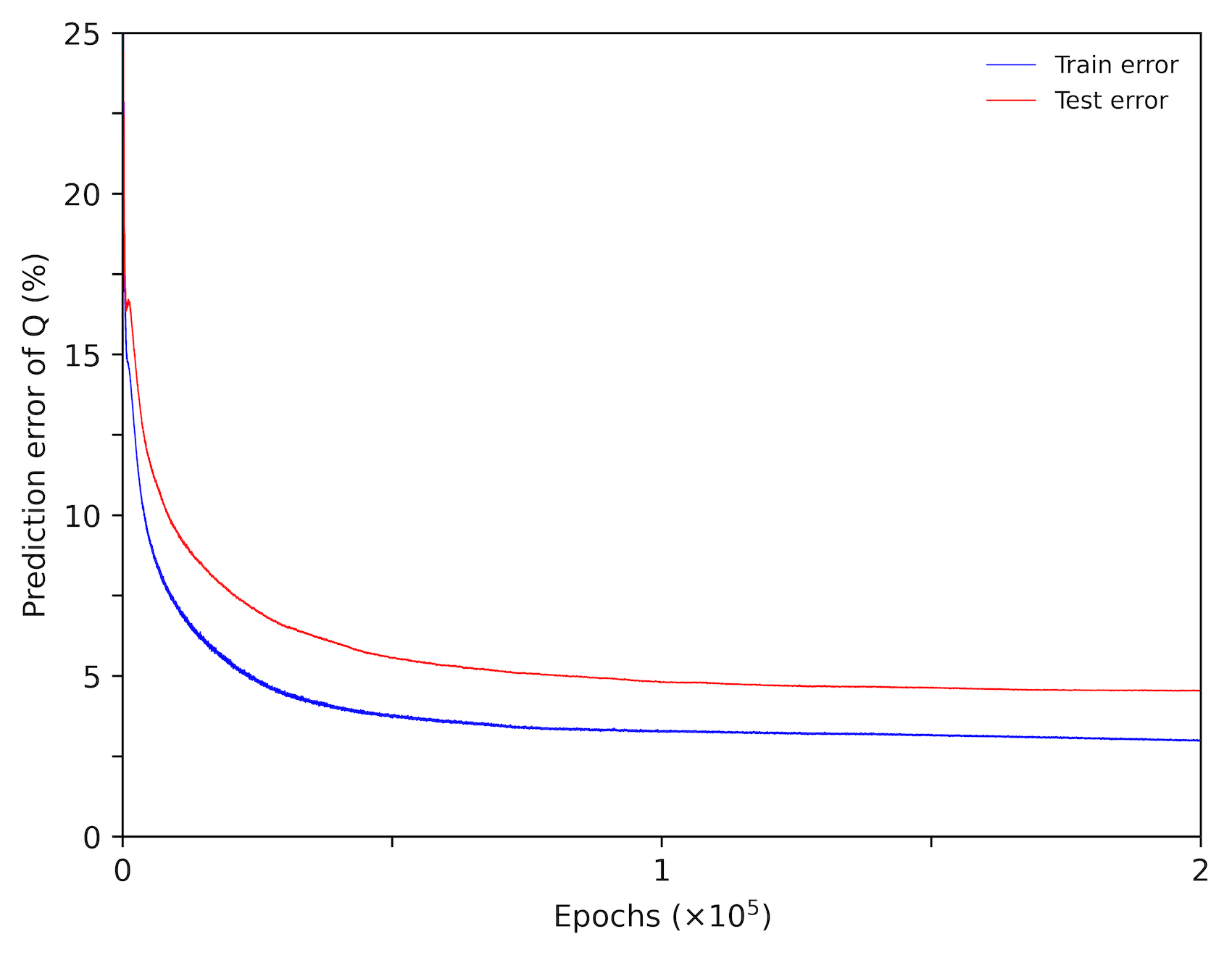}
        \caption{NN2: $\lambda = 0.001$, $\epsilon_{\mathrm{pred}} = 4.55\%\,(2.98\%)$}
        \label{fig:NC21_training_l0.001_slant}
    \end{subfigure}
    \caption{Training process, L2 with sidewall slant (NN (bold) responsible for optimizing L2 cavity 3)}
    \label{fig:NC_training_slant}
\end{figure}
\newpage

\begin{figure}[h!]
    \centering
    \begin{subfigure}{0.495\textwidth}
        \centering
        \includegraphics[width=\linewidth]{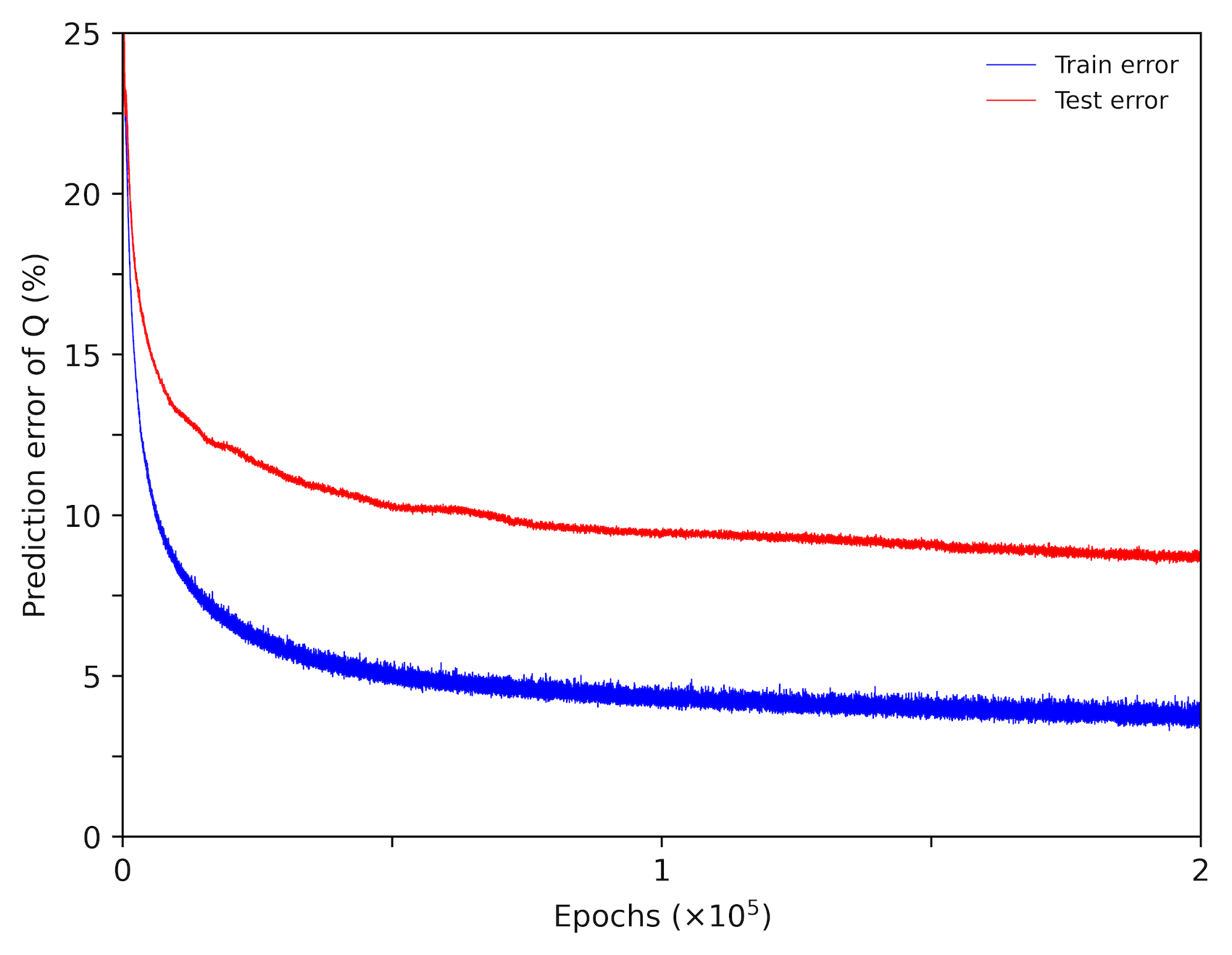}
        \caption{NN1: $\lambda = 0$, $\epsilon_{\mathrm{pred}} = 8.62\%\,(3.97\%)$}
        \label{fig:FB41_training_l0_slant}
    \end{subfigure}
    \begin{subfigure}{0.495\textwidth}
        \centering
        \includegraphics[width=\linewidth]{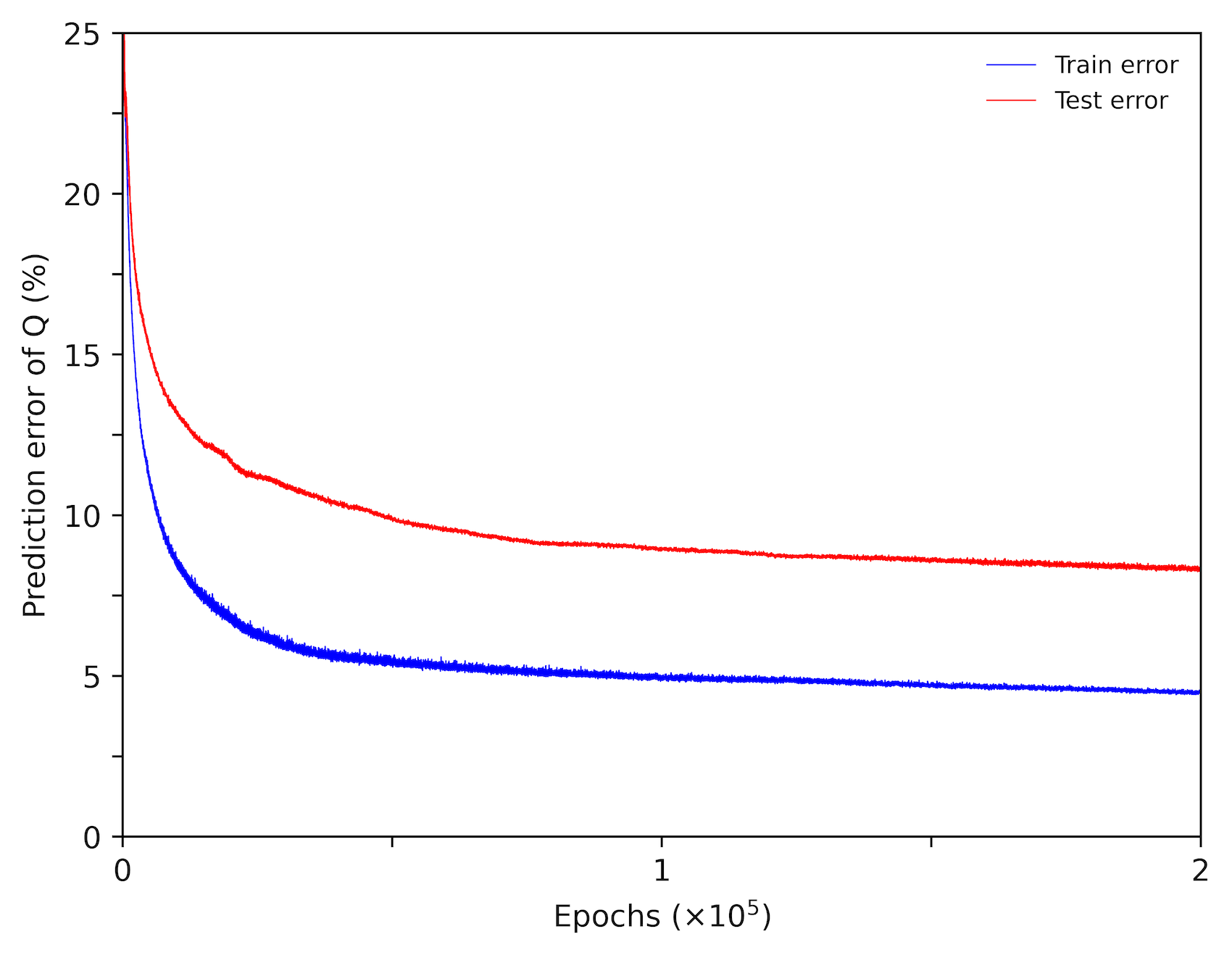}
        \caption{\textbf{NN1}: $\lambda = 0.001$, $\epsilon_{\mathrm{pred}} = 8.25\%\,(4.52\%)$}
        \label{fig:FB41_training_l0.001_slant}
    \end{subfigure}
    
    \begin{subfigure}{0.495\textwidth}
        \centering
        \includegraphics[width=\linewidth]{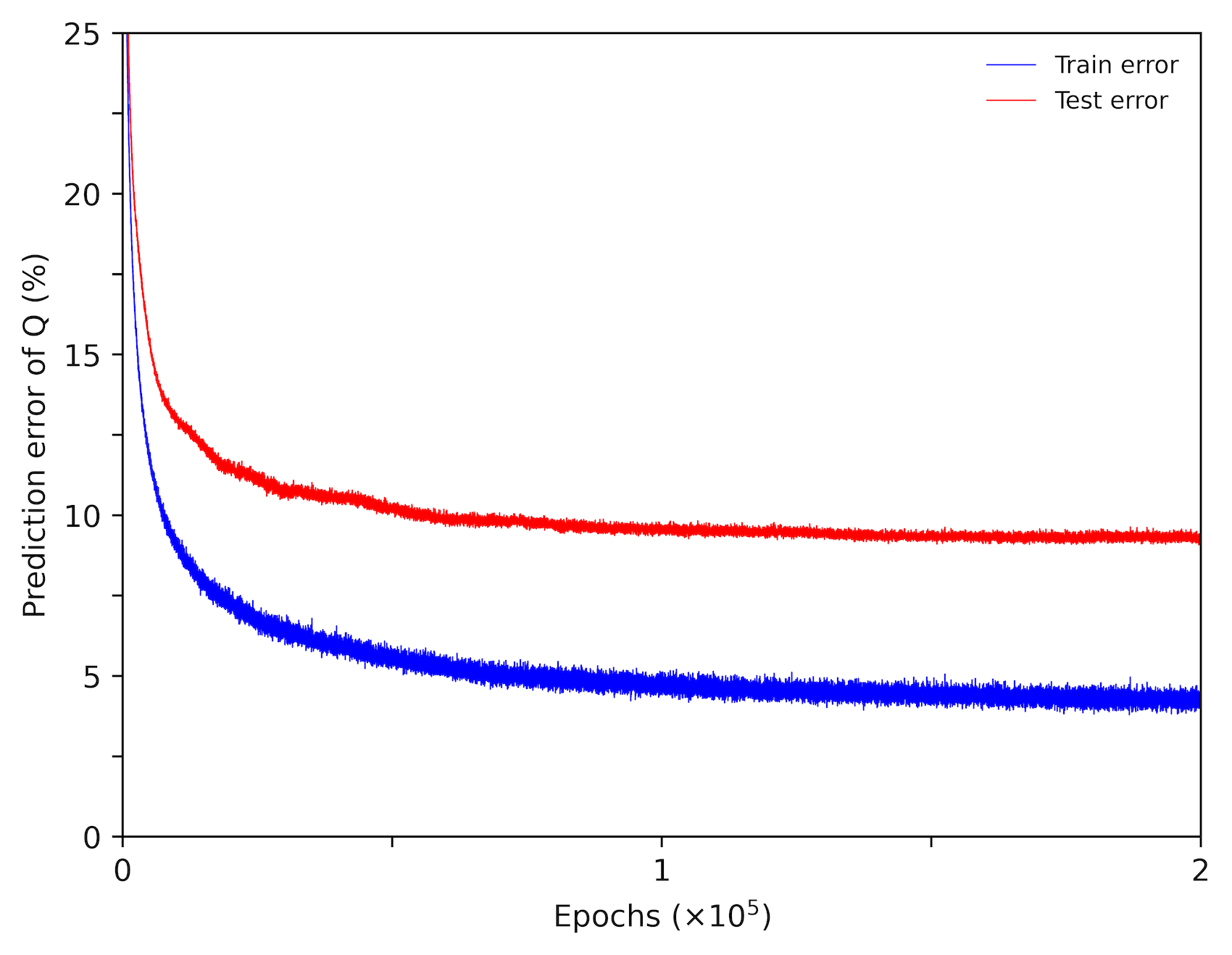}
        \caption{NN2: $\lambda = 0$, $\epsilon_{\mathrm{pred}} = 9.24\%\,(4.32\%)$}
        \label{fig:FB44_training_l0_slant}
    \end{subfigure}
    \begin{subfigure}{0.495\textwidth}
        \centering
        \includegraphics[width=\linewidth]{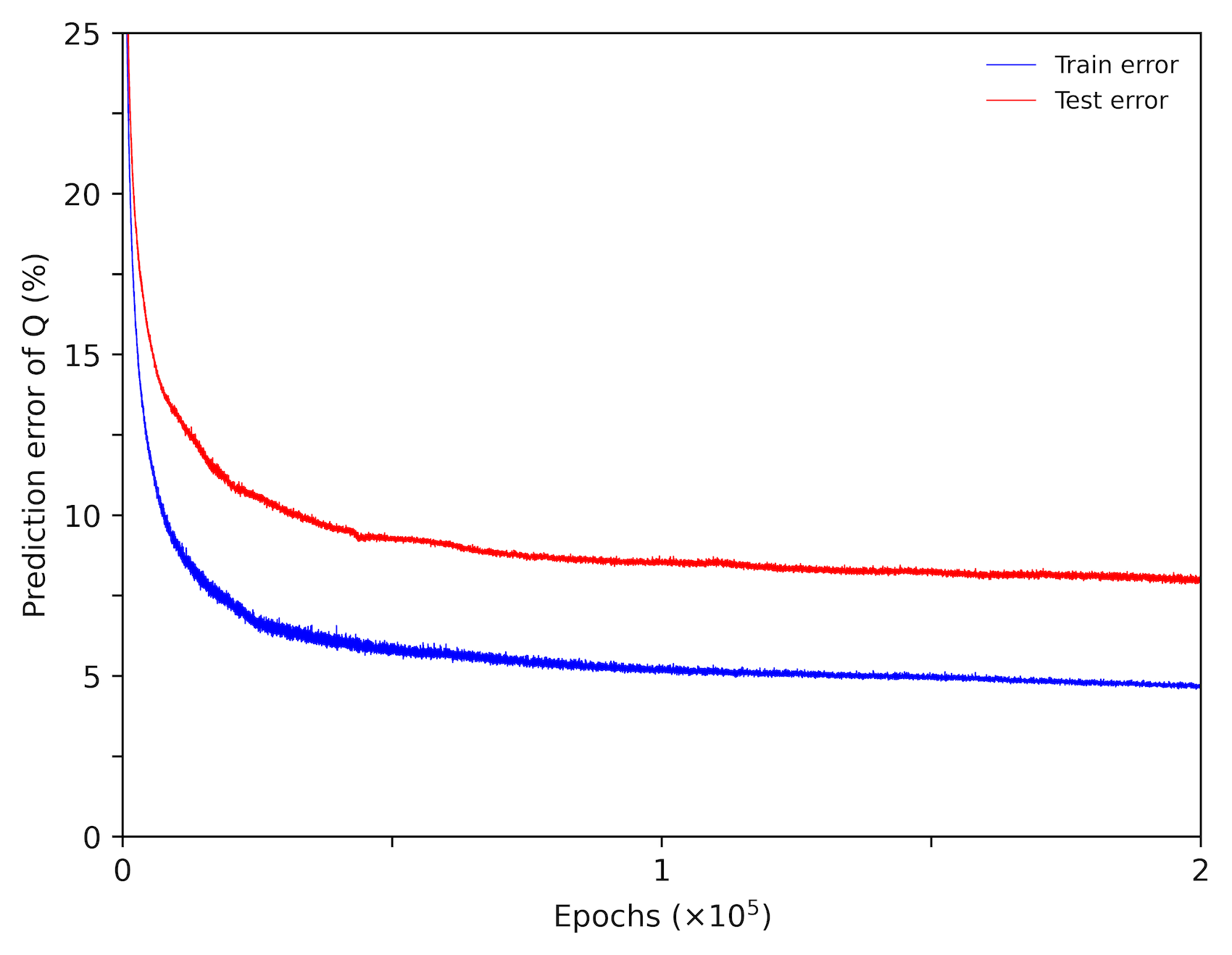}
        \caption{NN2: $\lambda = 0.001$, $\epsilon_{\mathrm{pred}} = 7.95\%\,(4.67\%)$}
        \label{fig:FB44_training_l0.001_slant}
    \end{subfigure}
    \caption{Training process, fishbone with sidewall slant (NN (bold) responsible for optimizing fishbone cavity 3)}
    \label{fig:FB_training_slant}
\end{figure}
\newpage

\begin{figure}[h!]
    \centering
    \begin{subfigure}{0.495\textwidth}
        \centering
        \includegraphics[width=\linewidth]{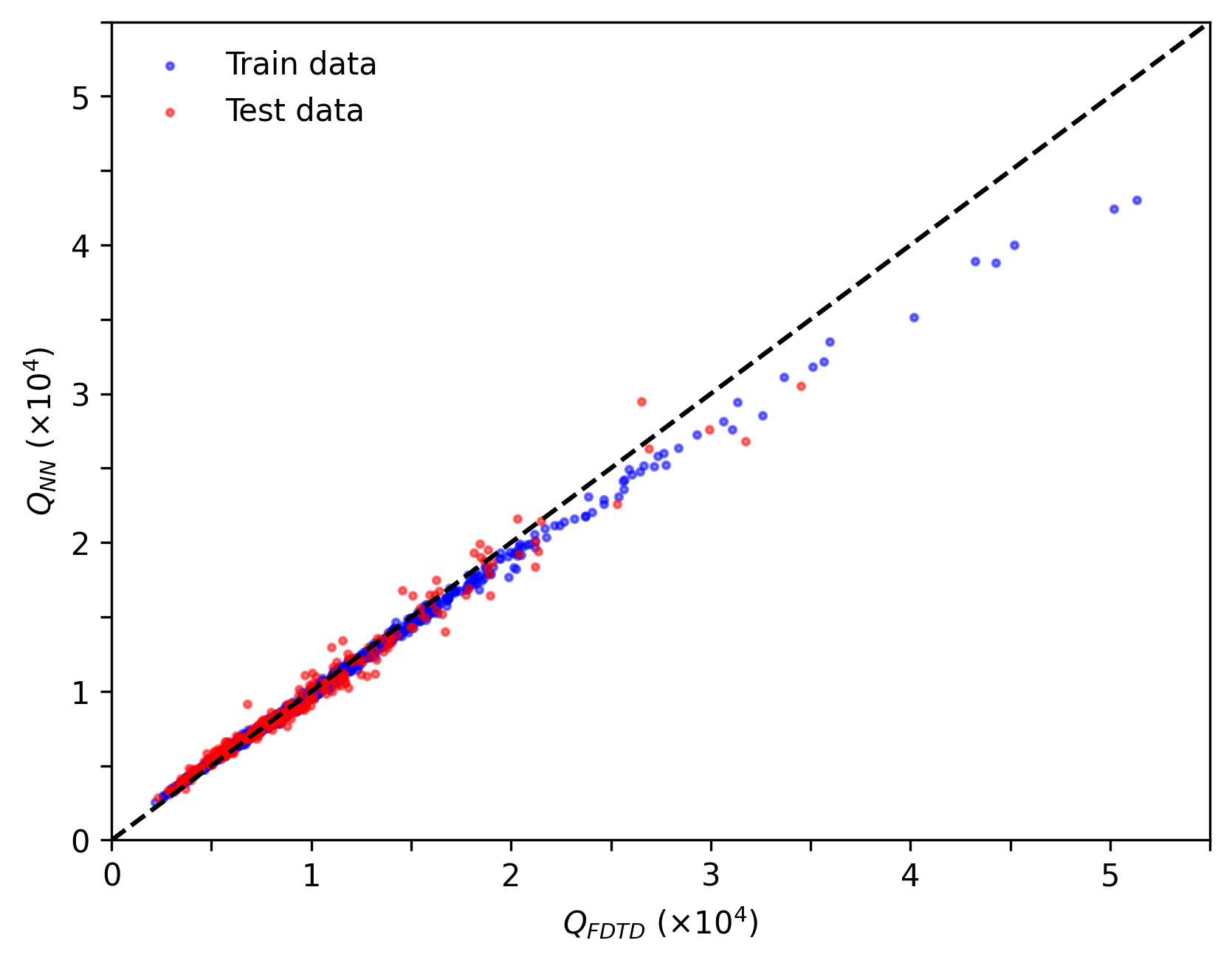}
        \caption{NN1: $\lambda = 0$, $R = 0.988\, (0.998)$}
        \label{fig:NC123_corr_l0_clean}
    \end{subfigure}
    \begin{subfigure}{0.495\textwidth}
        \centering
        \includegraphics[width=\linewidth]{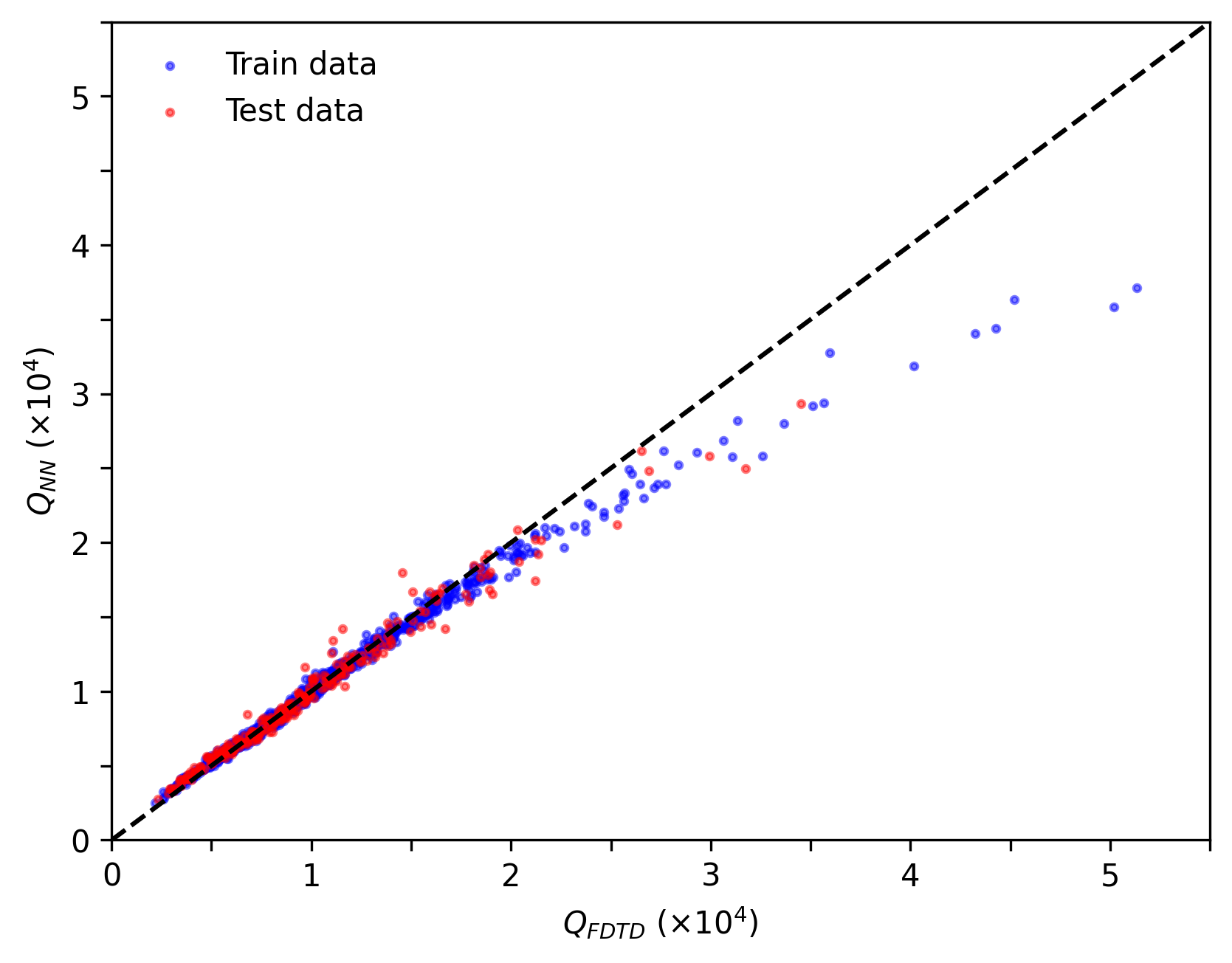}
        \caption{NN1: $\lambda = 0.001$, $R = 0.987\, (0.991)$}
        \label{fig:NC123_corr_l0.001_clean}
    \end{subfigure}
    
    \begin{subfigure}{0.495\textwidth}
        \centering
        \includegraphics[width=\linewidth]{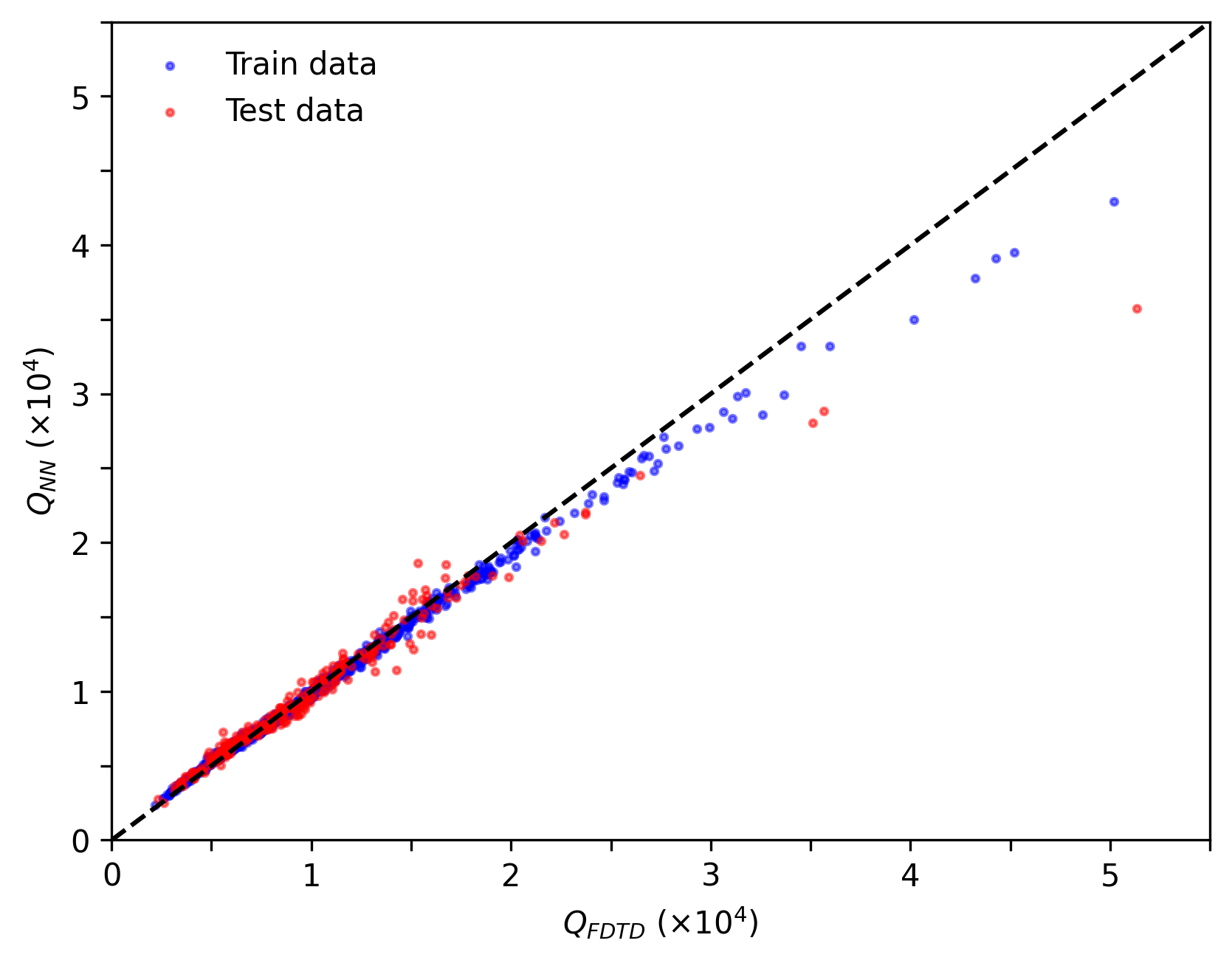}
        \caption{\textbf{NN2}: $\lambda = 0$, $R = 0.979\, (0.998)$}
        \label{fig:NC21_corr_l0_clean}
    \end{subfigure}
    \begin{subfigure}{0.495\textwidth}
        \centering
        \includegraphics[width=\linewidth]{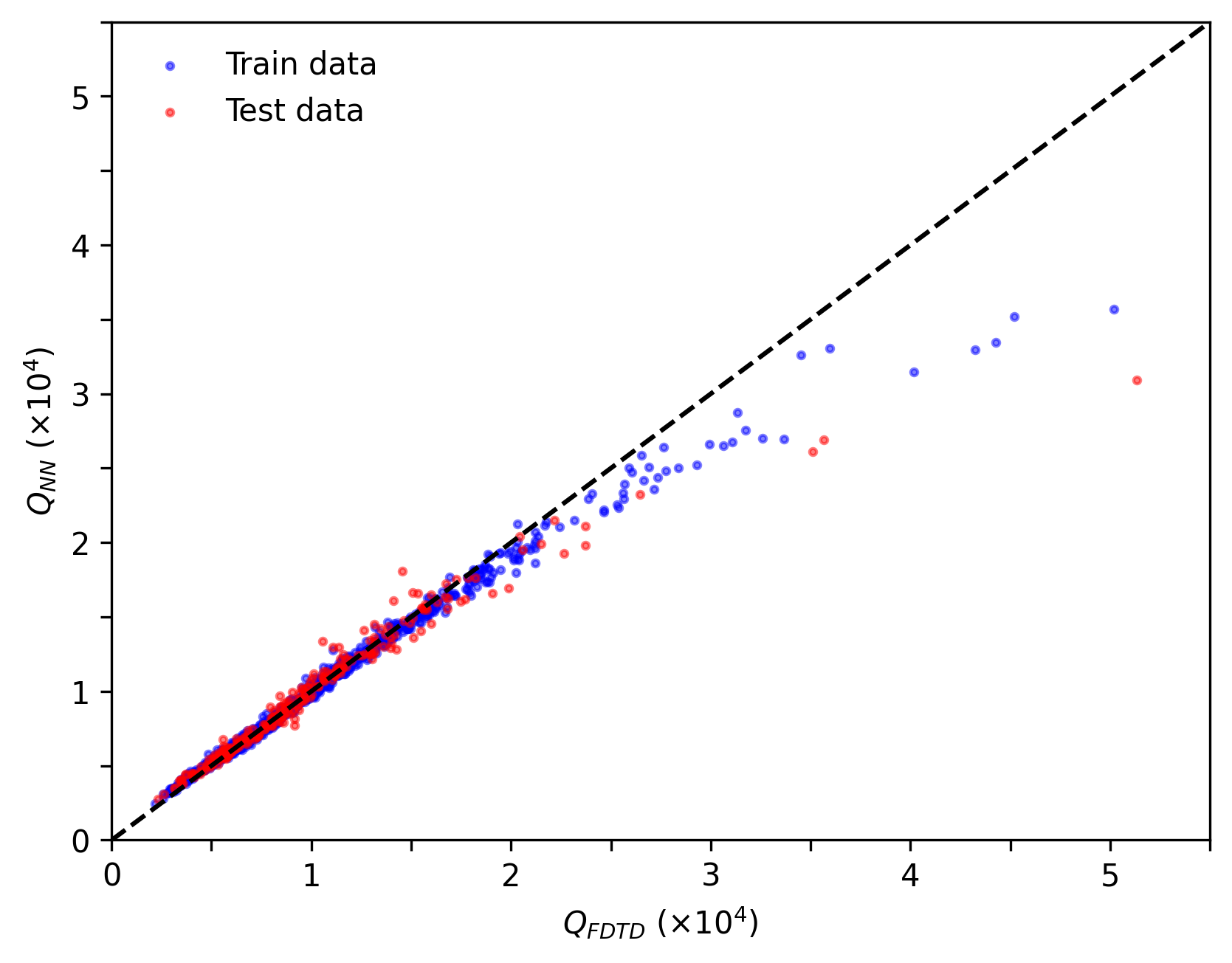}
        \caption{NN2: $\lambda = 0.001$, $R = 0.965\, (0.991)$}
        \label{fig:NC21_corr_l0.001_clean}
    \end{subfigure}
    \caption{Correlation graphs, L2 without imperfections (NN (bold) responsible for optimizing L2 cavity 1)}
    \label{fig:NC_corr_clean}
\end{figure}
\newpage

\begin{figure}[h!]
    \centering
    \begin{subfigure}{0.495\textwidth}
        \centering
        \includegraphics[width=\linewidth]{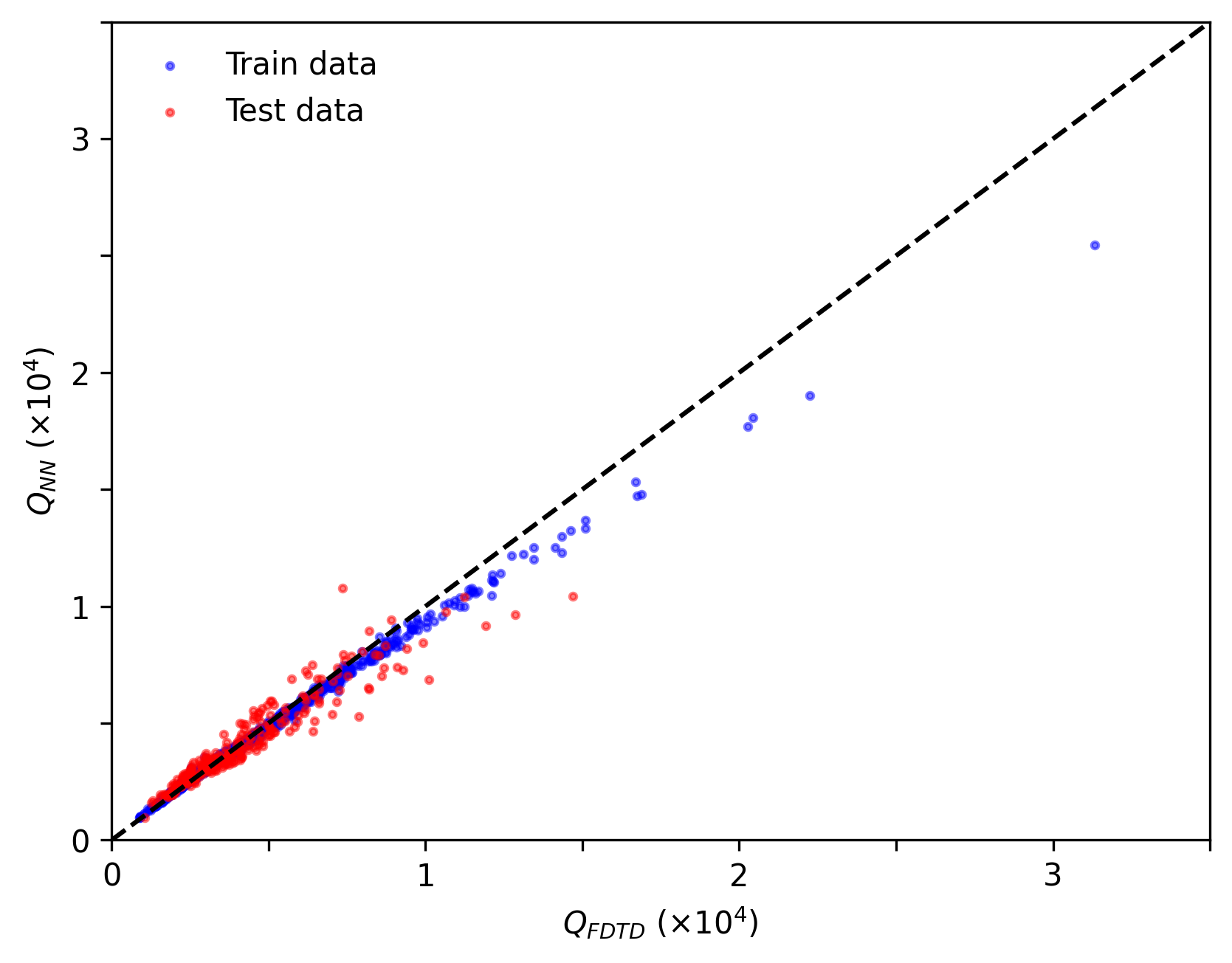}
        \caption{NN1: $\lambda = 0$, $R = 0.948\, (0.998)$}
        \label{fig:FB32_corr_l0_clean}
    \end{subfigure}
    \begin{subfigure}{0.495\textwidth}
        \centering
        \includegraphics[width=\linewidth]{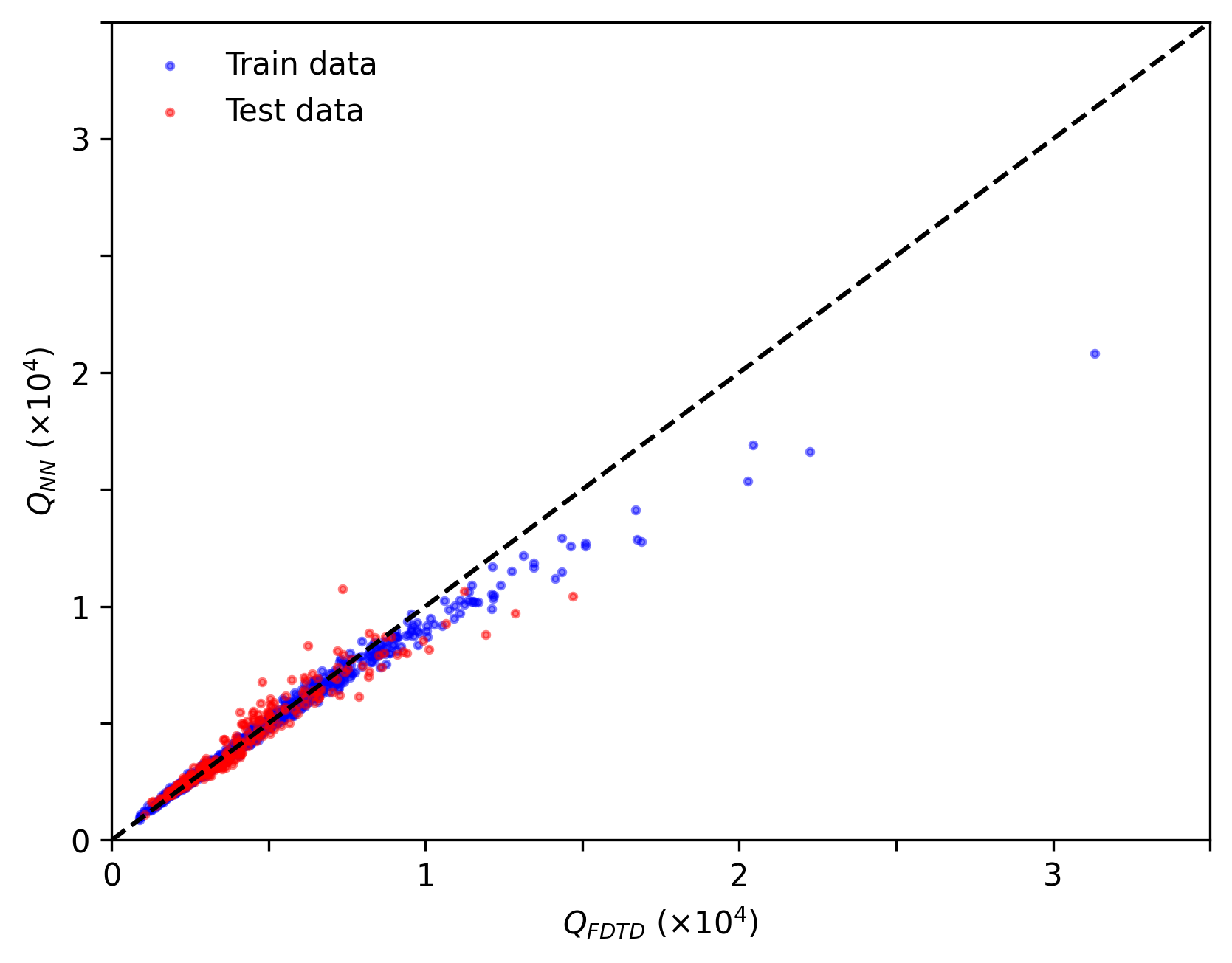}
        \caption{NN1: $\lambda = 0.001$, $R = 0.956\, (0.988)$}
        \label{fig:FB32_corr_l0.001_clean}
    \end{subfigure}
    
    \begin{subfigure}{0.495\textwidth}
        \centering
        \includegraphics[width=\linewidth]{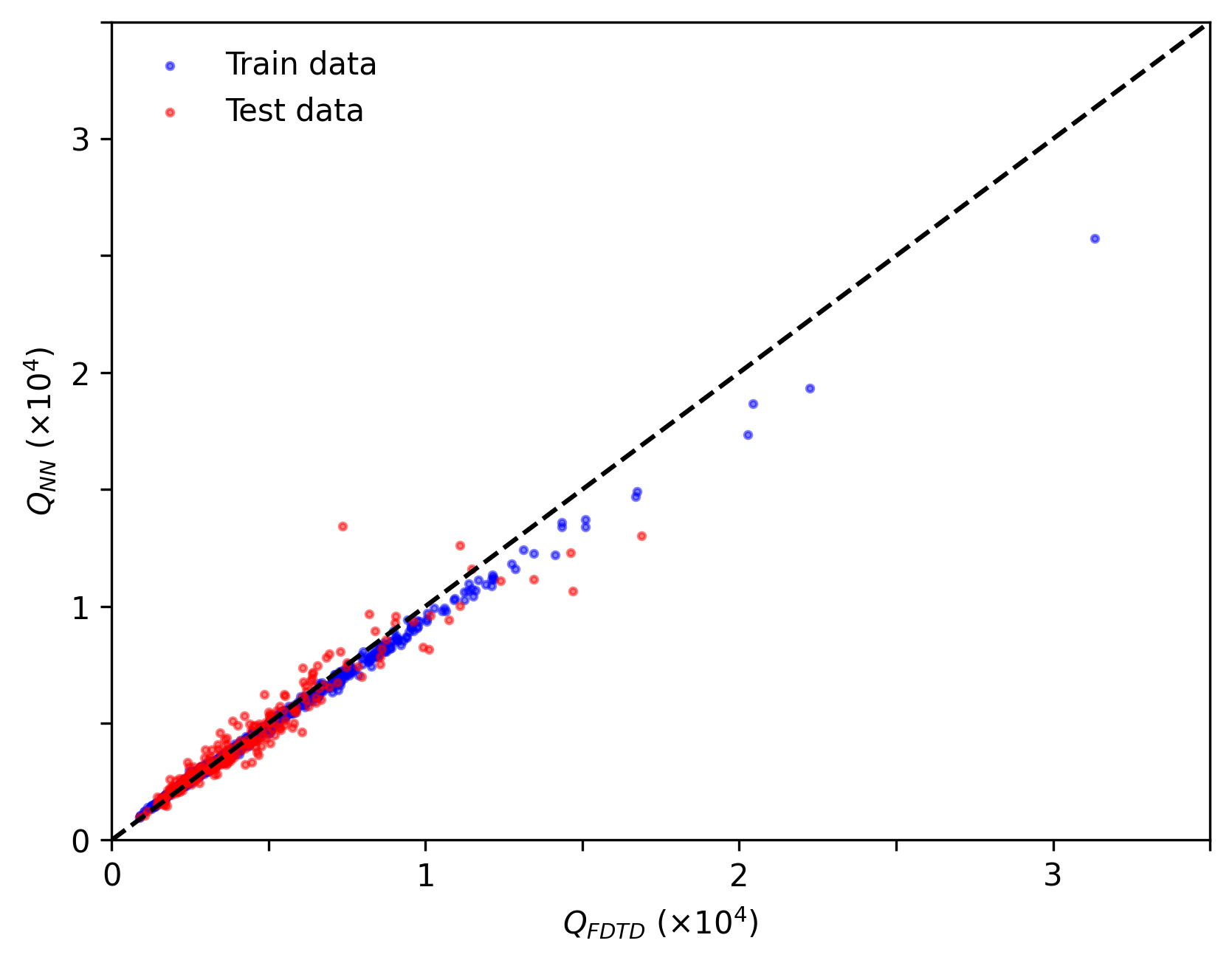}
        \caption{NN2: $\lambda = 0$, $R = 0.955\, (0.998)$}
        \label{fig:FB212_corr_l0_clean}
    \end{subfigure}
    \begin{subfigure}{0.495\textwidth}
        \centering
        \includegraphics[width=\linewidth]{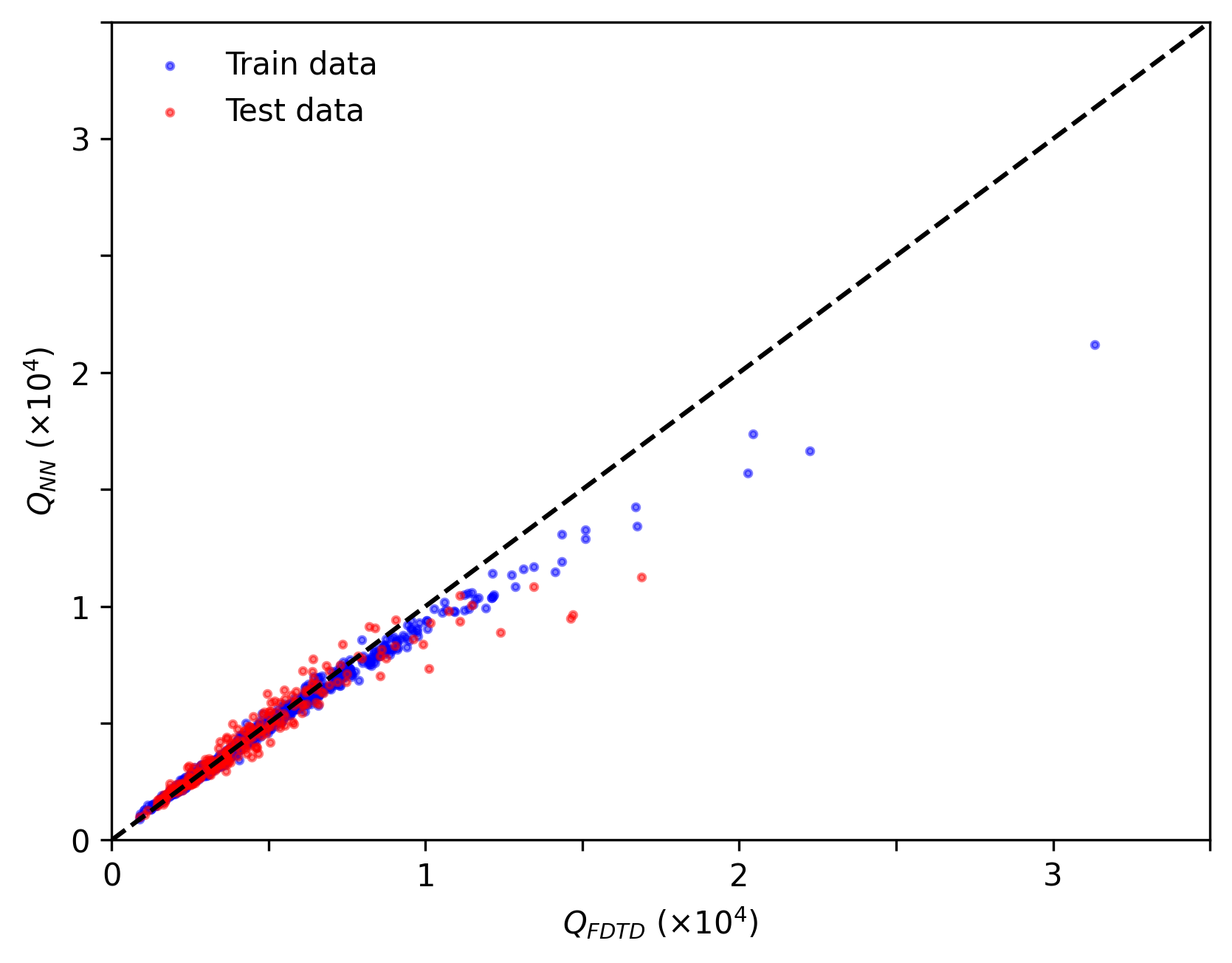}
        \caption{\textbf{NN2}: $\lambda = 0.001$, $R = 0.957\, (0.990)$}
        \label{fig:FB212_corr_l0.001_clean}
    \end{subfigure}
    \caption{Correlation graphs, fishbone without imperfections (NN (bold) responsible for optimizing fishbone cavity 1)}
    \label{fig:FB_corr_clean}
\end{figure}
\newpage

\begin{figure}[h!]
    \centering
    \begin{subfigure}{0.495\textwidth}
        \centering
        \includegraphics[width=\linewidth]{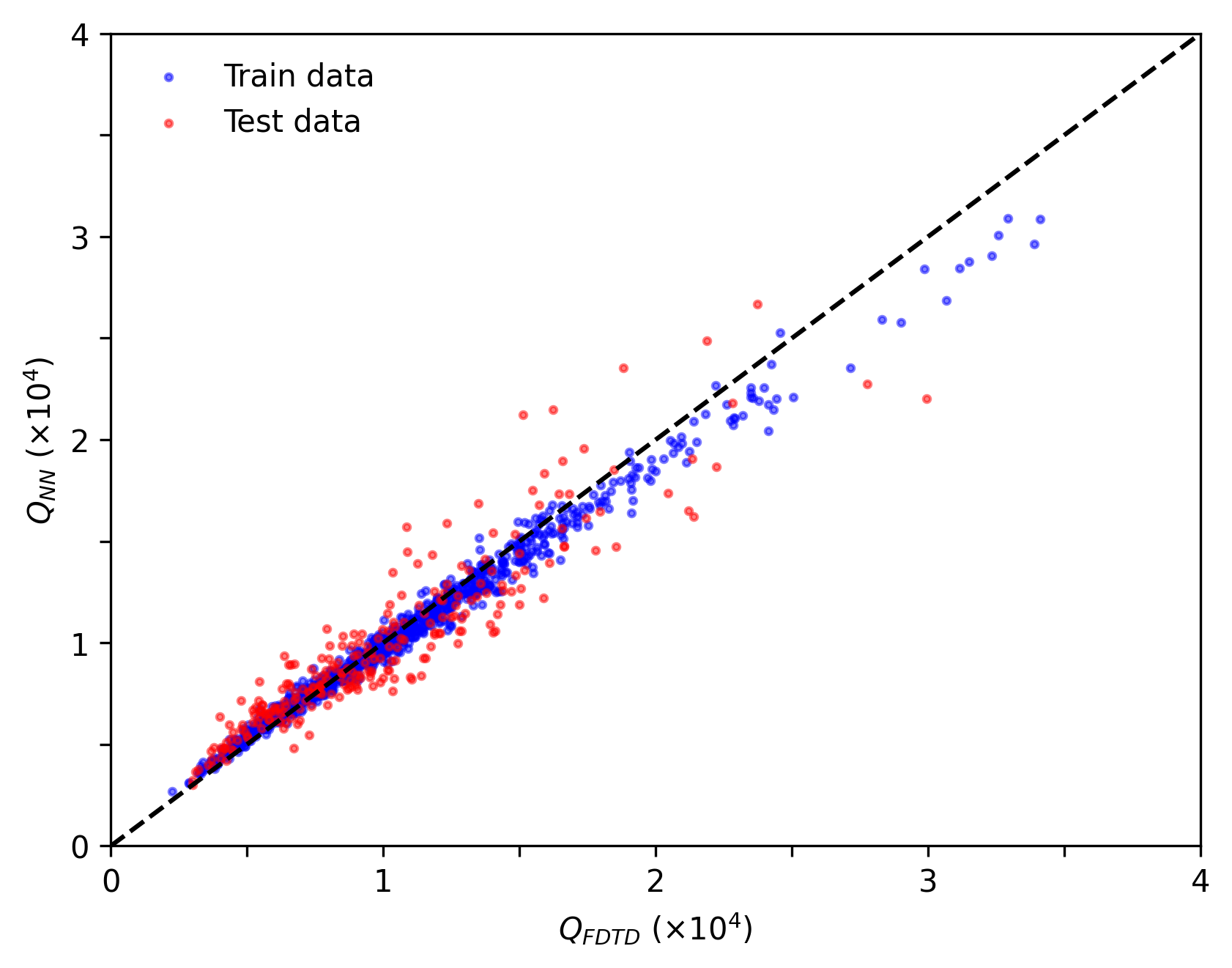}
        \caption{NN1: $\lambda = 0$, $R = 0.924\, (0.995)$}
        \label{fig:NC123_corr_l0_rough}
    \end{subfigure}
    \begin{subfigure}{0.495\textwidth}
        \centering
        \includegraphics[width=\linewidth]{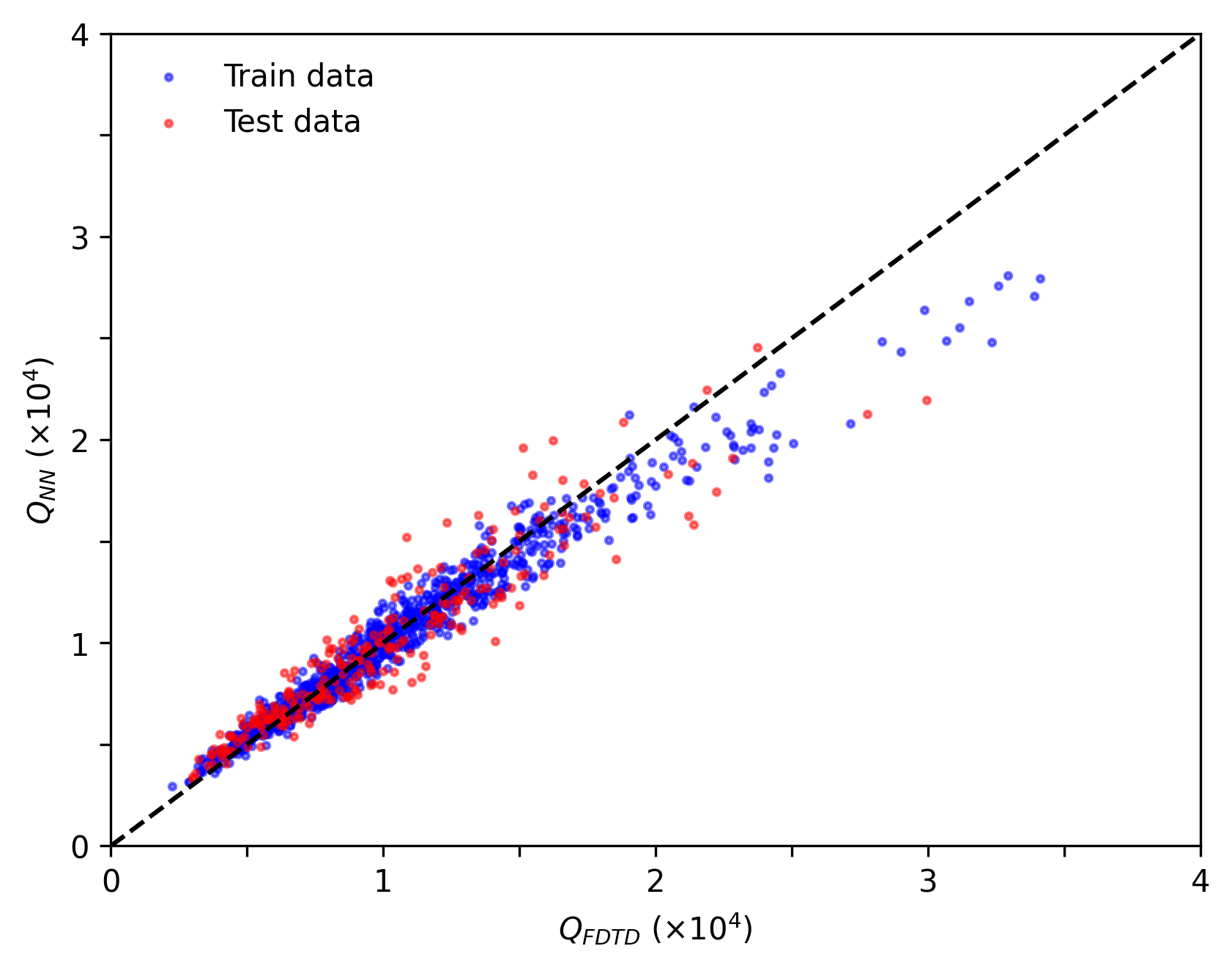}
        \caption{\textbf{NN1}: $\lambda = 0.001$, $R = 0.940\, (0.981)$}
        \label{fig:NC123_corr_l0.001_rough}
    \end{subfigure}
    
    \begin{subfigure}{0.495\textwidth}
        \centering
        \includegraphics[width=\linewidth]{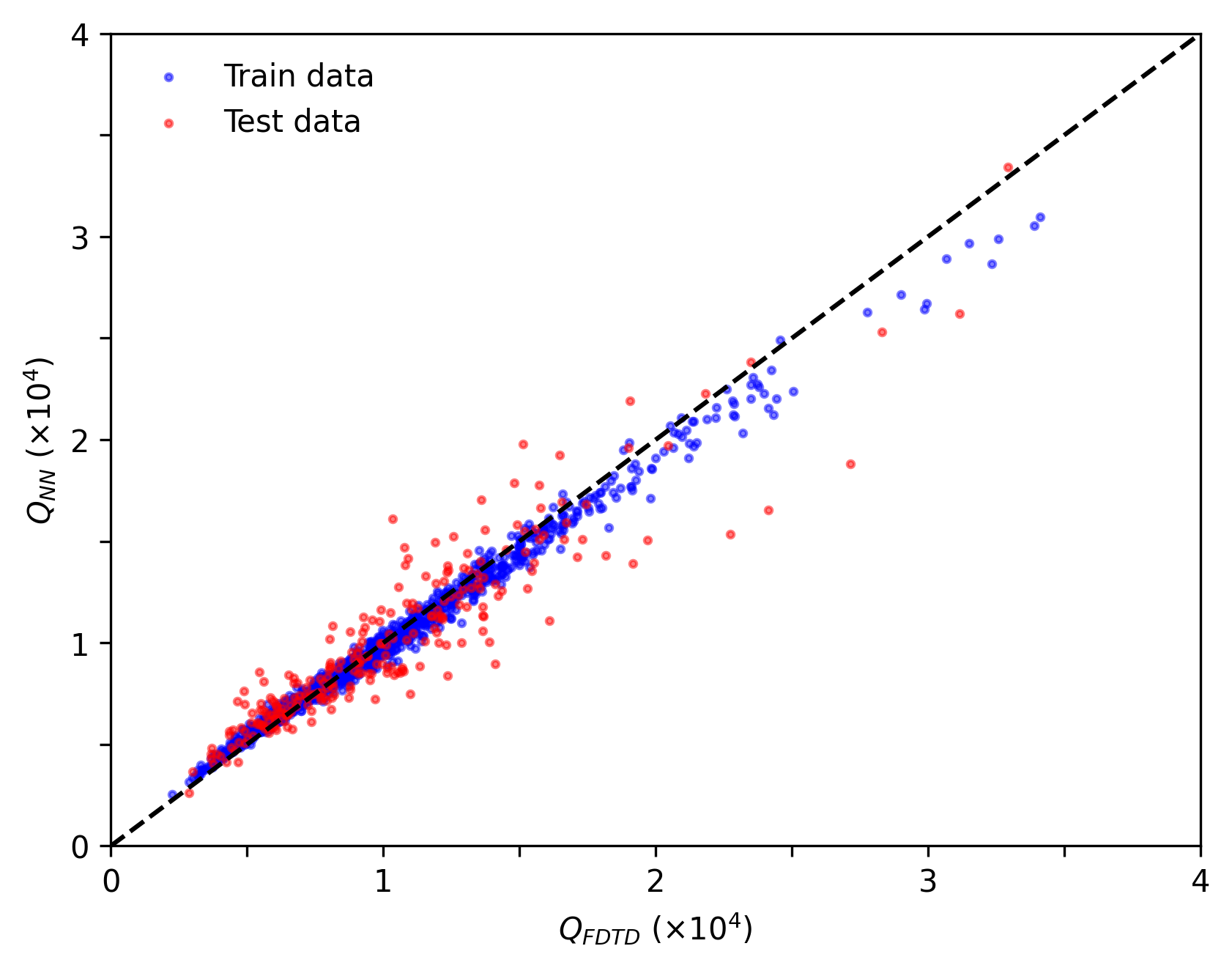}
        \caption{NN2: $\lambda = 0$, $R = 0.927\, (0.995)$}
        \label{fig:NC21_corr_l0_rough}
    \end{subfigure}
    \begin{subfigure}{0.495\textwidth}
        \centering
        \includegraphics[width=\linewidth]{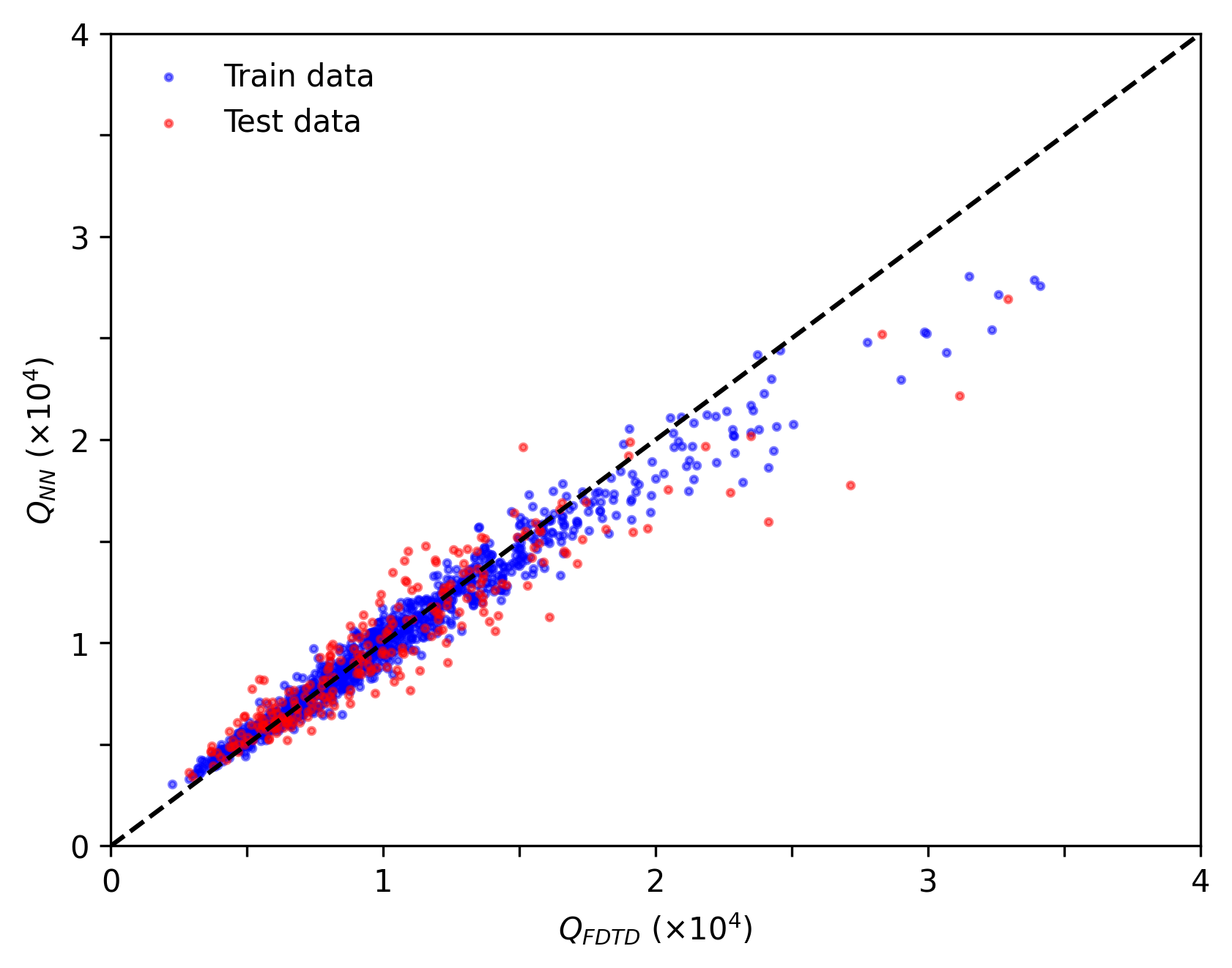}
        \caption{NN2: $\lambda = 0.001$, $R = 0.935\, (0.983)$}
        \label{fig:NC21_corr_l0.001_rough}
    \end{subfigure}
    \caption{Correlation graphs, L2 with surface roughness (NN (bold) responsible for optimizing L2 cavity 2)}
    \label{fig:NC_corr_rough}
\end{figure}
\newpage

\begin{figure}[h!]
    \centering
    \begin{subfigure}{0.495\textwidth}
        \centering
        \includegraphics[width=\linewidth]{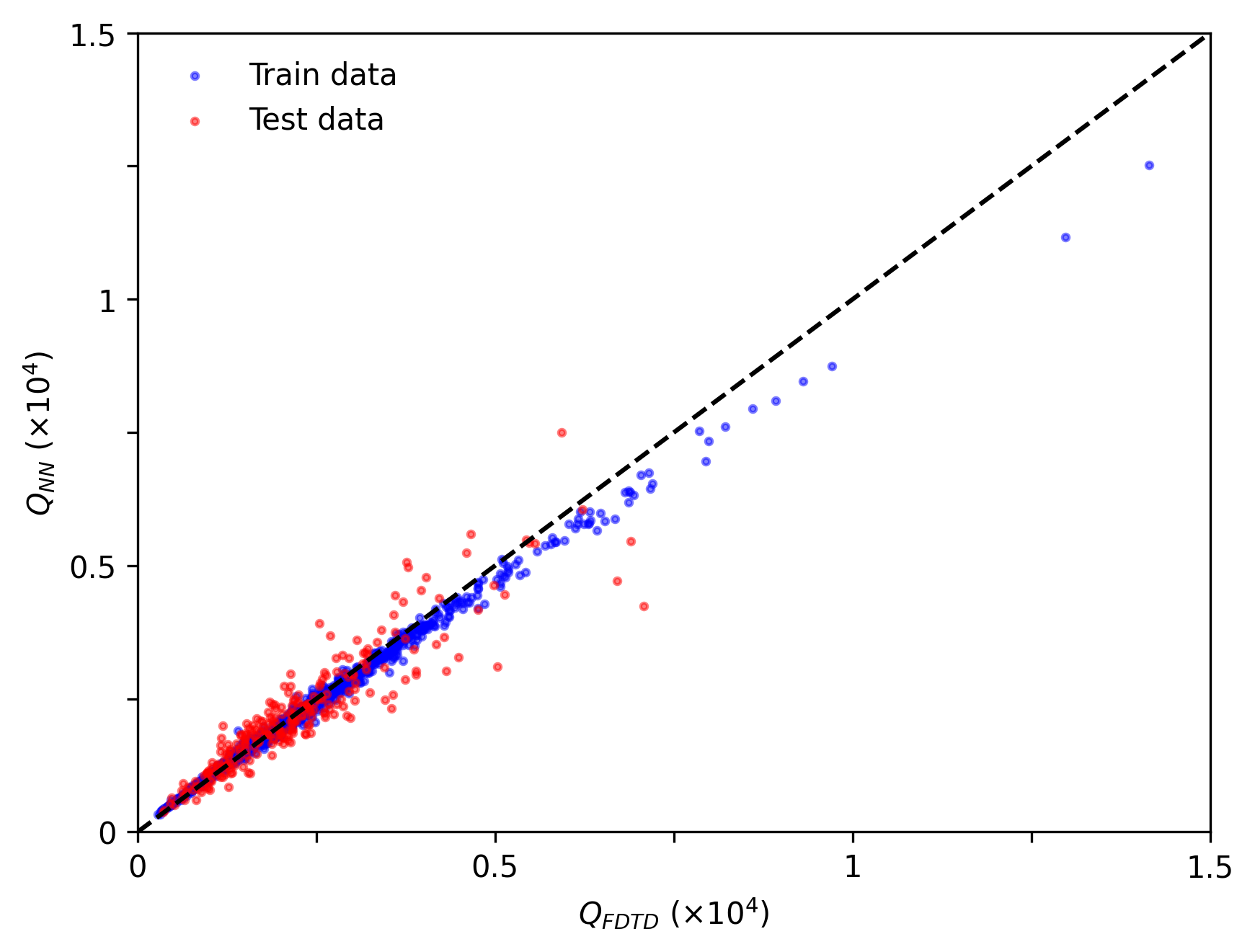}
        \caption{NN1: $\lambda = 0$, $R = 0.917\, (0.997)$}
        \label{fig:FB32_corr_l0_rough}
    \end{subfigure}
    \begin{subfigure}{0.495\textwidth}
        \centering
        \includegraphics[width=\linewidth]{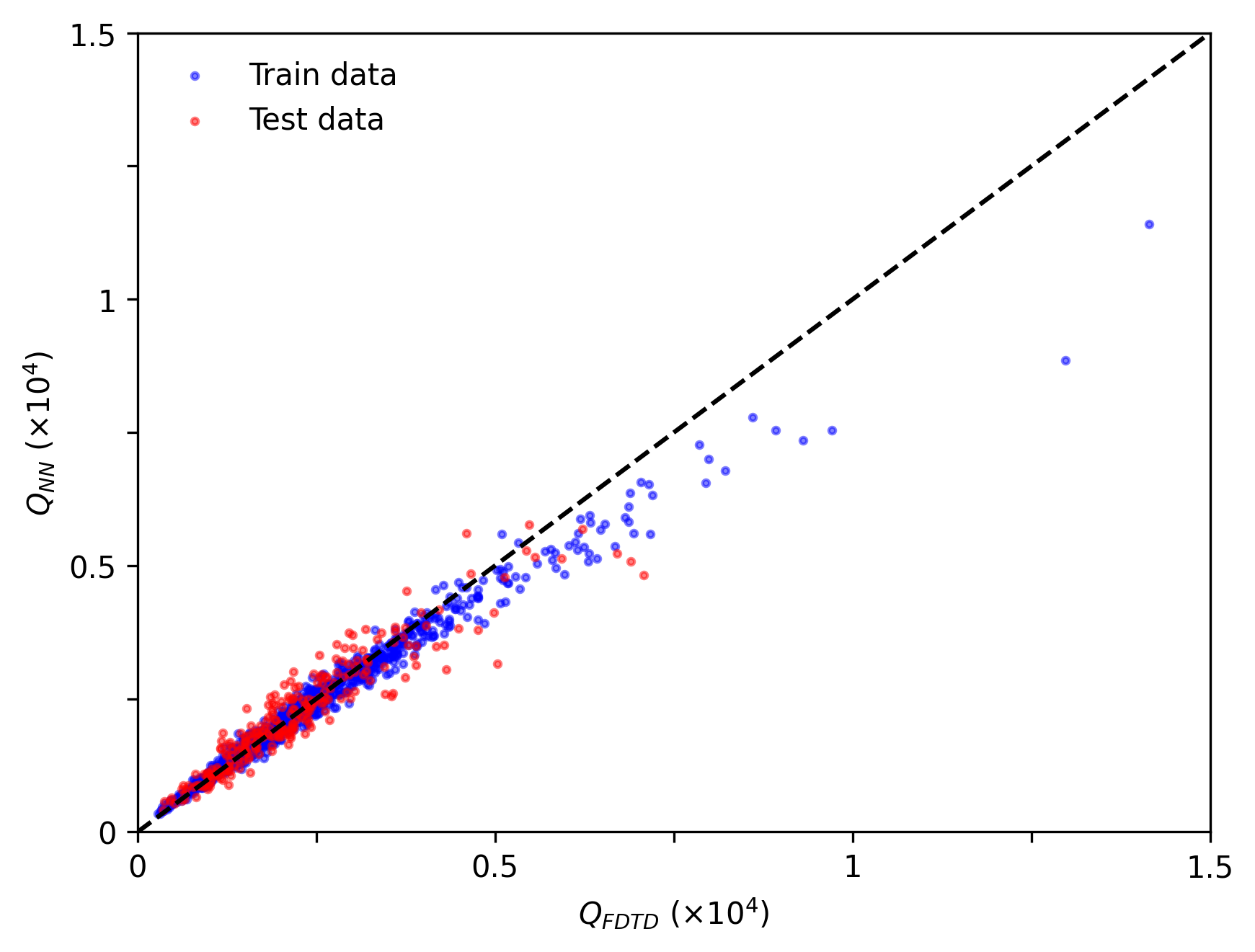}
        \caption{NN1: $\lambda = 0.001$, $R = 0.938\, (0.988)$}
        \label{fig:FB32_corr_l0.001_rough}
    \end{subfigure}
    
    \begin{subfigure}{0.495\textwidth}
        \centering
        \includegraphics[width=\linewidth]{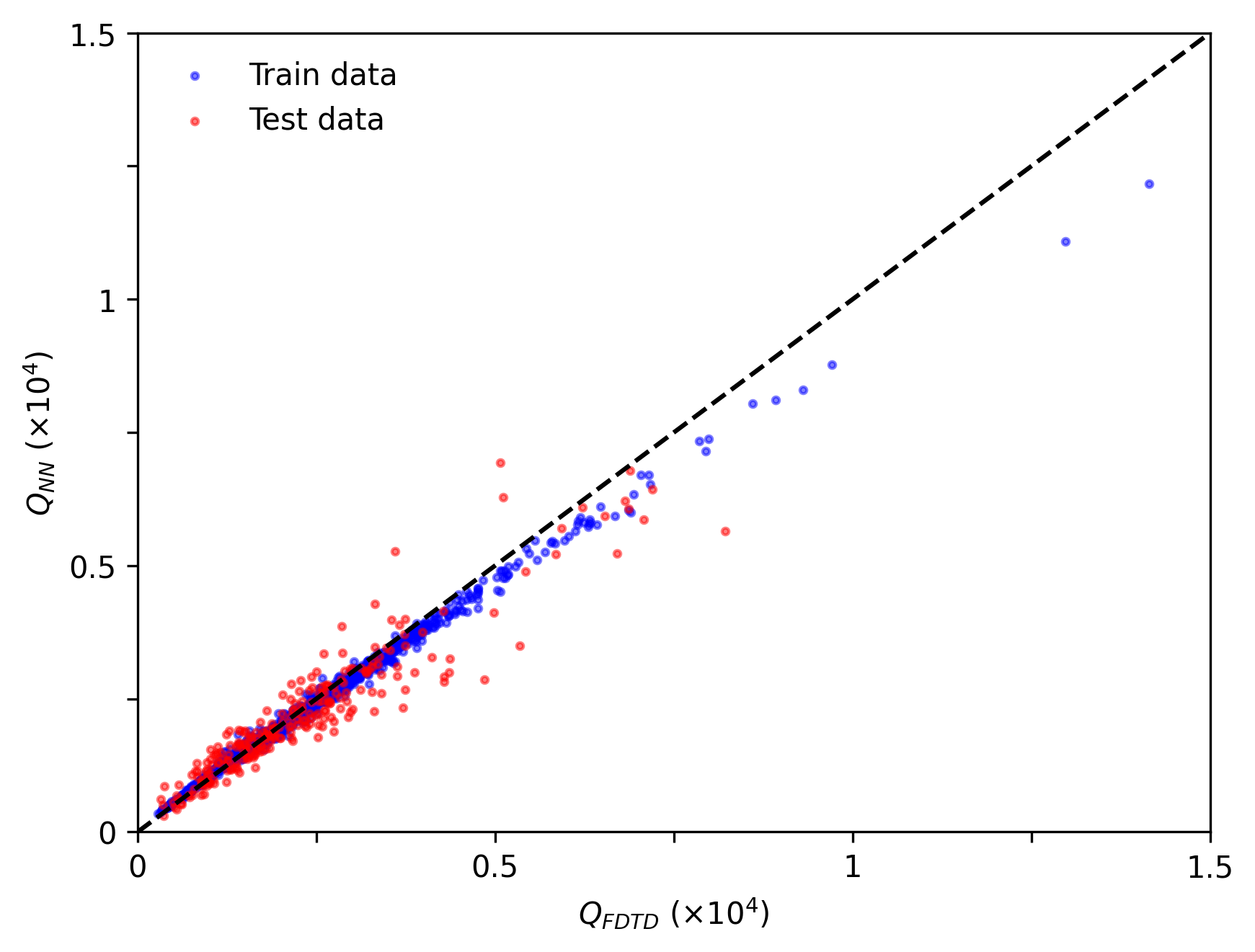}
        \caption{NN2: $\lambda = 0$, $R = 0.938\, (0.997)$}
        \label{fig:FB212_corr_l0_rough}
    \end{subfigure}
    \begin{subfigure}{0.495\textwidth}
        \centering
        \includegraphics[width=\linewidth]{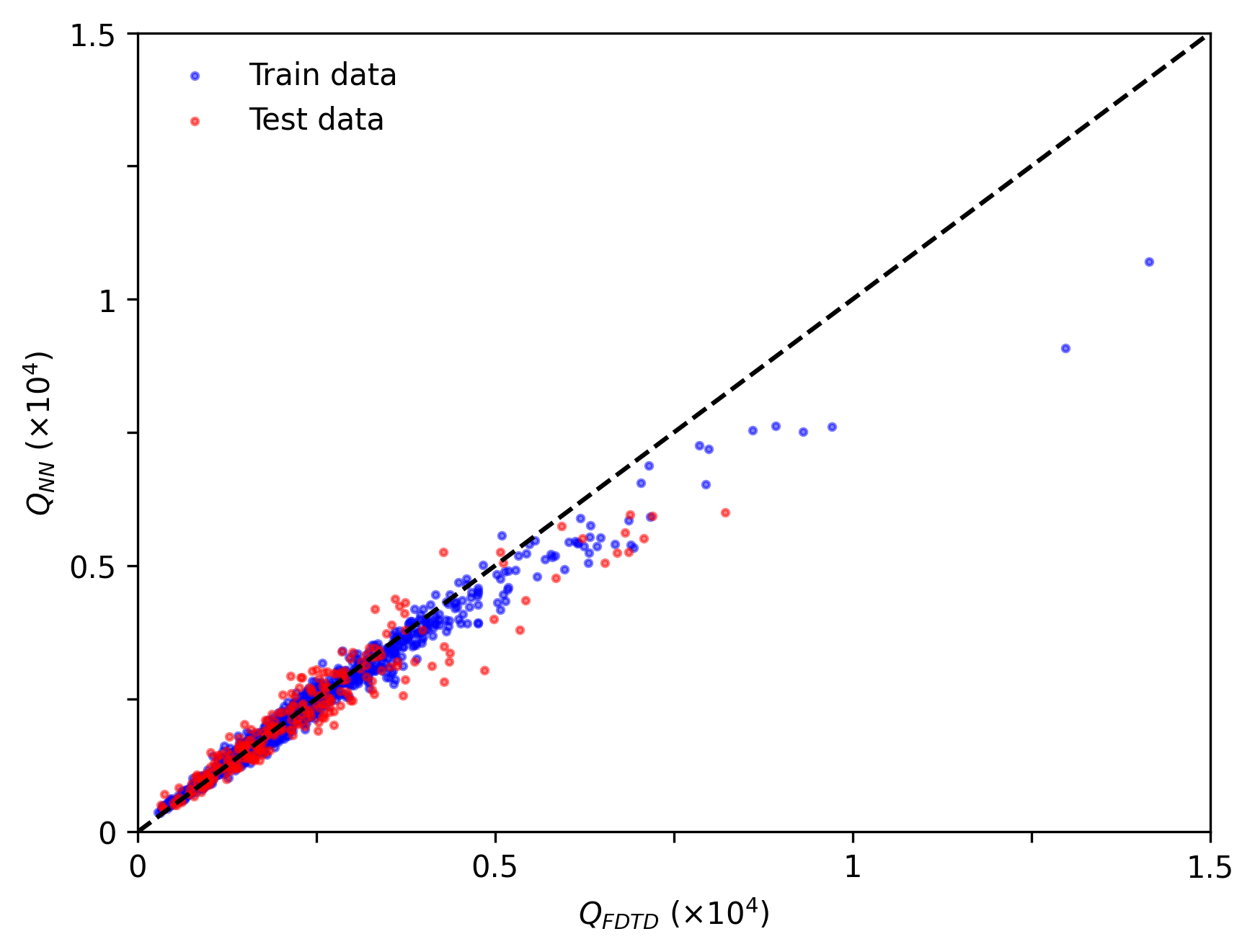}
        \caption{\textbf{NN2}: $\lambda = 0.001$, $R = 0.952\, (0.986)$}
        \label{fig:FB212_corr_l0.001_rough}
    \end{subfigure}
    \caption{Correlation graphs, fishbone with surface roughness (NN (bold) responsible for optimizing fishbone cavity 2)}
    \label{fig:FB_corr_rough}
\end{figure}
\newpage

\begin{figure}[h!]
    \centering
    \begin{subfigure}{0.495\textwidth}
        \centering
        \includegraphics[width=\linewidth]{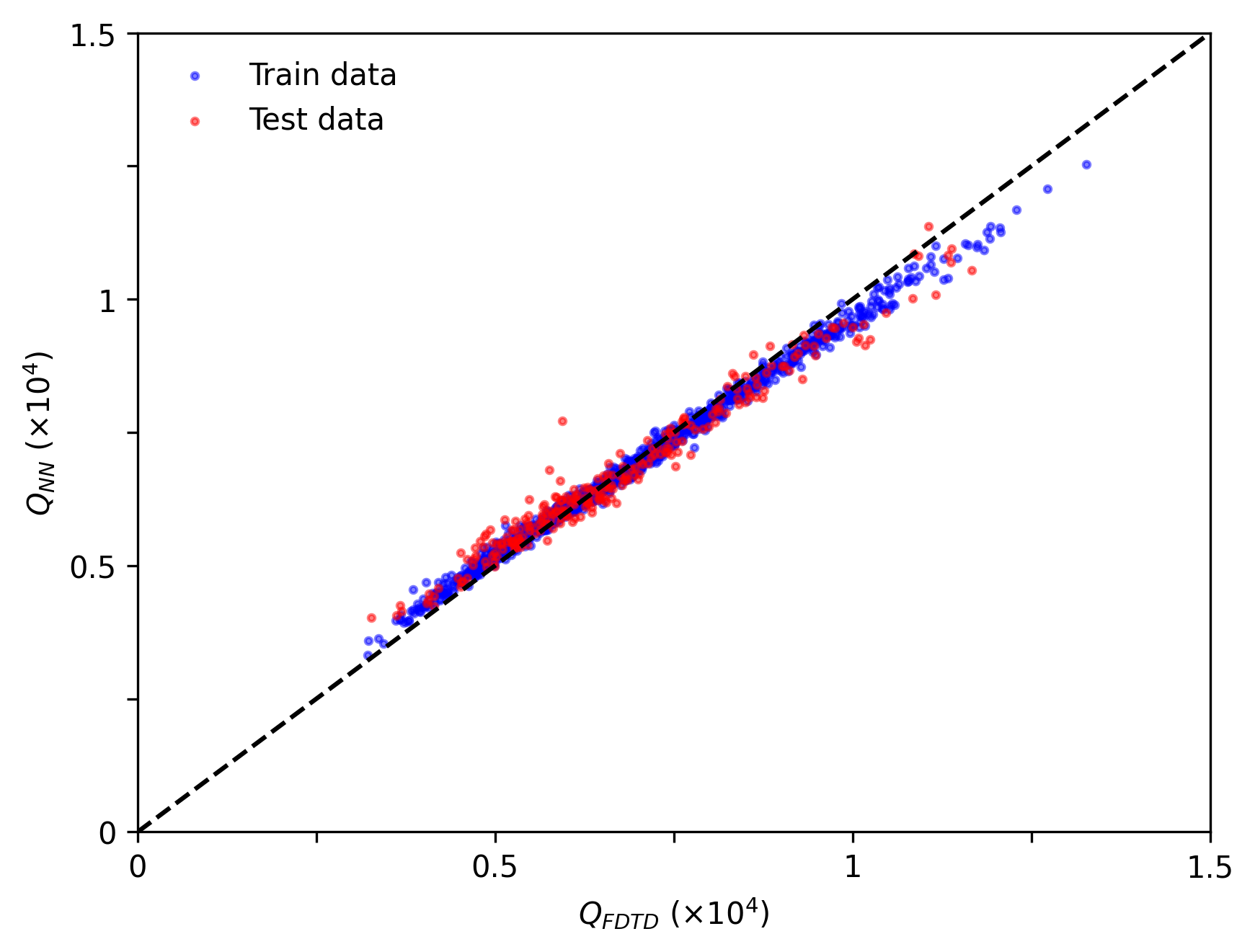}
        \caption{NN1: $\lambda = 0$, $R = 0.985\, (0.997)$}
        \label{fig:NC123_corr_l0_slant}
    \end{subfigure}
    \begin{subfigure}{0.495\textwidth}
        \centering
        \includegraphics[width=\linewidth]{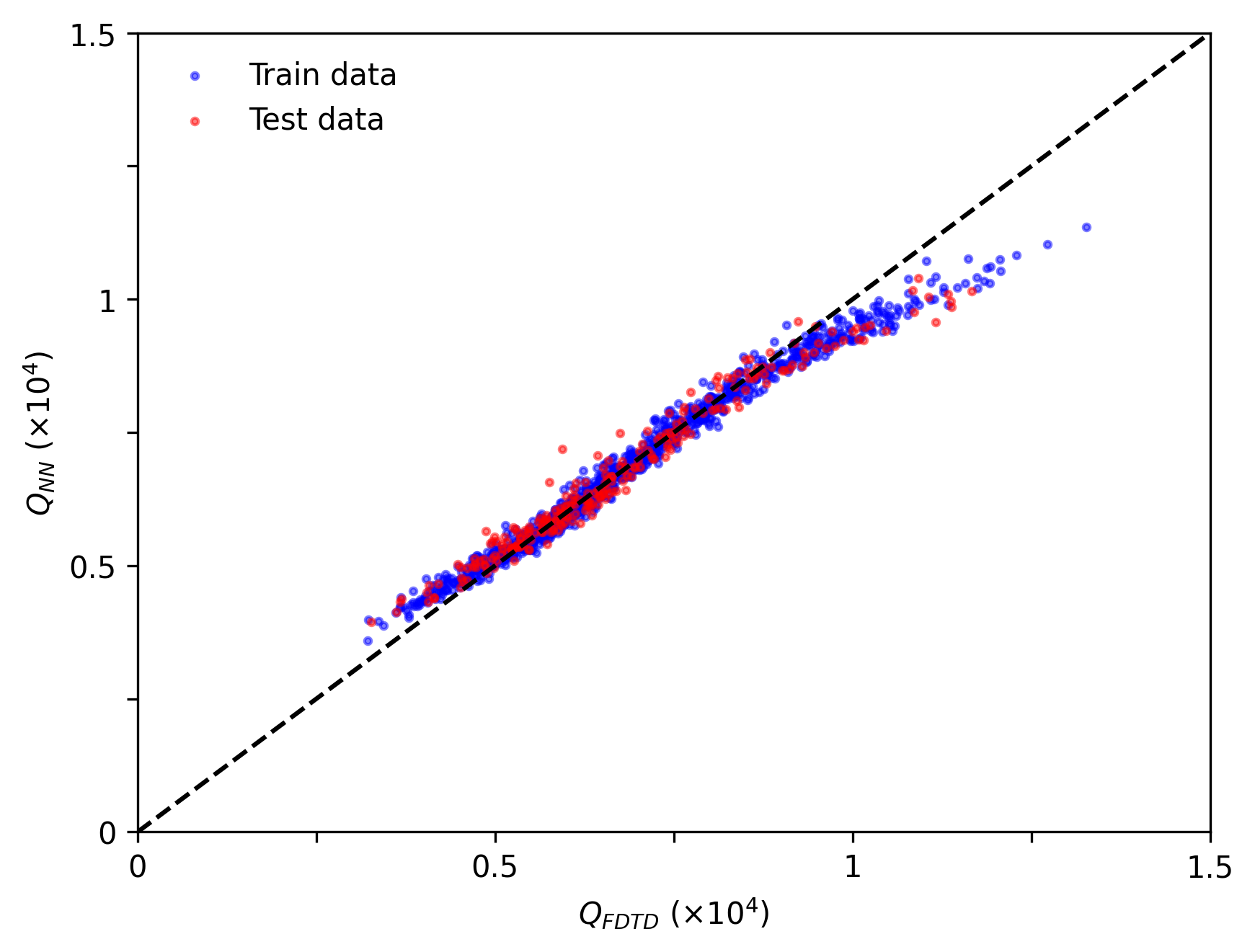}
        \caption{NN1: $\lambda = 0.001$, $R = 0.983\, (0.990)$}
        \label{fig:NC123_corr_l0.001_slant}
    \end{subfigure}
    
    \begin{subfigure}{0.495\textwidth}
        \centering
        \includegraphics[width=\linewidth]{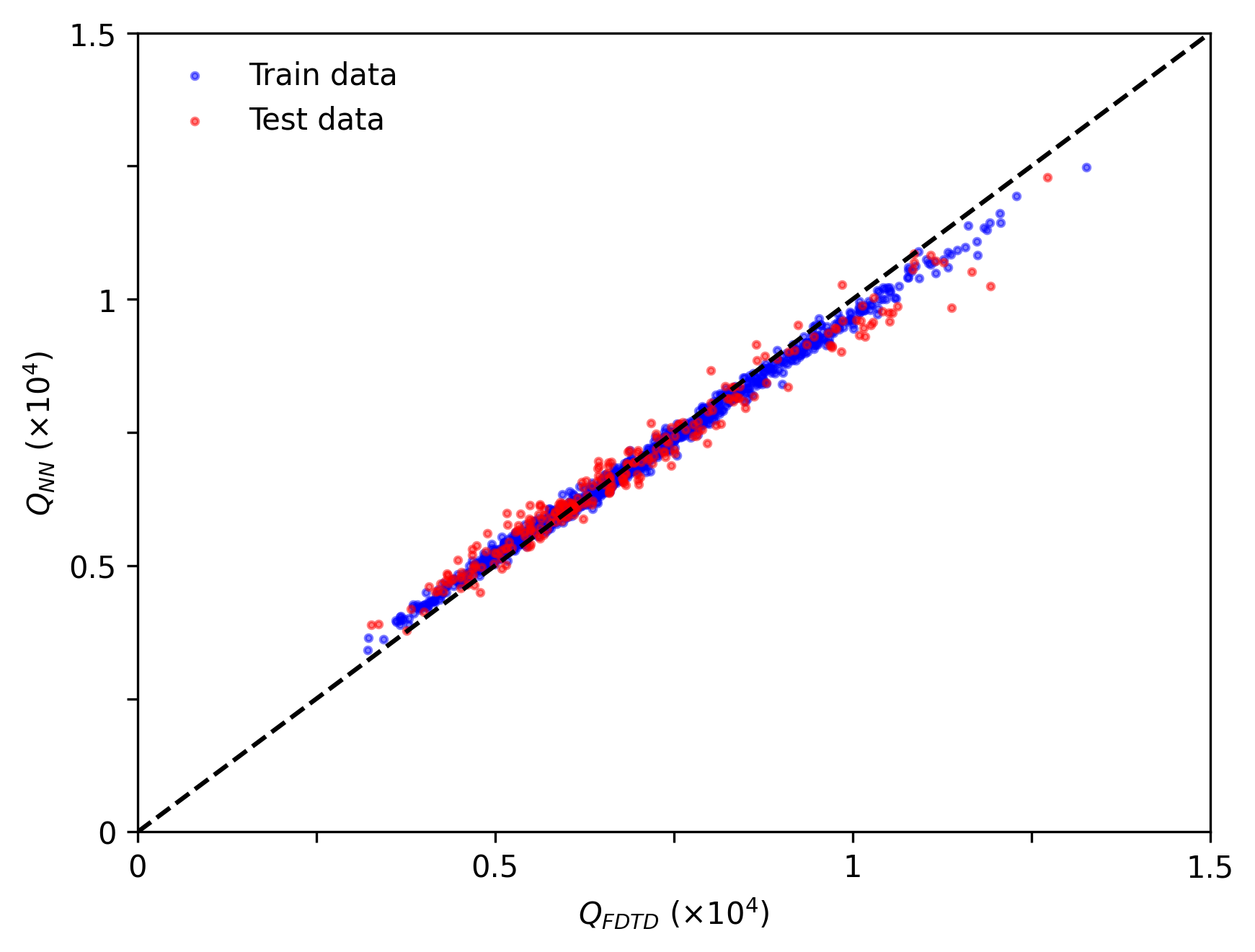}
        \caption{\textbf{NN2}: $\lambda = 0$, $R = 0.988\, (0.998)$}
        \label{fig:NC21_corr_l0_slant}
    \end{subfigure}
    \begin{subfigure}{0.495\textwidth}
        \centering
        \includegraphics[width=\linewidth]{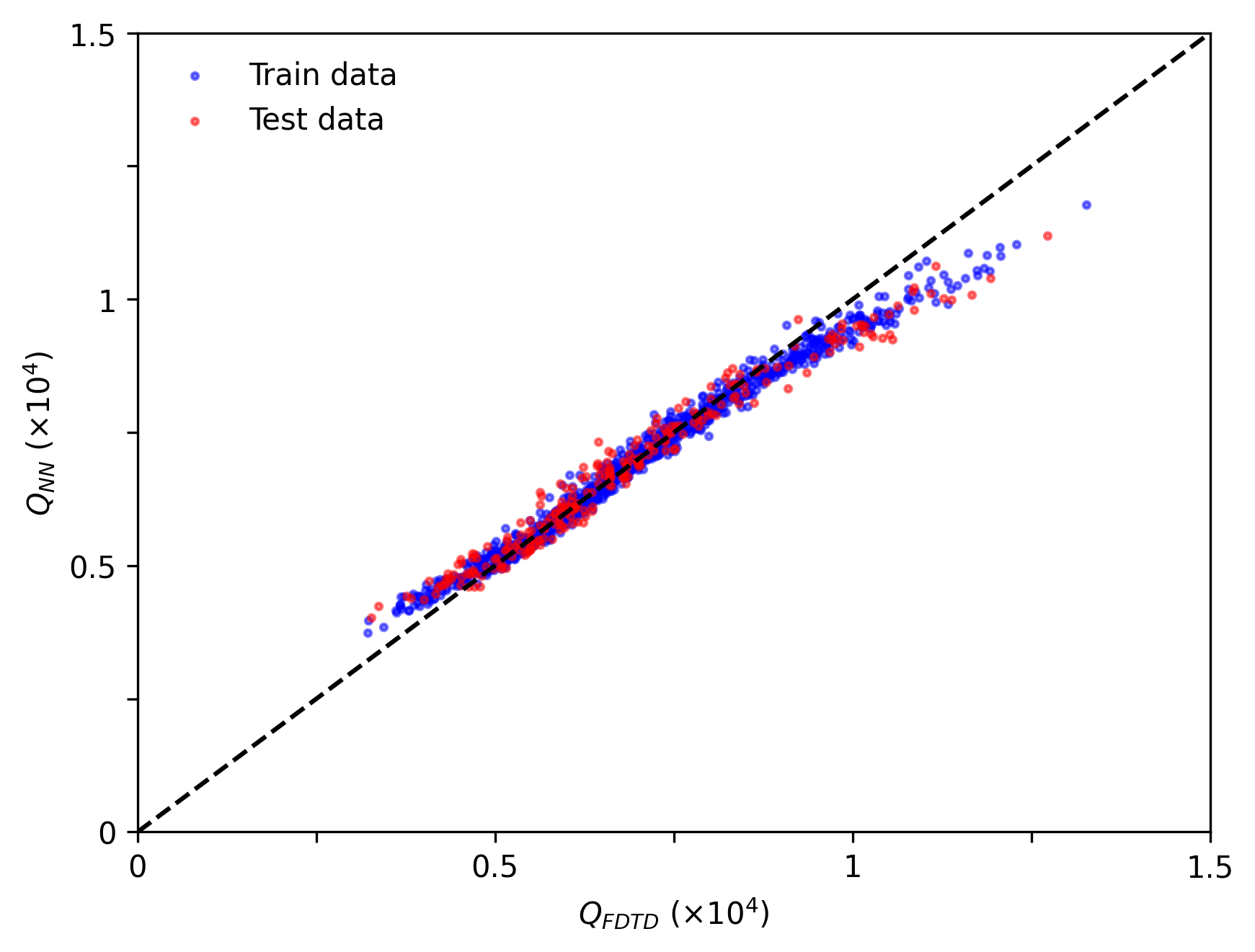}
        \caption{NN2: $\lambda = 0.001$, $R = 0.984\, (0.992)$}
        \label{fig:NC21_corr_l0.001_slant}
    \end{subfigure}
    \caption{Correlation graphs, L2 with sidewall slant (NN (bold) responsible for optimizing L2 cavity 3)}
    \label{fig:NC_corr_slant}
\end{figure}
\newpage

\begin{figure}[h!]
    \centering
    \begin{subfigure}{0.495\textwidth}
        \centering
        \includegraphics[width=\linewidth]{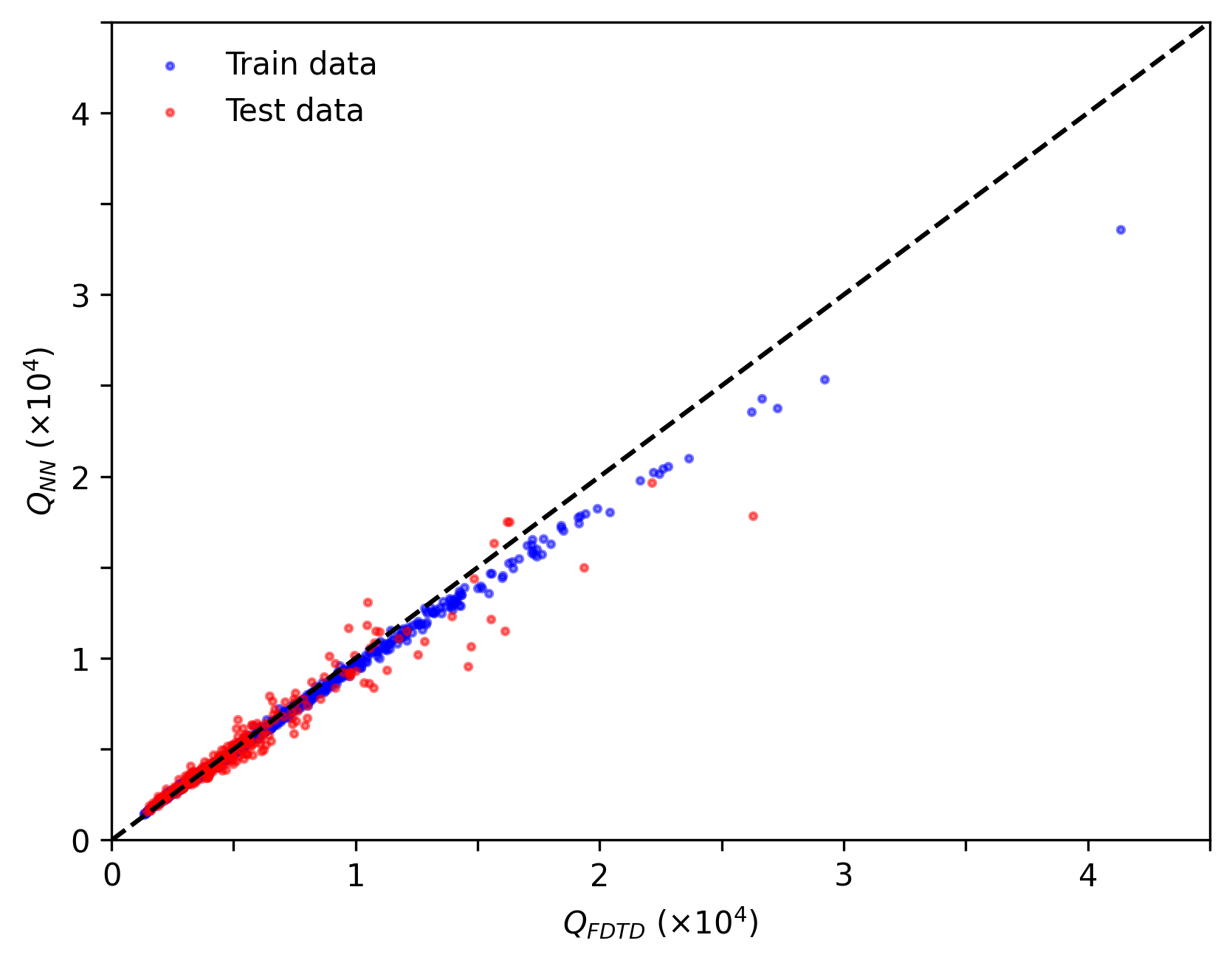}
        \caption{NN1: $\lambda = 0$, $R = 0.965\, (0.998)$}
        \label{fig:FB41_corr_l0_slant}
    \end{subfigure}
    \begin{subfigure}{0.495\textwidth}
        \centering
        \includegraphics[width=\linewidth]{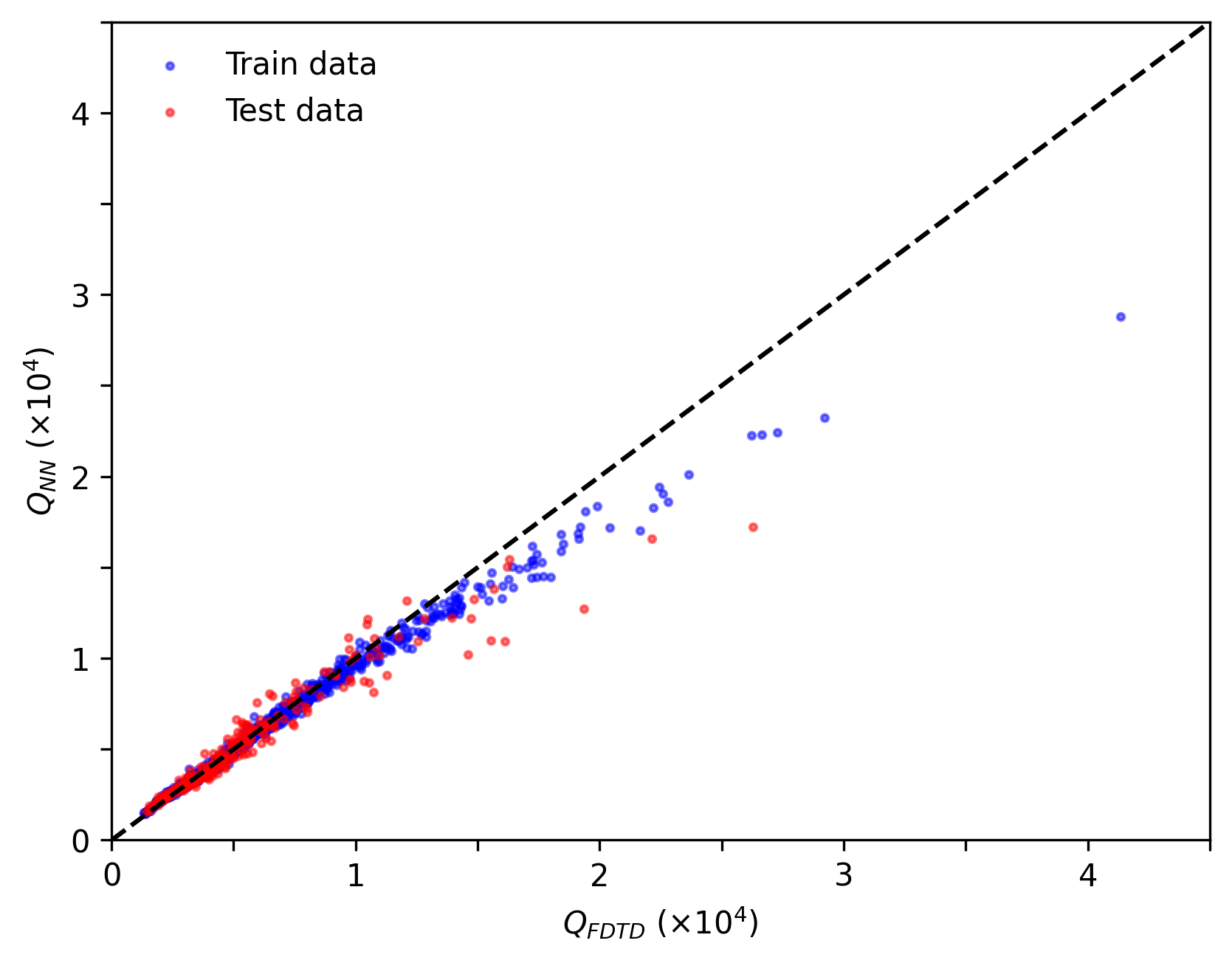}
        \caption{\textbf{NN1}: $\lambda = 0.001$, $R = 0.962\, (0.993)$}
        \label{fig:FB41_corr_l0.001_slant}
    \end{subfigure}
    
    \begin{subfigure}{0.495\textwidth}
        \centering
        \includegraphics[width=\linewidth]{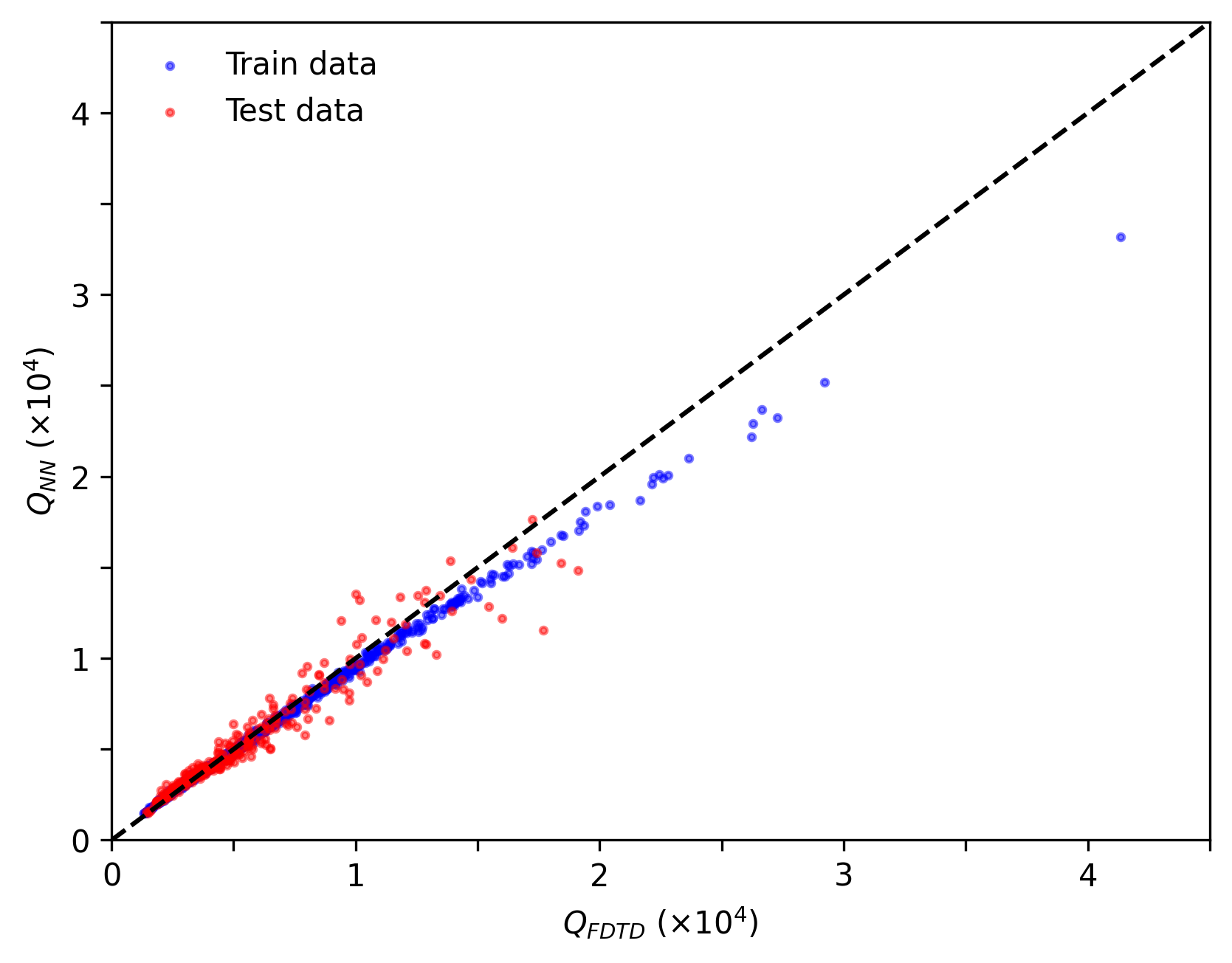}
        \caption{NN2: $\lambda = 0$, $R = 0.966\, (0.998)$}
        \label{fig:FB44_corr_l0_slant}
    \end{subfigure}
    \begin{subfigure}{0.495\textwidth}
        \centering
        \includegraphics[width=\linewidth]{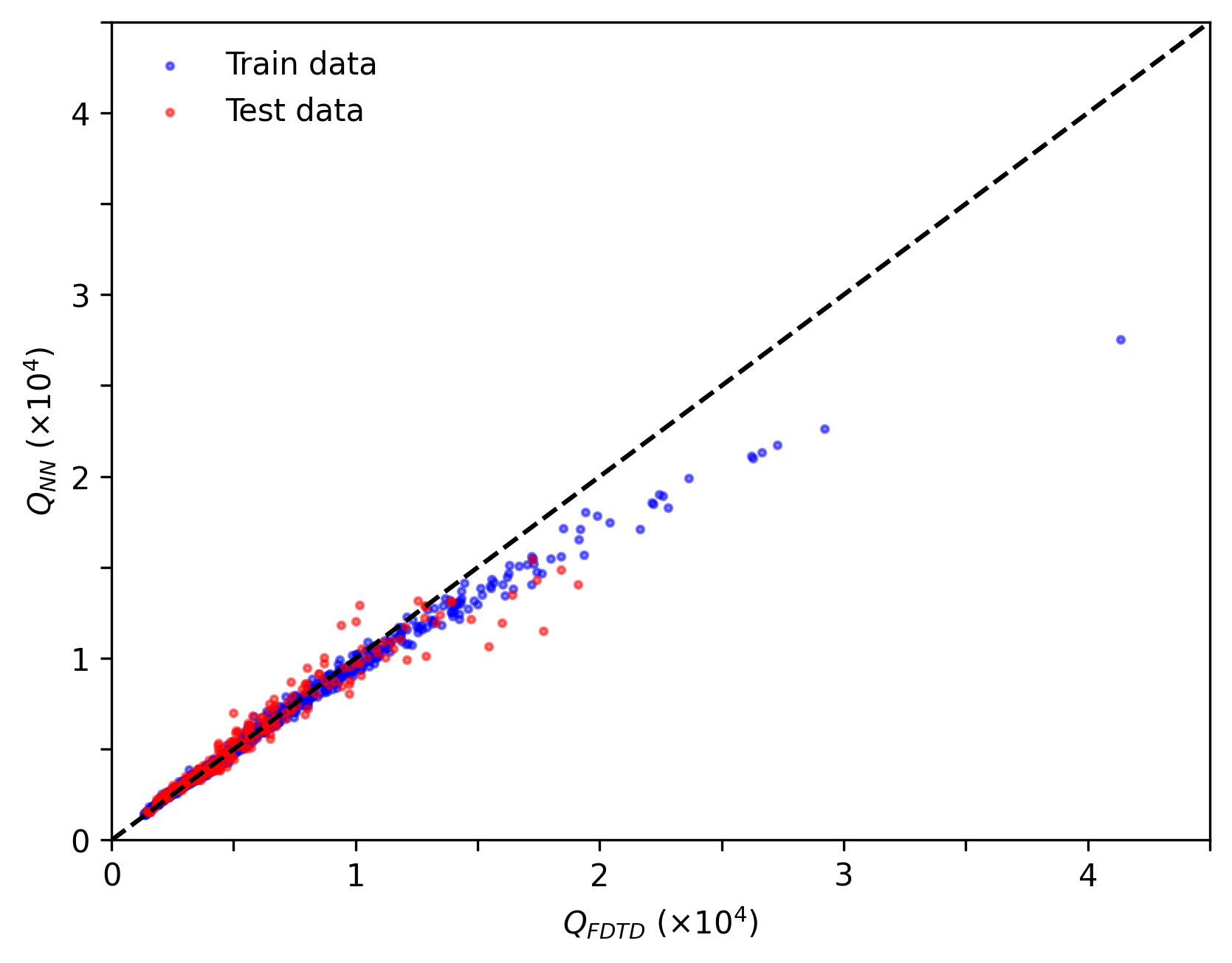}
        \caption{NN2: $\lambda = 0.001$, $R = 0.969\, (0.992)$}
        \label{fig:FB44_corr_l0.001_slant}
    \end{subfigure}
    \caption{Correlation graphs, fishbone with sidewall slant (NN (bold) responsible for optimizing fishbone cavity 3)}
    \label{fig:FB_corr_slant}
\end{figure}
\newpage